\documentclass[12pt]{report} %add 'twoside' for book layout
\usepackage{floatrow} %for tables spectral clustering
\newfloatcommand{capbtabbox}{table}[][\FBwidth] %for tables spectral clustering
\usepackage[utf8]{inputenc}
\usepackage{graphicx}
\graphicspath{{images/}}
\usepackage{caption}
\usepackage{subcaption}
\usepackage{epigraph} %for citations at the beginning of chapters
\usepackage{csquotes} %for quotes in the middle of text
\usepackage{amsmath,amsthm,amssymb,algorithmic,epsfig,graphicx, tikz}
\usepackage{amsfonts}
\usepackage{mathtools}
\usepackage{algorithm} %pseudo code
\usepackage{qcircuit} %quantum circuit
\usepackage{float} %to force figure positions with "H"
\usepackage{multirow} %for tables with multiple rows
\usepackage{multicol} %for tables with multiple columns
\usepackage{booktabs} %for tables
\usepackage[shortlabels]{enumitem} %enumerate with labels (a,b,c..)
\usepackage{hyperref} %for clickable reference, and table of content in acrobat reader

\usepackage[most]{tcolorbox}
\usepackage{amssymb}
\tcbuselibrary{theorems}

\newtcbtheorem
  [auto counter,number within=chapter]% init options
  {definition}% name
  {Definition}% title
  {%
    colback=blue!5,
    colframe=blue!75!black,
    fonttitle=\bfseries,
  }% options
  {def}% LABEL prefix

  \newtcbtheorem
  [auto counter,number within=chapter]% init options
  {theorem}% name
  {Theorem}% title
  {%
    colback=red!5,
    colframe=red!75!black,
    fonttitle=\bfseries,
  }% options
  {thm}% LABEL prefix

  \newtcbtheorem
  [auto counter,number within=chapter]% init options
  {claim}% name
  {Claim}% title
  {%
    colback=orange!5,
    colframe=orange!75!black,
    fonttitle=\bfseries,
  }% options
  {thm}% LABEL prefix

  \newtcbtheorem
  [auto counter,number within=chapter]% init options
  {result}% name
  {Result}% title
  {%
    colback=red!5,
    colframe=red!75!black,
    fonttitle=\bfseries,
  }% options
  {res}% LABEL prefix

\usepackage{afterpage}
\newcommand\blankpage{%
    \null
    \thispagestyle{empty}%
    \addtocounter{page}{-1}%
    \newpage}

\def\bra#1{\mathinner{\langle{#1}|}}
\def\ket#1{\mathinner{|{#1}\rangle}}
\newcommand{\braket}[2]{\langle #1|#2\rangle}
\DeclarePairedDelimiter\ceil{\lceil}{\rceil}
\DeclarePairedDelimiter\floor{\lfloor}{\rfloor}
\newcommand{\R}{\mathbb{R}}
\newcommand{\C}{\mathbb{C}}
\newcommand{\E}{\mathbb{E}}
\newcommand{\PP}{\mathbb{P}}
\newcommand{\Z}{\mathbb{Z}}

\newcommand{\errdist}{\epsilon_1}
\newcommand{\errmult}{\epsilon_2}
\newcommand{\errnorms}{\epsilon_3}
\newcommand{\errtom}{\epsilon_4}
\newcommand{\errkappa}{\epsilon_{\tau}}
\newcommand{\n}{N}

\newcommand{\norm}[1]{\left\lVert#1\right\rVert}

\makeatletter
\newcommand{\xMapsto}[2][]{\ext@arrow 0599{\Mapstofill@}{#1}{#2}}
\def\Mapstofill@{\arrowfill@{\Mapstochar\Relbar}\Relbar\Rightarrow}
\makeatother

\usepackage[a4paper,width=150mm,top=25mm,bottom=25mm,bindingoffset=6mm]{geometry}
\usepackage{fancyhdr}
\title{My PhD Thesis}
\author{Jonas Landman}
\date{April 2021}

\begin{document}

\begin{titlepage}
    \begin{center}
    
        \includegraphics[width=0.9\textwidth]{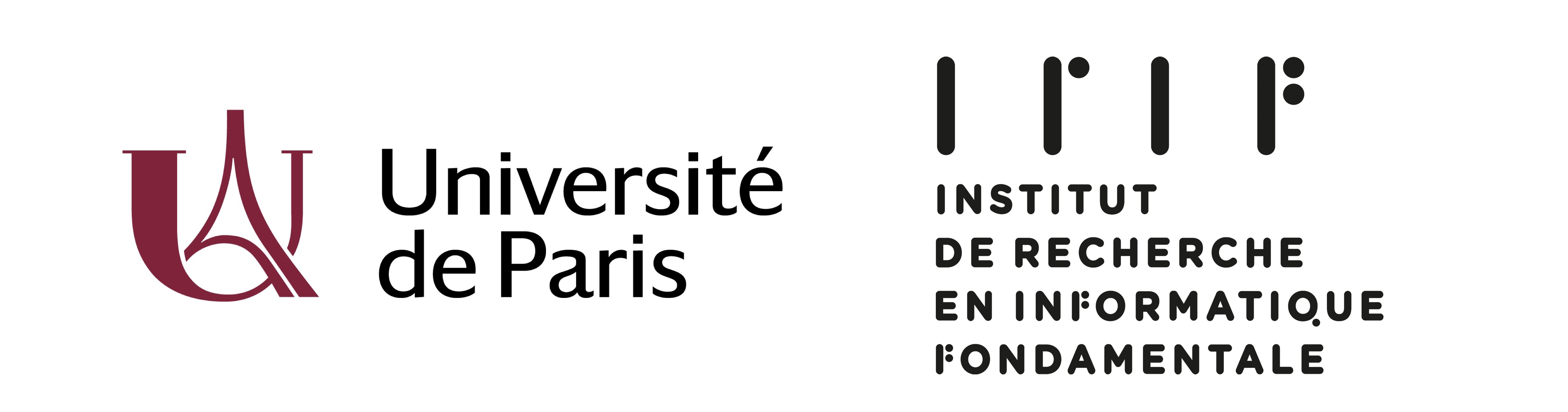}
        
        \vspace{1.3cm}
        \Large
        Université de Paris, CNRS\\
        Institut de Recherche en Informatique Fondamentale (IRIF)\\
        Paris, France\\
        
        \vspace*{1cm}

        \LARGE
        \textbf{Quantum Algorithms for Unsupervised Machine Learning and Neural Networks}
        
        \noindent\rule{8cm}{0.4pt}  

        \vspace{0.5cm}
        \LARGE
        By \textbf{Jonas Landman}

        \vspace{1.4cm}
        \Large
        A thesis presented for the degree of\\
        \emph{Doctor of Philosophy}

        \vspace{1.1cm}
        \Large
        Supervised by \textbf{Iordanis Kerenidis}

        \vspace{1.6cm}
        
        Publicly defended before a jury composed of
        \begin{table}[h]
        \begin{center}
        \begin{tabular}{ll}
        Frederic Magniez & Université de Paris\\
        Ashwin Nayak & University of Waterloo\\
        Xiaodi Wu & University of Maryland\\
        Iordanis Kerenidis & Université de Paris\\
        Elham Kashefi & Sorbonne Université \\ 
        & \& University of Edinburgh\\
        \end{tabular}
        \end{center}
        \end{table}
        
        October 2021
        
    \end{center}
\end{titlepage} %version En, online

%% add a blank page
\pagestyle{empty} % to not have the header etc.
\afterpage{\blankpage}
\pagestyle{fancy}
%%%%%%%%

%%%%-------------FOREWORDS---------------%%%%
\chapter*{Abstract}
%\addcontentsline{toc}{chapter}{Abstract}
%\input{chapters/Foreword/abstract_keyword} %abstract with keywords at the end (UP)
Combining two great scientific revolutions of the XX$^{th}$ century, quantum computing uses the strange properties of quantum physics to redefine the notion of information processing to solve computational problems. From the initial proposal to the present day, much work has been done to develop quantum algorithms for these machines to push the limits of what was thought achievable in the fields of computational physics, chemistry, optimisation, communication, cryptography, and many more. \\

In this thesis, we investigate whether quantum algorithms can be used in the field of artificial intelligence, or machine learning. In the last decade, this field has been revolutionizing our ability to predict, classify, and learn, for many applications.\\

We will first recall the fundamentals of machine learning and quantum computing, and then describe more precisely how to link them through linear algebra. By encoding vectors in the form of quantum states, we will present and introduce quantum algorithms to efficiently solve tasks such as matrix product or distance estimation. \\

These results are then used to develop new quantum algorithms for unsupervised machine learning, such as $k$-means and spectral clustering. This allows us to define many fundamental procedures, in particular in vector and graph analysis.
We will also present new quantum algorithms for artificial neural networks, or deep learning. For this we will introduce an algorithm to perform a quantum convolution product on images, as well as a new way to perform fast tomography on quantum states. \\

We prove that these quantum algorithms are faster compared to their classical version, but exhibit random effects due to the quantum nature of the computation. Many simulations have been carried out to study these effects and measure their learning accuracy on real data.\\

Finally, we will present a quantum orthogonal neural network circuit adapted to the currently available small and imperfect quantum computers. This allows us to perform real experiments to test our theory. \\

The quantum algorithms presented in this thesis give hope for the utility of an ideal quantum computer in the future. Indeed, we prove an asymptotic advantage for each algorithm in terms of complexity or running time, compared to the classical case. That being said, for this hope to become reality, many efforts remain to be realized in practice, from error correction to quantum data access. %just the abstract

\chapter*{Résumé}
%\addcontentsline{toc}{chapter}{Résumé}
%\input{chapters/Foreword/resume_long} %version université de paris (4 pages) (UP)
Combinant deux grandes révolutions scientifiques du XX$^{\text{ème}}$ siècle, l'ordinateur quantique utilise les étranges propriétés de la physique quantique pour redéfinir la notion d'ordinateur afin de résoudre des problèmes calculatoires. Depuis leur imagination jusqu'à nos jours, de nombreux travaux ont été réalisés pour développer des algorithmes pour ces machines afin de repousser les limites de ce que l'on pensait faisable par les ordinateurs dans les domaines de la physique, de la chimie, de l'optimisation, de la communication, de la cryptographie, et bien d'autres encore. 

Dans cette thèse, nous cherchons à savoir si des algorithmes quantiques pourront être utilisés dans le domaine du \emph{machine learning}, ou intelligence artificielle. Ce domaine révolutionne depuis une dizaine d'années notre approche de l'apprentissage, de la prédiction et de la classification pour d'innombrables applications.

Dans un premier temps nous rappellerons les fondamentaux du \emph{machine learning} et de l'ordinateur quantique, puis nous décrirons plus précisément comment faire le lien entre les deux, à travers l'algèbre linéaire. En encodant des vecteurs sous formes d'états quantiques, nous présenterons et introduirons des algorithmes quantiques permettant de résoudre rapidement des tâches telles que le produit matriciel ou l'estimation de distance. 

Ces résultats seront ensuite utilisés pour développer de nouveaux algorithmes quantiques en \emph{machine learning} non supervisés, tels le \emph{$k$-means} et le \emph{spectral clustering}. Cela nous permettra de définir de nombreuses procédures, en particulier dans l'analyse vectorielle et l'analyse de graphes.
Nous présenterons aussi de nouveaux algorithmes quantiques pour les réseaux de neurones artificiels, ou \emph{deep learning}. Pour cela nous introduirons un algorithme pour réaliser un produit de convolution quantique sur des images, ainsi qu'une nouvelle façon de réaliser une tomographie rapide sur les états quantiques. 

Nous prouverons que ces algorithmes quantiques sont des équivalents plus rapides que leur version classique, mais présentent des effets aléatoires dûs à la nature quantique du calcul. Afin d'étudier ces effets, de nombreuses simulations ont été faites pour tester la précision d'apprentissage sur des données réelles.

Enfin, nous présenterons un circuit quantique pour les réseaux de neurones orthogonaux, adapté aux ordinateurs quantiques petits et imparfaits actuellement disponibles. Cela nous permet de réaliser de vraies expériences afin de tester notre théorie. 

Les algorithmes quantiques présentés dans cette thèse donnent espoir quant à l'utilité d'un ordinateur quantique idéal dans le futur. En effet, pour chaque algorithme nous prouverons que sa complexité, ou temps de calcul, est asymtotiquement plus efficace que dans le cas classique. Ceci étant dit, pour que ces espoirs deviennent réalité, de nombreux efforts resteront à réaliser en pratique dans le domaine de la correction d'erreur et de l'accès aux données.

 %version online (short)

\chapter*{Remerciements}
%\addcontentsline{toc}{chapter}{Remerciements}
This Ph.D. was a long and wonderfully rewarding experience, half of which happened during the Covid-19 global pandemic. Despite this, I was able to travel, learn something new every day and meet many inspiring people. \\

I would like to thank Iordanis Kerenidis, my Ph.D. supervisor, who gave me this opportunity after we met at a NASA facility in California. I knew immediately that he would be an ideal supervisor and mentor, both scientifically and on a human level. His constant support and guidance over the years allowed me to go through this Ph.D. with confidence, determination, and joy. I look forward to working with Iordanis again and learn more from him.\\

I would like to thank the Jury, in particular Ashwin Nayak and Xiaodi Wu for kindly accepting to review and provide insights for this manuscript.\\

I have been fortunate to study, work, or just interact with great researchers who have truly inspired me, notably Frederic Magniez, Elham Kashefi, Sophie Laplante, Andre Chailloux, Ashwin Nayak, Ashley Montanaro, Seth Lloyd, Eleni Diamanti, Pascale Senellart, Philippe Grangier, Umesh Vazirani and John Preskill.\\

These years were enlightened by many new collaborations and friendships around the world, including Alessandro Luongo, Daniel Szilagyi, Anupam Prakash, Amine Cherrat, Sander Gribling, Alex Grillo, Yassine Hamoudi, and Yixin Shen from our research group at IRIF. All the LIP6-Edinburgh team including Niraj Kumar, Brian Coyle, Slimane Thabet, Constantin Dalyac and Pierre-Emmanuel Emeriau. And Natansh Mathur, Vincent Fortuin, Noah Berner, Juan Ignacio Adame, Federico Centrone, Alex Singh, Avinash Mocherla, Adam Bouland, Chris Cade, and Charles Hadfield. I would like to thank all members of PCQC and IRIF for their support, in particular Etienne Mallet and Eva Ryckelynck.\\

The quantum computing ecosystem has expended a lot since I started this Ph.D., and I am very glad to have met inspiring actors of the field such as Christophe Jurczak, Olivier Ezratty, Matt Johson, Loïc Henriet, Théau Peronnin, Matthieu Desjardins, Elvira Shishenina, and many more people that will shape the future of quantum computing in France and abroad. \\

À mes amis qui pourront enfin arrêter de me demander ``c'est quoi ta thèse déjà ?", merci pour tout. En particulier ceux qui m'ont convaincu de démarrer ce doctorat, Alexis Léautier et Pierre Fredenucci (2728 Grant St, Berkeley), Antoine Michon et Reda Agoumi (the Board), Ainsi que ceux dont les conseils sur le doctorat furent essentiels, Batiste Le Bars, Maxence Ernoult, Baptiste Louf et Geoffrey Negiar.
Merci à tous les Pototunes, à mes trois citrons de Polycool (et La Crampe) qui rendent ma vie si musicale et riche en couleur, en particulier mon niéseux Léon Vidal pour nos discussions du midi sur les ondes graviationnelles et les synthétiseurs analogiques. Merci à Romain Palmieri, Rafael Cohen, Dorian Perron et à toute l'équipe Groover, je suis fier de ce qu'est devenu notre projet aujourd'hui. À ceux et celles que je n'ai pas mentionnés, la thèse fait déjà presque 200 pages donc je vous remercie par télépathie, vous vous reconnaitrez. \\

Je dois tant à ma famille, mon père qui m'a donné le goût de la science pour le bien commun, ma mère qui m'a sans cesse répété ``\emph{quand on cherche on trouve}" (en parlant de mes affaires perdues, mais ça marche aussi pour le reste), Julia, Gabriel et Miko le chat de Schrödinger. Je suis si fier de rejoindre enfin le clan des ``Docteurs" Landman, même si je ne pourrai pas prescrire d'ordonnances. Une pensée pour mes grands parents qui m'accompagnent où qu'ils soient, Moshé qui m'a donné le goût du jeu d'échec, Maurice pour le goût de la musique, Adèle qui rigole toujours en m'appelant ``\emph{Zveinstein}" ou ``\emph{Knobel Price}", et Monique que je connais si bien sans l'avoir connue. \\

Enfin, Solène, merci et merci. Une thèse entière serait nécessaire pour exprimer à quel point la vie est belle avec toi.\\

Ci-dessous le ticket boisson de la NASA (Ames Research Center California), que j'avais sur moi le jour où j'ai su que je voulais faire une thèse en quantum machine learning.

\vspace{2,5cm}

\begin{figure}[h!]
    \centering
    \includegraphics[width=0.25\textwidth]{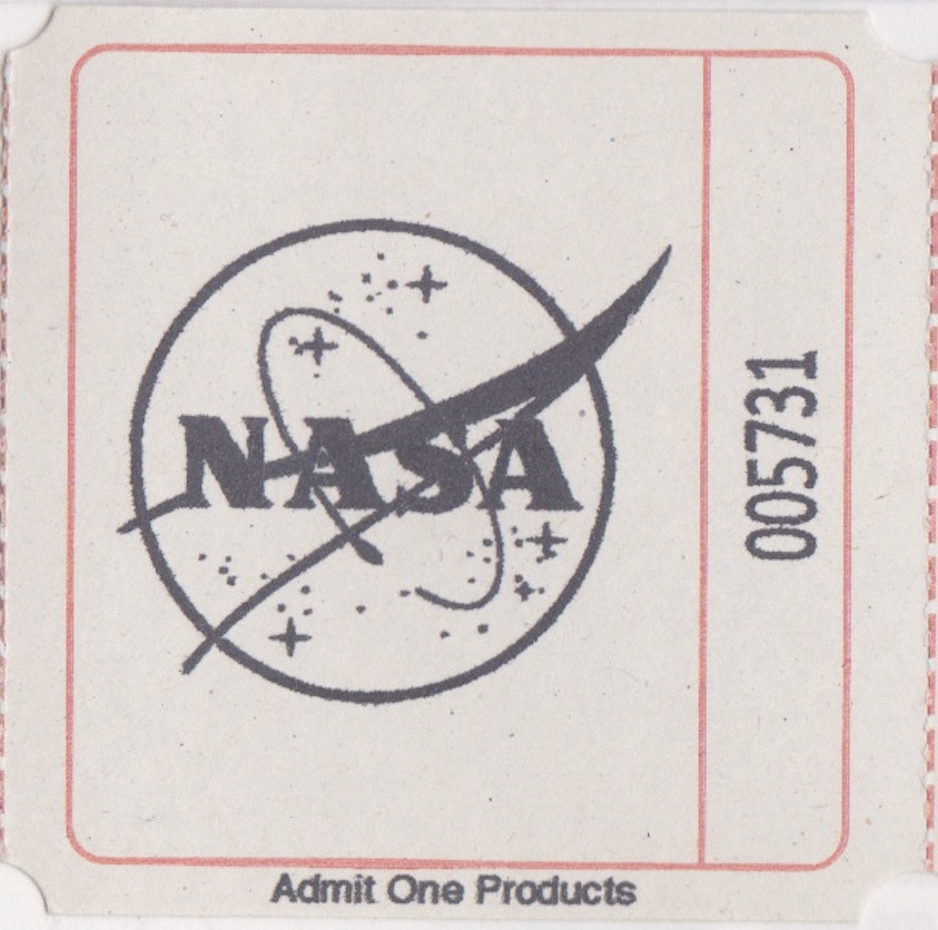}
    %\caption{Le ticket boisson de la NASA (Ames Research Center, California), le jour où j'ai su que je commencerai un thèse en quantum machine learning.}
\end{figure}

\chapter*{Avant-Propos}
%\addcontentsline{toc}{chapter}{Avant-Propos}
\epigraph{\textit{"La science à tout moment recule les limites du merveilleux."}}{Guy De Maupassant \\ \emph{La Peur} (1882)}

Je me permets d'introduire cette thèse de \emph{Doctor of Philosophy} (Ph.D.) par une touche de philosophie, une série de pensées qui m'ont accompagné durant ces années de Doctorat. 

J'aime à croire que l'émerveillement scientifique qu'il arrive à certaines et certains de ressentir dans leur vie est en partie dû à notre relation avec les \emph{limites}. Parmi les plus communes, celles qui hantent l'humanité depuis toujours, on peut citer ``d'où vient-on ?", ``qu'il y a-t-il au plus loin ?", ou bien ``quelle est la plus petite chose ?". De ces questions enfantines a jailli l'exploration du réel, emmenant nos connaissances toujours plus loin. Je ne sais pas ce qui est le plus surprenant entre les nouveaux savoirs acquis ou le fait même que nous ayons pu les acquérir. 
%On est ainsi en mesure de nous demander si cette exploration possède elle-même une limite ? Se faisant, on découvre l'élégance des limites qui portent sur le savoir lui-même: 
D'ailleurs, les questions les plus élégantes sont souvent celles qui portent sur les limites du savoir lui-même:
``pourra-t-on savoir un jour pourquoi l'univers existe ?", "puis-je me prononcer sur la conscience, étant moi-même conscient ?". D'un degré supérieur, ces limites sont la passion communes des philosophes, logiciens, mathématiciens et physiciens. 

La \emph{beauté} des mathématiques, c'est ce que je ressens devant la preuve d'une limite sur la connaissance elle-même, ou la démonstration habile de son absence. Les limites elles-mêmes sont devenues des objets mathématiques. D'ailleurs, une limite fait-elle partie de ce qu'elle délimite ? Qu'en est-il du Big Bang ? On voit que les limites ne sont pas des murs, mais bien des portes.

Si la quête des limites est l'archétype d'un \emph{hybris}, elle n'est pas forcément à confondre avec le désir d'utiliser ou de contrôler. Chercher les limites, c'est avant tout une méthode pour interroger notre compréhension du monde. La théorie de la physique quantique n'a pas seulement mieux délimité l'infiniment petit mais en a redéfini la notion même. Le principe d'incertitude de Heisenberg en est un bon exemple: il ne s'agit pas, au contraire de ce que l'on entend souvent, d'une limite de ce que l'on peut \emph{savoir} conjointement sur la position et la vitesse d'une particule, mais plutôt la découverte que ces notions ont un sens physique différent à cette échelle. Ainsi la quête de limite a modifié notre conception de la réalité.\\

Un sudoku à résoudre, l'équation du mouvement des planètes, la configuration d'une molécule. À partir de nombreux problèmes émergent des interrogations similaires sur les limites du possible: Est-ce que ce problème a une solution ? Si oui, puis-je la trouver avec une feuille et un stylo ? Est-ce qu'un ordinateur peut la trouver ? Si oui, peut-il la trouver rapidement, disons avant que le Soleil n'ait englouti la Terre ? Plus généralement, y a-t-il une limite à ce qu'un ordinateur peut résoudre rapidement ? Est-ce qu'un ordinateur peut simuler une conscience, ou un univers ? Les limites qui s'appliqueraient à ces résultats nous informent-elles sur la nature de l'information dans l'univers lui-même ? Par exemple, si j'autorise mon ordinateur à manipuler l'information sous forme quantique, possède-t-il les mêmes limites ? Si non, pourquoi ? 

Ainsi, des sciences de l'ordinateur ou \emph{Computer Science}, de nombreuses questions fondamentales émergent et passionnent les chercheurs. Se demander quelles sont les limites de ce qui peut être résolu ou encore \emph{résolu efficacement} permet de voir d'un autre angle de nombreuses questions philosophiques \cite{aaronson2013philosophers}. Il en va de même avec l'intelligence artificielle ou \emph{Machine Learning}. Se demander jusqu'à quel point une machine peut apprendre à différencier des images d'animaux, résoudre un problème de mécanique des fluides, ou détecter l'ironie dans un texte, c'est un chemin alternatif pour comprendre la nature de l'apprentissage, de la physique, ou du language. \\

Cette thèse, portant sur ce qu'il est théoriquement possible ou impossible de faire avec un ordinateur quantique dans le champ de l'intelligence artificielle, est d'une certaine façon motivée par ces considérations, ces émerveillements.\\

Enfin, il faut dire que sur le chemin de comprendre les limites se trouve assez fréquemment le désir de dépasser nos propres limites à travers la technique. Ces dernières années, j'ai pu assister en temps réel à l'émergence de l'intelligence artificielle et à l'apparition des premiers ordinateurs quantiques. Ces nouvelles technologies, part leur élégante universalité, sont ou seront probablement capables du meilleur comme du moins meilleur. J'espère que l'on fera attention à ne pas ériger le dépassement des limites en un principe supérieur aux principes naturels et humains. Car si nous sommes ici flottant dans l'espace, le principal est peut être simplement de comprendre et de prendre soin. \\

\vspace{2,5cm}

\begin{flushright}
Jonas Landman\\

Juin 2021
\end{flushright}

\tableofcontents

%%optional:
%\listoffigures
%\listoftables

%%%%-------------PART I---------------%%%%
\part{Introduction}\label{part:Introduction}
\chapter{Introduction}\label{chap:intro_intro}
\epigraph{\textit{"A mathematician is a blind man in a dark room looking for a black hat which isn’t there."}}{Charles Darwin}

\section{Context and Motivation}

\subsubsection{Quantum Physics}
Quantum physics is often considered as the most wonderful intellectual adventure of modern science. As Einstein, Bohr, Schrödinger, Dirac and others taught us, this theory is a new paradigm to our comprehension of the world. Small objects behave differently, by following specific equations and having the ability to be in multiple states before we look at them. Even though it concerns the tiniest objects such as atoms, electrons, or photons, the consequences are indeed macroscopic: without it, we wouldn't understand the Cosmic Microwave Background or photosynthesis, and we wouldn't have atomic clocks, lasers, computers and smartphones. Since the 1980s and the experimental realization of \emph{entangled} particles by Alain Aspect \cite{aspect1982experimental}, solving the Einstein-Podolsky-Rosen (EPR) paradox, we even started to manipulate information at a quantum level. We tend to forget it, but information is physical, and therefore it could also be \emph{quantum physical}. This second quantum revolution, paved the way to counterintuitive applications such as quantum teleportation \cite{bennett1993teleporting}. At the same time emerged the idea that handling quantum systems could help us performing computation, as Richard Feynman's famous quote \cite{feynmanquote1} puts it: 

\begin{displayquote}
``Trying to find a computer simulation of physics seems to me to be an excellent program to follow out [...] the real use of it would be with quantum
mechanics [...] Nature isn’t classical  dammit, and if you want to make a simulation of Nature, you’d better make it quantum mechanical, and by golly it’s a wonderful problem, because it doesn't look so easy."
\end{displayquote}

On year later, inspired by the work of Bennet and Fredkin, he added \cite{feynmanquote2}:

\begin{displayquote}
``We can in principle make a computing device in which the numbers are represented by a row of atoms with each atom in either of the two states. That’s our input. The Hamiltonian starts “Hamiltonianizing” the wave function [...] The ones move
around, the zeros move around [...] Finally, along a particular bunch of atoms, ones and zeros [...] occur that represent the answer."
\end{displayquote}

In addition, using quantum information processing to create a computer seemed to solve an impending problem faced by traditional classical computers: Moore's law. First stated in 1965, this empirical rule claims that the transistors, building blocks of computers and physical embodiment of bits (0's and 1's), will be twice smaller every 18 months. As transistors reach a size of few nanometers with few atoms per unit, quantum and thermodynamical effects will disturb their properties. Shrinking has its limits \cite{mooreslawend} and experts expect the end of Moore's law in the current decade.\\

\subsubsection{Quantum Computing}\label{sec:intro_quantum_computing}

So, what is a quantum computer, and why it may surpass classical computers? First of all it is a computer. It manipulates quantum objects, such as photons, electrons, or ions, as bits of information (see Section \ref{sec:preliminaries_quantum_computing} for mathematical formalism). Therefore the quantum bits or \emph{qubits}, representing the 0's and 1's, would inherit a quantum nature. It allows them to be in both states a the same time or to be \emph{entangled} with each other. But these properties are only available before any measurement is made, after which everything becomes classical again. In the meantime, the quantum computer would apply logical operations to the qubits so that the measurements would give the desired answer with less resource globally.     

Intuitively, the key difference lies in the exponential superposition of binary inputs. Indeed, one classical bit can be either in state 0 or 1 as a transistor can be opened or closed, but a qubit can simultaneously be in a combination of both states, informally:
\begin{equation}
    \ket{0} + \ket{1}
\end{equation}
where the Dirac notation $\ket{\cdot}$ reminds us that the bit is a quantum system. Similarly, two qubits can be in four states simultaneously, informally:
\begin{equation}
    \ket{00} + \ket{01} + \ket{10} + \ket{11}
\end{equation}
It follows that $n$ qubits can be in a superposition of $2^n$ states. Since $2^n$ classical bits would be necessary to encode the same amount of state. This gives the intuition of the exponential advantage quantum could offer, informally: 
\begin{equation}
    \ket{0\cdots00} + \ket{0\cdots01} + \ket{0\cdots010} + \cdots + \ket{10\cdots0}
\end{equation}

Let's consider a simplified, informal, and intuitive reasoning. We are required to solve the task of checking if a given name is "Albert", among a random list of $N = 10^{9}$ names (a billion). The best thing a classical computer can do is to repetitively instantiate bits to be each name of the list, one by one, and check if the name is "Albert" until it finds it. The time to find the right name, called the \emph{complexity} of the algorithm, would be on average $N/2$. We see that this algorithm would be \emph{linear} in $N$, and doubling $N$ would double the time. 
\begin{equation}
    \begin{split}
    \text{Niels} \mapsto 0\\
    \text{Marie} \mapsto 0\\
    \vdots\\
    \text{Albert} \mapsto 1\\
    \vdots\\
    \text{Erwin} \mapsto 0
    \end{split} 
\end{equation}

However, if a quantum computer could instantiate all names in superposition using only 30 qubits ($2^{30} \approx 10^9$), it would need to check only \emph{once} if the quantum state is "Albert" and get a superposition of all answers: 

\begin{equation}\label{albertquantum}
    \begin{rcases*}
    \ket{\text{Niels}}\\
    +\ket{\text{Marie}}\\
    \vdots\\
    +\ket{\text{Albert}}\\
    \vdots\\
    +\ket{\text{Erwin}}
\end{rcases*}
\mapsto
\begin{cases*}
    \ket{0}\\
    +\ket{0}\\
    \vdots\\
    +\ket{1}\\
    \vdots\\
    +\ket{0}
\end{cases*}
\end{equation}
One computation instead of one billion seems astonishing. Note however that the output in Eq.\ref{albertquantum} is still in a quantum state before any measurement. Therefore the answer is not directly accessible to us, classical beings. One would have to measure and therefore destroy this state to get only one of the output, most probably a $\ket{0}$. In fact, the effective method is called Grover's algorithm \cite{groveralgo} and is quite different. It has the benefit of a complexity of $\sim	\sqrt{N}$ instead of $\sim N$. 

In fact, during the 1990s, computer scientists and physicists tried to develop a theory on quantum computing and find specific problems where a quantum computer would be beneficial. The first algorithms made by Deutsch-Josza \cite{deutsch1992rapid}, Bernstein–Vazirani \cite{bernstein1997quantum}, and Simon \cite{simon1997power}, where simple but already showed provable exponential speedups. Later, the development of Phase Estimation, Quantum Fourier Transform led to the famous Shor's algorithm \cite{shor1999polynomial} for solving prime number factoring in 1994.

Behind these specific algorithms hides the field of \emph{Complexity theory}, and the question of what nature is able to compute efficiently, and what happens if we add quantum physics to it? The discovery of efficient quantum algorithms would invalidate the Church-Turing thesis which, in its modern complexity theoretical formulation, states:

\begin{displayquote}
``A probabilistic Turing machine can efficiently simulate any realistic model of computation."
\end{displayquote}

Consider a yes-no problem with an input of $n$ bits. We call \textbf{P} the class of problems \emph{solvable} in time polynomial in $n$ by a computer, or more precisely a Turing Machine. \textbf{NP} is the class of problems where, if a solution is given, we can \emph{verify} it in polynomial time in $n$. Some problems that are in \textbf{NP} but not in \textbf{P} are called \textbf{NP-complete} problems. Notably, the question of proving or not that \textbf{P}$\neq$\textbf{NP} is one of the million-dollar problems of the Clay Math Institute. The class of problems efficiently solvable by a quantum computer with some constant error allowed is called \textbf{BQP} for Bounded-Error Quantum Polynomial-Time. We could compare it to \textbf{P} but since quantum measurements are probabilistic, it is fairer to compare it to \textbf{BPP} for Bounded-error Probabilistic Polynomial time, the equivalent of \textbf{P} with the ability to give a solution with some constant probability. It is easy to show \textbf{BQP} contains \textbf{BPP}, meaning that every (probabilistic) classical circuit can be simulated by a quantum circuit. But is the reciprocal true? In fact, proving that \textbf{BQP}$\neq$\textbf{BPP} would invalidate the Church-Turing thesis cited above, and be key to understand the power of quantum computing. 
However, it is believed that \textbf{NP} $\nsubseteq$ \textbf{BQP}, meaning that some important problems hard to solve but easily checkable, would eventually not be solvable by a quantum computer. In conclusion, \textbf{BQP} is something else, a complexity class made stranger due to quantum nature.\\

\begin{figure}[h]
    \centering
    \includegraphics[width=0.4\textwidth]{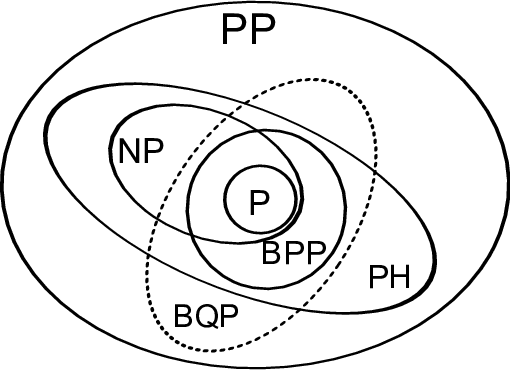}
    \caption{A map of fundamental complexity classes. Source: \cite{nakata2014diagonal}.}
    \label{fig:complexitclasses}
\end{figure}

Researchers realized quickly that qubit errors, due to various quantum effects such as decoherence or uncontrolled state perturbation, would be a major flaw for quantum computers. As for classical computing, a theory of \emph{error correction} has been developed \cite{knill1997theory, nielsen2002quantum}. In error correcting codes, a \emph{logical} qubit in state $\ket{0}$ or $\ket{1}$, is in fact composed of many \emph{physical} qubits. Current error correcting codes imply a strong overhead in the number of qubits required for a denoised device. 

We call \emph{universal fault tolerant} quantum computers (FTQC) the ideal quantum computers, with a universal set of gates, and a high number of logical qubits. Although they seem far away, conceptualizing and working on these ideal computers help us to understand theoretically what are the hopes and the limits. Proving a serious limitation could thus call into question the efforts currently deployed, or on the contrary provide even more excitement as Shor's algorithm did in 1994. 

Throughout this thesis, the majority of the quantum machine learning algorithms will be suited for such FTQC devices, except for Chapter \ref{chap:OrthoNN_nisq} where we propose a quantum circuit that we effectively implement on a real quantum computer.

\subsubsection{Machine Learning}

Machine learning is a subfield of artificial intelligence.  Its specificity is to perform tasks in a radically different way than what is usually considered as algorithms. Indeed, these algorithms are made to progressively learn how to solve a problem instead of being the most efficient solution by design. There exist plenty of algorithm families which all have their properties, their formalism and applications, while remaining very general. Modern developments include deep learning, or artificial neural networks, which are allegedly built to mimic neurons connectivity in the brain. These methods have become essential in all domains of science. 

\begin{figure}[h!]
    \centering
    \includegraphics[width=0.41\textwidth]{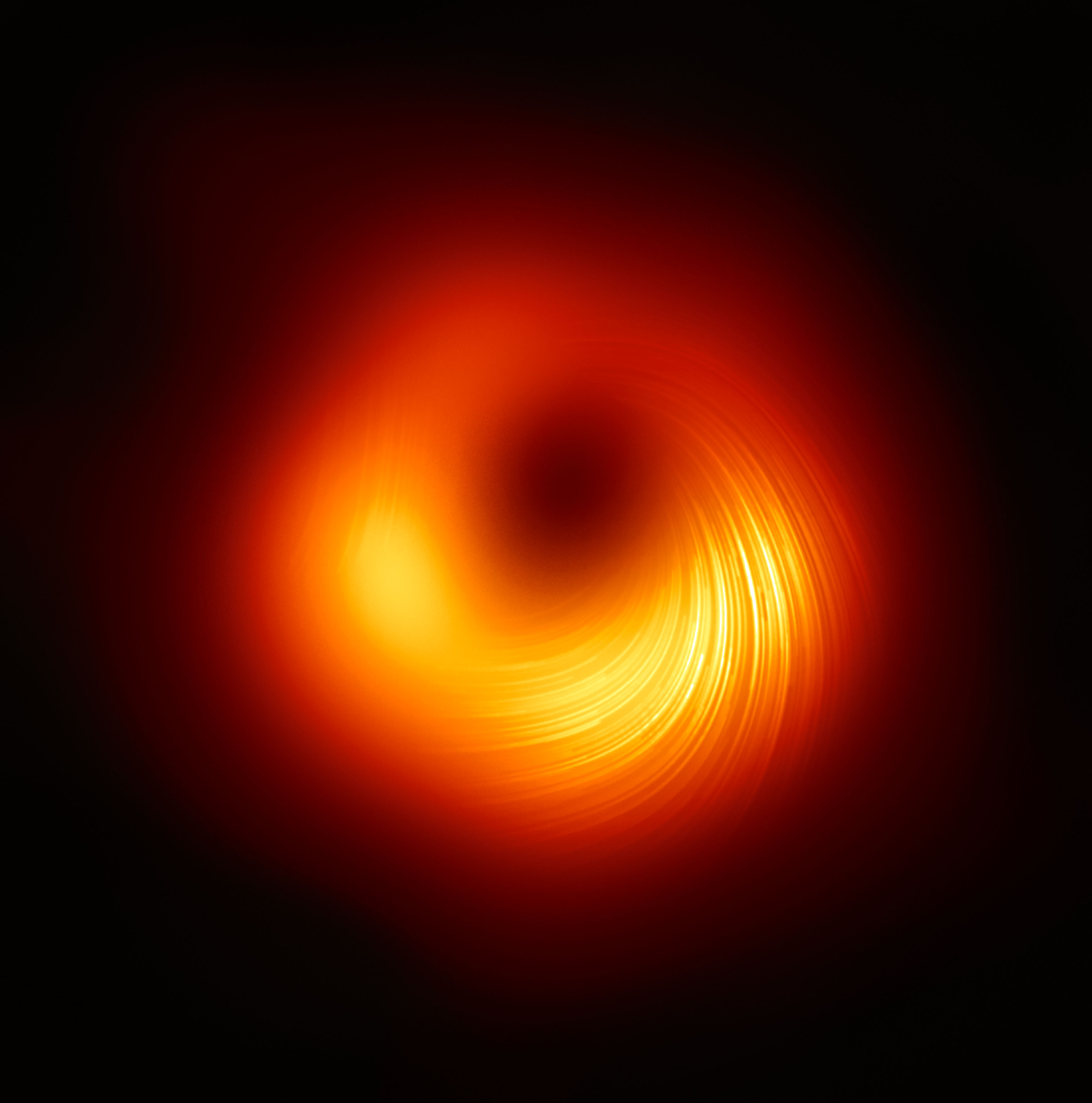}
    \caption{On April 2019, the first image of the blackhole M87$^*$ was produced using petabytes of data and machine learning algorithms \cite{M87}. Updated image from March 2021.}
    \label{fig:blakchole}
\end{figure}

In the last decade, machine learning algorithms have pushed the boundaries of science and information processing more than we could have imagined. 
Among the most exciting and recent discoveries, in 2019, the Event Horizon Telescope (EHT) reconstructed the first image of a black hole in the galaxy M87 using a machine learning algorithm \cite{M87}. It used a supercomputer to process an enormous amount of data, gathered from 10 telescopes around the world: more than 10 petabytes, equivalent to the number of pictures 100.000 people would take in their entire life. In 2021, 1200 gravitational lenses, another phenomenon predicted by Einstein (again) in the theory of general relativity, have been discovered thanks to a machine learning algorithm. This algorithm found them in an image of Space containing 10 trillion pixels, or 1 petabyte \cite{gravitationallenses}. This discovery could help us understand the expansion of the universe and discover new galaxies. Discoveries also concern biology since in 2020 researchers used machine learning algorithms \cite{deepmindalphafold} to solve the problem of protein folding for the first time at the CASP competition \cite{CASP}. This opens new paths for understanding life or discovering new medical treatments.

All these developments come at a cost of a tremendous amount of data processing, requiring the most powerful supercomputers available. Machine learning may reach its limit with the never-ending global data growth, associated with the increasing complexity of the algorithms used. Efficient computing will become mandatory since machine learning is now helping in many domains, from medical applications, image processing, social networks, experimental science, safety systems, and may even help for fighting against climate change \cite{climatechangeAI}.

\subsubsection{Recent Developments}
In recent years, a lot of efforts have been done by physicists all over the world to make the first experimental realizations of qubits, quantum logical gates, and now quantum computers. In 2017, anyone could access IBM's superconducting qubit quantum computer and launch a circuit. More recently, in 2019, Google demonstrated the first quantum ``supremacy" experiment \cite{googlesupremacy}. Their 53 qubit quantum chip was processed random quantum logical gates to create a complex quantum state and sample random outputs. Despite the noise of their device, they showed that for a classical computer to output random results following the same distribution it would require allegedly 10,000 years, versus in 200 seconds for the quantum computer. Note that this result was recently nuanced by \cite{pan2021simulating} who were able to classically simulate the same quantum circuit in 149 days. However, the ``supremacy" experiment remains a great achievement that proves the reality of exponential Hilbert spaces. 

Since this news made the front page, quantum computing has became more realistic and triggered a lot of hope. Many countries, universities, big and small private companies have started the race of building a fault tolerant universal quantum computer. The expectations are high, and the pressure on the achievements is rising. Due to the universality of quantum computing, every domain is now interested, including optimization, machine learning, chemistry, material science, health, and of course quantum physics in general.

But one question remains uncertain: will a fault tolerant quantum computer be useful? Despite the difficulty of physically building one, which would already be a fantastic scientific achievement for mankind, what useful task would we do with such a machine? And isn't there any fundamental limits to their power, and why? In the rest of this thesis, we will focus these questions on the field of unsupervised machine learning and neural networks and try to answer the following:

\begin{displayquote}
    Can a fault tolerant universal quantum computer provide an advantage in machine learning over classical computing?
\end{displayquote}

Both quantum computing and machine learning have universal properties, and the future will probably allow us to find unexpected results by combining these two fields.

\section{Quantum Machine Learning}\label{sec:intro_qml}

Combining quantum computing and machine learning is audacious but justified as they share a deep connection. Both theories are based on a common mathematical formalism, linear algebra, which makes a certain translation possible. Indeed, as we will see later, all machine learning can be written as vectors, matrices, vector spaces and transformations. Algorithms rely on linear algebra properties and theorems to classify, modify or create data points, seen as vectors (Chapter \ref{chap:intro_classical_ml} for details). They can also play with representations and map points from a vector space to another where the task is efficient. On the other hand, quantum physics formalism was built around the mathematical description of quantum states, represented as vectors in a complex vector space called the Hilbert space. Therefore, a set of qubits can always be seen as a vector in a high dimensional space, and any quantum gate or circuit as a matrix or linear operation in that space (see Section \ref{sec:preliminaries_quantum_computing} for details).

In short, both theories speak the same language, but they also differ in many aspects. To cite a few: quantum Hilbert spaces are exponentially bigger but don't allow non-linear transformations, which are common in machine learning. quantum Hilbert spaces are complex, whereas data in machine learning is mostly real numbers. quantum algorithms deal with quantum states but machine learning requires classical inputs and classical outputs. quantum vectors are normalized, which can be undesirable for representing data in machine learning. 

The goal of quantum machine learning (QML) \cite{biamonte2017quantum} is to find a common theory for developing quantum algorithms that implement known or unknown machine learning tasks. It is also about using the differences between the two fields to propose new algorithms or to enhance the existing ones.

This deep connection between the two fields become real in 2009 with the HHL algorithm \cite{HHL} that solve linear systems and matrix inversion with a proven exponential speedup on a quantum computer. Given an $N$-dimensional input vector $b\in\R^{N}$ and a Hermitian matrix $A \in\R^{N\times N}$, the task is to find a vector $x \in\R^{N}$ such that:
\begin{equation}
    Ax = b
\end{equation}
Solving this problem comes down to find the inverse $A^{-1}$ of the matrix $A$, as $x = A^{-1}b$. On a classical computer, this requires in general $\sim N$ iterations, but the HHL allows to solve it in only $\sim \log(N)$ steps. This represents an exponential speedup that could be a practical game changer, as this computational task appears all over science including fluid mechanics, optimization, physics in general, but also machine learning. This breakthrough also questioned our abilities to convert the inputs $b$ and $A$ in quantum states to be further processed by a quantum circuit, as well as the way of recovering a classical output from it \cite{readthefineprint}. 

\begin{figure}[h!]
    \centering
    \includegraphics[width=0.8\textwidth]{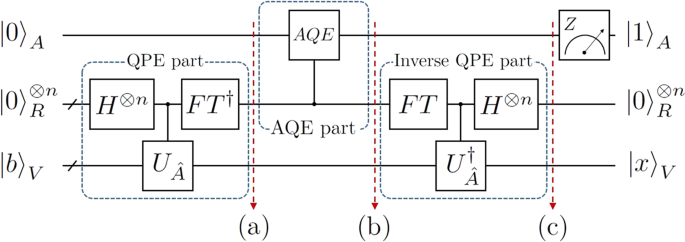}
    \caption{Quantum circuit for the HHL algorithm. Source: \cite{lee2019hybrid}}
    \label{fig:hhl_circuit}
\end{figure}

Using the HHL algorithm, the first proposals for precise quantum machine learning algorithms appeared a few years later. They concerned various tasks including simple linear regressions, the dimensionality reduction called Principal Component Analysis (PCA) \cite{Lloyd_PCA_quantum}, Nearest Neighbors algorithm \cite{wiebe_nearest_neigbhors}, topological data analysis \cite{Lloyd_topological_ml}, recommendation systems \cite{kerenidis_recommendation_system}, classification with Support Vector Machines (SVM) \cite{rebentrost2018quantum_svm}, unsupervised learning and clustering \cite{aimeur2013quantum, LMR13}. Later on, attempts to provide quantum algorithms for neural networks and deep learning were proposed \cite{wiebe2014quantum_deeplearning, Lloyd_hopfield_nn, farhi2018classification}.\\

\begin{figure}[h!]
    \centering
    \includegraphics[width=0.6\textwidth]{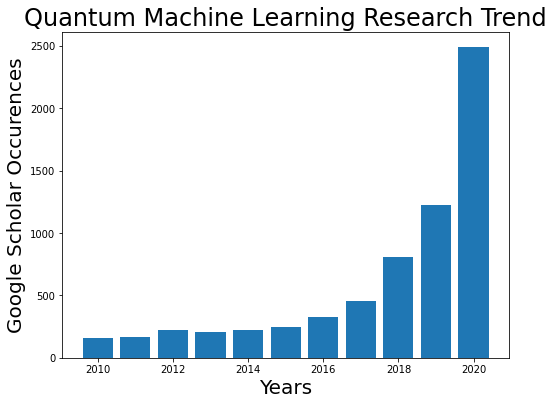}
    \caption{Occurrences of the basic keywords “quantum machine learning”, “quantum neural network(s)”, and “quantum deep learning” in Google Scholar’s articles 2010–2020.}
    \label{fig:qml_trends}
\end{figure}

In parallel to quantum machine learning, a concerted research effort has been made to find quantum algorithms for optimization problems. As we will see in Chapter \ref{chap:intro_classical_ml}, machine learning is intimately linked to optimization, in particular concerning the various way of performing gradient descent. Recent results include a quantum gradient descent algorithm \cite{kerenidis2020_gradient_descent}, as well for the interior point method \cite{kerenidis2020quantum_IPM} and more generally algorithms for solving LP and SDP problems \cite{van2017quantum, brandao2017quantum, van2018improvements}. For instance, a quantum solver for second order cone programming is directly applicable to support vector machine \cite{kerenidis2021quantum_SOCP}, a popular machine learning algorithm.\\

It is worth noticing that since the start of this thesis, an impressive amount of new QML algorithms were proposed. Some of them followed the initial works presented above. Others shifted to a new paradigm called variational quantum circuits, where ideas of machine learning (tunable parameters) were directly applied to quantum circuits themselves. In addition, the first implementations on actual quantum hardware were made possible by various institutions and companies allowing for real experimentation: IonQ, IBM, Google, Xanadu, Rigetti, Pasqal, and many more. Besides, many software projects were made to program quantum circuits easily: Qiskit, Pennylane, Cirq, Forge, Q\# and plenty more. In only three years, the evolution is noticeable at all levels and brings great hope for the future of quantum computing.

\section{Contributions}

The approach of this thesis is to pursue these works and find new quantum algorithms that correspond to existing machine learning methods that are used in practice. We focus our scope to clustering or unsupervised learning algorithms, as well as on neural networks or deep learning methods. Moreover, we always prove that using a quantum computer would benefit in some manner. Defining and proving a quantum \emph{advantage} is the key difficulty in most cases. It can be a theoretical complexity result for the running time, a quantum circuit with shorter depth, or with few qubits. It can also relate to the final accuracy of the algorithm, in theory with the control on the errors, but also practice on real datasets.\\

Complexity or running time results are presented with the $O(\cdot)$ notation, indicating asymptotic growth with the size of the problem. For instance, problems of size $N$ that require $N$, $1/2N + 3$, or $100N$ steps to be solved, have each a complexity of $O(N)$, indicating proportionality to $N$. This notation always emphasizes on most dominant asymptotic terms, and for instance we have $N^3 + 10\sqrt{N} + 4\log(N) = O(N^3)$.
We also use the $\widetilde{O}(\cdot)$ symbol to discard the terms that grow logarithmically slow. Therefore, $O(\log(N)N^2)$ can be written $\widetilde{O}(N^2)$. See Section \ref{sec:math_notations} for more details. 

\begin{figure}[h]
    \centering
    \includegraphics[width=0.8\textwidth]{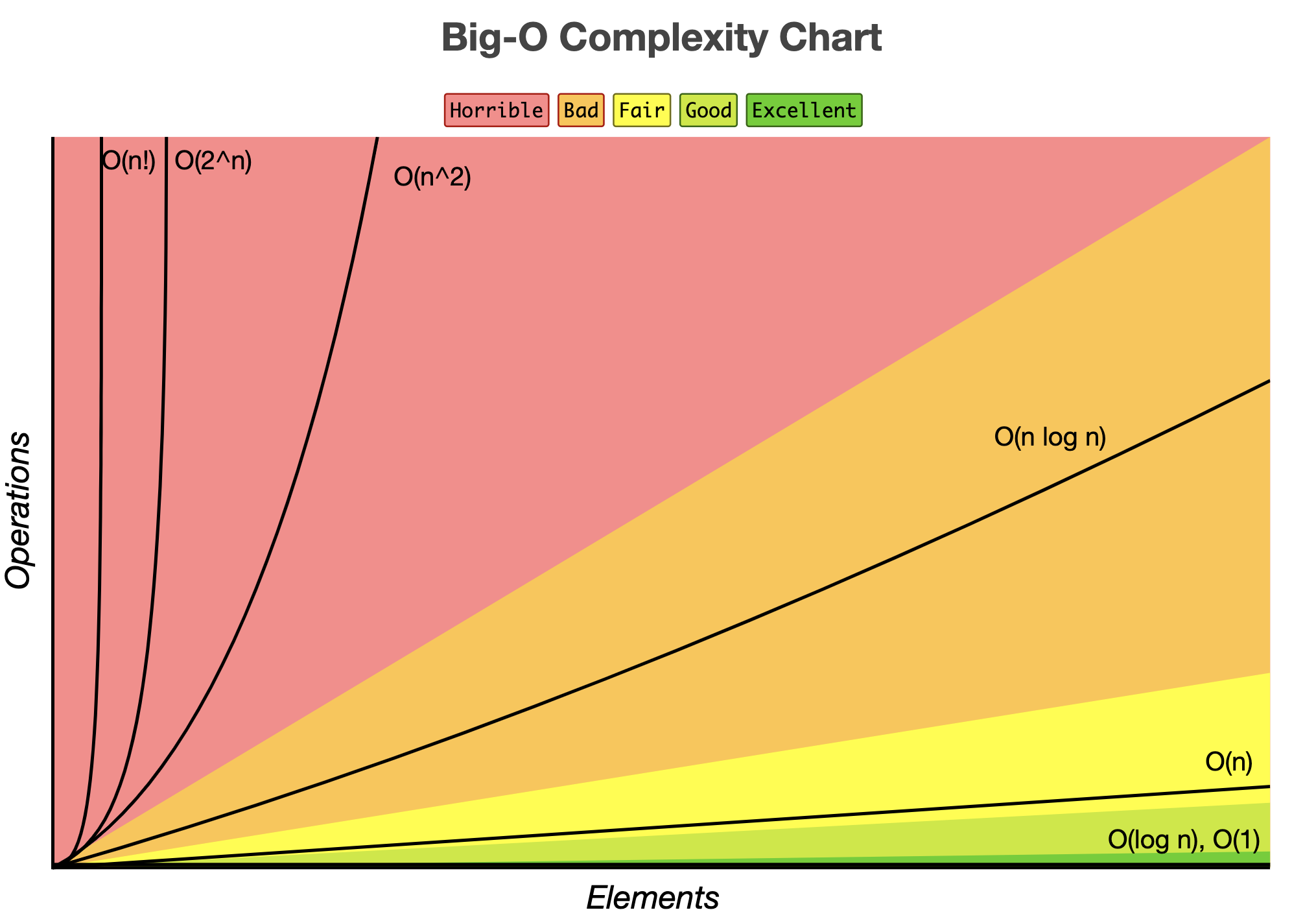}
    \caption{Representation of Big-O notation for different asymptotic growths. Source: Bigocheatsheet}
    \label{fig:bigO}
\end{figure}

To prove a speedup over a classical algorithm, the quantum algorithm must be comparable in some sort. To do so, we require it to be end-to-end and comparable: It should start from classical input, follow similar steps and returning a classical answer. With rigor, we tried to analyze any source of error due to quantum randomness during measurement, and include it in our final running time for a fair comparison. We also have simulated our quantum algorithms on real datasets to compare them in practice with their classical counterparts.\\

We now present the structure of this thesis and the corresponding results. for each result, we compare the complexity of the previous classical or quantum algorithm and the one from this thesis. Note that they often depend on some parameters define as
\begin{itemize}
    \item $N$: usually the number of points, size of the problem, input size, etc.
    \item $d$: dimension of the points or vectors for the problem to solve.
    \item $k$: number of classes or clusters to find in a dataset.
    \item $\epsilon$, $\delta$: error or precision parameters due to quantum effects. There is most of the time a trade-off between speed and accuracy.
    \item $\mu$, $\eta$: these are data dependant parameters specific to quantum linear algebra. See Definitions \ref{def:mu} and \ref{def:eta}. In a nutshell, $\mu$ is derived from a data matrix norm, and $\eta$ is the maximum norm of the vectors in the dataset. 
    \item Other parameters should be defined in Section \ref{sec:math_notations} or in the corresponding theorems. 
\end{itemize}

\paragraph{Part \ref{part:preliminaries_QML}} To create our quantum algorithms, we used existing quantum linear algebra tools and developed new ones as well. These tools were general enough to be reused across algorithms and are the common thread of most of this thesis. Among others, we propose  quantum algorithms for inner product and distance estimation (Section \ref{sec:inner_product_and_distance_quantum}, Theorem \ref{thm:distance_innpdct_quantum}). As we will see, they allow quantum computing to speak the same language as machine learning. Using them, we also provide an algorithm for a quantum convolution product (Section \ref{QCNNForward}, Theorem \ref{thm:convolution_layer}). As well as a quantum processing routine for graph-based machine learning including the fast creation of an adjacency graph and its Laplacian matrix (Section \ref{quantumspectralclusteringsection}, Theorem \ref{thm:quantumaccesstoL}). Finally, a new quantum tomography procedure with $\ell_{\infty}$-norm error bounds is introduced (Section \ref{sec:l_2_and_l_infinite_tomography}, Theorem \ref{thm:tomography_linfinity}), to retrieve a classical description of a quantum state faster, while keeping the meaningful information in the context of neural networks.

\begin{table}[h!]
\begin{center}
\begin{tabular}{|c|c|c|}
\hline
Algorithm             & Type      & Running Time      \\ \hline
Inner product or distance estimation (IPE) & Classical & $O(Nd)$          \\
\textbf{Quantum IPE \cite{qmeans}} & Quantum   & $\widetilde{O}(\eta/\epsilon)$ \\ \hline
Tensor convolution product & Classical & $O(N_oK)$          \\
\textbf{Quantum tensor convolution product \cite{QCNN}} & Quantum   & $\widetilde{O}(\eta/\epsilon)$ \\ \hline
Projected Laplacian matrix creation& Classical & $O(N^3)$          \\
\textbf{Quantum projected Laplacian matrix \cite{quantumspectralclustering}} & Quantum   & $\widetilde{O}(\mu\kappa/\epsilon)$ \\\hline
\end{tabular}
\caption{Summary of contributions for fundamental linear algebra routines. In the classical convolution product, $N_o$, and $K$ are respectively the output size and the kernel size. The quantum convolution product returns a quantum state.\label{table:linearalgebra_contribution}}
\end{center}
\end{table}

\begin{table}[h!]
\begin{center}
\begin{tabular}{|c|c|c|}
\hline
Algorithm             & Type      & Running Time      \\ \hline
$\ell_2$ tomography \cite{kerenidis2020quantum_IPM} & Quantum & $O(d\log(d)/\epsilon^2)$         \\
\textbf{$\ell_{\infty}$ tomography \cite{QCNN}} & Quantum   & $O(\log(d)/\epsilon^2)$ \\ \hline
\end{tabular}
\caption{Summary of contributions for quantum tomography. $d$ is the dimension or the number of elements in the quantum vector.\label{table:tomography_contribution}}
\end{center}
\end{table}

\paragraph{Part \ref{part:Unsupervised_QML}} We then focus on quantum algorithms for unsupervised machine learning. We propose q-means, a new quantum algorithm \cite{qmeans} providing a potential exponential speedup to one of the most basic and widely used clustering algorithms, the $k$-means algorithm (Chapter \ref{chap:qmeans}, Theorem \ref{thm:qmeans_main}). We build upon this result by introducing another quantum algorithm \cite{quantumspectralclustering}, an analog of the spectral clustering algorithm, which uses the $k$-means method on top of graph-based machine learning. (Chapter \ref{chap:Q_spectral_clustering}, Theorem \ref{thm:Qspectralclustering_main}). 

\begin{table}[h!]
\begin{center}
\begin{tabular}{|c|c|c|}
\hline
Algorithm             & Type      & Running Time      \\ \hline
$k$-means & Classical & $O(Nkd)$        \\
Quantum $k$-means \cite{LMR13} & Quantum & $O(Nk\log(d)/\epsilon)$        \\
\textbf{q-means \cite{qmeans}} & Quantum   & $\widetilde{O}(\log(N)k^2d\eta^{1.5}/\epsilon^3)$ \\ \hline
\end{tabular}
\caption{Summary of contributions for the $k$-means algorithms. $N$ is the size of the dataset, $d$ the dimension of each vector, and $k$ the number of clusters. The runtime for $q$-means is simplified and for the case of \emph{well-clusterable} datasets. The result from \cite{LMR13} outputs a quantum state and would become linear in $d$ to produce a classical output as in our work \cite{qmeans}.\label{table:kmeans_contribution}}
\end{center}
\end{table}

\begin{table}[h!]
\begin{center}
\begin{tabular}{|c|c|c|}
\hline
Algorithm             & Type      & Running Time      \\ \hline
Spectral clustering \cite{ng2002spectral} & Classical & $O(N^3)$          \\
\textbf{Quantum spectral clustering \cite{quantumspectralclustering}} & Quantum   & $O(\log(N)\mu)$ \\ \hline
\end{tabular}
\caption{Summary of contributions for the spectral clustering algorithms. $\mu$ in the quantum algorithm is $O(N)$ in the worst case and in our numerical experiments.\label{table:spectralclustering_contribution}}
\end{center}
\end{table}

\paragraph{Part \ref{part:Q_NeuralNet}} Next, the same tools are adapted to develop a framework for quantum neural networks, also called quantum deep learning. In particular, we introduce an algorithm for quantum convolution neural network \cite{QCNN} (Chapter \ref{chap:QCNN}, Theorem \ref{res:qcnn1}).  

We also propose a different type of quantum circuit, suited for Noisy Intermediate Scale Quantum computers, or NISQ \cite{NISQpreskill} (see Section \ref{sec:NISQ}), currently available. These quantum circuits have a specific pyramid shape and data encoding, allowing them to implement a neural network with orthogonal properties (Chapter \ref{chap:OrthoNN_nisq}). 

\begin{table}[h!]
\begin{center}
\begin{tabular}{|c|c|c|}
\hline
Algorithm             & Type      & Running Time      \\ \hline
Convolutional CNN layer \cite{lecun1998gradient} & Classical & $O(N_oK)$          \\
\textbf{Quantum CNN \cite{QCNN}} & Quantum   & $O(\sigma N_o \eta /\epsilon)$ \\ \hline
Orthogonal NN inference \cite{jia2019orthogonal} & Classical & $O(N^2)$          \\
Orthogonal NN training \cite{jia2019orthogonal} & Classical & $O(N^3)$          \\
\textbf{Pyramidal OrthoNN \cite{Quantum_OrthoNN} inference} & Quantum   & $O(N/\delta^2)$ \\
\textbf{Pyramidal OrthoNN \cite{Quantum_OrthoNN} training} & Classical   & $O(N^2)$ \\ \hline
\end{tabular}
\caption{Summary of contributions for neural networks algorithms. For CNN, $N_o$ and $K$ are respectively the output size and the kernel size. $\sigma$ is a ratio in [0,1]. \emph{OrthoNN} stands for Orthogonal Neural Network and $N$ is both the input and output size of a single layer.\label{table:cnn_contribution}}
\end{center}
\end{table}

\newpage
These contributions were the subject of scientific publications, which are listed below:
\begin{itemize}
    \item \cite{qmeans} ``q-means: A quantum algorithm for unsupervised machine learning". Published in \emph{Proceedings of the 33rd Conference on Neural Information Processing Systems (NeurIPS) - 2019, pp.4136-4146}. By I. Kerenidis, J. Landman, A. Luongo, A. Prakash.
    \item \cite{QCNN} ``Quantum Algorithms for Deep Convolutional Neural Networks" Published in \emph{Proceedings of the 8th International Conference on Learning Representation (ICLR) - 2020}. By I. Kerenidis, J. Landman, A. Prakash
    \item \cite{quantumspectralclustering} ``Quantum Spectral Clustering". Published in \emph{Physical Review A 103, 042415 - April 2021}. By I. Kerenidis, J. Landman.
    \item \cite{Quantum_OrthoNN} ``Classical and Quantum Algorithms for Orthogonal Neural Networks". By I. Kerenidis, J. Landman, N. Mathur. \emph{(Under submission)}
    \item \cite{QBNN} ``Quantum Inference Algorithm for Bayesian Neural Networks". By N. Berner, V. Fortuin, J. Landman. (Under submission)
    \item \cite{QMedicalImagingRoche} ``Medical Image Classification Via Quantum Neural Networks". By N. Mathur, J. Landman, Y. Li, M. Strahm, S. Kazdaghli, A. Prakash, I. Kerenidis. (Under submission)
\end{itemize}

\vspace{2cm}

\begin{figure}[h]
    \centering
    \includegraphics[width=\textwidth]{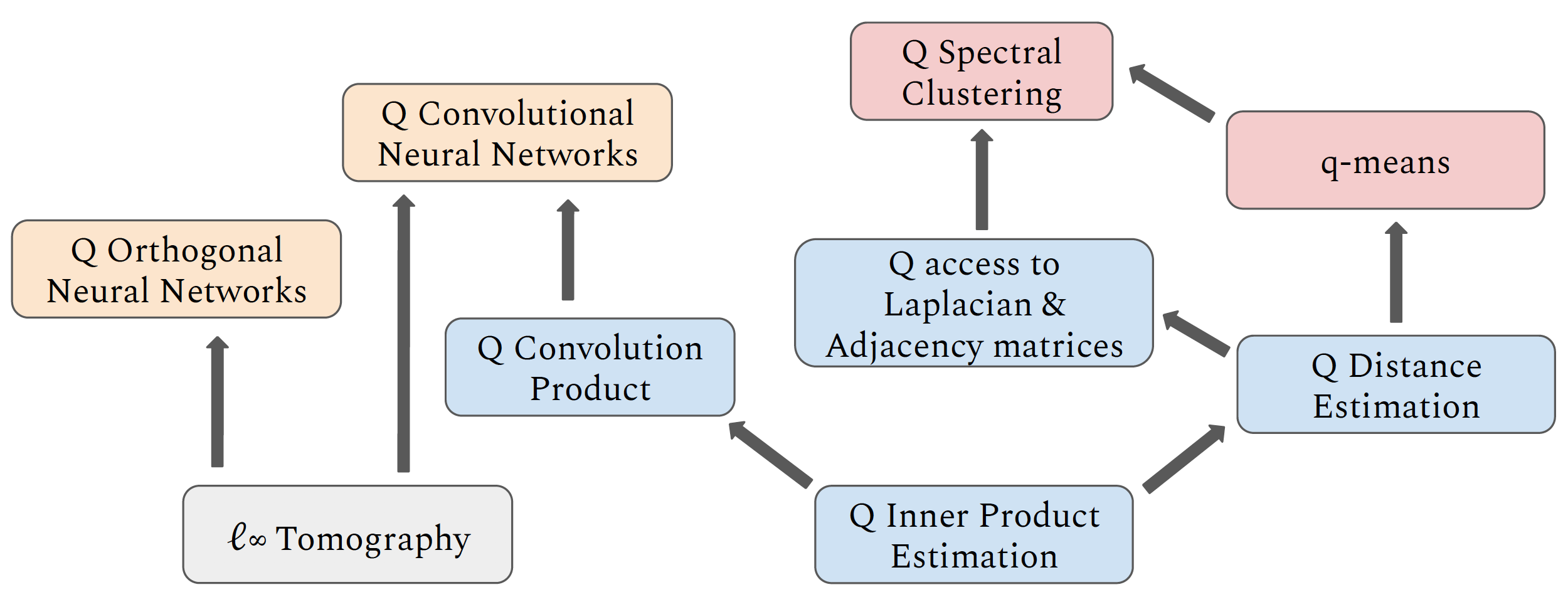}
    \caption{Diagram representation of the main contributions. Arrows denote dependencies between linear algebra (blue), tomography (grey), unsupervised machine learning (red), and neural networks (orange) algorithms. Q stands for ``Quantum".}
    \label{fig:plan_these}
\end{figure}

\newpage
\section{Mathematical Notations}\label{sec:math_notations}

We introduce basic notations and definitions for the understanding of this dissertation.  \\

$\mathbb{N}$, $\Z$, $\R$, $\R^+$, and $\C$ are respectively the integers, natural numbers, real numbers, positive real numbers, and complex numbers. For an integer $N>0$, $[N]$ denotes the set of integers between $1$ and $N$ included. $i$ is the imaginary number such that $i^2=-1$, but is also often used as the index of numbered elements. A Hilbert space is a real or complex vector space with an inner product.\\

Vectors are often written as $x$, $v$, $s$, or $y$ and are elements of (a subspace of) $\R^{d}$ (or $\C^{d}$), where $d>0$ is the dimension. Therefore, a vector $x\in \R^d$ vectors have $d$ real components, each in $\R$ (or $\C$). We write $x = (x_1,x_2,\cdots,x_d)$ or $x = \sum_{i=1}^d x_i e_i$ where $e_i$ is the $i^{th}$ vector of the standard or canonical basis. Note that in some cases, the vector's components will be indexed from 0 to $d-1$ instead. $x$ or $v$ will often denote the input vector of an algorithm, $y$ the output vector. $c$ often stands for a centroid vector (center of a cluster). 

The $\ell_p$ norm of a vector is $\norm{x}_p = (\sum_{i=1}^d |x_i|^p)^{1/p}$. In particular, we use the $\ell_2$ norm $\norm{x}_2 = \sqrt{\sum_{i=1}^d |x_i|^2}$, and we define the $\ell_{\infty}$ norm as $\norm{x}_{\infty} = \max_{i\in[d]} |x_i|$. If the subscript is not specified, $\norm{x}$ usually represents the $\ell_2$ norm.

The inner product between two $d$-dimensional real vectors $x$ and $y$ is written $(x,y)$, $x \cdot y$ or $x^Ty$ and is equal to $\sum_{i\in[d]}x_i y_i$. 
We have $\norm{x}_2 = \sqrt{(x,x)}$ and the euclidean distance between $x$ and $y$ is: 
\begin{equation}
d(x,y) = \norm{x-y}_2 = \sqrt{\norm{x}_2^2 + \norm{y}_2^2 - 2(x,y)}
\end{equation}
The normalized inner product is $\braket{x}{y}$ such that $(x,y) = \norm{x}_2\norm{y}_2\braket{x}{y}$. Two vectors $x$ and $y$ are orthogonal if $(x,y)=0$. We use $x^{\perp}$ to denote a vector orthogonal to $x$.\\
 
For a collection of $N$ vectors, also called \emph{dataset} of size $N$, we can number each vector in the set $\{x^i\}_{i\in[N]}$. Therefore, the $j^{th}$ component of the $i^{th}$ vector is written $x^i_j$. 
However, we often used $\{x_i\}_{i\in[N]}$ with each $x_i \in \R^d$ to denote the $i^{th}$ vector and not its component, which could lead to confusion. A dataset can also be represented as a matrix $A \in \R^{N\times d}$ (also $V$, $X$, $Y$, $M$ or $S$). Indeed, $N$ vectors of $d$ dimensions can compose the $N$ rows of a matrix. Therefore the $i^{th}$ vector can be written $A_i$ and its $j^{th}$ component is $A_{ij}$ or $A_{i,j}$. We denote its transpose $A^T \in \R^{d\times N}$, with elements $A_{ji}$.\\

Let $A$ be a \emph{square} matrix in $\C^{N\times N}$. If $A$ is said \emph{invertible}, we denote $A^{-1}$ the inverse of $A$ such that $AA^{-1}=I$ where $I$ is the identity matrix. The adjoint of $A$ is $A^{\dagger} = \overline{A}^T$, where $\overline{A}$ is the complex conjugate of $A$. The matrix $A$ is Hermitian if $A = A^{\dagger}$ and unitary if $AA^{\dagger} = A^{\dagger}A = I$. Note that for real matrices, being unitary is equivalent to being orthogonal.

The Singular Value Decomposition (SVD) of a \emph{rectangular} matrix $A\in\R^{N\times d}$ is of the form $A = U\Sigma V$, where $U\in\R^{N\times N}$, $V\in\R^{d\times d}$ and $\Sigma$ is a rectangular diagonal matrix with non negative elements $\sigma_i$ called the \emph{singular values}. If $r \leq \min(N,d)$ is the rank of $A$, we can write:

\begin{equation}
    A = \sum_{i\in[r]} \sigma_i u_i v_i^T
\end{equation}

where $u_i$ and $v_i$ are respective columns of $U$ and $V$. We can define the pseudo inverse of $A$ as $A^+ = \sum_{i\in[r]} \frac{1}{\sigma_i} u_i v_i^T$. The condition number $\kappa(A)$ is the ratio between the biggest and the smallest singular values $\kappa(A) = \frac{\sigma_{\max}}{\sigma_{\min}}$.

Let $A\in\C^{N\times N}$ be a \emph{diagonalizable} matrix. Then $A$ has $N$ eigenvectors $v_i$ and eigenvalues $\lambda_i$ such that $Av_i =\lambda_i v_i$. The vectors $v_i$ form a basis in $\R^N$. An unitary matrix is diagonalizable and its eigenvalues are such that $|\lambda_i|=1$. The sparsity $s$ of $A$ is the maximum number of non zero elements in a row of $A$.

A symmetric matrix $A\in\R^{N\times N}$ is said to be \emph{positive semidefinite} if, for any vector $x \in \R^N$, we have $x^TAx \geq 0$. Then all eigenvalues of $A$ are non negative.\\

We define possible norms for a matrix $A\in\C^{N\times d}$ with elements $A_{ij}$.
the Frobenius norm $\norm{A}_F$ if the generalization of the $\ell_2$ norm, $\norm{A}_F = \sqrt{\sum_{i,j}|A_{ij}|^2} = \sqrt{\sum_i \norm{A_i}_2^2}$. For a \emph{square} matrix we have $\norm{A}_F = \sqrt{\sum_i \lambda_i^2}$. The spectral norm of $A$ is written $\norm{A}_2$ or $\norm{A}$ and is the biggest singular value of $A$.\\

Tensors are the generalization of matrices with more than two dimensions. 3D tensors are indexed by $i$, $j$ and $d$.\\

For an algorithm depending on the variable $N$, its running time is stated in the standard asymptotic notation $O(f(N))$ which indicates a running time upper bounded by $cf(N)$ for a fixed $c \in \R^+$ and sufficiently large $N>0$. The notation $\widetilde{O}()$ hides
polylogarithmic factors (e.g. $\log^2(N)$), that is $O(f(N)polylog(N))$ is represented by $\widetilde{O}(f(N))$. \\

Note finally that across Chapters and algorithms, it is possible that notations switch due to the context e.g. in Chapter \ref{chap:Q_spectral_clustering} singular values are also written $\lambda_i$. For neural networks in Chapter \ref{chap:OrthoNN_nisq}, the input vectors are of size $n$ and the output of size $d$.

\chapter{Classical Machine Learning}\label{chap:intro_classical_ml}
\epigraph{\textit{"A computer would deserve to be called intelligent if it could deceive a human into believing that it was human."}}{Alan Turing \\ \emph{Computing Machinery and Intelligence} (1950)}

\section{Introduction}

The growing importance of machine learning in Science, but also in industry and in our society, is undeniable. Recent advances in signal processing, time series forecasting, medical predictions, image recognition, anomaly detection, or generative data, have surpassed most expectations. In 2019 deep learning inventors \cite{lecun2015deep} were awarded the Turing medal, and together with Learning Theory and Neuroscience, the scientific community is working to understand how the brain learns.

The sophistication of the algorithms used, and the amount of data necessary to train them seems staggering. The quantity of data generated by our society, from the web, various captors, open medical data, is expected to grow beyond comprehension. Most of this data will be multimodal, complex, and unlabelled. Therefore the help of unsupervised machine learning and neural networks (or deep learning) will become more necessary, but more powerful ways of making sense of this amount of data will be necessary as well.

In the following, we will introduce basic concepts and notations in machine learning, required for this thesis. Complete courses can be found in \cite{bishop2006pattern} and \cite{goodfellow2016deep} for deep learning. \\

A machine learning task aims to extract information from data. It usually consists of a parametrized function (or model). Its parameters are is progressively tuned (or trained) such that the task is done as efficiently as possible. We then say that the model has \emph{learned} to predict using the data provided. 

Data points could represent points in space, a set of numbers, or even images decomposed as pixels. In the general case, a dataset $\mathcal{D}$ consists of $N$ vectors $\{x_i\}_{i\in[N]}$, where each vector lies in a subspace of $\R^d$. Said differently, each vector has $d$ features, or is $d$-dimensional. The model to train can be written as a function $f(x|\theta)$, where $\theta$ are the parameters to tune, that should map each input $x_i \in \R^d$ to an output $y_i \in \R^{d'}$. The outputs can be of any sort as well. In the context of regression, it usually consists of a number, and $d'=1$. In classification, where the goal is to put on each input a \emph{label} (or \emph{class}), we have $d'=k$, where $k$ is the number of classes. \\

The most common machine learning branch is supervised learning, where the dataset is provided with labelled data. Namely we are given a dataset $\mathcal{D} = {x_i,y_i}_{i\in[N]}$. Training such algorithms boils down to being able to predict the right $y_i$ for each $x_i$ with a \emph{training set}, and then ensure that the model can also predict the right result for a \emph{testing set} of pairs $(x_i,y_i)$ that haven't been seen during the training. Supervised learning are usually trained  by adjusting the parameters $\theta = (\theta_1,\cdots,\theta_m)$ for a chosen number $m$, such that a \emph{loss} or \emph{cost} function $\mathcal{C}$ decreases. This loss is calculated from the accuracy of the predictions made on the training set. Then, we perform a \emph{gradient descent} to update each parameters $\theta_j$ with a \emph{learning rate} $\lambda>0$:
\begin{equation}
    \theta_i \gets \theta_i -\lambda \frac{\partial \mathcal{C}}{\partial \theta_i}
\end{equation}

In this thesis, supervised learning will be used as the framework for neural networks (see Section \ref{sec:preliminaries_classical_nn}). Before that, in the next section, we will introduce unsupervised machine learning and two specific algorithms.

\begin{figure}[h]
    \centering
    \includegraphics[width=0.8\textwidth]{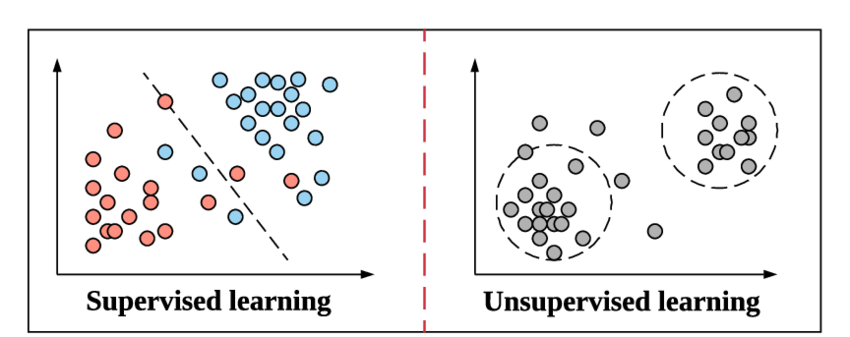}
    \caption{Schematic differences between supervised and unsupervised machine learning. Source: \cite{qian2020orchestrating}.}
    \label{fig:supervised_vs_unsupervised}
\end{figure}

Note that, for each algorithm presented, notations for inputs, outputs, parameters, matrices, and different numbers, may change but will remain consistent between classical and quantum versions.

\section{Unsupervised Learning}

As shown in Fig.\ref{fig:supervised_vs_unsupervised}, unsupervised learning deals only with unlabelled dataset $\mathcal{D} = \{x_i\}_{i\in [N]}$. The algorithms must find by themselves the labels $y_i$ to assign at each input $x_i$. In the context of classification or segmentation, this automatic process is often referred to as \emph{clustering}. It aims to find clusters among data points, each cluster can then be converted as a certain class or label $y_i$. Note that unsupervised learning can also include generative tasks such as generative adversarial neural networks \cite{goodfellow2014generative}, which have been studied as quantum algorithms as well \cite{dallaire2018quantum,lloyd2018quantum}. 

\begin{figure}[h!]
    \centering
    \includegraphics[width=150px]{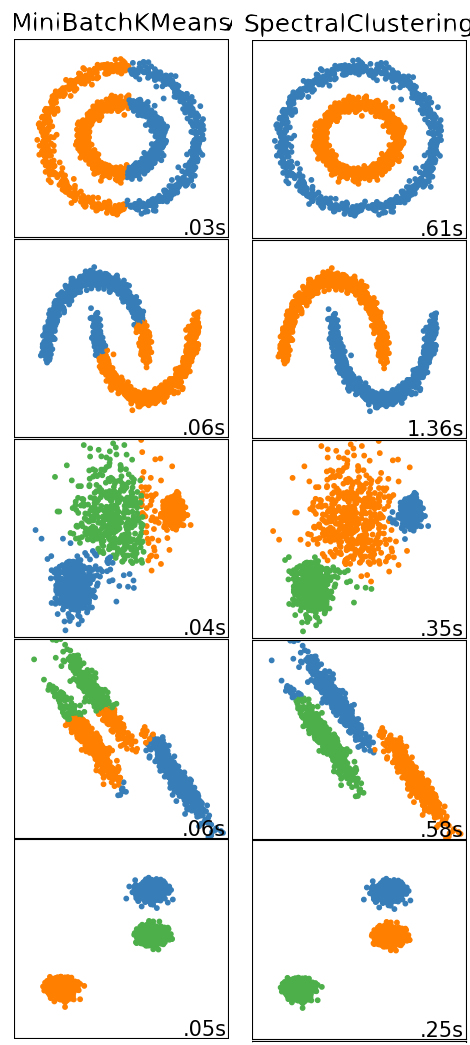}
    \caption{A comparison between $k$-means and spectral clustering on different types of toy datasets. We can see that spectral clustering distinguishes  nested datasets with more accuracy, but is unfortunately slower. Source: Scikit Learn \cite{scikit-learn}.}
    \label{fig:kmenasvsspectral}
\end{figure}

The next two algorithms are the $k$-means algorithm and the spectral clustering algorithm. They are closely linked, as the latter relies on the former. Given a distribution of points in a vector space, their goal is to identify clusters among them and further classify new points. This problem is known to be \text{NP-complete} \cite{vattani2009hardness} (see Section \ref{sec:intro_quantum_computing}). Both algorithms are iterative, non deterministic algorithms, or heuristic, that solve this problem with good accuracy on simple cases. $k$-means clustering has a complexity of $O(N)$ per iteration, where $N$ is the number of points, but suffers from poor flexibility and requires well-shaped datasets. Spectral clustering however uses properties of graph theory to distinguish complex data, at the cost of a higher complexity of $O(N^3)$ per iteration. This can be seen in simple examples showed in Fig.\ref{fig:kmenasvsspectral}.

\subsection{$k$-means Clustering}\label{sec:classical_kmeans_preliminaries}

The $k$-means algorithm was introduced in 1982 \cite{lloyd1982least}, and is extensively used for unsupervised problems. The inputs to $k$-means algorithm are vectors $v_{i} \in \R^{d}$ for $i \in [\n]$. These points must be partitioned in $k$ subsets according to a similarity measure,  which in k-means is the Euclidean distance between points. The output of the $k$-means algorithm is a list of $k$ cluster centers, which are called \textit{centroids}. 

The algorithm starts by selecting $k$ initial centroids randomly or using efficient heuristics like the $k$-means++ \cite{arthur2007k}. It then alternates between two steps: (i) Each data point is assigned the label of the closest centroid. (ii) Each centroid is updated to be the average of the data points assigned to the corresponding cluster. These two steps are repeated until convergence, that is until the change in the centroids during one iteration is sufficiently small. 

\begin{figure}[h]
    \centering
    \includegraphics[width=0.5\textwidth]{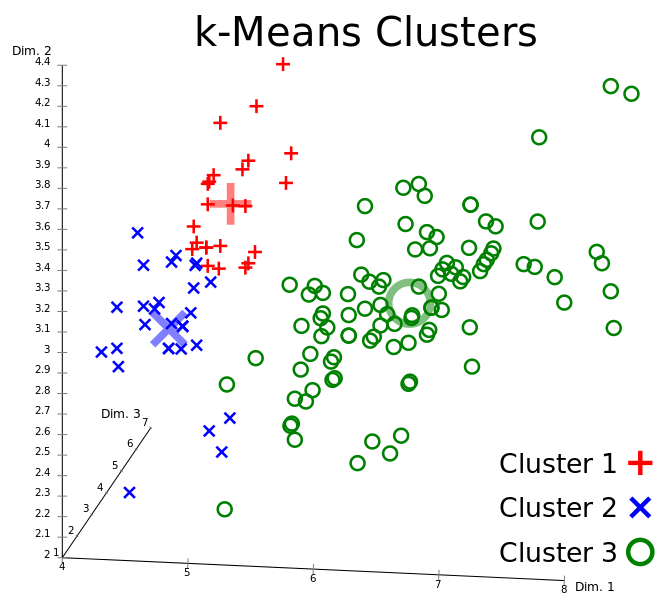}
    \caption{3D representation of the $k$-means clustering applied on the IRIS dataset of 3 types of flowers. Source: Wikipedia.}
    \label{fig:kmeans_iris}
\end{figure}

More precisely, we are given a dataset $V$ of vectors $v_{i} \in \R^{d}$ for $i \in [\n]$. At step $t$, we denote the $k$ clusters by the sets $C_j^t$ for $j \in [k]$, and each corresponding centroid by the vector $c_{j}^{t}$. At each iteration, the data points $v_i$ are assigned to a cluster $C_j^t$ such that $C_1^t \cup C_2^t \cdots \cup C_K^t = V$ and $C_i^t \cap C_l^t = \emptyset$ for $i \neq l$. Let $d(v_{i}, c_{j}^{t})$ be the Euclidean distance between vectors $v_{i}$ and $c_{j}^{t}$.  
The first step of the algorithm assigns each $v_{i}$ a label $\ell(v_{i})^t$ corresponding to the closest centroid, that is 
\begin{equation}
\ell(v_{i})^{t} = \text{argmin}_{j \in [k]}(d(v_{i}, c_{j}^{t}))
\end{equation}
The centroids are then updated, $c_{j}^{t+1} = \frac{1}{ |C_{j}^t|} \sum_{i \in C_{j}^t } v_{i},$
so that the new centroid is the average of all points that have been assigned to the cluster in this iteration. We say that we have converged if for a small threshold $\tau$ we have
\begin{equation}
\frac{1}{k}\sum_{j=1}^{k}{d(c_{j}^{t},c_{j}^{t-1}}) \leqslant \tau
\end{equation}
The loss function that this algorithm aims to minimize is the RSS (residual sums of squares), the sum of the squared distances between points and the centroid of their cluster. 
\begin{equation}
\text{RSS} := \sum_{j \in [k]}\sum_{i\in C_j} d(c_j, v_i)^2
\end{equation}
The RSS decreases at each iteration of the $k$-means algorithm, the algorithm therefore converges to a local minimum for the RSS. The number of iterations $T$ for convergence depends on the data and the number of clusters.  A single iteration has complexity of $O(k\n d)$ since the $\n$ vectors of dimension $d$ have to be compared to each of the $k$ centroids. 

%From a computational complexity point of view, we recall that it is NP-hard to find a clustering that achieves the global  minimum for the RSS. There are classical clustering algorithms based on optimizing different loss functions, however the k-means algorithm uses the RSS as the objective function. 
The algorithm can be super-polynomial in the worst case (the number of iterations is $2^{\omega(\sqrt{\n})}$  \cite{arthur2006slow}), 
but the number of iterations is usually small in practice. The $k$-means algorithm with a suitable heuristic like $k$-means++ to initialize the centroids finds a clustering such that the value for the RSS objective function is within a multiplicative $O(\log \n)$ factor of the minimum value \cite{arthur2007k}.\\

In Section \ref{sec:delta_k_means}, we will introduce a slightly different version of the algorithm, named $\delta$-$k$-means, which includes some noise and randomness to be fairly comparable to the quantum algorithm $q$-means presented in Section \ref{sec:qmeans}.

\subsection{Spectral Clustering}\label{sec:classical_spectral_clustering}

A summary of all variables along with their definition is given in Chapter \ref{chap:Q_spectral_clustering}, Table \ref{table:spectral_clustering_variable_summary}.

\subsubsection{Notations and Definitions}\label{notationsanddefinition}

Let $S \in \R^{N\times d}$ be the input of our clustering task. $S$ is the data matrix composed of $N$ vectors $s_i \in \R^{d}$, for $i \in[N]$. The spectral clustering method uses a graph derived from the data $S$, where similar points are connected. We define the distance between two points by $d_{ij}=\norm{s_i-s_j}$.\\

We consider the undirected graph for which each of the $N$ nodes corresponds to a data point.
The value of the edge connecting two nodes $i$ and $j$ is 1 if the two nodes are connected and 0 otherwise. More generally we will denote by $a_{ij} \in \{0,1\}$ the value of this edge. By convention we have $a_{ii}=0$. 
We define the Adjacency matrix $A \in\R^{N\times N}$ as the symmetric matrix with elements $a_{ij}$. \\

\begin{figure}[h!]
\centering
   \includegraphics[width=\textwidth]{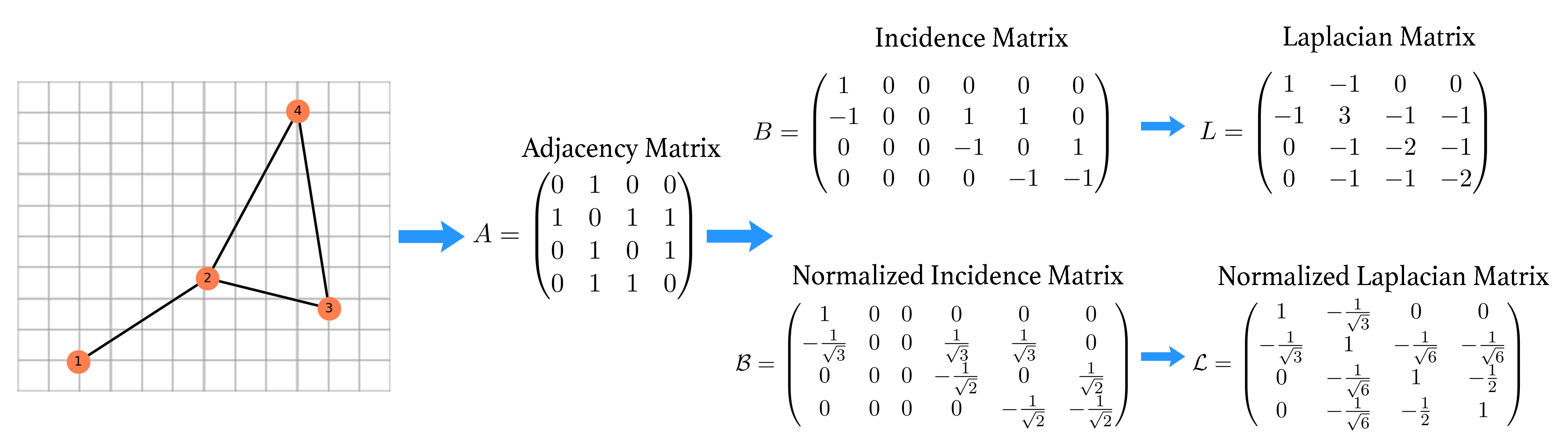}
\caption{Example of the Adjacency, Incidence, and Laplacian matrices of a N=4 nodes graph.}
\label{fig:spectral_clust_graph_example}
\end{figure} 

We will use the following construction rule for the graph: the value of an edge between two points $s_i$ and $s_j$ is equal to 1 if their distance satisfies $d_{ij} \leq d_{min}$ and 0 otherwise, for a given threshold $d_{min} >0$. This choice has been made for simplicity and to take into account constraints from quantum circuits that will be detailed later. \\

The Incidence matrix $B$ is another way of representing the graph. Each row of $B$ represents a node whereas a column represents a possible edge. An element of $B$ indicates if an edge is incident to a node. $B$ is not symmetric and has size $N \times \frac{N(N-1)}{2}$. 
%We introduce the following notation that will be useful when considering quantum circuit implementation: an edge can be indexed by the indices of its two nodes $p$ and $q$ ordered so that $p<q$. 
We index the elements of $B$ by three numbers $B_{i,(p,q)}$ where $i$ is the node and $(p,q)$ represents the edge connecting the nodes $p$ and $q$ ordered so that $p<q$. Even though the graph is undirected, the values of $B$ must follow an oriented convention. Therefore the rule for constructing $B$ is the following:
\begin{equation}\label{Bdefinition}
    B_{i,(p,q)} = \begin{cases}
    a_{pq} \quad \text{if } i=p \\
    -a_{pq} \text{ if } i=q \\
    0 \qquad \text{ if } i\notin \{p,q\}  
    \end{cases}
\end{equation}

We introduce the \emph{normalized} incidence matrix $\mathcal{B}$, with elements defined by $\mathcal{B}_{i,(p,q)}=\frac{B_{i,(p,q)}}{\norm{B_i}}$, where $B_i$ is the $i^{th}$ row of $B$. Therefore each row $\mathcal{B}_i$ has unit norm.

The Laplacian matrix is defined by $L=BB^T$. We introduce the normalized Laplacian matrix as $\mathcal{L} = \mathcal{B}\mathcal{B}^T$. It inherits the properties of the Laplacian matrix and will be used for classification. 
Note that the usual definition $\mathcal{L} = D^{-\frac{1}{2}}LD^{-\frac{1}{2}}$, with $D$ the Degree matrix, coincides if the edges are either 0 or 1. 

$\mathcal{L}$ is a symmetric and positive semidefinite matrix in $\R^{N\times N}$. The $n$ eigenvalues of $\mathcal{L}$ are real and positive. We denote them $\{\lambda_1, \cdots, \lambda_N\}$, and their corresponding eigenvectors are $\{u_1, \cdots, u_N\}$. The eigenvalues are ordered such that $\lambda_1 \leq \lambda_2 \leq \cdots \leq \lambda_N $. 
For a given integer $k \in [N]$, we will denote by $\widetilde{\mathcal{L}}^{(k)}$ %$L_{\leq k}$ 
the projection of $\mathcal{L}$ on its $k$ lowest eigenvalues.
 
Since $\mathcal{L}=\mathcal{B}\mathcal{B}^T$, the $N$ singular values $\lambda^\mathcal{B}_j$ of $\mathcal{B}$ are such that $\lambda_j = (\lambda^\mathcal{B}_j)^2$. Indeed, using the singular value decomposition (SVD), there exist two orthonormal matrices $U$ and $V$ and the diagonal matrix $\Sigma \in \R^{N \times N}$ with elements $(\lambda^\mathcal{B}_1,\cdots,\lambda^\mathcal{B}_N)$, such that $\mathcal{B}= U\Sigma V^T$. Therefore $\mathcal{L}=U\Sigma^2 U^T$, and the eigenvectors $u_j$ of $\mathcal{L}$ are the left singular vectors of $\mathcal{B}$.\\

In Chapter \ref{chap:Q_spectral_clustering}, to ensure a better running time for the quantum algorithm, we will slightly modify the incidence matrix $B$ by replacing the ``0'' elements with a small parameter $\epsilon_B > 0$. We will see in the experiments that it does not affect the accuracy of the clustering.

\subsubsection{Partitioning the Graph into $k$ Clusters}\label{classicalkpartition}
Once the graph's normalized incidence matrix $\mathcal{B}$ is computed, we can calculate its normalized Laplacian and find its eigenvalues and eigenvectors.

Let $\widetilde{\mathcal{L}}^{(k)} \in \R^{N\times k}$ be the \emph{projected} normalized Laplacian matrix on its $k$ lowest eigenvectors, i.e. The $j^{th}$ column of $\widetilde{\mathcal{L}}^{(k)}$ is $u_j$, the $j^{th}$ eigenvector of $\mathcal{L}$, for $j=1,\cdots, k$. 
The method developed by \cite{ng2002spectral} consists in applying the clustering algorithm $k$-means (see previous Section \ref{sec:classical_kmeans_preliminaries}) with input the $N$ rows $\widetilde{\mathcal{L}}^{(k)}_i$ of the projected normalized Laplacian $\widetilde{\mathcal{L}}^{(k)}$. 
Each row $\widetilde{\mathcal{L}}^{(k)}_i$ is a vector of dimension $k$, corresponding to an input vector $s_i$ in the initial input space.

\begin{figure}[h!]
\centering
   \includegraphics[width=\textwidth]{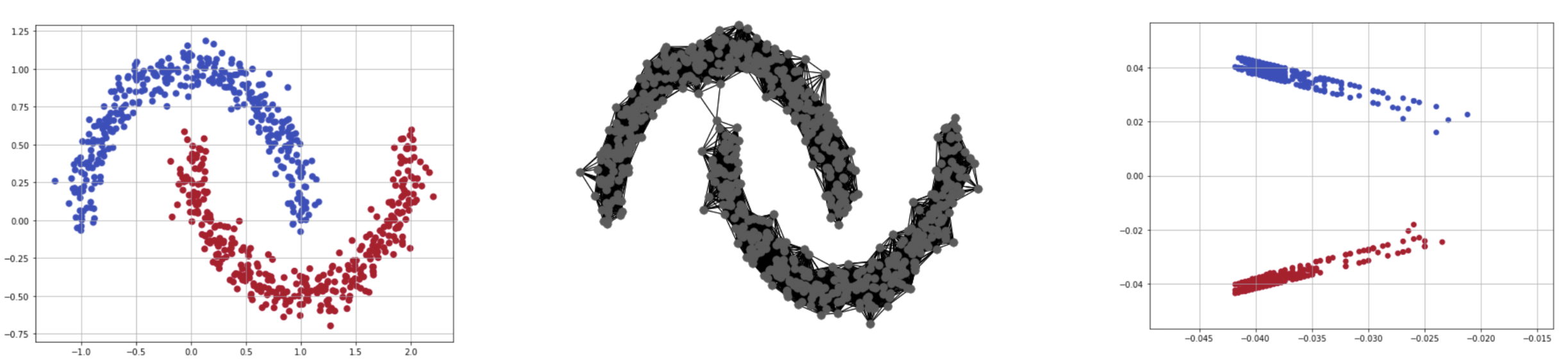}
\caption{Example of the spectral clustering algorithm on two nested half moons. (left) The two classes of $N$ points in two dimensions. (center) The similarity graph. (right) The points projected into the space spanned by the first two eigenvectors of the Laplacian matrix of the graph. In the spectral space (right) we see that the point are easy to separate.}
\label{fig:spectral_clust_half_moon}
\end{figure}

With this procedure, the $k$-means clustering takes place in a low-dimensional and appropriate space, ensuring an efficient clustering (see Fig.\ref{fig:spectral_clust_half_moon}). 
The output of the algorithm could be the label of each point (for example the label corresponding to the nearest centroid) or the $k$ centroids in the spectral space. \\

\subsubsection{Classical Running Time}

The classical algorithm can be decomposed into several steps: the distance calculation between points runs in time $O(dN^2)$, and the creation of the Laplacian matrix in $O(Nm)$ where $m$ is the number of edges in the graph, which is $O(N^2)$ in the worst case. Then, the extraction of eigenvalue and eigenvectors of the Laplacian matrix is done in $O(N^3)$. Finally, the $k$-means clustering runs in $O(Nk^2)$. The dominant term is in practice therefore $O(N^3)$ (we assume we have more points than dimensions), and the impractical running time of spectral clustering is due to the need for diagonalization of the Laplacian matrix \cite{li2011time}.

\subsubsection{Alternative Classical Algorithms}

Randomness naturally occurs during our quantum spectral clustering algorithm (see Chapter \ref{chap:Q_spectral_clustering}), and one may wonder for fair comparison if an alternative classical algorithm can efficiently spectral clustering with noisy or sampled methods as well. To circumvent the prohibitive running time of the classical algorithm, several approximations have indeed been proposed on different steps. A recent review of these techniques \cite{tremblay2020approximating} concludes that despite many efforts, methods with provable scalability are found limited or worse in practice, whereas other good empirical methods have no provable guarantees. 

Some methods aim to build the similarity graph using sampling \cite{choromanska2013fast,rahimi2008random,li2011time}. They present limitations \cite{wang2019scalable} and act by sampling partially the input data, which is not the case of our quantum algorithm. Their running time is often proportional to $O(Nm)$ or $O(Nm^2)$ where $m$ is the number of edges, which is $O(N^2)$ in the worst case. If one of these methods was empirically efficient, we could actually adapt it to our quantum algorithm, by first applying a similar sparsification and then using it as input in our quantum algorithm. Recently such techniques have been done in the quantum setting \cite{apers2019quantum}.

At the next step, it is possible to use Lanczos methods to compute the $k$ lowest eigenvalues and eigenvectors, with a running time of $O(N^2)$ for a fully connected graph. However, they seem to suffer from poor efficiency in practice since they strongly rely on the distribution of the eigenvalues and can require many iterations that would ruin the advantage \cite{bai2000templates}. Note that even with an effective application of this method, our quantum algorithm would still be advantageous. We can also cite the use of power methods \cite{boutsidis2015spectral} to solve clustering using approximated eigenvectors.

Some methods try to improve the clustering step itself, by modifying the $k$-means algorithm \cite{hamerly2015accelerating}. One should compare these methods directly with the quantum $k$-means \cite{qmeans} (Chapter \ref{chap:qmeans}). In most cases, such variations carry over to the quantum case as well. Finally, other attempts use solely preprocessing techniques \cite{yan2009fast} on the initial dataset. Again, one could simply use them before the quantum algorithm to similarly improve its practical efficiency.

\section{Neural Networks}\label{sec:preliminaries_classical_nn} % ex- \label{preliminaries_classical_nn}

Artificial neural networks may be the most impressive advance in contemporary computing. They were imagined to roughly mimic neuron connectivity in the brain. While being far from complete, this imitation already allowed for impressive advances in machine learning. Today, deep learning has become a state-of-the-art standard in most cases, be it speech recognition, image or video processing, disease detection, etc.  

The first attempts to create artificial neural networks go back to the mid XX$^{th}$ century, but the key paradigm for training them efficiently was developed in the 1980s with the \emph{backpropagation} algorithm \cite{rumelhart1986learning, lecun1998gradient}. Neural networks had to wait until the late 2000s to achieve their worldwide success, thanks to the impressive development of GPUs (Graphics Processing Units) allowing fast implementation of linear algebra routines.

To continue to improve, today's architectures are becoming increasingly complex, deep, and resource-intensive. Training deep networks requires large clusters of GPUs and an excessive amount of time and energy.\\ 

For this thesis, we will only review few basic types of neural networks for which quantum algorithms are proposed in Part \ref{part:Q_NeuralNet}. Emphasis is made on mathematical formalism which will be helpful for the quantum versions of these algorithms. Fully connected neural networks (FCNN) are the original and most basic ones, followed by the backpropagation algorithm, necessary to train all neural networks. We then outline the recent proposal of orthogonal neural networks (OrthoNN). OrthoNNs show special abilities for learning, but that is most interest is their orthogonality constraint that arises naturally in quantum computing. We then present the widely used convolutional neural networks (CNN) which are specialized in signal or image processing. Finally, we detail the backpropagation algorithm in the context of CNN.

\subsection{Fully Connected Neural Networks}\label{sec:classical_fcnn_intro}

\begin{figure}[h!]
\centering
   \includegraphics[width=0.55\textwidth]{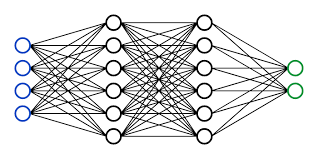}
\caption{A fully connected neural network for 4-dimensional inputs, 2 classes outputs, and two hidden layers. Each line represents a tunable weight.}
\label{fig:NNsmall}
\end{figure}

A neural network usually consists of layers of neurons or \emph{nodes}, each being a numerical value. Adjacent layers are connected through \emph{weights}. A network is said to be fully connected if all nodes of a layer are connected to all nodes of the next layer, as in Fig.\ref{fig:NNsmall}. The input layer has as many nodes as the input vector has dimensions. The output layer size is also the dimension of the output. A key feature of neural networks, that gives them the expressive power and universal abilities, is the presence of non-linear \emph{activation functions} at each layer.

Layers are numbered $\ell =1,\cdots,L$, from input to output. Between two layers $\ell$ and $\ell+1$, respectively of size $n_{\ell}$ and $n_{\ell+1}$, the weights can be embedded in a matrix $W^{\ell}\in \R^{n_{\ell+1} \times n_{\ell}}$. For instance, the first column of $W^{\ell}$ will correspond to the weights connecting each node of layer $\ell$ to the first node of layer $\ell+1$, that is $(w_{00}, w_{10},\cdots,w_{n^{\ell+1}0})$.

We denote by $a^{\ell} \in \R^{n_\ell}$ the vector of layer $\ell$. The 
\emph{feedforward} procedure consists of creating the next layer $a^{\ell+1}$ by first doing a matrix product with the weight matrix, and then applying a non-linear function $\sigma$. Usually, an extra parameter $b^\ell$ called \emph{bias} is added to the layer to ensure flexibility to the model. Note that this bias can be discarded in the formalism, as it is equivalent to adding an extra dimension to the value 1 at each layer.    

\begin{equation}\label{eq:nn_basic_equation_1}
    z^{\ell+1} = W^\ell a^\ell + b^\ell 
\end{equation}
    
\begin{equation}\label{eq:nn_basic_equation_2}
    a^{\ell+1} = \sigma(z^{\ell+1})
\end{equation}

This procedure is continued until we obtain the last layer $a^{L}$. As we will see in the rest of this thesis, it is important to notice that Eq.(\ref{eq:nn_basic_equation_1}) can be decomposed as several inner products between the input vector $a^\ell$ and the rows of $W^\ell$. 

The non-linearity $\sigma$ is usually taken to be the \emph{sigmoid} function, which has the property of pushing positive and negative values respectively towards +1 and 0.

\begin{equation}\label{eq:sigmoid}
    \sigma: x \mapsto \frac{1}{1 + e^{-x}}
\end{equation}

\begin{figure}[h!]
\centering
   \includegraphics[width=0.5\textwidth]{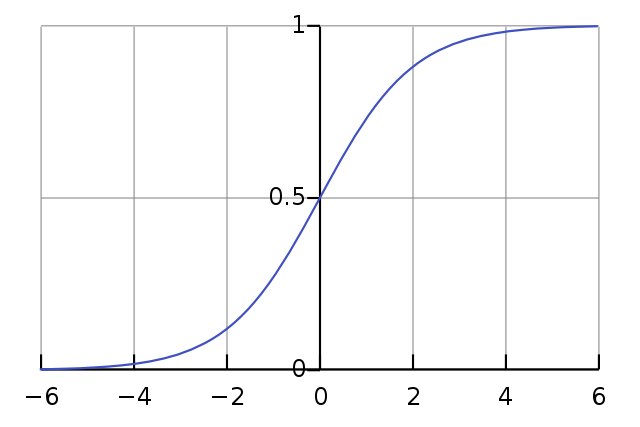}
\caption{The sigmoid function. Source: Wikipedia.}
\label{fig:sigmoid}
\end{figure}

It follows that the running time of a single fully connected is dominated by the matrix-vector multiplication at its core, which takes $O(n^{\ell}n^{\ell+1})$, or $O(n^2)$ in the case of \emph{square} layers.

\subsection{Backpropagation}\label{sec:bacpropagation_fcnn}

The backpropagation algorithm in a fully connected neural network is a well know and efficient procedure to update the weight matrix at each layer \cite{hecht1992theory, rojas1996backpropagation}. 

After the last layer $a^{L}$, one can define the cost function $\mathcal{C}$ that compares the output to the ground truth during the supervised training. The goal is to calculate the gradient of $\mathcal{C}$ with respect to each weight and bias, namely 
$\frac{\partial \mathcal{C}}{\partial W^{\ell}}$ and $\frac{\partial \mathcal{C}}{\partial b^{\ell}}$. In the backpropagation, we start by calculating these gradients for the last layer, then propagate back to the first layer. 

We will require to obtain the \emph{error} vector at layer $\ell$ defined by $\Delta^{\ell} = \frac{\partial \mathcal{C}}{\partial z^{\ell}}$. One can show the backward recursive relation 
\begin{equation}\label{eq:bakcprop_error_formula}
\Delta^{\ell} = (W^{\ell+1})^T\cdot \Delta^{\ell+1}\odot \sigma'(z^{\ell}),
\end{equation}

where $\odot$ symbolizes the Hadamard product, or entry-wise multiplication. If using the sigmoid function, we also have the property $\sigma'(x) = \sigma(x)(1-\sigma(x))$. Note that the previous computation requires simply to apply the layer (i.e. apply matrix multiplication) in reverse. We can then show that each element of the weight gradient matrix at layer $\ell$ is given by $\frac{\partial \mathcal{C}}{\partial W^{\ell}_{jk}} = \Delta^{\ell}_j\cdot a^{\ell-1}_k$. Similarly, the gradient with respect to the biases is easily defined as $\frac{\partial \mathcal{C}}{\partial b^{\ell}_{j}} = \Delta^\ell_j$. 

Once these gradients are computed, we update the parameters using the gradient descent rule, with learning rate $\lambda$ :
\begin{equation}\label{gradient_descent_intro}
\centering
W^{\ell}_{jk} \gets W^{\ell}_{jk} -\lambda \frac{\partial \mathcal{C}}{\partial W^{\ell}_{jk}} \quad;\quad
b^{\ell}_{j} \gets b^{\ell}_{j} -\lambda \frac{\partial \mathcal{C}}{\partial b^{\ell}_{j}}
\end{equation}

The task is repeated until the cost function stops decreasing, indicating a local minimum has been reached. We also refer as \emph{stochastic} gradient descent (SGD) when the loss and its gradients are estimated with one or few samples only, and not on the entire dataset. SGD allows for faster iterations in big datasets, and the imperfect gradient estimations along with good learning rate can help escape from local minima.

\begin{figure}[h!]
\centering
   \includegraphics[width=0.7\textwidth]{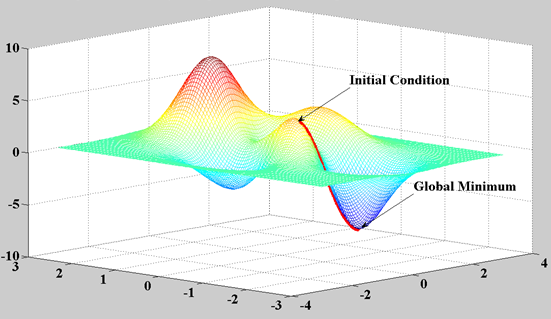}
\caption{Representation of gradient descent on two-dimensional parameter space. The z-axis is the cost function. Source: Matlab.}
\label{fig:gradient_descent_3D}
\end{figure}

The complexity of a single gradient descent update of one layer only is dominated by the time to compute all gradients. Since these are made using a similar matrix-vector multiplication as in the forward pass (see Eq.(\ref{eq:bakcprop_error_formula})), the complexity is also $O(n^{\ell}n^{\ell+1})$ for a layer with input size $n^{\ell}$ and output size $n^{\ell+1}$. This becomes $O(n^2)$ for a square layer.

\subsection{Orthogonal Neural Networks}\label{sec:classical_ortho_nn_preliminaries}

The idea behind Orthogonal Neural Networks (OrthoNNs) is to add a constraint to the weight matrices corresponding to the layers of a neural network. Imposing orthogonality to these matrices has theoretical and practical benefits in the generalization error \cite{jia2019orthogonal}. Orthogonality ensures a low weight redundancy and preserves the magnitude of the weight matrix's eigenvalues to avoid vanishing gradients. In terms of complexity, for a single layer, the feedforward pass of an OrthoNN is simply a matrix multiplication, hence has a running time of $O(n^2)$ if $n\times n$ is the size of the orthogonal matrix (input and output layers of size $n$). It is also interesting to note that OrthoNNs have been generalized to convolutional neural networks \cite{wang2020orthogonal}. 

The main drawback of OrthoNNs is to preserve the orthogonality of the matrices while updating them during gradient descent. Several algorithms have been proposed to this end \cite{wang2020orthogonal, bansal2018can, lezcano2019cheap}, but they all point that pure orthogonality is computationally hard to conserve. Therefore, previous works allow for approximations: strict orthogonality is no longer required, and the matrices are often pushed toward orthogonality using regularization techniques during weights update. \\

We present two algorithms from \cite{jia2019orthogonal} for updating orthogonal matrices. 

The first algorithm is an approximated one, called \emph{Singular Value Bounding} (SVB). It starts by applying the usual gradient descent update on the matrix, therefore making it not orthogonal anymore. Then, the singular values of the new matrix are extracted using Singular Value Decomposition (SVD), their values are manually pushed to be close to 1, and the matrix is recomposed hence enforcing orthogonality. This method shows less advantage on practical experiments \cite{jia2019orthogonal}. It has a complexity of $O(n^3)$ due to the SVD, which in practice is better than the next algorithm. Note that this running time is still longer than $O(n^2)$, the running time to perform standard gradient descent.

The second algorithm can be considered perfect since it ensures strict orthogonality by performing the gradient descent in the manifold of orthogonal matrices, called the Stiefel Manifold. In practice \cite{jia2019orthogonal}, this method showed advantageous classification results on standard datasets. This algorithm requires $O(n^3)$ operations, but is very prohibitive in practice. We give a very informal step-by-step detail of this algorithm:

\begin{enumerate}
    \item Compute the gradient $G$ of the weight matrix $W$.
    \item Project the gradient matrix $G$ in the tangent space,  (The space tangent to the manifold at this point $W$): multiply $G$ by some other matrices based on $W$: 
    \begin{equation}
    (I-WW^T)G+\frac{1}{2}W(W^TG-G^TW)
    \end{equation} 
    This requires several matrix-matrix multiplications. In the case of square $n\times n$ matrices, each has complexity $O(n^3)$. the result of this projection is called the \emph{manifold gradient} $\Omega$.
    \item update $W' = W - \eta\Omega$, where $\eta$ is the chosen learning rate. 
    \item Perform a \emph{retraction} from the tangent space to the manifold. To do so we multiply $W'$ by $Q$ factor of the \emph{QR decomposition}, obtained using Gram Schmidt orthonormalization, which has complexity $O(2n^3)$. 
\end{enumerate}

\subsection{Convolutional Neural Networks}\label{sec:classical_cnn_preliminaries}

Convolutional neural networks (CNN) are a specific type of neural networks, designed in particular for image processing or time series. They use the \emph{convolution product} as the main procedure for each layer. They were originally developed by Yann LeCun and others \cite{lecun1998gradient} in the 1980s. They are now the most widely used algorithms for image recognition tasks \cite{krizhevsky2012imagenet}. Their capacities have been used in various domains such as autonomous vision \cite{bojarski2016visualbackprop} or gravitational wave detection \cite{george2018deep}. Despite these successes, CNNs suffer from a computational bottleneck that makes deep CNNs resource expensive in practice.

In the following, we will focus on image processing with a tensor framework for all elements of the network. Our goal is to explicitly describe the CNN procedures in a form that can be translated in the context of quantum algorithms. 
As a regular neural network, a CNN should learn how to classify any input, in our case images. The training consists of optimizing parameters learned on the inputs and their corresponding labels.

\subsubsection{Tensor representation}
Images, or more generally layers of the network, can be seen as tensors. A tensor is a generalization of a matrix to higher dimensions. For instance, an image of height $H$ and width $W$ can be seen as a matrix in $\R^{H\times W}$, where every pixel is a greyscale value between 0 and 255 (8 bit). However, the three channels of color (RGB: Red Green Blue) must be taken into account, by stacking three times the matrix for each color. The whole image is then seen as a 3 dimensional tensor in $\R^{H\times W \times D}$ where $D$ is the number of channels. We will see that the Convolution Product in the CNN can be expressed between 3-tensors (input) and 4-tensors (convolution \emph{filters} or \emph{kernels}), the output being a 3-tensor of different dimensions (spatial size and number of channels).

\begin{figure}[h]
\centering
\includegraphics[width=0.95\textwidth]{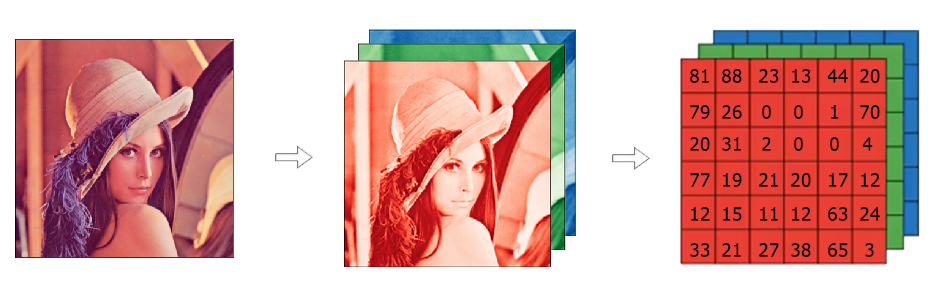} 
\captionsetup{justification=raggedright, margin=1cm}
\caption{RGB decomposition, a colored image is a 3-tensor.}\label{fig:CNN_RGB}%used to be \label{RGB}
\end{figure} 

\subsubsection{Architecture}

A CNN is composed of 4 main procedures, compiled and repeated in any order: Convolution layers, most often followed by an Activation Function, Pooling Layers, and some Fully Connected layers at the end. We will denote by $\ell$ the current layer.

\textbf{Convolution Layer :} The $\ell^{th}$ layer is convolved by a set of filters called \emph{kernels}. The output of this operation is the $(\ell+1)^{th}$ layer. A convolution by a single kernel can be seen as a feature detector, that will screen over all regions of the input. If the feature represented by the kernel, for instance a vertical edge, is present in some part of the input, there will be a high value at the corresponding position of the output. The output is called the \emph{feature map} of this convolution.

\textbf{Activation Function :} As in regular neural networks, we insert some non-linearities also called \emph{activation functions}. These are mandatory for a neural network to be able to learn any function. In the case of a CNN, each convolution is often followed by a Rectified Linear Unit function, or \emph{ReLu}. This is a simple function that puts all negative values of the output to zero, and lets the positive values as they are. 

\textbf{Pooling Layer :} This downsampling technique reduces the dimensionality of the layer, in order to improve the computation. Moreover, it gives the CNN the ability to learn a representation invariant to small translations. Most of the time, we apply a Maximum Pooling or an Average Pooling. The first one consists of replacing a subregion of $P\times P$ elements only by the one with the maximum value. The second does the same by averaging all values. Recall that the value of a pixel corresponds to how much a particular feature was present in the previous convolution layer.  

\textbf{Fully Connected Layer :} After a certain number of convolution layers, the input has been sufficiently processed so that we can apply a fully connected network. Weights connect each input to each output, where inputs are all elements of the previous layer. The last layer should have one node per possible label. Each node value can be interpreted as the probability of the initial image to belonging to the corresponding class.

\begin{figure}[h]
\centering
\includegraphics[width=0.95\textwidth]{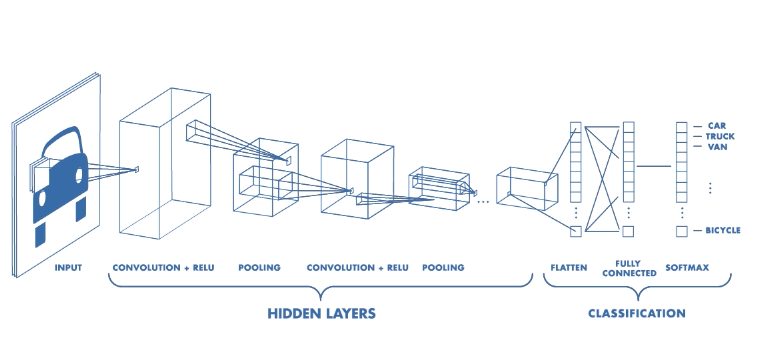} 
\captionsetup{justification=raggedright, margin=1cm}
\caption[Caption for LOF]{Representation of a CNN's layers and operations. Source: Mathworks}
\label{fig:CNN_image}% ex \label{CNN-image}
\end{figure}

\subsubsection{Convolution Product as a Tensor Operation}\label{tensors} %attention il y bcp de \ref{tensors} dans QCNN
\begin{figure}[h]
\centering
\includegraphics[width=0.8\textwidth]{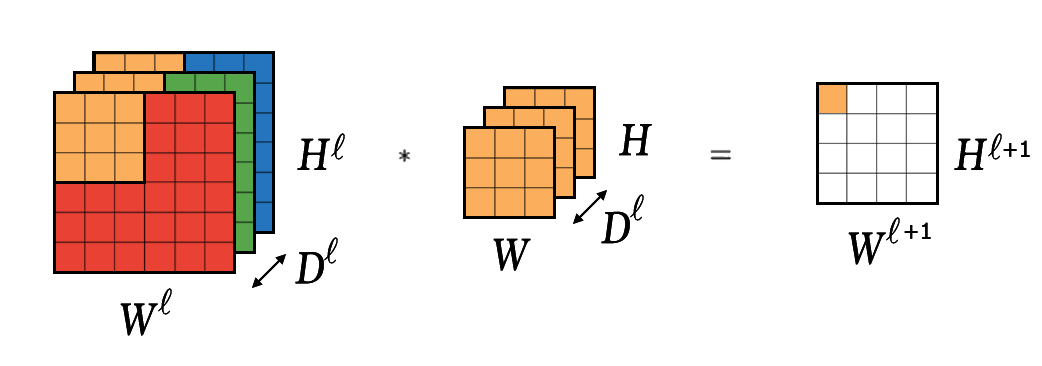} 
\captionsetup{justification=raggedright, margin=1cm}
\caption{Convolution of a 3-tensor input (Left) by one 3-tensor kernel (Center). The ouput (Right) is a matrix for which each entry is a inner product between the kernel and the corresponding overlapping region of the input.}\label{volume-convolution}
\end{figure}

Most of the following mathematical formulations have been very well detailed in \cite{CNNIntro}. At layer $\ell$, we consider the convolution of a multiple channels image, seen as a 3-tensor $X^{\ell} \in \R^{H^{\ell}\times W^{\ell} \times D^{\ell}}$. Let's consider a single kernel in $\R^{H\times W \times D^{\ell}}$. Note that its third dimension must match the number of channels of the input, as in Fig.\ref{volume-convolution}. The kernel passes over all possible regions of the input and outputs a value for each region, stored in the corresponding element of the output. Therefore the output is 2 dimensional, in $\R^{H^{\ell+1}\times W^{\ell+1}}$.

In a CNN, the most general case is to apply several convolution products to the input, each one with a different 3-tensor kernel. Let's consider an input convolved by $D^{\ell+1}$ kernels. We can globally see this process as a whole, represented by one 4-tensor kernel $K^{\ell} \in \R^{H\times W \times D^{\ell} \times D^{\ell+1}}$. As $D^{\ell+1}$ convolutions are applied, there are $D^{\ell+1}$ outputs of 2 dimensions, equivalent to a 3-tensor $ X^{\ell+1} \in \R^{H^{\ell+1}\times W^{\ell+1} \times D^{\ell+1}}$

\begin{figure}[h]
\centering
\includegraphics[width=0.9\textwidth]{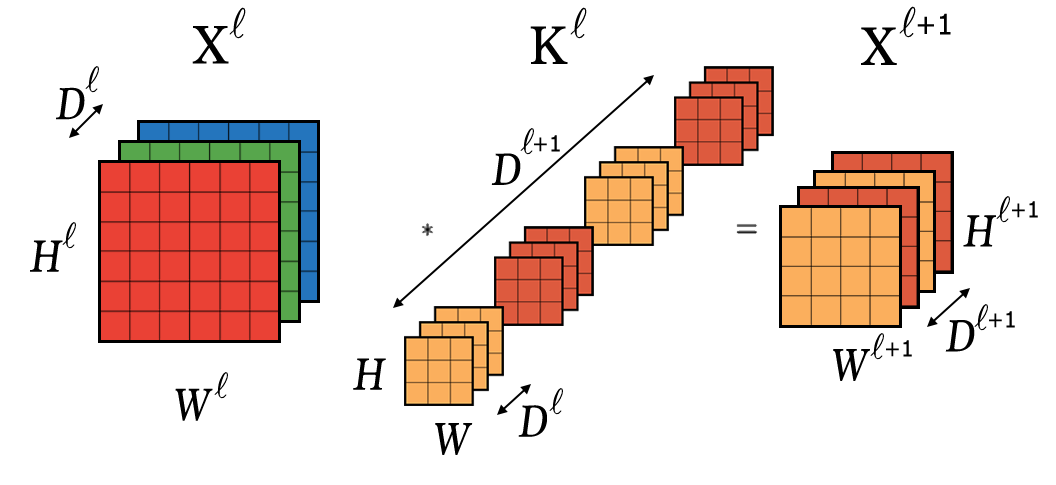} 
\captionsetup{justification=raggedright, margin=1cm}
\caption{Convolutions of the 3-tensor input $X^{\ell}$ (Left) by one 4-tensor kernel $K^{\ell}$ (Center). Each channel of the output $X^{\ell+1}$ (Right) corresponds to the output matrix of the convolution with one of the 3-tensor kernel.}
\label{fig:CNN_volume_convolution_tensor}%ex \label{volume-convolution-tensor}
\end{figure} 

This tensor convention explains why Fig.\ref{fig:CNN_image} is represented with layers as volumes of different shapes. Indeed we can see in Fig.\ref{fig:CNN_volume_convolution_tensor} that the output's dimensions are modified given the following rule:

\begin{equation}
  \begin{cases}
    H^{\ell+1} = H^{\ell}-H+1\\
    W^{\ell+1} = W^{\ell}-W+1\\
  \end{cases}
\end{equation}

We omit to detail the use of \emph{Padding} and \emph{Stride}, two parameters that control how the kernel moves through the input, but these can easily be incorporated in the algorithms.

An element of $X^{\ell}$ is determined by 3 indices $(i^{\ell},j^{\ell},d^{\ell})$, while an element of the kernel $K^{\ell}$ is determined by 4 indices $(i,j,d,d')$. For an element of $X^{\ell+1}$ we use 3 indices $(i^{\ell+1},j^{\ell+1},d^{\ell+1})$. We can express the value of each element of the output $X^{\ell+1}$ with the relation

\begin{equation}\label{analytic-expression}
X^{\ell+1}_{i^{\ell+1},j^{\ell+1},d^{\ell+1}} = \sum_{i=0}^{H}\sum_{j=0}^{W}\sum_{d=0}^{D^{\ell}}K^{\ell}_{i,j,d,d^{\ell+1}}X^{\ell}_{i^{\ell+1}+i, j^{\ell+1}+j, d}
\end{equation}

\subsubsection{Matrix Expression}
\begin{figure}[h]
\centering
\includegraphics[width=\textwidth]{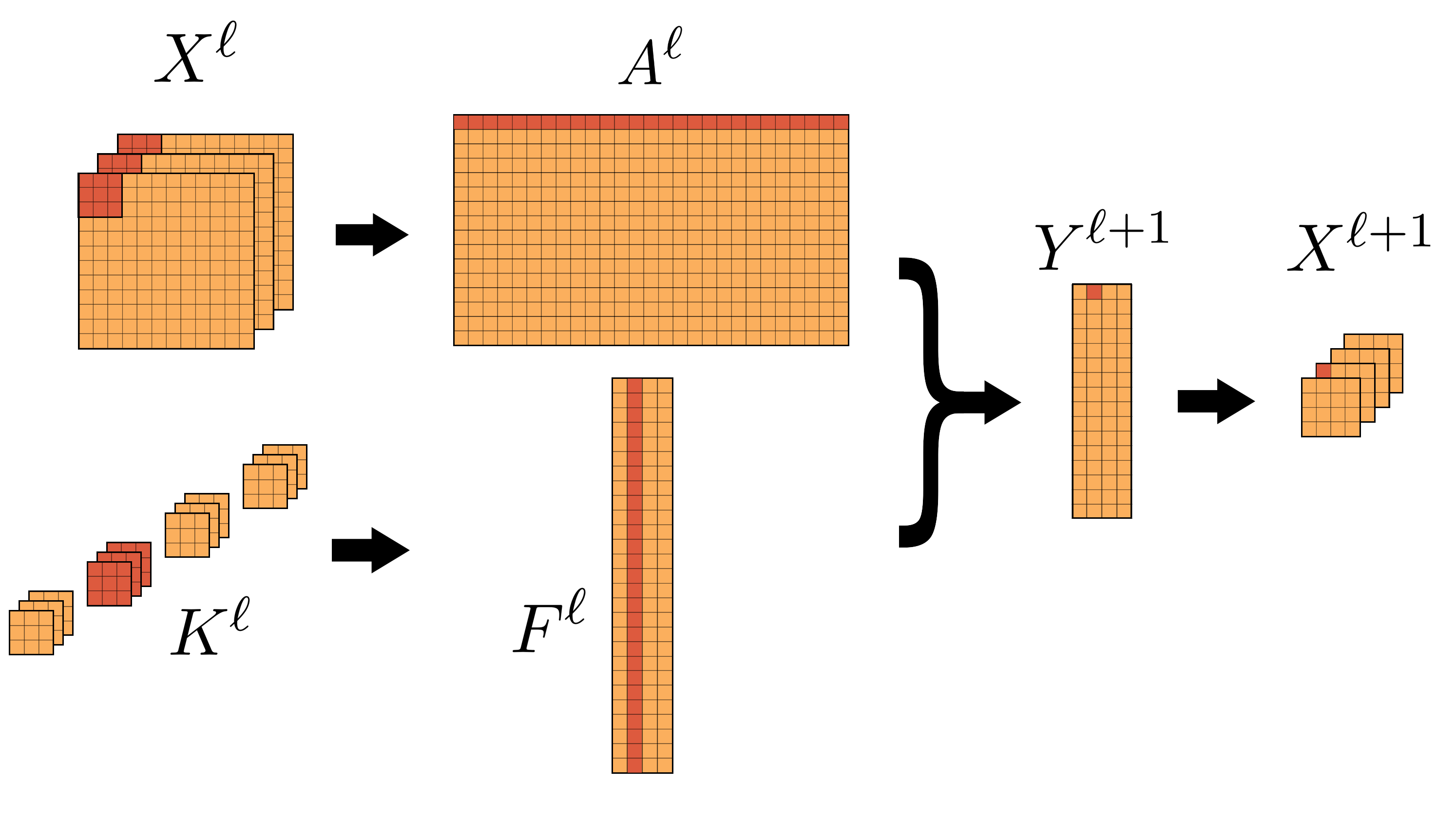} 
\captionsetup{justification=raggedright, margin=1cm}
\caption{A convolution product is equivalent to a matrix-matrix multiplication.}
\label{fig:CNN_tensors_figure}% ex \label{CNN-tensors-figure}
\end{figure}

It is possible to reformulate Eq.(\ref{analytic-expression}) as a matrix product. For this we have to reshape our objects. We expand the input $X^{\ell}$ into a matrix $A^{\ell} \in \R^{(H^{\ell+1}W^{\ell+1})\times(HWD^{\ell})}$. Each row of $A^{\ell}$ is a vectorized version of a subregion of $X^{\ell}$. This subregion is a volume of the same size as a single kernel volume $H\times W\times D^{\ell}$. Hence each of the $H^{\ell+1}\times W^{\ell+1}$ rows of $A^{\ell}$ is used for creating one value in $X^{\ell+1}$. Given such a subregion of $X^{\ell}$,  the rule for creating the row of $A^{\ell}$ is to stack, channel by channel, a column first vectorized form of each matrix. Then, we reshape the kernel tensor $K^{\ell}$ into a matrix $F^{\ell} \in \R^{(HWD^{\ell})\times D^{\ell+1}}$, such that each column of $F^{\ell}$ is a column first vectorized version of one of the $D^{\ell+1}$ kernels.

As proved in \cite{CNNIntro}, the convolution operation $X^{\ell} * K^{\ell} = X^{\ell+1}$ is equivalent to the following matrix multiplication 
\begin{equation}
A^{\ell}F^{\ell} = Y^{\ell+1},
\end{equation}
where each column of $Y^{\ell+1} \in \R^{(H^{\ell+1}W^{\ell+1})\times D^{\ell+1}}$ is a column first vectorized form of one of the $D^{\ell+1}$ channels of $X^{\ell+1}$. Note that an element $Y^{\ell+1}_{p,q}$ is the inner product between the $p^{th}$ row of $A^{\ell}$ and the $q^{th}$ column of $F^{\ell}$. It is then simple to convert $Y^{\ell+1}$ into $X^{\ell+1}$ 
The indices relation between the elements $Y^{\ell+1}_{p,q}$ and $X^{\ell+1}_{i^{\ell+1},j^{\ell+1},d^{\ell+1}}$ is given by:
\begin{equation}\label{YtoX}
  \begin{cases}
    d^{\ell+1} = q\\
    j^{\ell+1} = \floor{\frac{p}{ H^{\ell+1}}}\\
    i^{\ell+1} = p - H^{\ell+1}\floor{\frac{p}{ H^{\ell+1}}}\
  \end{cases}
\end{equation}

A summary of all variables along with their meaning and dimensions is given in Chapter \ref{chap:QCNN}, Table \ref{variable_summary_1}.\\

Finally, we can give a running time for one single convolutional layer. From all the routines, the convolution product is the most costly and dominates the rest. With an input tensor of size $H^{\ell}W^{\ell}D^{\ell}$ and $D^{\ell+1}$ kernels of size $HWD^{\ell}$, we produce an output of size $H^{\ell+1}W^{\ell+1}D^{\ell+1}$. We have seen that each pixel of the output was created by applying an inner product between one kernel and a same-size part of the input. Each output's pixel is therefore created in $O(HWD^{\ell})$, and the whole convolution product takes:

\begin{equation}
   O(H^{\ell+1}W^{\ell+1}D^{\ell+1} \cdot HWD^{\ell}) 
\end{equation}

We can summarize this complexity as:

\begin{equation}
O(\text{output size} \cdot \text{kernel size})
\end{equation}

\subsection{Backpropagation for Convolutional Neural Networks}\label{sec:backpropagaton_cnn}

After each forward pass, the outcome is compared to the true labels and a suitable loss function is computed. We can update our weights by gradient descent to minimize this loss, and iterate. The main idea behind the backpropagation is to compute the derivatives of the loss $\mathcal{L}$, layer by layer, starting from the last one.

At layer $\ell$, the derivatives needed to perform the gradient descent are $\frac{\partial \mathcal{L}}{\partial F^{\ell}}$ and $\frac{\partial \mathcal{L}}{\partial Y^{\ell}}$. The first one represents the gradient of the final loss $\mathcal{L}$ with respect to each kernel element, a matrix of values that we will use to update the kernel weights $F^{\ell}_{s,q}$. The second one is the gradient of $\mathcal{L}$ with respect to the layer itself and is only needed to calculate the gradient $\frac{\partial \mathcal{L}}{\partial F^{\ell-1}}$ at layer $\ell-1$.

\subsubsection{Convolution Product}
We first consider a classical convolution layer without non-linearity or pooling. Thus the output of layer $\ell$ is the same tensor as the input of layer $\ell+1$, namely $X^{\ell+1}$ or equivalently $Y^{\ell+1}$. Assuming we know $\frac{\partial \mathcal{L}}{\partial X^{\ell+1}}$ or equivalently $\frac{\partial \mathcal{L}}{\partial Y^{\ell+1}}$, both corresponding to the derivatives of the $(\ell + 1)^{th}$ layer's input, we will show how to calculate $\frac{\partial \mathcal{L}}{\partial F^{\ell}}$, the matrix of derivatives with respect to the elements of the previous kernel matrix $F^{\ell}$. This is the main goal to optimize the kernel's weights. 

The details of the following calculations can be found in \cite{CNNIntro}. We will use the notation $vec(X)$ to represents the vectorized form of any tensor $X$.

Recall that $A^{\ell}$ is the matrix expansion of the tensor $X^{\ell}$, whereas $Y^{\ell}$ is a matrix reshaping of $X^{\ell}$. By applying the chain rule $\frac{\partial \mathcal{L}}{\partial vec(F^{\ell})^T} = \frac{\partial \mathcal{L}}{\partial vec(X^{\ell+1})^T}\frac{\partial vec(X^{\ell+1})}{\partial vec(F^{\ell})^T}$, we can obtain (See \cite{CNNIntro} for calculations details):
\begin{equation}\label{updateF}
\frac{\partial \mathcal{L}}{\partial F^{\ell}} = (A^{\ell})^{T} \frac{\partial L}{\partial Y^{\ell+1}}
\end{equation}
Eq.(\ref{updateF}) shows that, to obtain the desired gradient, we can just perform a matrix-matrix multiplication between the transposed layer itself ($A^{\ell}$) and the gradient with respect to the previous layer ($\frac{\partial L}{\partial Y^{\ell+1}}$).

Eq.(\ref{updateF}) explains also why we will need to calculate $\frac{\partial \mathcal{L}}{\partial Y^{\ell}}$ in order to backpropagate through layer $\ell-1$. To calculate it, we use the chain rule again for $\frac{\partial \mathcal{L}}{\partial vec(X^{\ell})^T} = \frac{\partial \mathcal{L}}{\partial vec(X^{\ell+1})^T}\frac{\partial vec(X^{\ell+1})}{\partial vec(X^{\ell})^T}$. Recall that a point in $A^{\ell}$, indexed by the pair $(p,r)$, can correspond to several triplets $(i^{\ell},j^{\ell},d^{\ell})$ in $X^{\ell}$. We will use the notation $(p,r) \leftrightarrow ({i^{\ell},j^{\ell},d^{\ell}})$ to express formally this relation. One can show that $\frac{\partial \mathcal{L}}{\partial Y^{\ell+1}}(F^{\ell})^T$ is a matrix of same shape as $A^{\ell}$, and that the chain rule leads to a simple relation to calculate $\frac{\partial \mathcal{L}}{\partial Y^{\ell}}$ (See \cite{CNNIntro} for calculations details):

\begin{equation}\label{updateX}
\left[\frac{\partial \mathcal{L}}{\partial X^{\ell}}\right]_{i^{\ell},j^{\ell},d^{\ell}} = \sum_{(p,r) \leftrightarrow ({i^{\ell},j^{\ell},d^{\ell}})}\left[\frac{\partial \mathcal{L}}{\partial Y^{\ell+1}}(F^{\ell})^T\right]_{p,r}
\end{equation}

We have shown how to obtain the gradients with respect to the kernels $F^{\ell}$ and to the layer itself $Y^{\ell}$ (or equivalently $X^{\ell}$).

\subsubsection{Non Linearity}

The activation function has also an impact on the gradient. In the case of the ReLu, we should only cancel the gradient for points with negative values. For points with a positive value, the derivatives remain the same since the function is the identity. A formal relation can be given by

\begin{equation}\label{relubackpropagation}
\left[\frac{\partial \mathcal{L}}{\partial X^{\ell+1}}\right]_{i^{\ell+1},j^{\ell+1},d^{\ell+1}} = 
\begin{cases}
    \left[\frac{\partial \mathcal{L}}{\partial f(X^{\ell+1})}\right]_{i^{\ell+1},j^{\ell+1},d^{\ell+1}} \text{ if } X^{\ell+1}_{i^{\ell+1},j^{\ell+1},d^{\ell+1}} \geq 0\\
    0 \text{ otherwise}\\
\end{cases}
\end{equation}

\subsubsection{Pooling}

If we take into account the pooling operation, we must change some of the gradients. Indeed, a pixel that hasn't been selected during pooling has no impact on the final loss, thus should have a gradient equal to 0.  We will focus on the case of Max Pooling (Average Pooling relies on similar idea). To state a formal relation, we will use the notations of Section \ref{pooling}: an element in the output of the layer, the tensor $f(X^{\ell+1})$, is located by the triplet $(i^{\ell+1},j^{\ell+1},d^{\ell+1})$. The tensor after pooling is denoted by $\tilde{X}^{\ell+1}$ and its points are located by the triplet $(\tilde{i}^{\ell+1},\tilde{j}^{\ell+1},\tilde{d}^{\ell+1})$. During backpropagation, after the calculation of $\frac{\partial \mathcal{L}}{\partial \tilde{X}^{\ell+1}}$, some of the derivatives of $f(X^{\ell+1})$ should be set to zero with the following rule:

\begin{equation}\label{poolingbackpropagation}
\left[\frac{\partial \mathcal{L}}{\partial f(X^{\ell+1})}\right]_{i^{\ell+1},j^{\ell+1},d^{\ell+1}} = 
\begin{cases}
    \left[\frac{\partial \mathcal{L}}{\partial \tilde{X}^{\ell+1}}\right]_{\tilde{i}^{\ell+1},\tilde{j}^{\ell+1},\tilde{d}^{\ell+1}} \text{ if } (i^{\ell+1},j^{\ell+1},d^{\ell+1}) \in \mathcal{P}\\
    0 \text{ otherwise}\\
\end{cases}
\end{equation}
where $\mathcal{P}$ is the set of indices selected during pooling.

\chapter{Quantum Computing}\label{chap:intro_quantum_computing}
\epigraph{\textit{"Où finit le télescope, le microscope commence. Lequel des deux a la vue la plus grande? Choisissez."}}{Victor Hugo \\ \emph{Les Misérables} (1862)}

\section{Preliminaries in Quantum Computing}\label{sec:preliminaries_quantum_computing}

We introduce a basic and succinct quantum information background necessary for this thesis. For a more detailed introduction we recommend \cite{nielsen2002quantum, kaye2007introduction, de2019quantum, childs2017lecture}.

%\subsection{Quantum Information}%% \label{qubitstuto}

\subsection{Quantum Bits and Quantum Registers} The bit is the basic unit of classical information. It can be either in state 0 or 1. Similarly, a quantum bit or \emph{qubit}, is a quantum system that can be in state $\ket{0}$, $\ket{1}$ (the \emph{braket} notation $\ket{\cdot}$ is a reminder that the bit considered is a quantum system) or in a superposition of both states 
\begin{equation}
    \alpha\ket{0}+\beta\ket{1}
\end{equation}

The coefficients $\alpha,\beta \in \mathbb{C}$, named \emph{amplitudes}, are such that $|\alpha|^2 + |\beta|^2 = 1$. It is also convenient to see this qubit as a unit norm, complex, vector of dimension two, in the \emph{computational basis} (see Fig.\ref{fig:blochsphere}):
\begin{equation}
\ket{0} =
\begin{pmatrix}
1 \\
0 \\
\end{pmatrix}
,\quad
\ket{1} =
\begin{pmatrix}
0 \\
1 \\
\end{pmatrix}
,\quad
\alpha\ket{0}+\beta\ket{1} =
\begin{pmatrix}
\alpha \\
\beta  \\
\end{pmatrix}
\end{equation}

The amplitudes are linked to the probabilities of observing either 0 or 1 when \emph{measuring} the qubit, since 
\begin{equation}
    P(0)=|\alpha|^2, \quad P(1)=|\beta|^2.
\end{equation}

Before the measurement, any superposition is possible, which gives quantum information special abilities in terms of computation. With $n$ qubits, the $2^n$ possible binary combinations can exist simultaneously, each with a specific amplitude. For instance we can consider an uniform distribution $\frac{1}{\sqrt{n}}\sum_{i=0}^{2^n-1}\ket{i}$ where $\ket{i}$ represents the $i^{th}$ binary combination (e.g.  $\ket{01\cdots1001}$). Multiple qubits together are often called a \emph{quantum register}. 

In its most general formulation, a quantum state with $n$ qubits can be seen as a vector in a complex Hilbert space of dimension $2^n$. This vector must be normalized under $\ell_2$-norm, to guarantee that the squared amplitudes sum to 1, to respect the probabilities of measuring each possible state. 

With two quantum states or quantum registers $\ket{p}$ and $\ket{q}$, the whole system is written as a tensor product $\ket{p}\otimes\ket{q}$, often simplified as $\ket{p}\ket{q}$ or $\ket{p,q}$.

\subsection{Quantum Computation}
To process qubits and therefore quantum registers, we use quantum gates. These gates are \emph{unitary operators} in the Hilbert space as they should map unit-norm vectors to unit-norm vectors. Formally, we can see a quantum gate acting on $n$ qubits as a Hermitian matrix $U \in \mathbb{C}^{2^n}$ such that $UU^{\dagger}=U^{\dagger}U=I$, where $U^{\dagger}$ is the conjugate transpose of $U$. 

There exist plenty of quantum logical gates. For a single qubit, living in a complex 2-dimension space, it is worth mentioning first the Pauli matrices $\sigma_x$, $\sigma_y$, and $\sigma_z$. They have core importance in quantum physics, and together with the identity matrix, they form a basis for all single qubit quantum gates:
\begin{equation}
\sigma_x =
\begin{pmatrix}
0 & 1 \\
1 & 0 \\
\end{pmatrix}
,\quad
\sigma_y =
\begin{pmatrix}
0 & -i \\
i & 0 \\
\end{pmatrix}
,\quad
\sigma_z =
\begin{pmatrix}
1 & 0 \\
0 & -1 \\
\end{pmatrix}
\end{equation}

$\sigma_x$ is often referred to as the NOT gate, also written $X$, that inverts $\ket{0}$ and $\ket{1}$. The Hadamard gate, written $H$, which truly captures the nature of quantum information processing:
\begin{equation}
H = \frac{1}{\sqrt{2}}\begin{pmatrix}
1 & 1 \\
1 & -1 \\
\end{pmatrix}
\end{equation}
Indeed, we can see that $H$, applied to the computational basis, creates the uniform quantum superposition:
\begin{equation}
H\ket{0} = 
\frac{1}{\sqrt{2}}
\begin{pmatrix}
1 & 1 \\
1 & -1 \\
\end{pmatrix}
\begin{pmatrix}
1 \\
0 \\
\end{pmatrix}
=
\begin{pmatrix}
\frac{1}{\sqrt{2}}\\
\frac{1}{\sqrt{2}}\\
\end{pmatrix}
=\frac{1}{\sqrt{2}}(\ket{0}+\ket{1})
\end{equation}

\begin{equation}
H\ket{1} = 
\frac{1}{\sqrt{2}}
\begin{pmatrix}
1 & 1 \\
1 & -1 \\
\end{pmatrix}
\begin{pmatrix}
0 \\
1 \\
\end{pmatrix}
=
\begin{pmatrix}
\frac{1}{\sqrt{2}}\\
-\frac{1}{\sqrt{2}}\\
\end{pmatrix}
=\frac{1}{\sqrt{2}}(\ket{0}-\ket{1})
\end{equation}

\begin{figure}[h]
\centering
\begin{subfigure}{.35\textwidth}
  \centering
  \includegraphics[width=\linewidth]{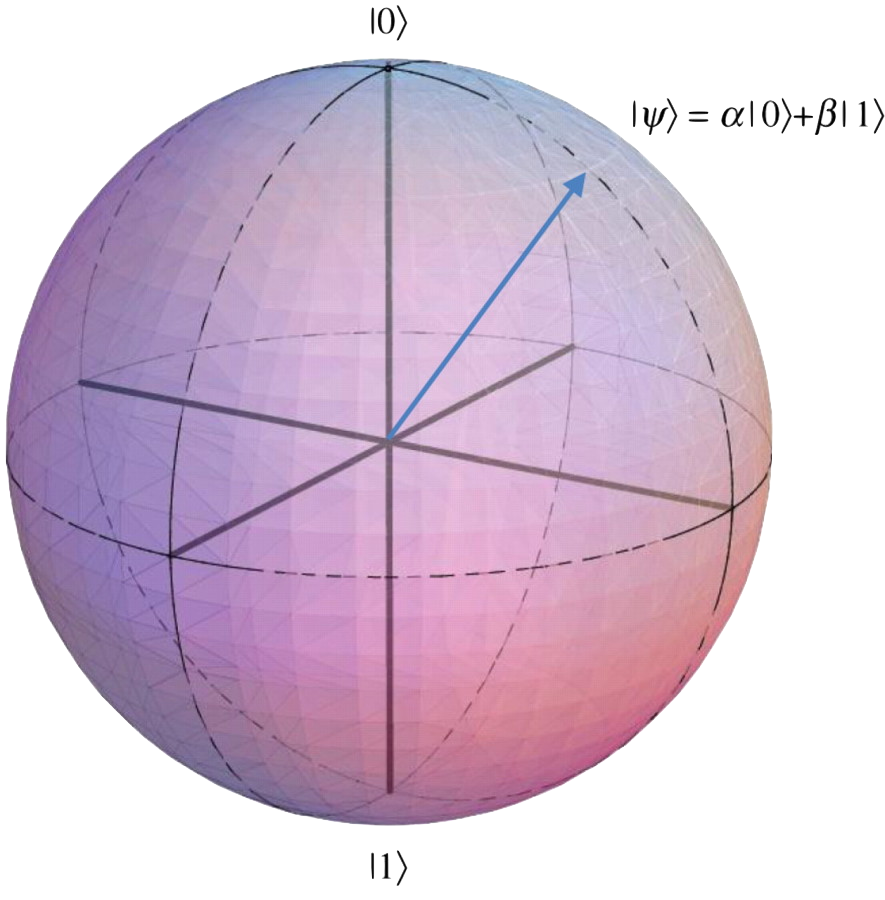}
  \caption{}
  \label{fig:blochsphere}
\end{subfigure}\hspace{0.15\textwidth}%add horizontal space
\begin{subfigure}{.35\textwidth}
  \centering
  \includegraphics[width=\linewidth]{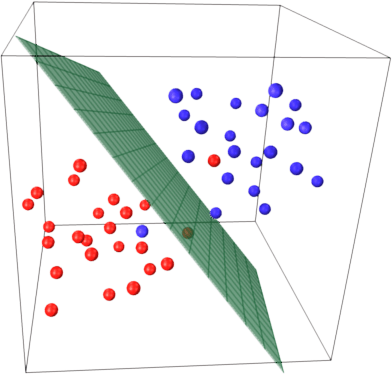}
  \caption{}
  \label{fig:svm_linearalgebra}
\end{subfigure}
\caption{ (a) A quantum state (blue vector) of one qubit is a unit vector in the Bloch sphere, embedded in a 2 dimensional Hilbert Space. Source: \cite{dorner2012towards} (b) Classical machine learning also manipulates data as vectors (blue and red dots) in vector spaces to classify or transform them. Source: \emph{Machine Learning in Action}.}
\label{fig:linearalgebra_quatum_vs_classical}
\end{figure}

Rotation gates rotate a qubit vector inside the Bloch sphere (see Fig.\ref{fig:blochsphere}), each on a respective plan. These gates take a real value angle as parameter:
\begin{gather}\nonumber
R_x(\theta) =
\begin{pmatrix}
\cos(\theta/2) & -i\sin(\theta/2) \\
-i\sin(\theta/2) & \cos(\theta/2) \\
\end{pmatrix}
,\quad
R_y(\theta) =
\begin{pmatrix}
\cos(\theta/2) & -\sin(\theta/2) \\
\sin(\theta/2) & \cos(\theta/2) \\
\end{pmatrix}
,\quad\\
R_z(\theta) =
\begin{pmatrix}
1 & 0 \\
0 & e^{i\theta} \\[1ex]
\end{pmatrix}
\end{gather}

Multiple qubit gates exist, such as the Controlled-NOT, or $CNOT$, that applies a NOT gate on a target qubit conditioned on the state of a control qubit. As well, the Controlled-$Z$ or $CZ$ gate, flips the phase of the amplitude of the target qubit ($\sigma_z$), if the controlled qubit is in state $\ket{1}$:
\begin{equation}
CNOT = \begin{pmatrix}
1 & 0 & 0 & 0 \\
0 & 1 & 0 & 0 \\
0 & 0 & 0 & 1 \\
0 & 0 & 1 & 0 \\
\end{pmatrix}, 
\quad
CZ = \begin{pmatrix}
1 & 0 & 0 & 0 \\
0 & 1 & 0 & 0 \\
0 & 0 & 1 & 0 \\
0 & 0 & 0 & -1 \\
\end{pmatrix}
\end{equation}

The same controlled gates exist for Rotations and other gates. They can be controlled by more qubits as well. If these gates are not native to a quantum device, they often come at the cost of being decomposed in practice into other gates, adding some depth to the circuit. 

An other fundamental and useful gate is the $SWAP$ gate, that swaps two qubits, such that $SWAP \ket{p}\ket{q} = \ket{q}\ket{p}$:
\begin{equation}
SWAP = \begin{pmatrix}
1 & 0 & 0 & 0 \\
0 & 0 & 1 & 0 \\
0 & 1 & 0 & 0 \\
0 & 0 & 0 & 1 \\
\end{pmatrix}
\end{equation}

The main advantage of quantum gates is their ability to be applied to a superposition of inputs. Indeed, given a gate $U$ on a quantum state $\ket{x}$, such that:
\begin{equation}
U\ket{x} \mapsto \ket{f(x)}
\end{equation}
we can apply it to all possible combinations of $\ket{x}$ at once: 
\begin{equation}
U\left(\frac{1}{C}\sum_{x}\ket{x}\right) \mapsto \frac{1}{C}\sum_{x}\ket{f(x)}
\end{equation}
where $C$ is is a normalization factor to respect the fact that quantum states must be unit vectors.

\subsection{Quantum Measurements}

Before any intervention, quantum states evolve according to unitary transformations. But there is a moment where one needs to get a result in the classical world. Measuring a quantum state is an action that transfers some information from quantum to classical, but the output can remain quantum if the measurement was made on a part of the state. Measuring a quantum state $\ket{\phi}$ can often result in different outcomes indexed $m$, symbolized by the set of measurement \emph{operators} $\{M_m\}$. These operators act on the quantum Hilbert space and the probability of obtaining the outcome $m$ is given by:

\begin{equation}
    p(m) = \bra{\phi}M^{\dagger}_mM_m\ket{\phi}
\end{equation}

and the quantum state of the system \emph{after} the measurement is:

\begin{equation}
    \frac{M_m\ket{\phi}}{\sqrt{\bra{\phi}M^{\dagger}_mM_m\ket{\phi}}}
\end{equation}

The denominator of this fraction appears to renormalize the output state. A \emph{complete} set of measurement operators is such that $\sum_m M^{\dagger}_mM_m = I$, where $I$ is the identity operator. 

The most common measurement operators are the ones in the \emph{computational basis}, where we want to know in which computational state qubits are. For a single qubit and its basis $\{\ket{0}$,$\ket{1}\}$, we write $M_0 = \ket{0}\bra{0}$, $M_1 = \ket{1}\bra{1}$, and the associated probability is the respective square of the amplitude. \\

More generally one can define \emph{POVM} (Positive Operator Valued Measure) and extend their applications to mixtures of quantum states known as density matrices. 

And in practice, we most commonly use \emph{Projective} measurements, which are derived from \emph{observables}. An Hermitian measurement operator $M$ can be decomposed in $m$ projections $P_m$. Each $P_m$ is a projection into an eigenspace of $M$ with eigenvalue $m$. The projectors are orthogonal between each other, complete, positive definite, and Hermitian.

\begin{equation}
    M = \sum_m m P_m
\end{equation}

Since for a projector $P$ we have $P = P^{\dagger}$ and $P^2=P$, the probability of obtaining the state $m$ is now given by:

\begin{equation}
    p(m) = \bra{\phi}P_m\ket{\phi}
\end{equation}

and the remaining state is:

\begin{equation}
    \frac{P_m\ket{\phi}}{\sqrt{\bra{\phi}P_m\ket{\phi}}}
\end{equation}

Notably, the average outcome of a projective measurement $M$ is the expectation value:

\begin{equation}
 \begin{split}
    \E (M) &= \sum_m mp(m)\\
    & = \sum_m m \bra{\phi}P_m\ket{\phi}\\
    & = \bra{\phi} (\sum_m m P_m) \ket{\phi}\\
    & = \bra{\phi}M\ket{\phi}
 \end{split}   
\end{equation}

Finally, we will use the following facts:
\begin{itemize}
    \item When asking the question `` what is the probability of measuring $\ket{\psi}$ from quantum state $\ket{\phi}$?" we use $p(\psi) = |\braket{\phi}{\psi}|^2$. To prove this, we use the projector $\ket{\psi}\bra{\psi}$.
    \item Two quantum states $\ket{\phi}$ and $\ket{\psi}$ are said \emph{orthogonal} is $\braket{\phi}{\psi} = 0$. 
    \item For tensor product of quantum states, linearity prevails: $(\bra{\phi}\otimes\bra{\phi'})(\ket{\psi}\otimes\ket{\psi'}) = \braket{\phi}{\psi} \braket{\phi'}{\psi'}$. We wan write $\ket{\psi^{\perp}}$ to denote a quantum state orthogonal to $\ket{\psi}$.
    \item We will often have state in the form $\ket{\phi} = \alpha\ket{y}\ket{0} + \beta\ket{y^{\perp}}\ket{1}$. The probability of measuring $\ket{0}$ on the last qubit is then given by: 
\begin{equation}
    \bra{\phi}(\ket{0}\bra{0})\ket{\phi} = \alpha^2
\end{equation}
and the remaining state after the measurement is simply $\ket{y}$.
\end{itemize}

\section{Quantum Algorithms}\label{sec:basic_quantum_algo}

\begin{figure}[h]
    \centering
    \includegraphics[width=0.6\textwidth]{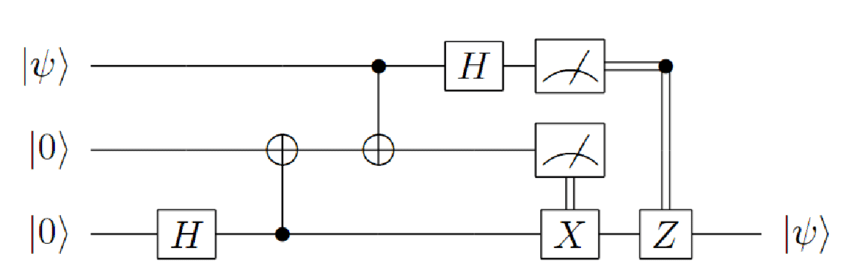}
    \caption{Quantum circuit for the teleportation of any quantum state $\ket{\psi}$ \cite{bennett1993teleporting}. Each wire corresponds to one or several qubits. The input on the left is processed through the circuit. The circuit uses Hadamard, CNOT, Pauli $\sigma_x$ and $\sigma_z$ gates. It also requires measurements and classical information flow.}
    \label{fig:itro_quantum_teleportation}
\end{figure}

Joined together, quantum gates form quantum circuits, also called quantum algorithms, and are often represented as in Fig.\ref{fig:itro_quantum_teleportation}. The complexity of a quantum circuit can be expressed as the relationship between its depth and the size of the problem. The complexity also takes into account the number of time the circuit must be run to obtain the desired output.\\

We already presented the first meaningful quantum algorithms in Section \ref{sec:intro_quantum_computing}. Among them, the Deutsch-Josza algorithm \cite{deutsch1992rapid}, the Bernstein-Vazirani algorithm \cite{bernstein1997quantum}, Simon's algorithm \cite{simon1997power}, and the famous Grover \cite{groveralgo} and Shor's algorithms \cite{shor1999polynomial}. In this section, we will introduce a few more quantum algorithms that will be used as subroutines in this thesis. The algorithms are presented in the form of theorems specifying the error and running time guarantees. We omit the proof of the theorems of this section (see \cite{nielsen2002quantum, kaye2007introduction, de2019quantum, childs2017lecture} for details).

\subsection{Phase Estimation}\label{sec:phase_estimation}

Phase estimation \cite{kitaev1995quantum} is an important quantum algorithm that creates the link with linear algebra. It is also a subroutine used in Shor's algorithm \cite{shor1999polynomial}, the HHL algorithm\cite{HHL}, and amplitude estimation (see Section \ref{sec:amplitude_amplification_estimation}). Phase estimation is itself based on the Quantum Fourier Transform (QFT) algorithm. The QFT provides an exponential speedup in the task of mapping a vector, encoded as a quantum state, into the Fourier space. 

\begin{figure}[h]
    \centering
    \includegraphics[width=0.6\textwidth]{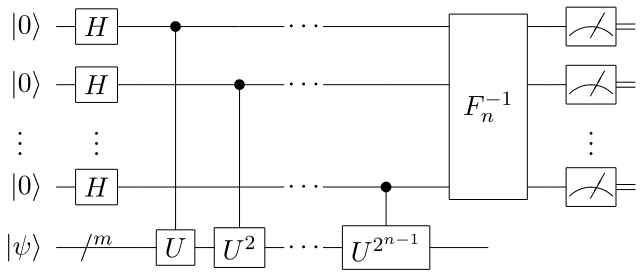}
    \caption{The phase estimation circuit for a unitary $U$ and any input $\ket{\psi}$, with measurements at the end. $F^{-1}_n$ is the inverse QFT. Source: Wikipedia}
    \label{fig:circuit_phase_estimation}
\end{figure}

Basically, the goal of phase estimation is to extract the eigenvalues of a matrix $U$. In fact, this matrix must be a quantum circuit $U$, therefore it should be an unitary operator. We denote its eigenvectors by $\ket{u_j}$ and eigenvalues $e^{i\theta_j}$, hence $U\ket{u_j} = e^{i\theta_j}\ket{u_j}$. Given a eigenvector and an extra register $\ket{u_j}\ket{0}$ as input, the algorithm should return $\ket{u_j}\ket{\theta_j}$. Since the eigenvectors compose a basis, any state $\ket{\psi}$ can be written as $\ket{\psi} = \sum_{j\in[n]}\alpha_j\ket{u_j}$. Therefore, the interesting feature of phase estimation is to apply in superposition. the circuit is shown in Fig.\ref{fig:circuit_phase_estimation}.

\begin{theorem}{Phase Estimation}{phase_estimation}
Let $U$ be a unitary operator than runs in time $T(U)$, with eigenvectors $\ket{u_j}$
and eigenvalues $e^{i\theta_j}$ for $\theta_j\in [-\pi,\pi]$, for $j\in [N]$. For a precision
parameter $\epsilon > 0$, there exists a quantum algorithm that runs in time $O(T(U)\log(N)/\epsilon)$
and with probability $1-1/poly(N)$ performs the mapping 
\begin{equation}
\ket{\psi} = \sum_{j\in[N]}\alpha_j\ket{u_j}\ket{0} \mapsto \sum_{j\in[N]}\alpha_j\ket{u_j}\ket{\overline{\theta_j}}
\end{equation}
where $\overline{\theta_j}$ is approximating $\theta_j$ with guarantee $|\overline{\theta_j} - \theta_j| \leq \epsilon$ for all $j\in [N]$.
\end{theorem}

One can decide to measure the result at the end to recover a classical description of the eigenvalues, or to leave the output as a quantum state for further processing. Note however that this circuit requires to apply sequentially the unitary $U$ in a controlled fashion. For certain circuits $U$, this can become a costly operation.

\subsection{Amplitude Amplification and Amplitude Estimation}\label{sec:amplitude_amplification_estimation}

As stated before, one of the major difficulties in quantum computing is to manipulate the amplitudes of the quantum states. These amplitudes often carry important information, in particular in quantum linear algebra (see Chapter \ref{chap:quantum_data}). It also happens that a part of the quantum superposition is considered as ``garbage" and that one wants to discard it. Finally, one may need to recover a classical value of a target amplitude. 

Following the Grover algorithm \cite{groveralgo}, a generalisation was proposed in \cite{brassard1997exact} and then \cite{brassard2002quantum} to amplify a target amplitude of a quantum state. This has proven to be very useful in many cases. It can also output an amplitude or writing it in a quantum register, and become very powerful when applied in superposition. As Grover algorithm, these algorithms usually provides a quadratic speedup (see Table \ref{table:amp_est_quantum_vs_classical}).

\begin{theorem}{Amplitude Amplification}{amplitude_amplification}
    Given the ability to implement a quantum unitary $U$ and $U^{-1}$,  such that $U\ket{0} = \sin(\theta) \ket{x, 1} + \cos(\theta) \ket{G, 0}$, where $\ket{G}$ is a garbage state, then we can instead create the state $\ket{x}$ in time $\widetilde{O}(\frac{T(U)}{\sin(\theta)})$, where $T(U)$ is the time to implement $U$ and $U^{-1}$.
\end{theorem}

\begin{theorem}{Amplitude Estimation (1)}{amplitude_estimation_other}
    Given the ability to implement a quantum unitary $U$ and $U^{-1}$, such that $U\ket{0} = \sin(\theta) \ket{x, 1} + \cos(\theta) \ket{G, 0}$, where $\ket{G}$ is a garbage state, then $\sin(\theta)$ can be estimated to multiplicative error $\epsilon>0$ in time $\widetilde{O}(\frac{T(U)}{\epsilon \sin(\theta)})$, or to additive error in $\widetilde{O}(\frac{T(U)}{\epsilon})$, where $T(U)$ is the time to implement $U$ and $U^{-1}$.
\end{theorem}

In this thesis, we will also use a specific version of this algorithm \cite{grover2005fixed,yoder2014fixed} where the amplitudes don't have to be known in advance to be estimated.

\begin{theorem}{Amplitude Estimation (2)}{amplitude_estimation}%ex- \label{theoremamplitudeamplification}
	Given the ability to implement a quantum unitary $U$ and $U^{-1}$, such that $U:\ket{0} \to \sqrt{p}\ket{y,1} + \sqrt{1-p}\ket{G,0}$ where $\ket{G}$ is a garbage state, then for any positive integer $P$, the amplitude estimation algorithm outputs $\tilde{p}$ $(0 \le \tilde p \le 1)$ such that
	\begin{equation}
	|\tilde{p}-p|\le 2\pi \frac{\sqrt{p(1-p)}}{P}+\left(\frac{\pi}{P}\right)^2
	\end{equation}
	with probability at least $8/\pi^2$. It uses exactly $P$ iterations of the algorithm $U$ and $U^{-1}$. 
	If $p=0$ then $\tilde{p}=0$ with certainty, and if $p=1$ and $P$ is even, then $\tilde{p}=1$ with certainty.
\end{theorem}

Proper proofs of these theorems are given in \cite{kaye2007introduction}. Briefly, let $U$ be the unitary that creates the state $\sqrt{p}\ket{y,1} + \sqrt{1-p}\ket{G,0}$. Amplitude amplification or estimation is phase estimation (Section \ref{sec:phase_estimation}) applied on a the unitary $Q = U^{-1}O_{{\perp}}UO_f$ where $O_{{\perp}}$ and $O_f$ are the Grover phase shift operators. One can show that the eigenvalues of the operator $Q$ which are estimated by phase estimation are linked to the desired amplitudes.

\begin{figure}[h]
    \centering
    \includegraphics[width=0.7\textwidth]{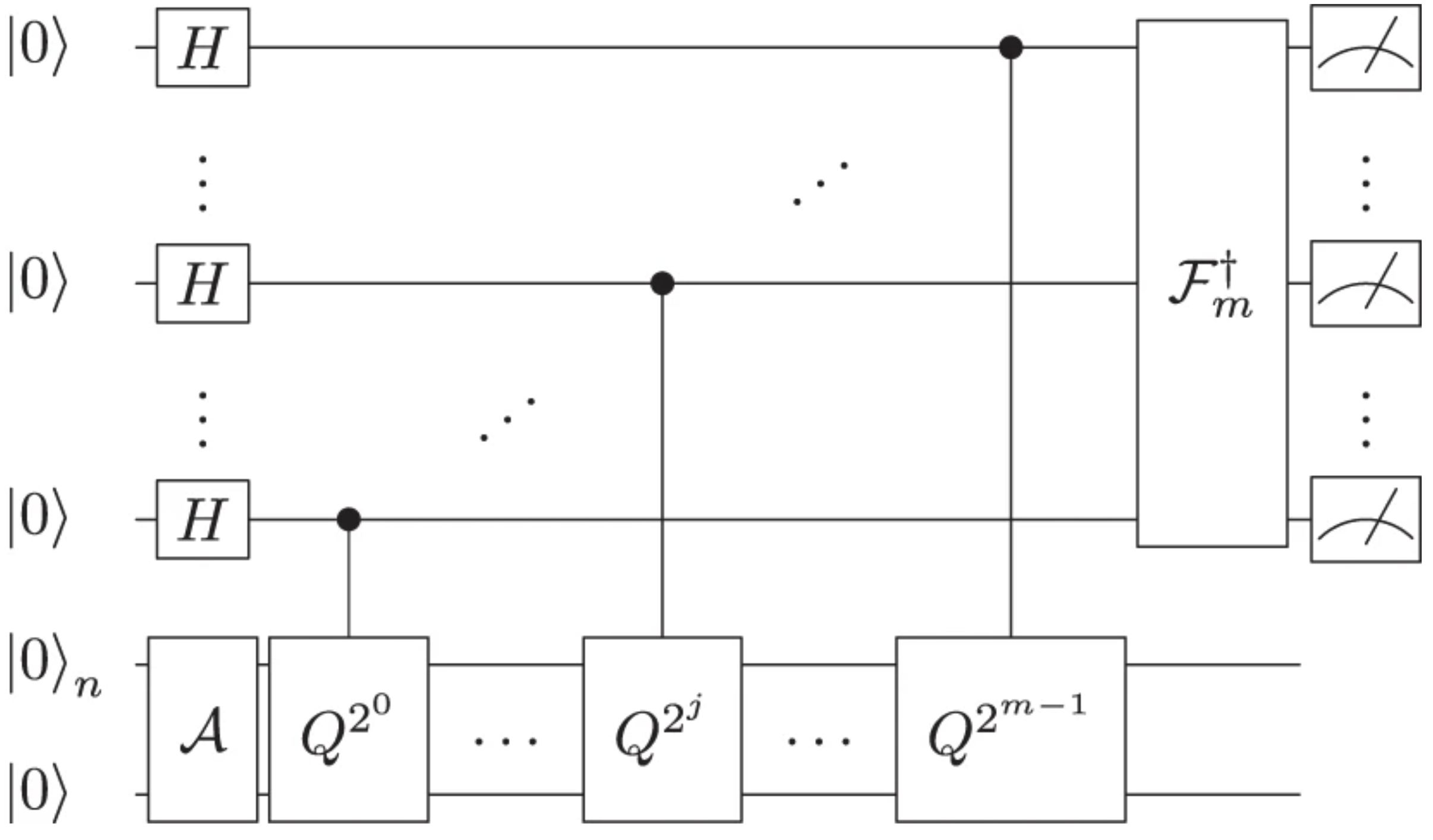}
    \caption{The amplitude estimation circuit, where $\mathcal{F}$ is the Quantum Fourier Transform, and $\mathcal{A}$ is the initial unitary (denoted $U$ in above Theorems). Source: \cite{grinko2021iterative}.}
    \label{fig:amplitude_estimation_circuit}
\end{figure}

It is also possible to obtain the amplitude as a quantum state, written in binary with some precision $\epsilon$, as $\ket{\sqrt{p}}$ or $\ket{\sin(\theta)}$ with some amplitude $\sqrt{\alpha}$ such that $\alpha > 8/\pi^2$. For this it suffices to not perform the measurement at the end of the phase estimation (see Section \ref{sec:phase_estimation}). We will refer indistinctly to ``amplitude estimation" and Theorem \ref{thm:amplitude_estimation} for both usages, and often for the additive error case (see Table \ref{table:amp_est_quantum_vs_classical}). Later on, we will also use Theorem \ref{thm:median_evaluation} to boost the amplitude $\sqrt{\alpha}$ and have a state arbitrary close to $\ket{\sqrt{p}}$ or $\ket{\sin(\theta)}$. \\

%\begin{table}[h]
%\begin{tabular}{|c|c|c|c|}
%\hline
%Type           & Guarantee                  & Quantum                                         & Classical                                  \\ \hline
%additive error & $|\tilde{p}-p|<\epsilon$   & $O\left(\frac{1}{\epsilon}\right)$             & $O\left(\frac{1}{\epsilon^2}\right)$      \\ \hline
%relative error & $|\tilde{p}-p|<\epsilon p$ & $O\left(\frac{1}{\epsilon \sqrt{P(0)}}\right)$ & $O\left(\frac{1}{\epsilon^2 P(0)}\right)$ \\ \hline
%\end{tabular}
%\caption{Comparison of classical and quantum amplitude estimation in the additive and relative cases.\label{table:amp_est_quantum_vs_classical}}
%\end{table}

\begin{table}[h]
\begin{tabular}{|c|c|c|c|}
\hline
Type           & Guarantee                  & Quantum                                         & Classical                                  \\ \hline
additive error & $|\tilde{p}-p|<\epsilon$   & $O\left(1/\epsilon\right)$             & $O\left(1/\epsilon^2\right)$      \\ \hline
relative error & $|\tilde{p}-p|<\epsilon p$ & $O\left(1/\epsilon \sqrt{P(0)}\right)$ & $O\left(1/\epsilon^2 P(0)\right)$ \\ \hline
\end{tabular}
\caption{Comparison of classical and quantum amplitude estimation in the additive and relative cases.\label{table:amp_est_quantum_vs_classical}}
\end{table}

Note that both amplitude amplification and estimation rely on phase estimation (see Section \ref{sec:phase_estimation}) and therefore can suffer from the same constraints. However recently, proposals for amplitude estimation without phase estimation have been made \cite{aaronson2020quantum,suzuki2020amplitude} and could better suit short term implementations for which shallow circuits are required \cite{giurgica2020low}.

\subsection{Other Subroutines}

\subsubsection{Classical Boolean Circuits}
In the following claim we state some primitive quantum circuits, which we will use in our algorithm. They are basically quantum circuits with a reversible version of the classical boolean ones.

Using quantum circuits, one can perform the following operations in time linear in the number of qubits used to encode the input values :
\begin{claim}{Classical Boolean Circuits}{boolean_circuits} %ex \label{claimcircuits}
    \begin{itemize}
    	\item For two integers $i$ and $j$, we can check their equality with the mapping $\ket{i}\ket{j}\ket{0} \mapsto \ket{i}\ket{j}\ket{[i=j]}$. 
    	\item For two real numbers $a>0$ and $b>0$, we can compare them using $\ket{a}\ket{b}\ket{0} \mapsto \ket{a}\ket{b}\ket{[a\leq b]}$. 
    	\item For a real number $a>0$, we can obtain its square $\ket{a}\ket{0} \mapsto \ket{a}\ket{a^2}$.
    \end{itemize}
\end{claim}

This can be extended since any classical boolean function can be embedded in a quantum circuit (see Section \ref{chap:intro_quantum_computing}). In particular, non linear functions ($\arcsin(x)$, $\sqrt{x}$, $\text{sigmoid}(x)$ etc.) can be applied to the value encoded in binary quantum registers, as we will do in Part \ref{part:Q_NeuralNet}. These functions can be implemented using Taylor decomposition or any other technique. Note however that non linear transformations are impossible on quantum amplitudes directly, which require unitary, thus linear, transformations.

\subsubsection{Conditional rotation} 
Conditional rotation is a convenient and short procedure, used in the HHL algorithm \cite{HHL} for instance, and throughout this thesis. In contrast to amplitude estimation (Theorem \ref{thm:amplitude_estimation}), the goal is to map a value, binary encoded in a quantum register, to the amplitude of an extra qubit. Therefore, this value should be in $[-1,1]$. 

\begin{theorem}{Conditional Rotation}{conditionrotation} %ex- \label{conditionrotationthm}
Given the quantum state $\ket{a}$ encoded in $q$ qubits, with $a \in [-1,1]$, There is a quantum circuit to perform $\ket{a}\ket{0} \mapsto \ket{a}(a\ket{0}+\sqrt{1-a^2}\ket{1})$. 
\end{theorem} 

\begin{proof}
    Let $\gamma = \arcsin(a)$. The controlled rotation starts by writing the state $\ket{\gamma}$
    in a $q$ qubits register. This can be done using a quantum implementation of the  $\arcsin$ function, as in Claim
    \ref{thm:boolean_circuits} or any other \cite{haner2018optimizing}. Quantum circuits are classical boolean operations, which usually apply part of the arcsine polynomial decomposition from the Taylor's series. This can require $O(poly(q))$ or less, depending on the solution adopted. 
    \begin{equation}
        \ket{a}\ket{0}\ket{0} \mapsto \ket{a}\ket{0}\ket{\gamma}
    \end{equation}
    The second step is the controlled rotation itself by performing a series of controlled rotation gates along the $y$-axis, for each one of the $q$ qubits of $\ket{\gamma}$. If we write the binary expansion $\gamma = 0.\gamma_1.\cdots.\gamma_q$, we can write the unitary that performs the rotation (see Fig.\ref{fig:conditional_rotation}) $R_y(2\gamma) = \prod_{j=1}^q R^{\gamma_j}_y(2^{1-j})$:
    \begin{equation}
        \ket{a}\ket{0}\ket{\gamma} \mapsto \ket{a}\left(\sqrt{1-a^2}\ket{0} + a\ket{1}\right)\ket{\gamma}
    \end{equation}
    Finally, we can get rid of $\ket{\gamma}$ by reverting the circuit, and we can switch $\ket{0}$ and $\ket{1}$ using a $NOT$ gate.
\begin{figure}[h]
    \centering
    \includegraphics[width=0.6\textwidth]{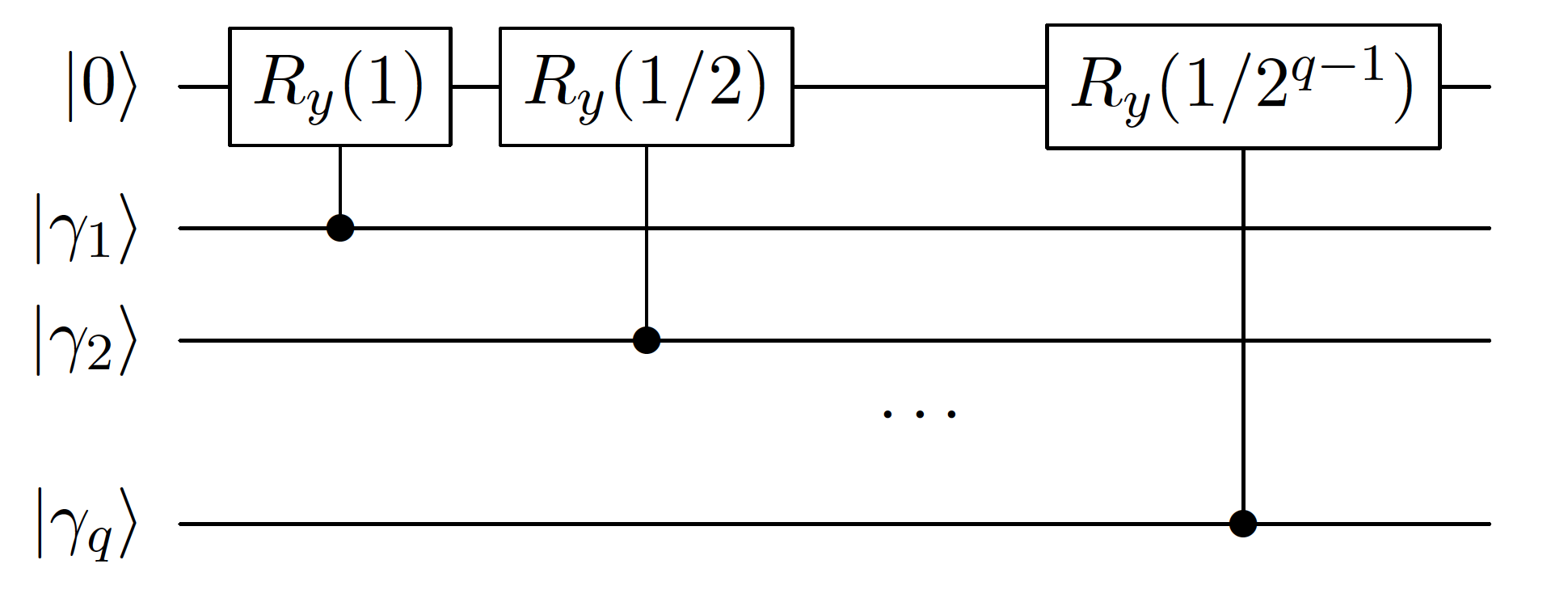}
    \caption{First half of a circuit implementing conditional rotation.}
    \label{fig:conditional_rotation}
\end{figure}
\end{proof}%\Hsquare

In addition, knowing in advance or computing an upper bound $\max(a)$ for the value of $a$ would allow applying the conditional rotation to values whose absolute value is bigger than 1, that is:

\begin{equation}
    \ket{a}\ket{0} \mapsto \ket{a}\left(\frac{a}{\max(a)}\ket{0}+\sqrt{1-\frac{a}{\max(a)}}\ket{1}\right)
\end{equation}

Using Theorem \ref{thm:conditionrotation} followed by Theorem \ref{thm:amplitude_estimation}, it then possible to transform the state $\frac{1}{\sqrt{d}}\sum_{j=0}^{d-1} \ket{x_j}$ into $\frac{1}{\norm{x}}\sum_{j=0}^{d-1} x_{j}\ket{x_j}$ and therefore alternate between the encoding types of a vector $x\in \R^d$ (see Chapter \ref{chap:quantum_data}).

\section{Noisy Intermediate Scale Quantum Computing (NISQ)}\label{sec:NISQ}

The algorithms presented in the previous Section are the continuation of the first results proving the theoretical superiority of quantum computing. Therefore, they all assume access to an ideal quantum computer, without decoherence, gate noise, and qubit errors. 

In recent years, we witnessed the advent of the first noisy quantum computers, up to the first quantum \emph{supremacy} experiment \cite{googlesupremacy}. Computer scientists and physicists tried to develop quantum algorithms that would suit these ``noisy intermediate scale quantum" devices, or NISQ for short \cite{NISQpreskill}. 

Several approaches exist, but the one that has attracted the most attention of researchers is called \emph{variational} quantum circuits (VQC) \cite{cerezo2020variational, bharti2021noisy}. Inspired by classical machine learning, it was proposed for quantum chemistry with the variational quantum eigensolver (VQE) algorithm \cite{peruzzo2014variational}, and for optimization with the quantum approximate optimization algorithm (QAOA) \cite{farhi2014quantum}. Later, a lot of derived applications in machine learning \cite{biamonte2017quantum,cong2019quantum, coyle2020born}.

\begin{figure}[h]
    \centering
    \includegraphics[width=0.7\textwidth]{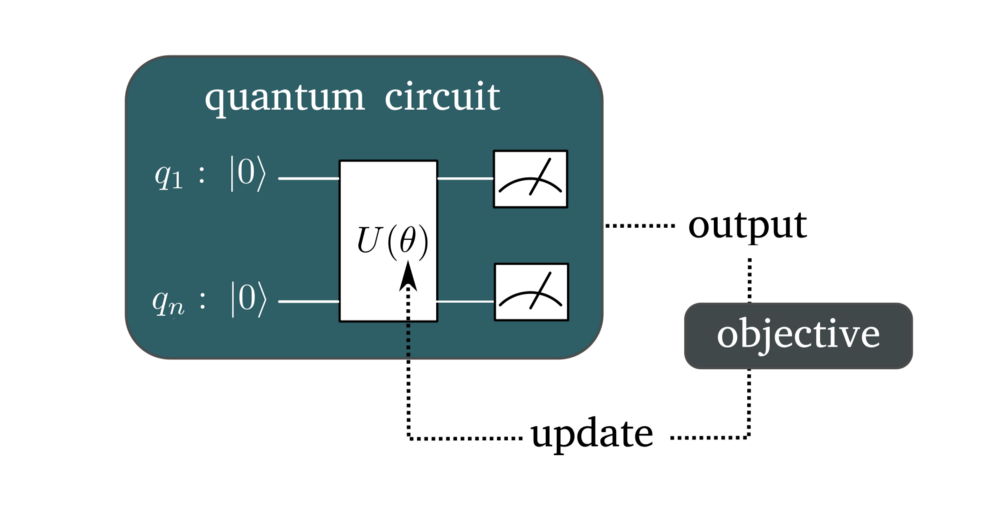}
    \caption{Representation of variational quantum circuit optimization scheme. Source: Xanadu}
    \label{fig:variational_circuit}
\end{figure}

VQC have universal properties \cite{biamonte2021universal} and already show encouraging results on real experiments \cite{abbas2020power}, however they are very different in nature. They are based on the following scheme (see Fig.\ref{fig:variational_circuit}): One defines a small circuit, called the \emph{ansatz}, made of many gates with tunable parameters, such as the angle of a rotation gate. Then, measurements of the resulting quantum state are performed and should give the right answers to the desired task (classification, regression). At first, the results are bad because the parameters are almost random. This metric is called the Objective Function or the Loss. Finally, optimization is done on a classical computer to propose a new and hopefully better set of parameters to try. And we repeat this loop until the circuit gives good results.\\

The main difference with previous quantum algorithms is that the circuit is not implementing a known classical ML algorithm. One would simply hope that the heuristic will converge and successfully classify data or predict values, and even more when quantum machines will become larger. 

Researchers hope that VQC would project data in large enough Hilbert space, to perform classically inaccessible correlations or separations.
Notably, research on variational quantum machine learning is less focused on proving computational speedups. The main interest is to reach a more expressive or complex state of information processing. 
Despite the excitement, VQC also suffers from theoretical disturbance. It is proven that when the number of qubits or the number of gates becomes too big, the optimization landscape will be flat and hinder the ability to optimize the circuit. Many efforts are made to circumvent this issue, called barren plateaus \cite{mcclean2018barren}, by using specific circuits \cite{pesah2020absence} or smart initialization of the parameters \cite{grant2019initialization}. These barren plateaus may be very fundamental in quantum information, as a deep link has been recently proven with quantum information scrambling and limitations for the Hayden-Preskill thought experiment on black holes information loss \cite{holmes2020barren}.\\

in a VQC, the gradients of a cost function with respect to each parameter have to be estimated. In classical neural networks, this is usually done using the backpropagation algorithm \ref{sec:bacpropagation_fcnn} over analytic operations. With VQC, operations become too complex, and we cannot access intermediate quantum states, without measuring them.
The current state-of-the-art solution is called the parameter shift rule \cite{mitarai2018quantum,schuld2019evaluating} and requires applying the circuit and measure its result 2 times for each parameter. By comparison, in classical deep learning, the network is applied just once forward and once backward to obtain all thousand or millions gradients. Hopefully, we could parallelize the parameter shift rule on many simulators or quantum devices, but this could be limited for a large number of parameters. \\

Finally, researchers tend to focus more and more on the importance of data loading into a quantum state \cite{johri2020nearest_dataloaders}, also called feature map \cite{schuld2021quantum}. Without the ideal amplitude encoding obtained with the QRAM (see Chapter \ref{chap:quantum_data}), there are doubts that we will be able to load and process high dimensional classical data with an exponential or high polynomial factor. \\

Note that the expression “Quantum Neural Networks” has been used to show the similarities with classical Neural Networks (NN) training. However they are not equivalent, since the VQC don’t have the same hidden layers architecture, and neither have natural non-linearities, unless a measurement is performed. And there’s no simple rule to convert any neural network to a VQC or vice versa.

In Chapter \ref{chap:OrthoNN_nisq}, we will propose an alternative NISQ algorithm for neural network implementation, and backpropagation, with exact equivalence.

\section{How to Test a Quantum Algorithm?}
Throughout this thesis, quantum algorithms will be developed theoretically. But actual experiments or simulations are necessary to judge, or at least gain intuition about the results of the algorithms. This becomes even more crucial when comparing to equivalent classical algorithms. Indeed, by the nature of quantum information, precision, noise, and randomness often arise, which can provide undesirable effects. 

However, as detailed in Section \ref{sec:basic_quantum_algo}, we will often consider algorithms that current and near term quantum computers would not support, due to the lack of error correction and qubit number. In the next sections, we will explain the three methods we used for testing our quantum algorithms and provide meaningful results.

\subsection{Real Quantum Computers and Emulators}\label{sec:real_qc_experiment_rbs}
The most conclusive experiment will always be to run the quantum circuit on actual hardware. For a few years, it becomes possible via cloud access to run quantum circuits on real quantum computers. The emergence of open-source quantum software from various institutions and companies allow to program easily quantum circuit and launch experiments. 

\begin{figure}[h]
    \centering
    \includegraphics[width=0.8\textwidth]{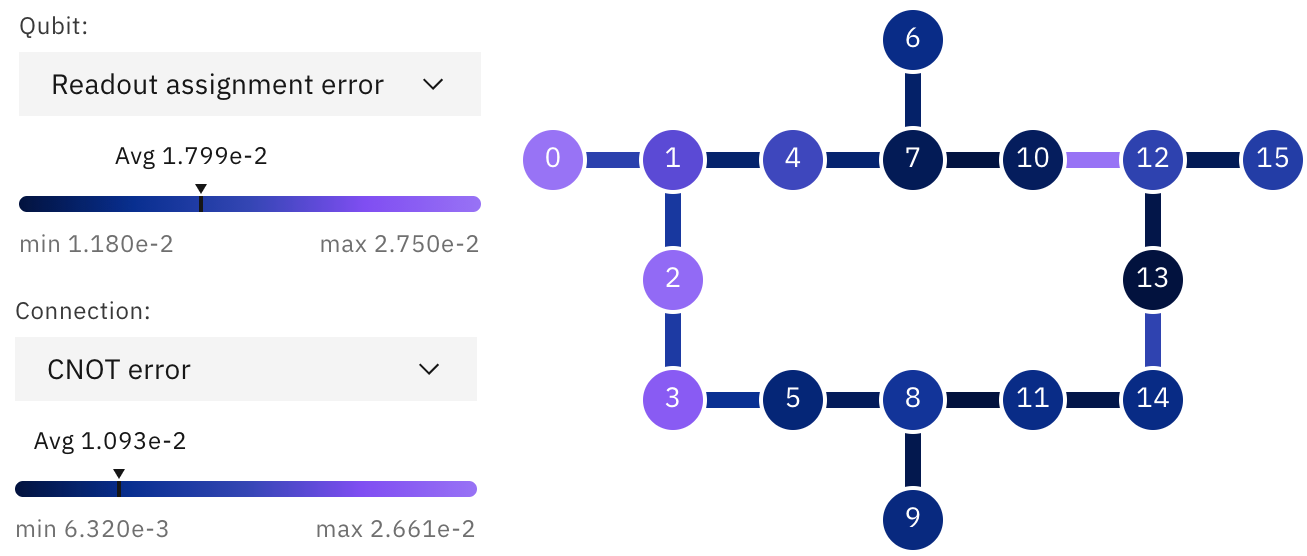}
    \caption{The \emph{IBM Guadalupe} quantum computer of 16 qubits. Visualization of the connectivity (right), and the calibrated error parameters(left). Qubits are connected with their adjacent neighbors, and only 3 qubits have 3 connections. May 2021.}
    \label{fig:ibm_guagalupe}
\end{figure}

In Chapter \ref{chap:OrthoNN_nisq}, we used several quantum computers made by IBM, ranging from 5 to 16 qubits (see Fig.\ref{fig:ibm_guagalupe}), for our orthogonal neural network algorithm. Indeed, this algorithm has the advantage of being shallow, repetitive, and requires only adjacent connectivity between qubits.

%\begin{figure}[h]
%    \centering
%    \includegraphics[width=0.7\textwidth]{images/RBS_qiskit         circuit.png}
%    \caption{?}
%    \label{fig:variational_circuit}
%\end{figure}

\begin{figure}[h]
\centering
\begin{subfigure}[b]{0.35\textwidth}
   \includegraphics[width=\linewidth]{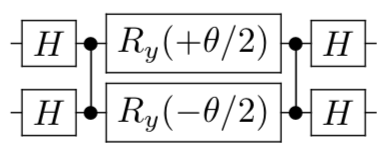}
   \caption{}
   \label{fig:RBS_decomposition} 
\end{subfigure}
\centering
\begin{subfigure}[b]{0.78\textwidth}
   \includegraphics[width=\linewidth]{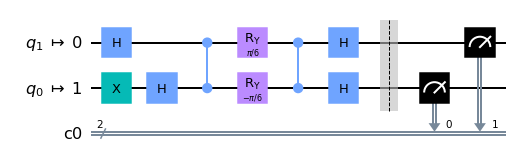}
   \caption{}
   \label{fig:RBS_qiskit_circuit}
\end{subfigure}
\caption{(a) Decomposition of the $RBS(\theta)$ gate (see Chapter \ref{chap:OrthoNN_nisq}) with Hadamard, y-axis Rotation, and $CZ$ gates. (b) Quantum circuit implementation using Qiskit \cite{Qiskit}, with additional $X$ gate to start in '10' and measurement on both qubits at the end.}
\end{figure}

However, the current state of these quantum computers makes them only interesting for proof of concepts and reality check. In practice, the qubits are noisy and prone to errors when applying gates and measuring qubit states. Still, it allows to learn the real constraints of quantum computing and to take into account the connectivity between the qubits, their quality, the noxious depth, and of course the monetary cost of such experiments. Moreover, current standard access is limited to few qubits, which is certainly too little to achieve quantum advantage.

It is however possible to move to a more ideal world, using emulators. Emulators are classical computers implementing the quantum circuit, by actually storing the exponentially large number of amplitudes and transforming it gate by gate. The main advantage of this method is to get rid of real hardware noise and augment the number of qubits. However, this number is often limited to ~30 or ~40 qubits for a general use case, since $2^{40}$ amplitudes with 32 bit floating point precision is already several Terabytes of data to process. Note that for some specific quantum circuits, efficient classical emulations can be found, which increases the number of qubits emulated. \\

\begin{figure}[h]
\centering
\begin{subfigure}{.45\textwidth}
  \centering
  \includegraphics[width=\linewidth]{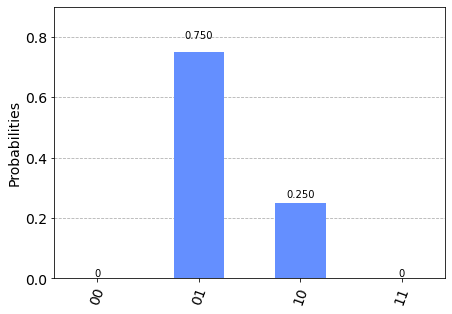}
  \caption{}
%  \label{fig:}
\end{subfigure}\hspace{0.05\textwidth}%add horizontal space
\begin{subfigure}{.45\textwidth}
  \centering
  \includegraphics[width=\linewidth]{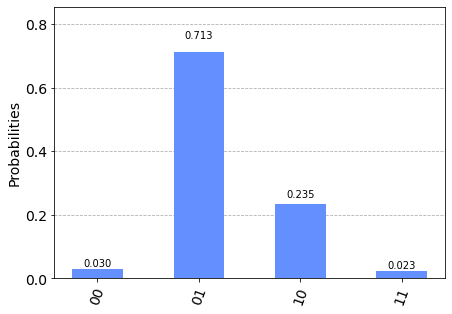}
  \caption{}
%  \label{fig:}
\end{subfigure}
\caption{Measurements results between the ideal (a) and the real (b) experiment of the RBS gate. This experiment took place on the 5 qubits \emph{IBM Santiago} quantum computer in May 2021.}
\label{fig:RBQ_measurements_results}
\end{figure}

In Fig.\ref{fig:RBQ_measurements_results}, we see the results of 8192 measurements of the circuit shown in Fig.\ref{fig:RBS_qiskit_circuit}. We see in Fig.\ref{fig:RBQ_measurements_results}a, the theoretical emulated result, that states '00' and '11' are not present in the quantum superposition and therefore should not be measured. However, on the real 5 qubits \emph{IBM Santiago} quantum computer, these states are effectively measured because of errors in gates, qubits, or readout.

\begin{figure}[h]
\centering
\begin{subfigure}[b]{0.6\textwidth}
   \includegraphics[width=\linewidth]{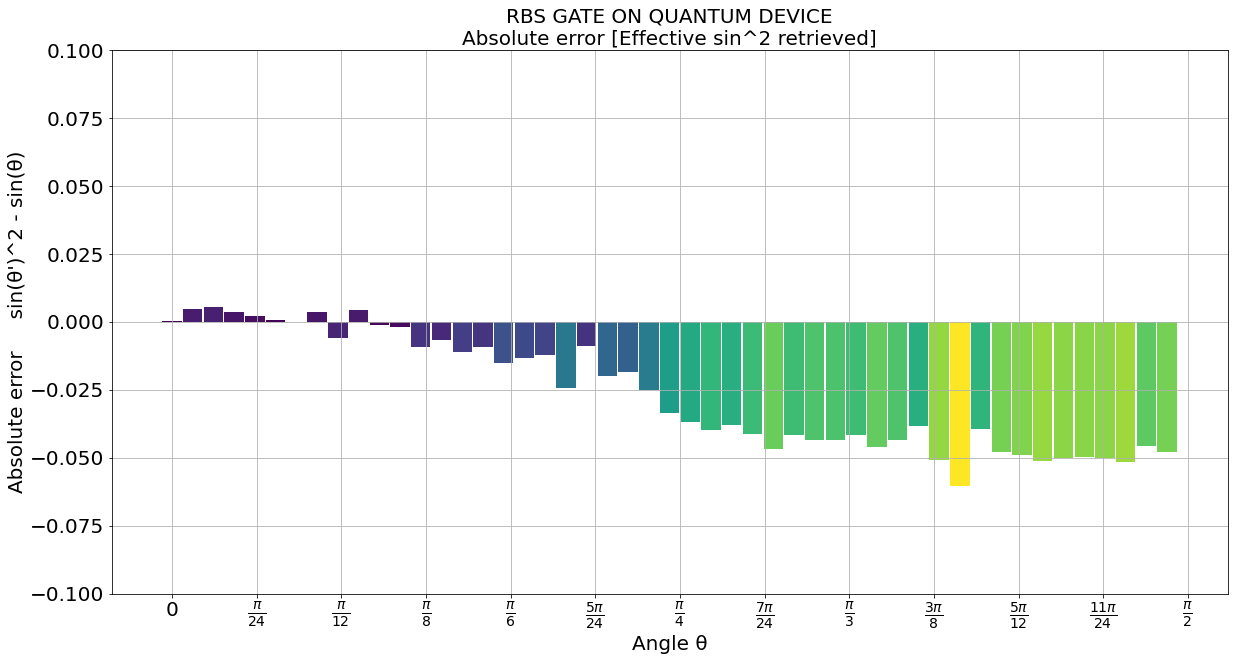}
   \caption{}
%   \label{fig:RBS_decomposition} 
\end{subfigure}
\centering
\begin{subfigure}[b]{0.6\textwidth}
   \includegraphics[width=\linewidth]{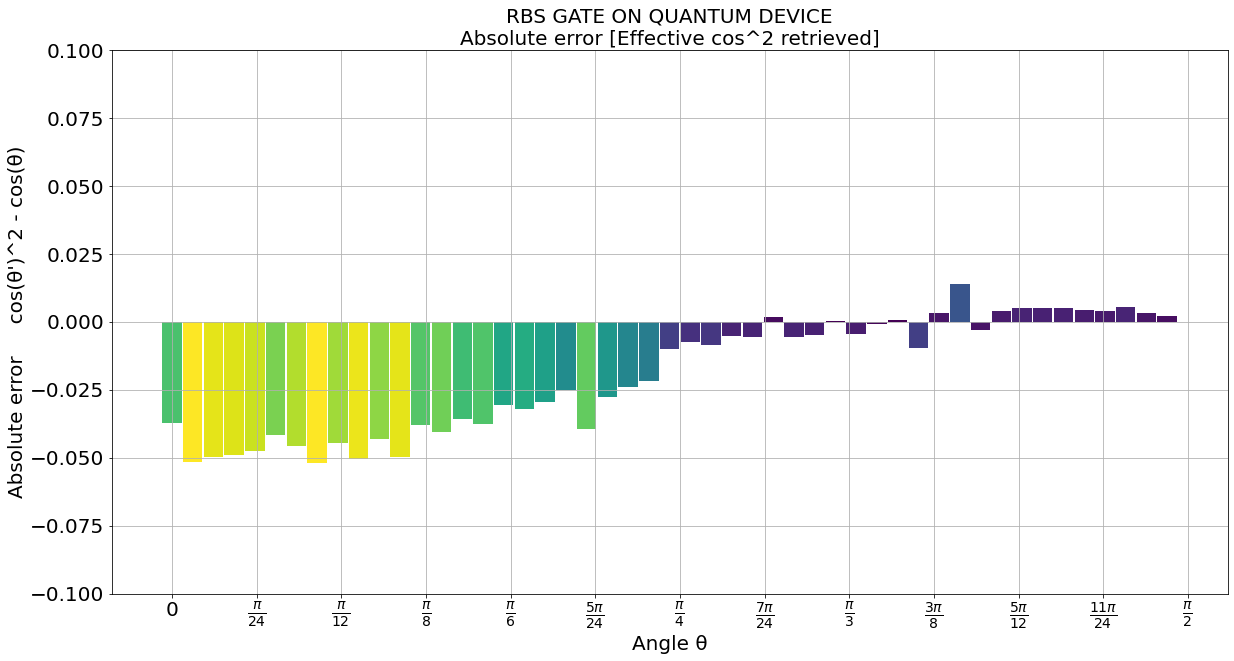}
   \caption{}
%   \label{fig:RBS_qiskit_circuit}
\end{subfigure}
\caption{Absolute error in the estimation of $\cos^2(\theta)$ and $\sin^2(\theta)$ from the measurements of the output of the $RBS(\theta)$ gate. We see that this error is correlated to the parameter $\theta$, making error mitigation more complex. This experiment took place on the 5 qubits \emph{IBM Santiago} quantum computer in May 2021.}
\label{fig:RBQ_error_cos_sin}
\end{figure}

As a result, it adds imprecision to the estimation of the amplitudes. In the case of the $RBS(\theta)$ gate, one expects to recover respectively $\cos(\theta)$ and $\sin(\theta)$ by measuring the relative size of the histogram bar '01' and '10'. In Fig.\ref{fig:RBQ_error_cos_sin}, we have computed this error, for different angles $\theta \in [0,\pi/2]$,

\subsection{Classically Simulating Quantum Algorithms}\label{sec:classically_simulating_an_algo}

Whether it is with 16 qubits on real hardware or 40 qubits on an emulator, we often require more. Indeed, for complex algorithms as most of those presented in this thesis, or for bigger size problems, a small number of qubits is often far from being useful. 

The advantage of the quantum machine learning algorithms we propose is their equivalence with their classical version. Indeed, as we will see, our algorithms follow the same steps, use the same inputs and outputs. The quantum circuits differ a lot, but we can mathematically describe their effects and we control the error committed, or the randomness during measurement. 

It is then possible to simply adapt programs, for instance in Python, that implement the classical machine learning algorithms. Since we control how and where the differences will occur, we can as well modify the classical programs to include them. With this, we can simulate our ideal fault-tolerant quantum algorithms in Python.

%%%%--------------PART II--------------%%%%
\part{Quantum Linear Algebra For Machine Learning}\label{part:preliminaries_QML}
\chapter{Quantum Data}\label{chap:quantum_data}
\epigraph{\textit{"There is no difference between Theory and Practice, except in Practice."}}{Benjamin Brewster (1882)}

In this chapter, we will cover the numerous techniques to interface quantum algorithms with classical data. As seen in Chapter \ref{chap:intro_classical_ml}, most machine learning relies on datasets containing samples $\{x_i\}_i\in[N]$, with $x_i\in\R^d$ for all $i$. These samples come from external experiments or data mining. As well, one needs classical outputs such as classes or labels $\{y_i\}_i\in[N]$, with $y_i\in\R^{d'}$ for all $i$. It seems that, for most \emph{real life} applications of quantum machine learning or optimization, it will be mandatory to have a way to handle classical data \cite{cortese2018loading}. 

In Section \ref{sec:classical_to_quantum_data}, we will see different propositions for encoding classical data as quantum states, and how we can generate them efficiently using \emph{quantum memory models}. Then, in Section \ref{sec:quantum_to_classical_to_data}, we will present the inverse task: recovering classical data from a quantum state.

Being able to propose such methods, even though some are only suited for perfect quantum computers (FTQC), is key to understand the potential benefit from near term and long term quantum computers.\\

Note that we will not cover the field on quantum machine learning on \emph{quantum data}. This is the case when the input of an algorithm is already a quantum state, as it could be in quantum communication or cryptography \cite{coyle2020variational}. As well, efforts are made for developing classical machine learning for quantum data \cite{dunjko2018machine}. We will not cover Grover based quantum memory models, such as the \emph{Quantum Associative Memory} \cite{de2019quantum_associativememory}.\\

Finally, as the translation between classical and quantum data remains a challenge in practice, there is interest in looking for ``dataless" problems. They could be problems that only involve an environment, such as solving partial differential equations, training a reinforcement learning algorithm, generate data or sampling from peculiar distributions, or even some chemistry applications.

\section{From Classical to Quantum Data}\label{sec:classical_to_quantum_data}

\subsection{Quantum Encodings}\label{sec:encoding_data}

In the following, numbers are given in their decimal ($N_{10}$) or binary ($N_2$) basis. If not specified, the basis shall be the decimal one. Unless otherwise specified, we will denote by $\ket{i}$ the $i^{th}$ quantum state in the computational basis, e.g. the state $\ket{0110\cdots10}$ that corresponds to the binary representation of the number $i$. 

Note that in the rest of this thesis, except Chapter \ref{chap:OrthoNN_nisq}, the \emph{amplitude encoding} described below will be used by default.

\subsubsection{Bit Encoding}

Classical data is naturally encoded as bits, e.g. $x=5_{10}$ and $y=3.25_{10}$ can be respectively written in binary as $x=101_2$ and $y=11.01_2$.
Therefore, it seems natural to first propose a similar simple \emph{bit encoding} using qubits, e.g. $\ket{x}=\ket{101}$. Similarly, a vector $x=
\begin{pmatrix}
3_{10} \\
2_{10}  \\
\end{pmatrix} 
= 
\begin{pmatrix}
11_{2} \\
10_{2}  \\
\end{pmatrix}$, can be encoded as the quantum state $\ket{x} = \ket{11}\otimes\ket{10}$ or $\ket{1110}$. 

For a small number of data points, or for low precision values, the loading of classical data known in advance is simple, using simple $NOT$ gates on qubits where $\ket{1}$ should be. This encoding is however poorly efficient as it requires as many qubits as bits, and has a limited precision. \\

In the quantum regime, it possible to use quantum superposition to handle multiple numbers at the same time. We can then propose the following bit encoding for a vector:

\begin{definition}{Bit Encoding}{bit_encoding}
    $m+\ceil{\log(d)}$ qubits can encode a vector $x = (x_1,\cdots,x_d) \in \R^d$ as a quantum superposition of bit strings:
    \begin{equation}
    \frac{1}{\sqrt{d}}\sum_{j=1}^{d}\ket{j}\ket{x_j}
    \end{equation}
    where $m$ is the number of qubits used for the precision, and $\ket{j}$ the $j^{th}$ quantum state in the computational basis. 
\end{definition}

\subsubsection{Amplitude Encoding}\label{sec:amplitude_encoding}

To use qubits to represent classical vectors or matrices, the most efficient encoding is by far the amplitude encoding scheme. It is the theoretical link between quantum computing and linear algebra that exploits quantum properties to the maximum. 

We will see that obtaining the amplitude encoding is usually the main bottleneck to our quantum algorithms (see Sections \ref{sec:quantum_memory_models} and \ref{sec:l_2_and_l_infinite_tomography}), while the rest consists in playing with the amplitudes. In fact, amplitude encoding was used in the pioneering work of \cite{HHL}, and later in many quantum machine learning and linear algebra works.

\begin{definition}{Amplitude Encoding}{ampltiude_encoding}
     $\ceil{\log(d)}$ qubits can encode a vector $x = (x_1,\cdots,x_d) \in \R^d$ using the amplitudes of the quantum state:
     \begin{equation}
    \ket{x} = \frac{1}{\norm{x}_2}\sum_{j=1}^{d}x_j\ket{j}
    \end{equation}
    where $\ket{j}$ is the $j^{th}$ quantum state in the computational basis
%\label{def:ampltiude_encoding}
\end{definition}

Since the quantum state $\ket{x}$ must be of unit norm in the Hilbert space, we use the normalization factor $1/\norm{x}_2$, which is equivalent to having a normalized input such that $\norm{x}_2=1$. Amplitude encoding uses only $\ceil{\log(d)}$ qubits: high dimensional data can be encoded with a small number of qubits, hence the exponential advantage. 

To go further, one can encode $N$ such vectors simultaneously, which is equivalent to encoding a matrix $X \in \R^{N\times d}$, as:

\begin{equation}
    \ket{X} 
    = \frac{1}{\norm{X}_F}\sum_{i=1}^{N}\norm{X_i}_2\ket{X_i}\ket{i}
\end{equation}

As in Definition \ref{def:ampltiude_encoding}, $\ket{X_i}$ is the quantum state of $X_i$, the $i^{th}$ row of $X$. $\norm{X}_F$ is the Frobenius norm of the matrix $X$. Note that this state is still normalized as 
$\norm{X}_F = \sqrt{\sum_{i=1}^{N}\norm{X_i}_2^2}$, which is permitted since all states in the superposition are orthogonal to each other thanks to the registers $\ket{i}$ at the end.\\

Finally, we present a methodology to switch from a superposition using bit encoding to a superposition using amplitude encoding. 

\begin{claim}{From Bit Encoding to Amplitude Encoding}{from_ampltide_to_bit_encoding}%label is reversed... but ok
    %Given the ability to create in time $T$ the bit encoding quantum state of the vector $x = (x_1,\cdots,x_d)$, namely we can perform the following mapping:
    Given an unitary $U$ which takes a ground state and creates in time $T_U$ the bit encoding quantum state of the vector $x = (x_1,\cdots,x_d)$, we can perform the following mapping:
    \begin{equation}
    \ket{0} \xmapsto[]{U} \frac{1}{\sqrt{d}}\sum_{j=1}^{d}\ket{j}\ket{x_j}
    \mapsto     
    %\frac{1}{\norm{x}_2}\sum_{j=1}^{d}x_j\ket{j}\ket{x_j}
    \frac{1}{\norm{x}_2}\sum_{j=1}^{d}x_j\ket{j}
    \end{equation}
    in time $\widetilde{O}(T_U\eta^2/\mathbb{E}(x_j^2))$, where $\eta \geq \max(x_j)$, and $\mathbb{E}(x_j^2)$ is the average value of the square components of $x$. 
\end{claim}

\begin{proof}
We start with a conditional rotation on the bit encoding state, see Theorem \ref{thm:conditionrotation}, and obtain:
\begin{equation}
    \frac{1}{\sqrt{d}}\sum_{j=1}^{d}\ket{j}\ket{x_j}\ket{0}
    \mapsto     
    \frac{1}{\sqrt{d}}\sum_{j=1}^{d}\ket{j}\ket{x_j}\left(\frac{x_j}{\eta}\ket{0} + \sqrt{1-\frac{x_j^2}{\eta^2}}\ket{1}\right)
\end{equation}
where $\eta$ is an upper bound of $\{x_j\}_{j\in[d]}$, or simply $\eta=1$ if $x$ is normalized. It would then suffice to measure the ancilla bit in the state $\ket{0}$ to end up with the desired state, with probability $P(0)$. This can also be done using amplitude amplification (Theorem \ref{thm:amplitude_amplification}). This second step has a complexity of $O(1/\sqrt{P(0)})$: 
\begin{equation}
    P(0) = \frac{1}{d}\sum_{j=1}^{d}\frac{x_j^2}{\eta^2}
    \leq \frac{\mathbb{E}(x_j^2)}{\eta^2}
\end{equation}
where $\mathbb{E}(x_j^2)$ is the expectation value, or average, of $\{x_j\}_{j\in[d]}$. We finally obtain the state
\begin{equation}
    \frac{1}{\sqrt{d}}\sum_{j=1}^{d}\alpha_j\ket{j}\ket{x_j}\ket{0}
\end{equation}
%from which we discard the last two quantum registers. (NO, they're entangled)
The new amplitudes $\alpha_j$ must be proportional to $x_j$, i.e. $\alpha_j = c x_j$. To respect the normalization, we must have $\frac{1}{d}\sum_{j}c^2x_j^2 = 1$, therefore $c=d/\sum_{j}x_j^2 = d/\norm{x}_2^2$. This shows that the remaining state is the amplitude encoding version of the vector $x$:
\begin{equation}
    \frac{1}{\norm{x}_2}\sum_{j=1}^{d}x_j\ket{j}\ket{x_j}
\end{equation}
Note that the output is close but not exactly the amplitude encoding of $x$ as defined in Defintion \ref{def:ampltiude_encoding}, but can be used in a similar manner. The last two registers are entangled and thus the last one cannot be simply discarded. The only solution is to apply $U^{\dagger}$, the reversed unitary that was used to create the bit encoding state. This adds another time $T_U$ in the computation but allows to obtain the exact amplitude encoding state $\frac{1}{\norm{x}_2}\sum_{j=1}^{d}x_j\ket{j}$. 

\end{proof}%\Hsquare

\subsubsection{Unary Encoding}\label{sec:unary_encoding}
As the previous encoding, unary encoding takes advantage of the amplitudes of quantum states in superposition to encode the components of a vector. However, a unary encoded quantum vector can be loaded with smaller circuits, as shown in Section \ref{sec:unary_data_loaders}.

The key feature of unary encoding is to use only the amplitude of unary states: the states that have one and only $\ket{1}$, e.g. $\ket{100\cdots0}$, $\ket{010\cdots0}$, etc.

\begin{definition}{Unary Encoding}{unary_encoding}
Given a vector $x=(x_0,\cdots,x_d) \in \R^d$, such that $\norm{x}_2=1$. We can encode it in a superposition of unary states:
\begin{equation} %ex \label{eq:x_unary_state}
    \ket{x} = 
    x_0\ket{10\cdots0} + x_1\ket{010\cdots0} + \cdots + x_{d-1}\ket{0\cdots01}
\end{equation}
We can also rewrite the previous state as:
\begin{equation}\label{eq:x_unary_state_notation}
\ket{x} = \sum^{d-1}_{i=0} x_i\ket{e_i}
\end{equation}
where $\ket{e_i}$ is the i$^{th}$ unary state with a $\ket{1}$ in the i$^{th}$ position e.g. $\ket{0\cdots010\cdots0}.$
\end{definition}

Note that if $x$ is not normalized, it is still possible to load it and each amplitude will naturally be divided by the norm. \\

This encoding is suited for short term quantum computers (NISQ) that are prone to errors. A very convenient consequence is the ability to perform \emph{error mitigation} while measuring the quantum states. Indeed,  Indeed, as we expect to obtain only quantum superposition of unary states, we can post-process our measurements and discard the ones that present non unary states (i.e. states with more that one qubit $\ket{1}$, or the ground state). The most expected error is a bit-flip between $\ket{1}$ and $\ket{0}$. The case where two bit-flips happened, which would pass through our error mitigation, is even less probable.

\subsubsection{Other Encodings}

Other encodings exist in the literature but are not used in this thesis. It is worth mentioning the \emph{gate encoding} used in variational quantum circuits (VQC), where the value of the input vector is directly put as the angle of rotation gate (note that it is close to unary loaders, see Section \ref{sec:unary_data_loaders}). \emph{Data reuploading}, the fact of repeating the gates that encode a vector, seems to add efficiency to such VQCs \cite{perez2020data}. Trying to understand how vectors are mapped into the high dimensional Bloch sphere, and what can be done to process the data \cite{schuld2021effect}.\\

Finally, there is also the Hamiltonian encoding which differs a lot since it consists of encoding the problem we desire to solve in a Hamiltonian form, and then try to perform the Hamiltonian evolution of an initial quantum state. This method is used in optimization, quantum annealing, QAOA type of VQCs \cite{farhi2014quantum}, but most importantly in Hamiltonian simulation, as used originally in the HHL algorithm \cite{HHL} where the matrix to be inverted is encoded as a Hamiltonian.

\subsection{Quantum Memory Models}\label{sec:quantum_memory_models}

A quantum memory model, or \emph{data loader}, is the link between classical data and quantum states. It is a classical structure, such as a table, a tree, or a list, where classical information is written. For each type of encoding we desire for the quantum state (see Section \ref{sec:encoding_data}), there exists one or several quantum circuits associated. In each case, it is important to differentiate the time to create the classical data structure, which should be done only once, and the time to load the quantum state. 

\begin{definition}{Quantum memory model}{}
    For a given type of quantum encoding, a quantum memory model is a classical data structure that stores vectors $X_i\in\R^{d}$ for $i\in[N]$. Along with a quantum circuit, it can perform the mapping:
    \begin{equation}
        \ket{i}\ket{0} \mapsto \ket{i}\ket{X_i}
    \end{equation}
\end{definition}

Note that starting with $\log(N)$ qubits in uniform superposition, all quantum states can be loaded as:
\begin{equation}
    \frac{1}{\sqrt{N}}\sum_{i=0}^{N-1}\ket{i}\ket{0} \mapsto \frac{1}{\sqrt{N}}\sum_{i=0}^{N-1}\ket{i}\ket{X_i}
\end{equation}

In the following, the number $i$ will be referred as the \emph{index} or sometimes the \emph{address}.

\subsubsection{Quantum Random Access Memory (QRAM)}

The QRAM, or Quantum Random Access Memory, is often used as a quantum memory model for amplitude encoding (see Definition \ref{def:ampltiude_encoding}), its name is derived from the classical RAM, for their equivalent underlying tree structure.   

\begin{figure}[h]
    \centering
    \includegraphics[width=0.8\textwidth]{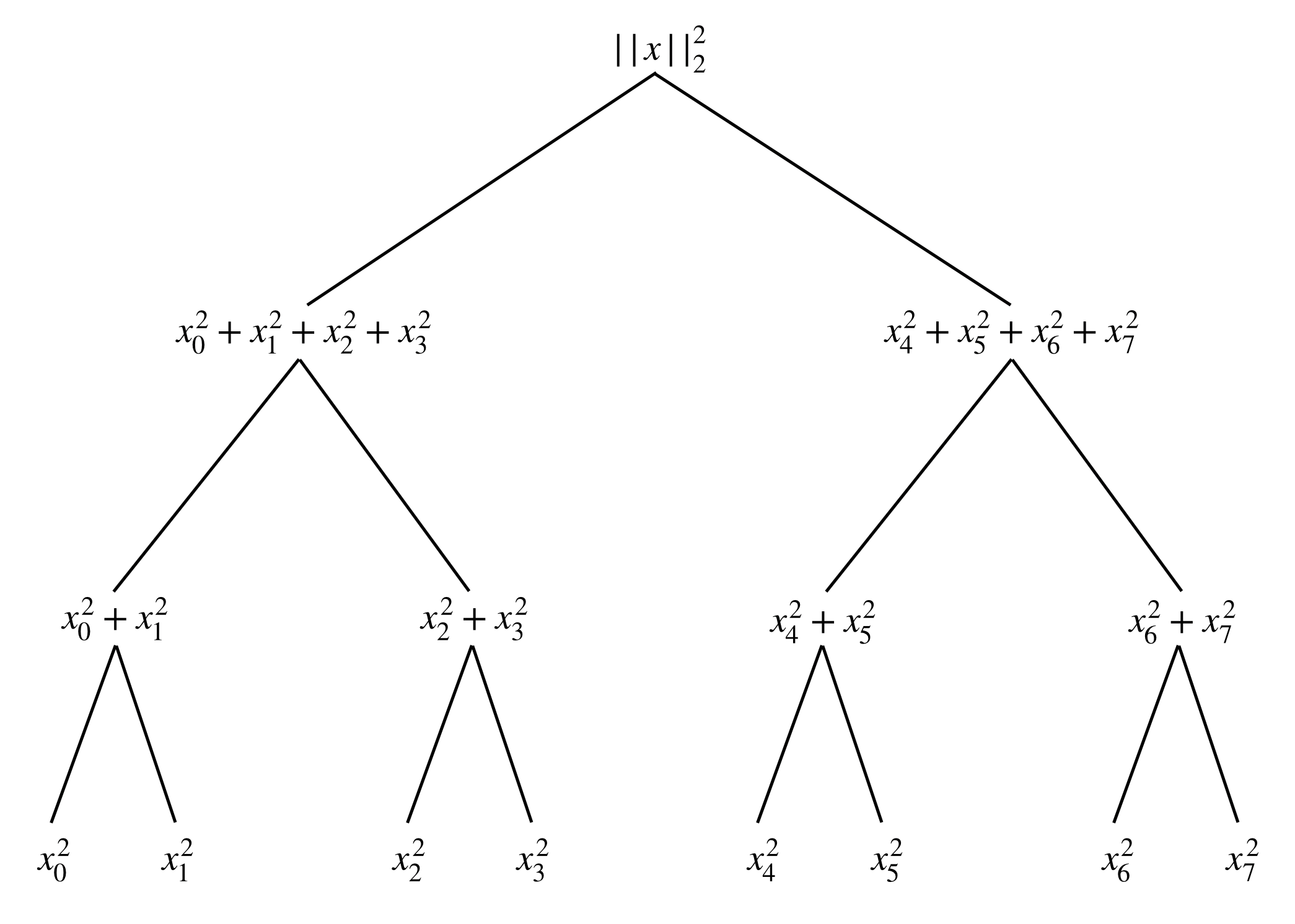}
    \caption{KP-tree structure for QRAM model on a 8-dimensional vector $x$.}
    \label{fig:QRAM_tree}
\end{figure}

For instance, an 8-dimensional vector $x = (x_0,\cdots,x_7)$ would be stored using a tree structure, also known as KP-tree \cite{prakash2014quantum, kerenidis_recommendation_system}, as shown in Fig.\ref{fig:QRAM_tree}. Compared to the number of dimensions, it has a linear number of leaves, but most importantly a logarithmic depth.

The time to create such a data structure is linear in the dimension $d$, and in the number $N$ of vectors to store, if applicable (see Theorem \ref{thm:QRAM}). 

\begin{figure}[h!]
    \centering
    \includegraphics[width=\textwidth]{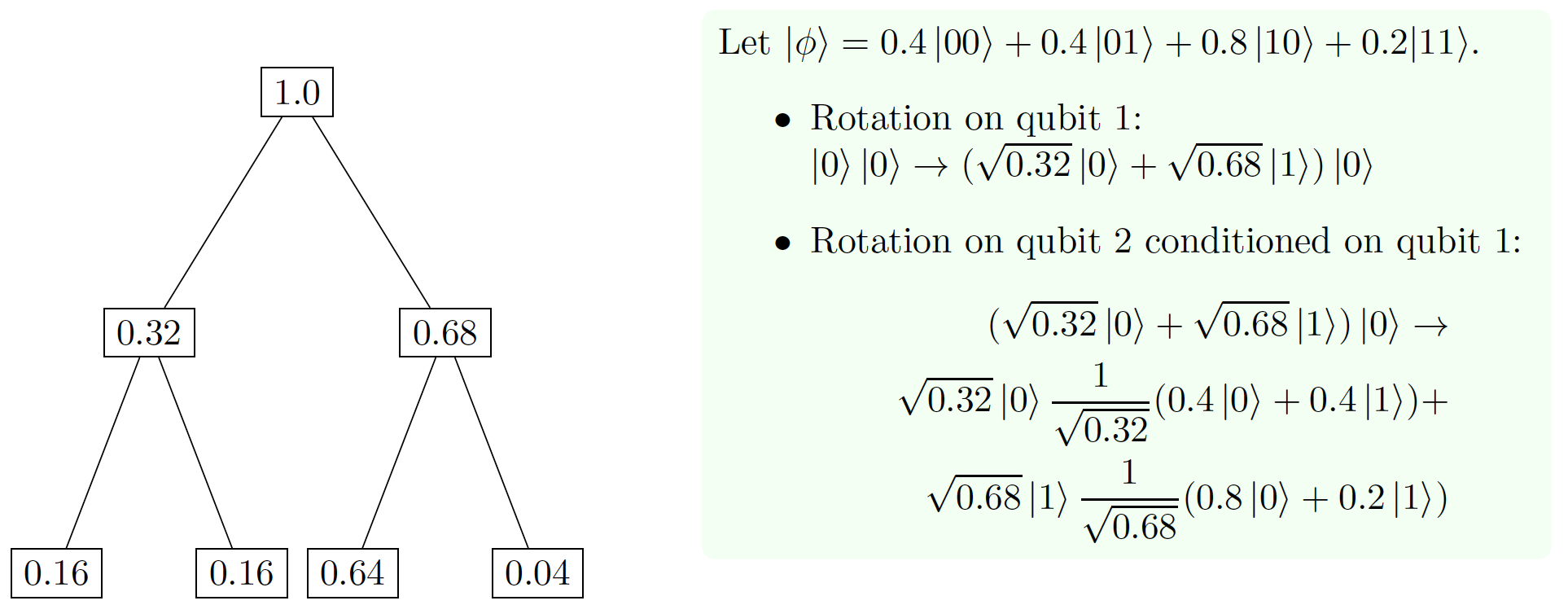}
    \caption{Example of sequential rotations to load the vector $\phi = (0.4, 0.4, 0.8, 0.2)$. Source: \cite{prakash2014quantum}.}
    \label{fig:QRAM_example_anupam}
\end{figure}

The interest of this framework is to compute angles that will be sequentially used to create $\ket{x}$, the amplitude encoding of $x$. Indeed, as shown in Fig.\ref{fig:QRAM_example_anupam} on a small example, each node of the tree has a relative weight that corresponds to the amount of rotation that should be applied. As a result, due to depth of the tree, the time to create the quantum state $\ket{x}$ is only $O(polylog(d))$ or, for $N$ such vectors, $O(polylog(Nd))$ (see Theorem \ref{thm:QRAM}). Note that some additional details must be taken care of to handle the signs of the components \cite{prakash2014quantum}.

For any matrix, we call \emph{quantum access} the ability to prepare, or \emph{load}, the amplitude encoding of the rows of the matrix. 

\begin{definition}{Quantum Access to Data}{quantum_access}
We say that we have quantum access to a matrix $X\in \R^{N\times d}$ if there exists a procedure to perform the following mapping, for $i \in [N]$:
\begin{itemize}
    \item $\ket{i}\ket{0} \mapsto \ket{i}\ket{X_i}$
    \item $\ket{0} \mapsto \frac{1}{\norm{X}_F}\sum_{i} \norm{X_i}_2\ket{i}$
\end{itemize}
\end{definition}

\begin{theorem}{QRAM}{QRAM}
	Let $X \in \mathbb{R}^{N \times d}$ be a matrix $N$ vectors of dimension $d$. There is a data structure to store the rows of $X$ such that, 
\begin{enumerate}
    \item The size of the structure is $O(Nd\log^2(Nd))$. 
    \item The time to store a row $X_i$ is $O(d\log^2(Nd))$, and the time to store the whole matrix $X$ is thus $O(Nd \log^2(Nd))$.
	\item The time to insert, update or delete a single entry $X_{ij}$ is $O(\log^{2}(Nd))$. 	
	\item A quantum algorithm with access to the data structure can perform the following unitaries (in superposition if necessary) in time $O(\log^{2}(Nd))$. 
	\begin{enumerate} 
		\item $\ket{i}\ket{0} \to \ket{i}\ket{X_{i}} $ for $i \in [n]$. 
		\item $\ket{0} \to \frac{1}{\norm{X}_F}\sum_{i \in [n]} \norm{X_{i}}_2\ket{i}$. 
	\end{enumerate}
\end{enumerate}
\end{theorem}

In practice, each component of the vectors $X_i$ will have to be stored using classical bits. The number of bits $k$ used for the precision of each value is not shown in Theorem \ref{thm:QRAM} but should appear simply as a multiplicative factor. 
%Finally, note that the polylogarithmic time to load the vectors, and to update a single entry in the data structure, is the best one could hope for.\\ 

Recently, using the QRAM data structure, the \emph{Block Encoding} framework was introduced by \cite{chakraborty2018power, gilyen2019quantum}, which allows us further improvements on quantum linear algebra subroutines (see Section \ref{sec:matrix_multiplication_quantum}). 

\begin{definition}{Block Encoding of a Matrix}{block_encoding}
    For a symmetric matrix $X\in\R^{n\times n}$, the $q$ qubits unitary $U\in\mathbb{C}^{2^q\times 2^q}$ is a $(\zeta,q)$-block encoding of $M$ if $U = \begin{pmatrix}
    X/\zeta & \cdot \\
    \cdot & \cdot
    \end{pmatrix}$. For a general matrix $M \in \R^{n \times m}$, we use a symmetrized version $\overline{P} = \begin{pmatrix}
    0 & M \\
    M^T & 0
    \end{pmatrix}$ to construct a block encoding for it.  
\end{definition}

QRAM then allows us to store and load such block encoding. In detail, for a matrix $X\in\R^{n\times n}$ it can implement in $\widetilde{O}(\log(n))$ a $(\zeta(X),2\log(n))$-block encoding with $\zeta(X) = \frac{1}{\norm{X}_2}\min(\norm{X}_F,s_1(X))$ where $s_1(X) = \max_i(\sum_j|X_{ij}|)$. As before, the storing takes a single pass over the matrix $X$, but a single update takes only $O(\log^2(n))$.\\

In the first proposals of a QRAM \cite{giovannetti2008quantum,giovannetti2008architectures}, the authors assumed access to a hardware platform that could naturally encode data into amplitudes. The circuit requires $O(d)$ qubits and $\log(d)$ depth, but necessitates specific hardware with light-matter interaction. On the other hand, one could also compose a controlled-$NOT$ based multiplexer with only $O(\log(d))$ qubits, which remains impractical since it requires performing a sequence of $d\log(d)$ controlled gates \cite{park2019circuit}. New ideas were suggested in \cite{arunachalam2015robustness_qram} with $O(d)$ depth and $O(d)$ qubits, with strong tolerance to noise and quantum errors. This last proposal encodes the state in $\log(d)$ qubits but requires $d$ additional classical bits to load the $d$ values. Resource estimation and resilience against noise have been studied in depth \cite{di2020fault, hann2021resilience_qram, zhao2018note}. They reveal that the task of creating such a circuit and use it efficiently remains a strong challenge. However, in theory, this should be no more difficult than building the fault-tolerant quantum computer itself. 

In the next section, we will introduce memory model that require only $\log(d)$ depth for $d$ qubits, but using unary encoding this time.\\

Finally, note that different methods were proposed for loading amplitude encoding for near term devices, with the help of variational circuits \cite{zhang2020toward}. However, they remain imperfect and too costly for now.

\subsubsection{Unary Data Loaders}\label{sec:unary_data_loaders}

The QRAM model for amplitude encoding might only be available for long term quantum computers. Therefore, shorter term data loaders were proposed in \cite{johri2020nearest_dataloaders}, suited for unary encoding (see Definition \ref{def:unary_encoding}). Since these quantum states are a superposition of unary states, i.e. states with one and only qubit in state $\ket{1}$, circuits with $O(d)$ qubits and depth ranging from $O(\log(d))$ to $O(d)$ can be proposed. 

They rely on the Reconfigurable Beam Splitter gate, or $RBS$ gate for short. This two-qubit gate is parametrizable with one angle $\theta \in [0,2\pi]$. Its matrix representation is given as:
\begin{equation}\label{eq:RBSgate} %ex \label{RBSgate}
RBS(\theta) = \begin{pmatrix}
1 & 0           & 0            & 0 \\
0 & \cos\theta & \sin\theta & 0 \\
0 & -\sin\theta & \cos\theta  & 0 \\
0 & 0           & 0            & 1
\end{pmatrix}
\end{equation}
We note that this gate leaves the states $\ket{00}$ and $\ket{11}$ unaffected. For the two other states, it equivalent to a planar rotation with angle $\theta$:
\begin{equation}
RBS(\theta) : 
 \begin{cases}
\ket{01} \mapsto \cos\theta\ket{01}-\sin\theta\ket{10}\\
\ket{10} \mapsto \sin\theta\ket{01}+\cos\theta\ket{10}\\
\end{cases}
\end{equation}
We can think of this gate as a rotation in the two-dimensional subspace spanned by the basis $\{\ket{01},\ket{10}$, while it acts as the identity in the remaining subspace $\{\ket{00},\ket{11}\}$. Or equivalently, starting with two qubits, one in the $\ket{0}$ state and the other one in the state $\ket{1}$, the qubits can be swapped or not in superposition. The qubit $\ket{1}$ stays on its wire with amplitude $\cos\theta$ or switches with the other qubit with amplitude $+\sin\theta$ if the new wire is below ($\ket{10}\mapsto\ket{01}$) or $-\sin\theta$ if the new wire is above ($\ket{01}\mapsto\ket{10}$). Note that in the two other cases ($\ket{00}$ and $\ket{11}$) the $RBS$ gate acts as identity.

\begin{figure}[h]
    \centering
    \includegraphics[width=0.85\textwidth]{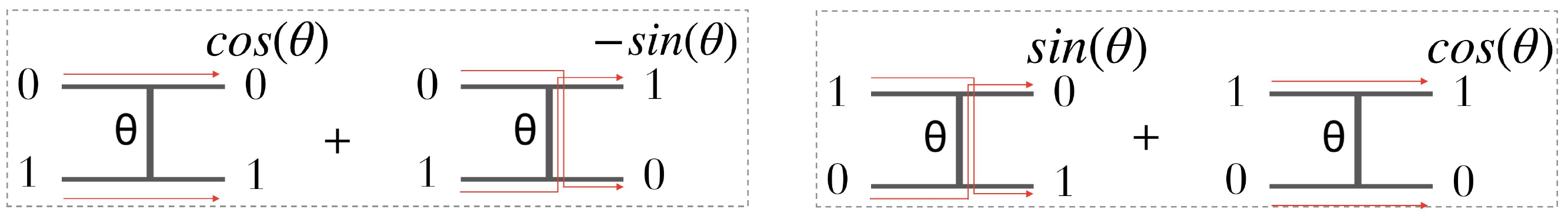}
    \caption{Representation of the quantum mapping from Eq.(\ref{eq:RBSgate}) on two qubits.}
    \label{fig:RBS_circuit_details}
\end{figure}

\begin{figure}[h]
\centering
\begin{subfigure}{.42\textwidth}
  \centering
  \includegraphics[width=\linewidth]{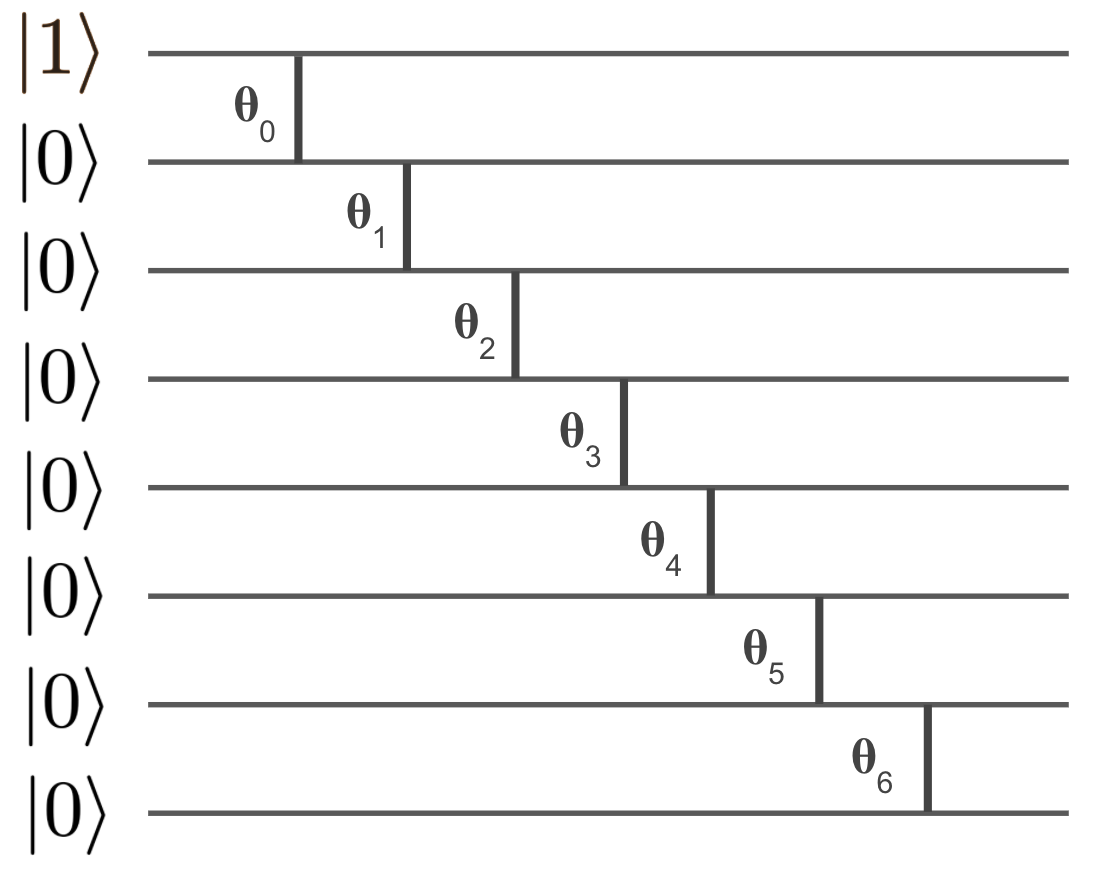}
  \caption{}
  \label{fig:unary_loader_diagonal}
\end{subfigure}\hspace{0.10\textwidth}%add horizontal space
\begin{subfigure}{.35\textwidth}
  \centering
  \includegraphics[width=\linewidth]{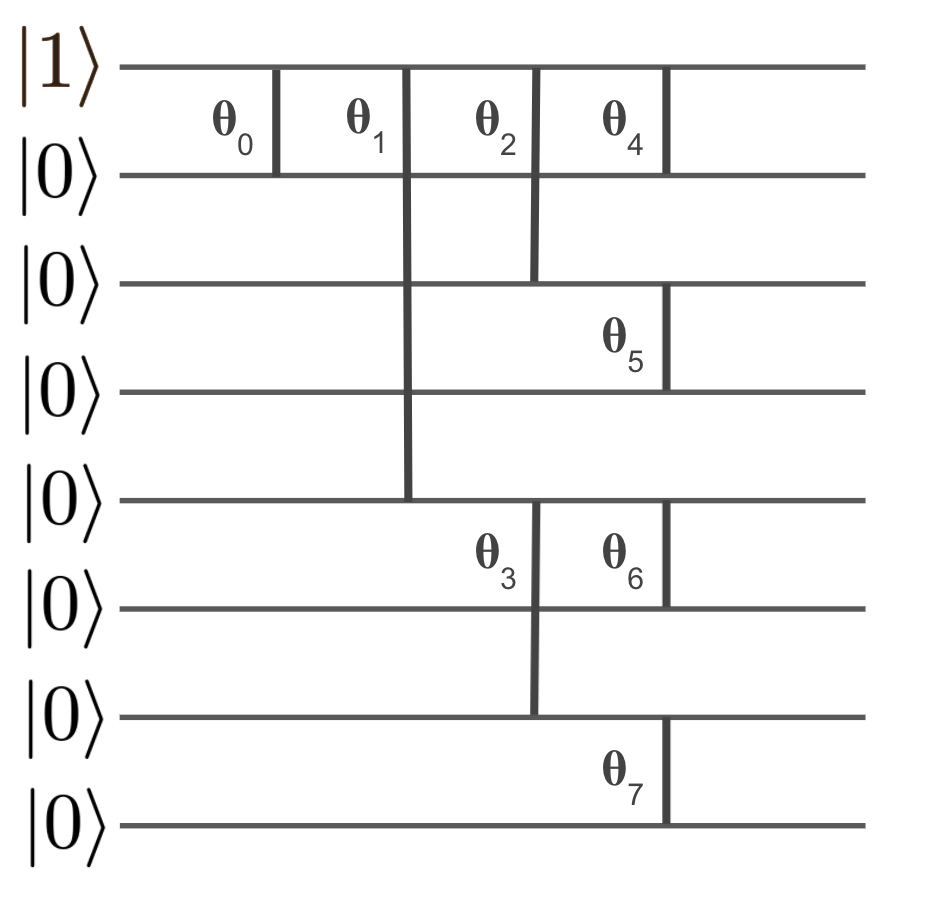}
  \caption{}
  \label{fig:unary_loader_parallel}
\end{subfigure}
\caption{Two quantum circuits for unary data loaders with (a) diagonal and (b) parallel structure. Each vertical bar is a $RBS$ gate applied on the two connected qubits, with angle $\theta_i$.}
\label{fig:unary_loader}
\end{figure}

Given a vector $x = (x_1,\cdots,x_d)\in\R^d$, the associated unary data loaders is a simple circuit using $d$-1 $RBS$ gates, along with the same number of precomputed angles $\theta_i$. In Chapter \ref{chap:OrthoNN_nisq}, we will use a diagonal circuit of depth $O(d)$, as shown on Fig.\ref{fig:unary_loader_diagonal}. This circuit has longer depth but has the property of using only adjacent qubits, which is convenient when using the current quantum computers available. The other circuit, a \emph{parallel} loader, shown on Fig.\ref{fig:unary_loader_parallel}, has a depth of $O(\log(d))$. An important advantage compared to the classical inner that uses $O(d)$ steps.

A trade-off is possible between the number of qubits and the depth, the product of the two remaining constant. The optimal solution would be a circuit of $\sqrt{d}\log(d)$ depth and $2\sqrt{d}$ qubits. This could also be extended to loading matrices. These unary loaders can be combined to apply linear algebra tasks such as inner product \cite{johri2020nearest_dataloaders} or matrix multiplication (see Chapter \ref{chap:OrthoNN_nisq}). 

These unary data loaders can be easily implemented by classical emulator as they consist of planar rotations, and are therefore not exploring an exponentially large Hilbert space.\\

The creation of the angles $\theta_i$ for $i\in[d-1]$ is an easy task, requiring $O(d)$ classical precomputations. For instance, for the diagonal unary loader (Fig.\ref{fig:unary_loader_diagonal}), we recursively obtain $d$-1 loading angles with: 

\begin{equation}
\centering
\begin{cases}
      \theta_0 = \arccos(x_0)\\
      \theta_1 = \arccos\left(\frac{x_1}{\sin(\theta_0)}\right)\\
      \theta_2 = \arccos\left(\frac{x_2}{\sin(\theta_0)\sin(\theta_1)}\right) \\
      \cdots
    \end{cases} 
\label{eq:data_loader_angles}
\end{equation}

Indeed, the diagonal unary loader starts in the all $\ket{0}$ state and flips the first qubit using an $x$ gate, in order to obtain the unary state $\ket{10\cdots 0}$ as shown in Fig.\ref{fig:data_loader}. Then a cascade of $RBS$ gates allows creating the state $\ket{x}$. The first gate will propagate the amplitude of the first qubit to the second one: 
\begin{equation}
x_0\ket{100\cdots} 
+ \sin(\theta_0)\ket{010\cdots}
\end{equation}
The second gate will create in turn the state :
\begin{equation}
x_0\ket{100\cdots} 
+ x_1\ket{010\cdots}
+ \sin(\theta_0)\sin(\theta_1)\ket{001\cdots}
\end{equation}
and so on, until obtaining $\ket{x}$ as in Eq.(\ref{eq:x_unary_state}).\\

\begin{figure}[h]
\centering
\begin{subfigure}{.48\textwidth}
  \centering
  \includegraphics[width=\linewidth]{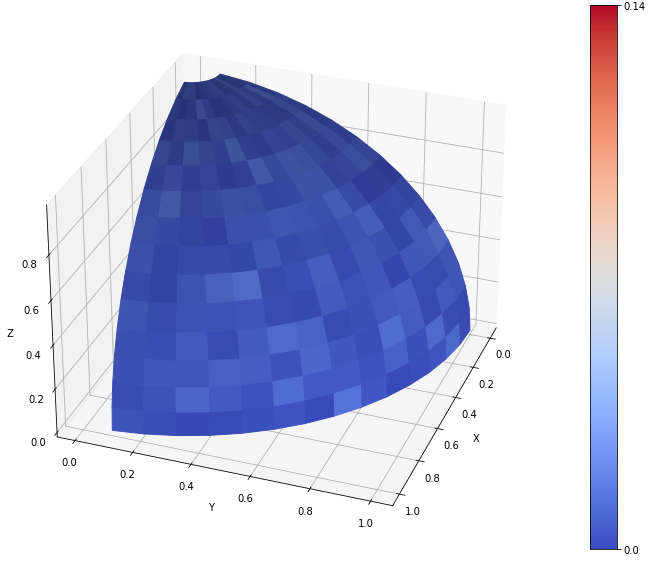}
  \caption{}
  \label{fig:ibm_diag_loader_simu}
\end{subfigure}\hspace{0.02\textwidth}%add horizontal space
\begin{subfigure}{.48\textwidth}
  \centering
  \includegraphics[width=\linewidth]{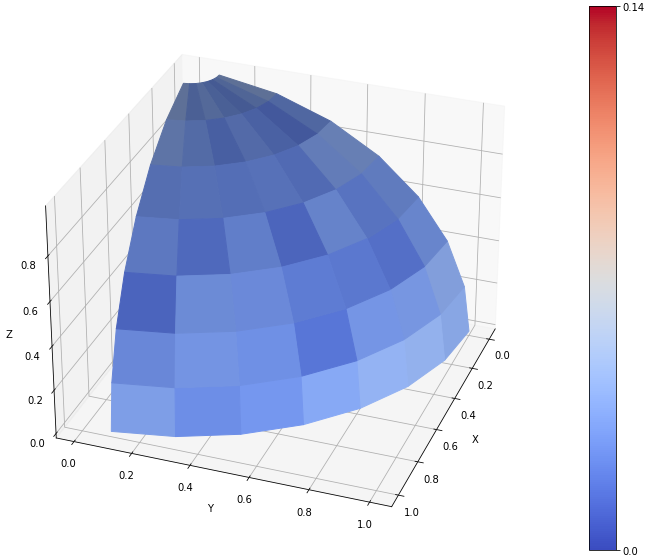}
  \caption{}
  \label{fig:ibm_diag_loader_real}
\end{subfigure}
\caption{$\ell_2$ norm errors when comparing vectors of the 3D unit sphere and their quantum version by using diagonal unary loaders. (a) Emulated results and (b) Actual hardware experiment on the $IBM Santiago$ quantum computer. May 2021.}
\label{fig:ibm_diag_loader}
\end{figure}

Finally, to verify the accuracy of these encoding, we performed a real hardware implementation of the unary diagonal loader. We compared the results with quantum circuit simulations. Using a 5 qubits superconducting quantum computer $IBM Santiago$, we were able to load unary encoded vectors on the 3-dimensional unit sphere. For this, we used vectors of the form $(x_0,x_1,x_2) \in [0,1]^3$. For several points, we created the quantum state, measured 8192 samples of it to recover its position, and calculated the euclidean error compared to the actual vector. Results are shown on Fig.\ref{fig:ibm_diag_loader}. Naturally, we see more errors (light blue coloring) for the real experiment, but the results seem consistent with the simulations.

\section{From Quantum to Classical data}\label{sec:quantum_to_classical_to_data}

For quantum linear algebra and machine learning applications, recovering classical data from quantum states is the other side of the coin. This process, called tomography, is usually very costly and is the second bottleneck of quantum algorithms \cite{readthefineprint}, after the data loading (see Section \ref{sec:encoding_data}). 

In this section we will present two different tomography procedures, both for amplitude encoded quantum vectors, the second one being a contribution of this thesis.

\subsection{$\ell_2$ and $\ell_{\infty}$ Tomography}\label{sec:l_2_and_l_infinite_tomography}

We consider a final state of a quantum circuit, expected to be $\ket{x} = \frac{1}{c}\sum^{d-1}_{i=0} x_i\ket{i}$, the quantum version of an unknown vector $x=(x_0,\cdots,x_{d-1})\in\R^d$. Measuring this state will randomly result in one of the binary strings $\ket{i}$. Performing a sufficiently large number of measurements would allow us to guess the underlying probability distribution of the binary strings, and therefore the amplitudes $(|x_0|^2,\cdots,|x_{d-1}|^2)$. As the number of measurements is not infinite, the probability distribution is only approximated and therefore an error is committed on the recovered vector, in particular for small value components. Depending on the guarantee we put on this error, the number of queries is modified. 

We first present an $\ell_2$-norm guarantee tomography, introduced in \cite{kerenidis2020quantum_IPM}. Informally, for some parameter $\delta>0$, we require that the resulting vector $\tilde{x}$ is $\delta$-close to the actual vector $x$. 

\begin{theorem}{$\ell_2$ Vector State Tomography}{tomography_l2} 
Given access to unitary $U$ such that $U\ket{0} = \ket{x}$ and its controlled version in time $T(U)$, there is a tomography algorithm with time complexity $O(T(U) \frac{ d \log d  }{\delta^{2}})$ that produces unit vector $\widetilde{x} \in \R^{d}$ such that $\norm{\widetilde{x}  - x }_{2} \leq \delta$ with probability at least $(1-1/poly(d))$. 
\end{theorem} 

Next, we introduce a new procedure, the $\ell_{\infty}$-norm guarantee tomography, where now each recovered component $\overline{x}_i$ must be  $\delta$-close to the actual vector's component $x_i$. Noticeably, this tomography requires exponentially fewer resources than the previous one.

\begin{theorem}{$\ell_{\infty}$ Vector State Tomography}{tomography_linfinity} % ex- \label{thm:tom}
Given access to unitary $U$ such that $U\ket{0} = \ket{x}$ and its controlled version in time $T(U)$, there is a tomography algorithm with time complexity $O(T(U) \frac{ \log d  }{\delta^{2}})$ that produces unit vector $\widetilde{x} \in \R^{d}$ such that $\norm{\widetilde{x}  - x }_{\infty} \leq \delta$ with probability at least $(1-1/poly(d))$. 
\end{theorem}

In some contexts, the $\ell_{\infty}$ tomography turns out to be more meaningful. 
In fact, when the quantum state $\ket{x}$ does not represent a vector per se, as a position on a mesh or some precise embedding in a latent space, but rather a collection of values like the pixels of an image or a time series, it is in practice less critical to have low precision on some components.
In some cases as well, we only care about the high-value elements of $x$, for instance in neural networks, where non-linearities are applied to push high values even higher, and downsize small values (see Section \ref{sec:classical_fcnn_intro}). 

For instance, in the case of visual neural networks (see Chapter \ref{chap:QCNN}), we will use this tomography to recover the highest valued pixels in an image.  

\begin{figure}[h]
    \centering
    \includegraphics[width=0.85\textwidth]{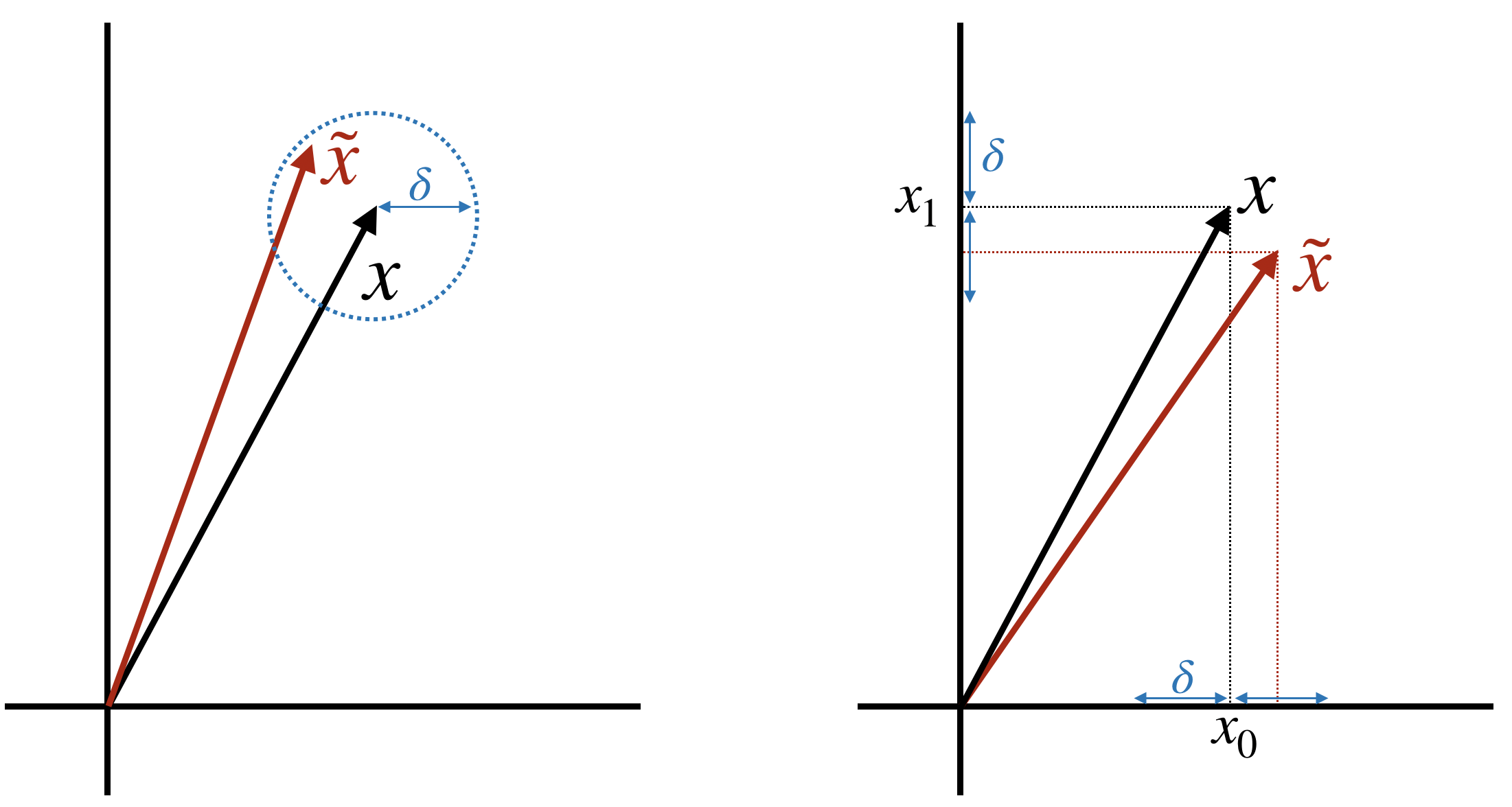}
    \caption{Representation of $\ell_{2}$ (left) and $\ell_{\infty}$ (right) error guarantees.}
    \label{fig:l2_vs_linfnty_error}
\end{figure}

\subsection{$\ell_{\infty}$ Tomography Details}\label{sec:l_infinite_proof}

To prove the Theorem \ref{thm:tomography_linfinity} introduced in this thesis, we follow the method from \cite{kerenidis2020quantum_IPM}. In the following we consider a quantum state $\ket{x}$ such that $x \in \R^d$ and $\norm{x}_2=1$.

\begin{algorithm} [h]
\caption{$\ell_{\infty}$ norm tomography} \label{alg:tom}
\begin{algorithmic}[1]
\REQUIRE   Error $\delta > 0$, access to unitary $U : \ket{0} \mapsto \ket{x} = \sum_{i \in [d]}x_i\ket{i}$, the controlled version of $U$, QRAM access. 
\ENSURE Classical vector $\widetilde{x} \in \R^{d}$, such that $\norm{\widetilde{x}}=1$ and $\norm{\widetilde{x}-x}_{\infty} < \delta$.
\vspace{10pt} 
\STATE Measure $N = \frac{36\ln{d}}{\delta^2}$ copies of $\ket{x}$ in the standard basis and count $n_i$, the number of times the outcome $i$ is observed. Store $\sqrt{p_i} = \sqrt{n_i/N}$ in QRAM data structure. 
\STATE Create $N = \frac{36\ln{d}}{\delta^2}$ copies of the state $\frac{1}{\sqrt{2}}\ket{0}\sum_{i \in [d]}x_i\ket{i} + \frac{1}{\sqrt{2}}\ket{1}\sum_{i \in [d]}\sqrt{p_i}\ket{i}$. 
\STATE Apply an Hadamard gate on the first qubit to obtain\\
\begin{equation}\label{eq:phi_state_tomography_proof}
\ket{\phi} = \frac{1}{2}\sum_{i \in [d]}\left( (x_i + \sqrt{p_i})\ket{0,i} + (x_i - \sqrt{p_i})\ket{1,i}\right)
\end{equation}
 \STATE Measure both registers of each copy in the standard basis, and count $n(0,i)$ the number of time the outcome $(0,i)$ is observed. 
 \STATE Set $\sigma(i)=+1$ if $n(0,i) > 0.4Np_i$ and $\sigma(i)=-1$ otherwise. 
 \STATE Output the unit vector $\widetilde{x}$ such that $\forall i \in [N], \widetilde{x}_i = \sigma_i\sqrt{p_i}$
\end{algorithmic}
\end{algorithm}

\noindent The following version of the Chernoff Bound will be used for analysis of algorithm \ref{alg:tom}. 

\begin{theorem}{Chernoff Bound}{}
Let $X_j$, for $j \in [N]$, be independent random variables such that $X_j \in [0,1]$ and let $X = \sum_{j \in [N]}X_j$. We have the three following inqualities:
\begin{enumerate}
  \item For $0<\beta<1,  \PP[ X < (1-\beta)\E[X]] \leq e^{-\beta^2\E[X]/2}$
  \item For $\beta>0,  \PP[ X > (1+\beta)\E[X]] \leq e^{-\frac{\beta^2}{2+\beta}\E[X]}$
  \item For $0<\beta<1,  \PP[ |X - \E[X]| \geq \beta\E[X]] \leq e^{-\beta^2\E[X]/3}$, by composing $1.$ and $2.$\\
\end{enumerate}
\end{theorem} 

\begin{theorem}{}{}
Algorithm \ref{alg:tom} produces an estimate $\widetilde{x} \in \R^{d}$ such that $\norm{\widetilde{x}-x}_{\infty} < (1+\sqrt{2}) \delta$ with probability at least $1 - \frac{1}{ d^{0.83}}$. 
\end{theorem} 

\begin{proof} 
\noindent Proving $\norm{x-\widetilde{x}}_{\infty} \leq O(\delta)$ is equivalent to showing that for all $i \in [d]$, we have $|x_i - \widetilde{x}_i| = |x_i - \sigma(i)\sqrt{p_i}| \leq O(\delta)$. Let $S$ be the set of indices defined by $S = \{i \in [d] ; |x_i|>\delta\} $. We will separate the proof for the two cases where $i \in S$ and $i \notin S$.

\paragraph{Case 1 : $i \in S$. \\}
We will show that if  $i \in S$, we correctly have $\sigma(i) = sgn(x_i)$ with high probability. Therefore  we will need to bound $|x_i - \sigma(i)\sqrt{p_i}| = ||x_i| - \sqrt{p_i}|$.

We suppose that $x_i>0$. The value of $\sigma(i)$ correctly determines $sgn(x_i)$ if the number of times we have measured $(0,i)$ at Step 4 is more than half of the measurements, i.e. $n(0,i) > \frac{1}{2}\E[n(0,i)]$. If $x_i<0$, the same arguments holds for $n(1,i)$. We consider the random variable that represents the outcome of a measurement on state $\ket{\phi}$. The Chernoff Bound (part 1) with $\beta=1/2$ gives

\begin{equation}\label{eq:exp}
\PP[n(0,i)\leq \frac{1}{2}\E[n(0,i)]]\leq e^{-\E[n(0,i)]/8}
\end{equation}
\noindent From the definition of $\ket{\phi}$, see Eq.(\ref{eq:phi_state_tomography_proof}),  we have $\E[n(0,i)] = \frac{N}{4}(x_i +\sqrt{p_i})^2$. We will lower bound this value with the following argument.

For the $k^{th}$ measurement of $\ket{x}$, with $k \in [N]$, let $X_k$ be a random variable such that $X_k=1$ if the outcome is $i$, and $0$ otherwise. We define $X = \sum_{k \in [N]}X_k$. Note that $X = n_i = N p_i$ and $\E[X] = N x_i^2$. We can apply the Chernoff Bound, part 3 on $X$ for $\beta = 1/2$ to obtain, 

\begin{equation}\label{eq:chernoff3}
\PP[|X - \E[X]| \geq  \E[X]/2] \leq e^{-\E[X]/12}
\end{equation}

\begin{equation}
\PP[|x_i^2 - p_i| \geq x_i^2/2] \leq e^{-Nx^2_i/12}
\end{equation}

We have $N = \frac{36 \ln{d}}{\delta^2}$ and by assumption $x_i^2 > \delta^2$ (since $i \in S$). Therefore,

\begin{equation}
\PP[|x_i^2 - p_i| \geq x_i^2/2] \leq e^{-36\ln{d}/12} = 1/d^3
\end{equation}

This proves that the event $|x_i^2 - p_i| \leq x_i^2/2$ occurs with probability at least $1-\frac{1}{d^3}$ if $i \in S$. This previous inequality is equivalent to $\sqrt{2p_i/3} \leq |x_i| \leq \sqrt{2p_i}$. Thus, with high probability we have 
$\E[n(0,i)] = \frac{N}{4}(x_i +\sqrt{p_i})^2 \geq 0.82 N p_i$, since $\sqrt{2p_i/3} \leq |x_i|$. Moreover, since $|p_i| \leq x_i^2/2$, $\E[n(0,i)] \geq 0.82 Nx_i^2/2 \geq 14.7\ln{d}$. Therefore, equation \eqref{eq:exp} becomes

\begin{equation}
\PP[n(0,i)\leq 0.41 N p_i]\leq e^{-1.83\ln{d}} = 1/d^{1.83}
\end{equation}
We conclude that for $i \in S$, if $n(0,i) > 0.41 N p_i$, the sign of $x_i$ is determined correctly by $\sigma(i)$ with high probability $1- \frac{1}{d^{1.83}}$, as indicated in Step 5. 

We finally show $|x_i - \sigma(i)\sqrt{p_i}| = ||x_i| - \sqrt{p_i}|$ is bounded. Again by the Chernoff Bound (3.) we have, for $0<\beta<1$:

\begin{equation}
\PP[|x_i^2 - p_i| \geq \beta x_i^2] \leq e^{\beta^2Nx^2_i/3}
\end{equation}

By the identity $|x_i^2 - p_i| = (|x_i| - \sqrt{p_i})(|x_i| + \sqrt{p_i}) $ we have

\begin{equation}
\PP\left[\Big||x_i| - \sqrt{p_i}\Big| \geq \beta \frac{x_i^2}{|x_i| + \sqrt{p_i}}\right] \leq e^{\beta^2Nx^2_i/3}
\end{equation}

Since $\sqrt{p_i} > 0$, we have $\beta \frac{x_i^2}{|x_i| + \sqrt{p_i}} \leq \beta \frac{x_i^2}{|x_i|} = \beta |x_i|$, therefore $\PP\left[\Big||x_i| - \sqrt{p_i}\Big| \geq \beta |x_i|\right] \leq e^{\beta^2Nx^2_i/3}$. Finally, by chosing $\beta = \delta/|x_i| < 1$ we have

\begin{equation}
\PP\left[\Big||x_i| - \sqrt{p_i}\Big| \geq \delta \right] \leq e^{36\ln{d}/3} = 1/d^{12}
\end{equation}

We conclude that, if $i \in S$, we have $|x_i - \tilde{x}_i| \leq \delta$ with high probability. 

Since $|S| \leq d$, the probability for this result to be true for all $i \in S$ is $1 - \frac{1}{d^{0.83}}$. This follows from the Union Bound on the correctness of $\sigma(i)$.

\paragraph{Case 2 : $i \notin S$. \\}
If $i \notin S$, we need to separate again in two cases. When the estimated sign is wrong, i.e. $\sigma(i) = -sgn(x_i)$, we have to bound $|x_i - \sigma(i)\sqrt{p_i}| = ||x_i| + \sqrt{p_i}|$. On the contrary, if it is correct, i.e. $\sigma(i) = sgn(x_i)$, we have to bound $|x_i - \sigma(i)\sqrt{p_i}| = ||x_i| - \sqrt{p_i}| \leq ||x_i| + \sqrt{p_i}|$. Therefore only one bound is necessary. 

We use Chernoff Bound (2.) on the random variable X with $\beta > 0$ to obtain

\begin{equation}
\PP[ p_i > (1+\beta)x_i^2] \leq e^{\frac{\beta^2}{2+\beta}Nx_i^2}
\end{equation}

We chose $\beta = \delta^2/x_i^2 $ and obtain $\PP[ p_i > x_i^2 + \delta^2] \leq e^{\frac{\delta^4}{3\delta^2}N} = 1/d^{12}$. Therefore, if $i \notin S$, with very high probability $1 - \frac{1}{d^{12}}$ we have  $p_i \leq x_i^2 + \delta^2 \leq 2\delta^2$. We can conclude and bound the error:
\begin{equation}
|x_i - \tilde{X}_i| \leq ||x_i| + \sqrt{p_i}| \leq \delta+\sqrt{2}\delta = (1+\sqrt{2})\delta
\end{equation}
Since $|\overline{S}| \leq d$, the probability for this result to be true for all $i \notin S$ is $1 - \frac{1}{d^{11}}$. This follows from applying the Union Bound on the event $p_i > x_i^2 + \delta^2$. 
\end{proof}%\Hsquare 

\chapter{Quantum Linear Algebra}\label{chap:quantum_linalg}
\epigraph{\textit{"The purpose of computation is insight, not numbers."}}{Richard Hamming \\ \emph{Numerical Methods for Scientists and Engineers} (1962)}

Throughout this thesis, we will use previously developed quantum linear algebra subroutines. These fundamental algorithms harness quantum encoding seen in Chapter \ref{chap:quantum_data} to perform their tasks with substantial speedup. 

We will present succinctly two of them: the singular value estimation (Section \ref{sec:singular_value_estimation_quantum}), and the matrix multiplication (Section \ref{sec:matrix_multiplication_quantum}).

\section{Singular Values Estimation and Projection}\label{sec:singular_value_estimation_quantum}

\subsubsection{Singular Value Decomposition (SVD)}

We recall some properties from Section \ref{sec:math_notations}. For any matrix $A \in \R^{m\times n}$, the singular value decomposition (SVD) of $A$ is 
\begin{equation}
A = U\Sigma V^T
\end{equation}
where $U \in \R^{m\times m}$ and $V \in \R^{n\times n}$ are unitary matrices, and $\Sigma \in \R^{m\times n}$ is a rectangular diagonal matrix with non-negative elements $\sigma_i$. With $r$ being the rank of $A$, the SVD can be also expressed as:
\begin{equation}
A = \sum_{i=1}^r \sigma_i u_i v_i^T
\end{equation}
where the left and the right singular vectors $u_i$ and $v_i$ are the columns of U and V.

The singular values $\sigma_i$ of a matrix $A$ of great importance to understand its properties, especially when the matrix is a transformation from one space to another, or when it is representing a graph (see Section \ref{sec:classical_spectral_clustering}). 

The singular values of a $m\times n$ matrix $A$ are the square roots of the eigenvalues of the $n\times n$ matrix $A^TA$. Thus, if $A$ is a $n\times n$, real and positive semidefinite matrix, the singular values and the eigenvalues are the same, which is not generally the case due to negative signs.

\subsubsection{Quantum Singular Value Estimation (SVE)}

A quantum algorithm for singular value estimation (SVE) was developed in \cite{kerenidis_recommendation_system} for solving the recommendation system problem, and later improved in \cite{kerenidis2020_gradient_descent}, inspired by the method of \cite{childs2010relationship} for solving eigenvalue estimation. They extended to non unitary matrices the method used in \cite{HHL} from extracting eigenvalues.

Given a matrix $A$ stored in the appropriate quantum memory model (see Section \ref{sec:quantum_memory_models}), the algorithm can map the quantum state of any right singular vector $\ket{v_i}$ to its singular value $\sigma_i$, with some precision $\epsilon>0$. The interesting feature comes when this is applied in superposition over all right singular vectors. And since they form a complete basis, any quantum vector $\ket{x}$ can be written as $\ket{x} = \sum_i \alpha_i \ket{v_i}$, for some coefficients $\alpha_i$. We begin by introducing the parameter $\mu(A)$ in Definition \ref{def:mu}.

\begin{definition}{Parameter $\mu(\cdot)$}{mu} %ex-\label{definitionmu} from spectral clust
For a matrix $A$, the parameter $\mu(A)$ is defined by 
\begin{equation}
\mu(A) = \min_{p\in[0,1]}\left(\norm{A}_F, \sqrt{s_{2p}(A)s_{2(1-p)}(A^T)}\right)
\end{equation}
 where $s_p(A) = \max_{i}(\norm{A_i}_p^p)$. 
%In general it is convenient to see it as $\mu(A) \leq \norm{A}_F$.
\end{definition}

\begin{theorem}{Quantum Singular Value Estimation}{SVE} %ex- \label{svethm} spectral clust
Given quantum access in time $T$ to a matrix $A\in\R^{m\times n}$ with singular value decomposition $A = \sum_i \sigma_i u_i v_i^T$, there is a quantum algorithm that performs the mapping 
\begin{equation}
\sum_i\alpha_i\ket{v_i}\ket{0} \mapsto \sum_i\alpha_i\ket{v_i}\ket{\overline{\sigma_i}}
\end{equation}
such that for any precision $\epsilon>0$, we have for all singular values $|\overline{\sigma_i} - \sigma_i|\leq \epsilon$, in time $\widetilde{O}(T\mu(A)/\epsilon)$, with probability at least $1-1/poly(n)$.
\end{theorem}

The parameter $\mu(A)$ will appear frequently in the running time of algorithms that use quantum linear algebra subroutines. For dense matrices, $\mu(A)$ can be taken to be the ratio Frobenius Norm / Spectral Norm of $A$. In some sense, it replaces the explicit dependence on the matrix dimension. Note that with $p=1/2$ we have $\mu(A) \leq s_1(A) = \max_i\norm{A_i}_1$. For sparse matrices, $\mu(A)$ can then be seen as the sparsity.\\

We informally give the details of the initial quantum SVE algorithm \cite{kerenidis_recommendation_system}. One starts by finding two matrices $P\in\R^{mn\times m}$ and $Q\in\R^{mn\times n}$ that form the decomposition $A/\norm{A}_F = P^TQ$, such that we have fast quantum access to $P$ and $Q$. Then we apply phase estimation (Theorem \ref{thm:phase_estimation}) on the unitary $W = (2PP^T - I)(2QQ^T-I)$. The input should be $\ket{Qx}$ that can be created from $\ket{x} = \sum_i \alpha_i \ket{v_i}$.
Since the eigenvectors of $W$ are $Qv_i$, with eigenvalues $e^{i\theta_i}$ such that $\cos(\theta_i/2) = \sigma_i/\norm{A}_F$, phase estimation allows us to obtain the state $\sum_i \alpha_i\ket{Qv_i}\ket{\overline{\theta_i}}$ and then $\sum_i \alpha_i\ket{Qv_i}\ket{\overline{\sigma_i}}$ with simple processing. Inverting the whole computation will yield to $\sum_i \alpha_i\ket{v_i}\ket{\overline{\sigma_i}}$. We also see that the error in the estimation of $\sigma_i$ comes directly from the error of phase estimation. The running time at this point is $O(polylog(mn)/\epsilon)$. 

In the improved version \cite{kerenidis2020_gradient_descent}, an additional qubit is used, and instead of choosing $P$ and $Q$ such that $A/\norm{A}_F = PQ$, we know require $A/\mu = P\circ Q$ for some parameter $\mu$. It is shown that the best parameter to choose is $\mu(A)$ from Definition \ref{def:mu}.

\subsubsection{Projections using SVE}\label{sec:quantum_projection_SVE}

It is a common task to project vectors in lower-dimensional space. If these space are spanned by a specific basis that results from singular vector or eigenvector decomposition, the projection is meaningful. This technique is widely used in machine learning to reduce the inputs, and quantum algorithms have been proposed to solve this in several ways \cite{Lloyd_PCA_quantum, kerenidis_recommendation_system, kerenidis2020classification_QSFA}. Often, one wants to select either the highest or lowest singular values and project all vectors in the subspace spanned by the corresponding singular vectors. 
We will also require such projections in Chapter \ref{chap:Q_spectral_clustering}. 

Using the quantum algorithm for SVE (Theorem \ref{thm:SVE}), it becomes easier since, starting from any vector $\ket{x}$ decomposed in the singular basis, we obtain a superposition of all singular vectors along with their singular values. 
\begin{equation}
    \ket{x}=\sum_i \alpha_i\ket{v_i}
    \mapsto
    \sum_i \alpha_i\ket{v_i}\ket{\sigma_i}
\end{equation}
It then suffices to separate the retained singular values from the others. For instance, if the singular values to select are the ones smaller than a parameter $\delta>0$, using a boolean comparison circuit (see Claim \ref{thm:boolean_circuits}) and a marked ancilla qubit, we obtain:
\begin{equation}
    \sum_{i|\sigma_i\leq\delta} \alpha_i\ket{v_i}\ket{\sigma_i} + \sum_{i|\sigma_i>\delta} \alpha_{i}\ket{v_i}\ket{\sigma_i}
    \mapsto
    \sum_{i|\sigma_i\leq\delta} \alpha_i\ket{v_i}\ket{\sigma_i}\ket{0} + \sum_{i|\sigma_i>\delta} \alpha_{i}\ket{v_i}\ket{\sigma_i}\ket{1}
\end{equation}
Then, as in Claim \ref{thm:from_ampltide_to_bit_encoding}, we start by applying a conditional rotation (Theorem \ref{thm:conditionrotation}) to have:
\begin{equation}\label{eq:projection_sve}% ex- \label{LaplacianCR}
    \sum_{\substack{i | \sigma_i \leq \delta}}\alpha_{i}\ket{v_i}\ket{\sigma_i}\ket{0}
    \left(\frac{\sigma_{i}}{\delta}\ket{0} + \sqrt{1-\frac{\sigma^2_i}{\delta^2}}\ket{1}\right)
    +
    \sum_{\substack{i | \sigma_i > \delta}}\alpha_{i}\ket{v_i}\ket{\sigma_i}\ket{1}\ket{0}
\end{equation}
and then we only have to measure the state $\ket{00}$ on the last two qubits. This could also be done using amplitude amplification on the state $\ket{00}$ (Theorem \ref{thm:amplitude_amplification}). In both cases we obtain the desired projection:
\begin{equation}% ex- \label{LaplacianCR}
    \sum_{\substack{i | \sigma_i \leq \delta}}\alpha'_{i}\ket{v_i}
\end{equation}
The new amplitudes $\alpha'_i$ will correspond to the component of the projection of vector $x$ in the new subspace.\\

The running time of this operation is driven by the amplitude amplification that involves $O(1/\sqrt{P(00)})$ queries to the SVE circuit itself, where $P(00)$ is the probability to measure '00' in the last two qubits of state (\ref{eq:projection_sve}), given by:
\begin{equation}
    P(00) = \sum_{\substack{i | \sigma_i \leq \delta}}\alpha_{i}^2\frac{\sigma_{i}^2}{\delta^2}
\end{equation}

\section{Matrix Multiplication and Inversion}\label{sec:matrix_multiplication_quantum}

As mentioned in Section \ref{sec:intro_qml}, the field of quantum machine learning was ignited by the pioneering work of \cite{HHL} giving exponential speedup for matrix multiplication and inversion. this construction was based on phase estimation (Theorem \ref{thm:phase_estimation}) and Hamiltonian simulation. 
Later, improvements were proposed in \cite{childs2017quantum,kerenidis2020_gradient_descent}, until \cite{chakraborty2018power,gilyen2019quantum} improved it again error wise, by introducing the \emph{Block Encoding} framework (see Definition \ref{def:block_encoding}). This ensures a complexity sublinear in the dimension. Recall that for a matrix $M$, $\kappa(M)$ is its condition number (the ratio between the biggest and the smallest singular values), and $\mu(M)$ is defined in Definition \ref{def:mu}. 

\begin{theorem}{Quantum Matrix Multiplication and Inversion}{quantum_matrix_multiplication_inversion}
Let $M \in \R^{d \times d}$ and $x \in \R^d$. Let $\delta_1,\delta_2>0$. If $M$ is stored in appropriate QRAM data structures and the time to prepare $\ket{x}$ is $T_{x}$, then there exist quantum algorithms that with probability at least $1-1/poly(d)$ return
    \begin{enumerate}
        \item A state $\ket{z}$ such that $\norm{ \ket{z} - \ket{Mx}}_{2} \leq \delta_1$ in time $\widetilde{O}((\kappa(M)\mu(M) + T_{x} \kappa(M)) \log(1/\delta_1))$.  \\
        Note that this also implies $\norm{ \ket{z} - \ket{Mx}}_{\infty} \leq \delta_1$
        \item A state $\ket{z}$ such that $\norm{\ket{z} - \ket{M^{-1}x}} \leq \epsilon$ in time $\widetilde{O}((\kappa(M)\mu(M) + T_{x} \kappa(M)) \log(1/\epsilon))$.
        \item Norm estimate $z \in (1 \pm \delta_2)\norm{Mx}_{2}$, with relative error $\delta_2$, in time $\widetilde{O}(T_{x} \frac{\kappa(M)\mu(M)}{\delta_2}\log(1/\delta_1))$.
    \end{enumerate}
\end{theorem}

Notably, these routines for quantum linear algebra can be applied to products of matrices. If the matrix $M$ is the product of $k$ matrices, i.e. $M = M_1 . . .M_k$, the resulting factor in the runtime is $\widetilde{O}(\kappa(M)\sum_k\mu(M_k)\log(1/\epsilon))$.

\section{Quantum Inspired Algorithms}\label{sec:quantum_inspired_algo}

A recent breakthrough by Tang et al. \cite{gilyen2018quantum, tang2018quantum, tang2019quantum}, proposed several classical machine learning algorithms obtained by \emph{dequantizing} the quantum
recommendation systems algorithm \cite{kerenidis_recommendation_system} and low rank linear system solvers. Like the quantum algorithms, the running time of these classical algorithms is $O(poly(k)polylog(mn))$, which is polylogarithmic in the dimension of the dataset and polynomial in the rank.  However, the polynomial dependence on the rank of the matrices is significantly worse than the quantum algorithms and in fact renders these classical algorithms highly impractical. For example, the classical algorithm for stochastic regression inspired by the HHL algorithm \cite{HHL} has a running time of $\tilde{O}(\kappa^{6} k^{16} \norm{A}_F^6 / \epsilon^6)$, which is impractical even for a rank-10 matrix. This running time has then been improved by \cite{shao2021faster} under some assumptions, to obtain a running time of $\widetilde{O}(\kappa_F^6\kappa^2/\epsilon^2)$ where $\kappa_F = \norm{A}_F\norm{A^{-1}}$ and $\kappa = \norm{A}_2\norm{A^{-1}}$, if $A$ is the matrix to invert or to multiply with. 

The extremely high dependence on the rank and the other parameters implies not only that the quantum algorithms are substantially faster, since their dependence on the rank is sublinear, but more importantly that in practice there exist much faster classical algorithms for these problems. While these new quantum inspired classical algorithms are based on the FKV methods \cite{frieze2004fast}, in classical linear algebra, algorithms based on the CUR decomposition (a low rank approximation of the SVD) that have a running time linear in the dimension and quadratic in the rank are preferred to the FKV methods \cite{frieze2004fast, drineas2004clustering, achlioptas2003fast}. Experimental comparisons with the usual classical algorithms have been done in \cite{arrazola2019quantum} but don't seem in favor of the new quantum inspired algorithms.

It remains an open question to find classical algorithms for these machine learning problems that are polylogarithmic in the dimension and are competitive with respect to the quantum or the classical algorithms for the same problems. This would involve using significantly different techniques than the ones presently used for these algorithms.\\

That being said, these results remain theoretically important as they reduce the exponential separation between quantum and classical linear algebra algorithms for low rank problems. In fact, the algorithms presented in this thesis could have their own quantum inspired versions (as we show for the quantum convolutional neural network algorithm in Chapter \ref{chap:QCNN}), but would not be usable in practice. This still forces us to find where quantum computations must draw to find an unrivaled advantage. Recent results indicate that sparsity-based algorithm, where the matrices and vectors are high dimensional but sparse, is a better candidate against quantum inspired classical algorithms.

\chapter{Inner Product and Distance Estimation}\label{chap:QIPE_and_distance}
\epigraph{\textit{"Quantum mechanics describes nature as absurd from the point of view of common sense. And yet it fully agrees with experiment. So I hope you can accept nature as She is - absurd."}}{Richard P. Feynman\\ \emph{QED: The Strange Theory of Light and Matter} (1985)}

\section{Related Work}

\subsection{SWAP Test}
In this section, we detail the seminal idea of \cite{LMR13} of using the SWAP test to compute the distance between two vectors. We assume quantum access to the vectors and their norms, using the amplitude encoding framework (Definition \ref{def:ampltiude_encoding}). For two vectors $v_i \in \R^d$ and $v_j \in \R^d$, respectively indexed by $i$ and $j$, we can query them in quantum registers in time $T$ using the mapping: 
\begin{equation}
\ket{i}\ket{0} \mapsto \ket{i}\ket{v_i}, ~~~  \ket{j}\ket{0} \mapsto \ket{j}\ket{v_j}
\end{equation}

With a QRAM data structure (Theorem \ref{thm:QRAM}) the query time is $O(\log d)$ 
where $d$ if the dimension of the vectors. We can also query their norms in a similar manner, 
\begin{equation}
\ket{i}\ket{0} \mapsto \ket{i}\ket{\norm{v_i}}, ~~~ \ket{j}\ket{0} \mapsto \ket{j}\ket{\norm{v_j}}
\end{equation}
We can compute the distance $d(v_i,v_j)$ in the amplitude of the ancillary register by performing a Swap Test between two states that were introduced in \cite{LMR13}. Define, 
\begin{equation}
\ket{\psi_{ij}} = \frac{1}{\sqrt{2}}(\ket{v_i}\ket{0} - \ket{v_j}\ket{1})
\end{equation} 
\begin{equation}
\ket{\phi_{ij}} = \frac{1}{\sqrt{Z_{ij}}}(\norm{v_i}\ket{0}+\norm{v_j}\ket{1})
\end{equation} 
Where $Z_{ij} = \norm{v_i}^2+\norm{v_j}^2$. 
Note that, in Section \ref{sec:inner_product_and_distance_quantum}, compared to \cite{LMR13}, we will exchange the minus sign between $\ket{\psi}$ and $\ket{\phi}$, and the two registers in $\ket{\psi}$, in order to avoid an extra quantum arithmetic operation to the conditional rotation step for $\ket{\phi_{ij}}$. Note that these states have been chosen for retrieving the distance between $v_i$ and $v_j$, but it suffices to replace $\ket{\phi_{ij}}$ by the $\ket{+}$ state to obtain the inner product $|\braket{v_i}{v_j}|^2$ instead. 

We now describe the preparation procedures for $\ket{\psi_{ij}}$ and $\ket{\phi_{ij}}$.
In order to create $\ket{\psi_{ij}}$, we first create 
$\ket{i}\ket{j} \ket{-} \ket{0}$ and then perform controlled queries as indicated below, 

\begin{equation}
	\left\lbrace
		\begin{array}{ccc}
			\ket{i}\ket{j}\ket{0}\ket{0} \mapsto \ket{i}\ket{j}\ket{0}\ket{v_i} \\
			\ket{i}\ket{j}\ket{1}\ket{0} \mapsto \ket{i}\ket{j}\ket{1}\ket{v_j}  
		\end{array}
	\right.
\end{equation}

\begin{equation}
\ket{i}\ket{j} \frac{1}{\sqrt{2}} (\ket{0} - \ket{1}) \ket{0} \to \ket{i}\ket{j}\frac{1}{\sqrt{2}} (\ket{0} \ket{v_{i}}  - \ket{1} \ket{v_j}) = \ket{i}\ket{j} \ket{\psi_{ij}} 
\end{equation}
A different procedure is used for creating $\ket{\phi_{ij}}$. We start with $\ket{i}\ket{j}$ as well and query the two norms in ancilla registers to obtain $\ket{i}\ket{j}\ket{\norm{v_i}}\ket{\norm{v_j}}$. We then add an extra qubit $\ket{0}$ and apply a controlled rotation,
to directly create $\ket{i}\ket{j}\ket{\norm{v_i}}\ket{\norm{v_j}}\ket{\phi_{ij}}$. We can then undo the first step to remove the third and fourth registers. 

Having prepared the states $\ket{\psi_{ij}}$ and $\ket{\phi_{ij}}$, we can apply the Swap Test circuit given below. The swap test circuit introduces an ancilla qubit on which we first apply a Hadamard gate, we then use the ancilla qubit to perform a controlled swap on two quantum registers, this is followed by a Hadamard gate and a measurement on the ancilla.
\begin{center}
	\begin{figure}[h]
		\[
		\Qcircuit @C=.7em @R=1.5em{
			\lstick{\ket{0}}   			 	& \gate{H} 	&\ctrl{1} \qwx[2]   & \gate{H}   		& \qw      \\
			\lstick{\ket{a}}    			& \qw		&  \qswap             & \qw        		& \qw      \\
			\lstick{\ket{b}}  			& \qw		&  \qswap             & \qw         		& \qw      
		}
		\]
		\caption{ Swap Test circuit on two quantum states $\ket{a}$ and $\ket{b}$}\label{swap_circuit}
	\end{figure}
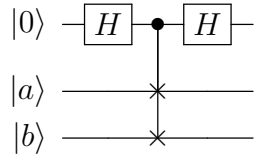 
\end{center} 
The action of the swap test over two registers is:
\begin{equation}
 \ket{0} \ket{a}\ket{b} \mapsto \left(\frac{1}{2}\ket{0}(\ket{a}\ket{b}+\ket{b}\ket{a})+\frac{1}{2}\ket{1}(\ket{a}\ket{b}-\ket{b}\ket{a}) \right)
\end{equation} 
It follows that after the swap test, the probability of measuring $\ket{0}$ in the ancilla register is equal to $\frac{1 + |\braket{a}{b}|^{2}}{2}$.

In our case, for the swap test between $\ket{\phi_{ij}}$ and the second register of $\ket{\psi_{ij}}$ which we denote as $\braket{\phi_{ij}}{\psi_{ij,2}}$. 
the probability $p_{ij}$ of measuring $0$ in the final ancilla register is $\frac{1 + |\braket{\phi_{ij}}{\psi_{ij,2}}|^2}{2}$.
In order to compute $p_{ij}$ we first compute $|\braket{\phi_{ij}}{\psi_{ij,2}}|^2$, 

\begin{align} \nonumber
 |\braket{\phi_{ij}}{\psi_{ij,2}}|^2 &= \frac{1}{2Z_{ij}}\left|(\norm{v_i}\bra{0}+\norm{v_j}\bra{1})(\ket{v_i}\ket{0}-\ket{v_j}\ket{1})\right|^2 \\ \nonumber
&= \frac{1}{2Z_{ij}} |\norm{v_i}\ket{v_i}-\norm{v_j}\ket{v_j}|^2 \\\nonumber
&= \frac{1}{2Z_{ij}}(\norm{v_i}^2 + \norm{v_j}^2 - 2 \norm{v_i}\norm{v_j}\braket{v_i}{v_j}) \\
&=\frac{d(v_i,v_j)^2}{2Z_{ij}}
\end{align}

Thus, the probability encodes the squared distance between $v_{i}$ and $v_j$, such that:
\begin{equation}
    p_{ij}= \frac{1}{2} + \frac{d(v_i,v_j)^2}{4Z_{ij}}
\end{equation}
The quantum state before measuring the ancilla qubit in the swap test circuit can be written as, 
\begin{equation}\label{eq:swap_test_dist_estimation_result}
\ket{i}\ket{j}(\sqrt{p_{ij}}\ket{y_{ij}}+\sqrt{1-p_{ij}}\ket{y_{ij}^{\perp}})
\end{equation}

Finally, we can apply this transformation over a superposition of indices $i$ and $j$, thus allowing to compute many pairwise distances simultaneously.\\

In Section \ref{sec:inner_product_and_distance_quantum}, we will propose a different method to obtain a similar quantum state as Eq.(\ref{eq:swap_test_dist_estimation_result}), and show how to extract the value $d(v_i,v_j)^2$ from the amplitude.

\subsection{Unary Inner Products}

Recently, \cite{johri2020nearest_dataloaders} used the unary encoding (Definition \ref{def:unary_encoding}) and unary data loaders (Section \ref{sec:unary_data_loaders}) to perform the inner product between two vectors.\\

\begin{figure}[h]
    \centering
    \includegraphics[width=0.6\textwidth]{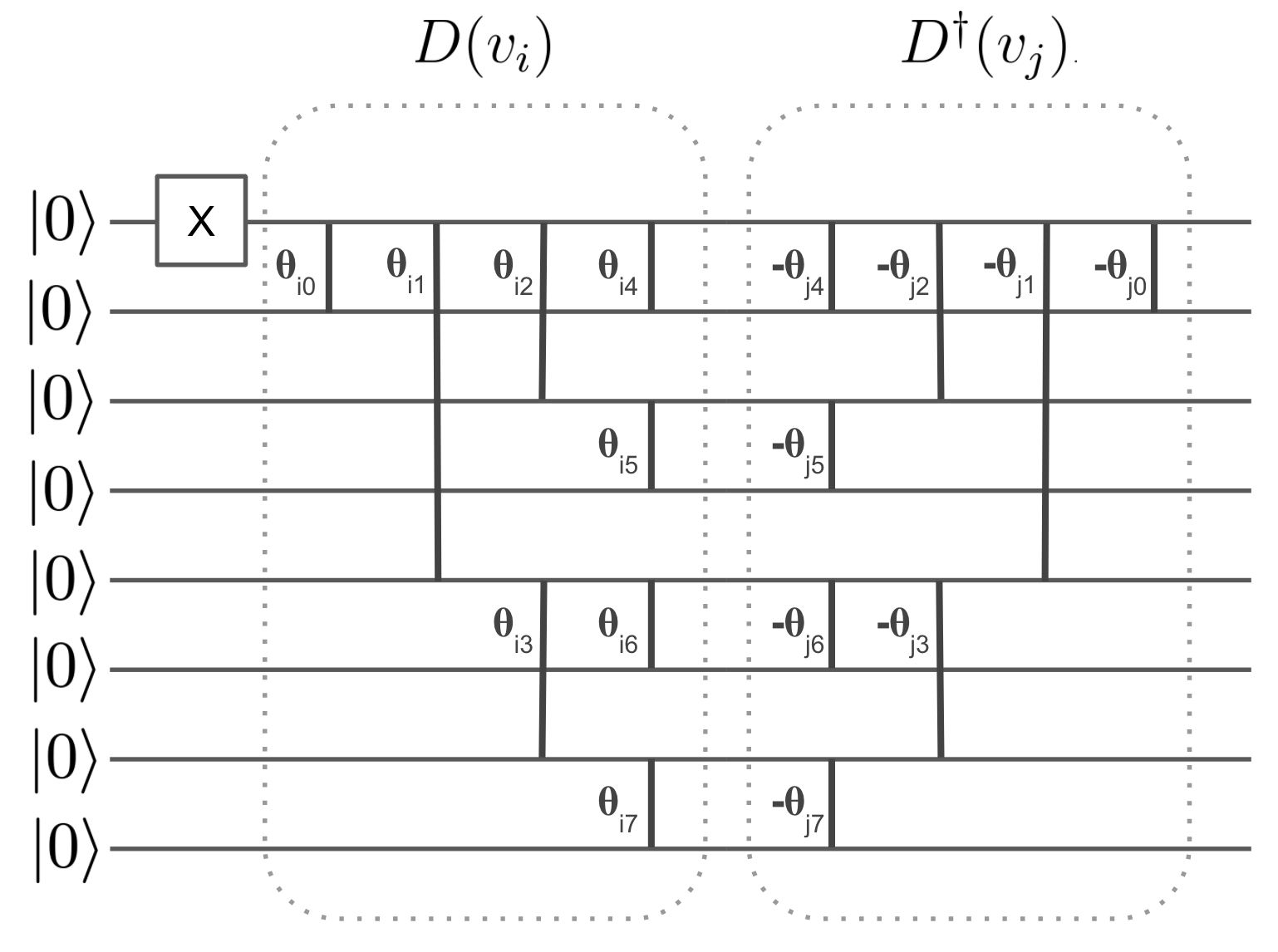}
    \caption{The inner product circuit for two unary quantum vectors of dimension 8 is the concatenation of the two data loaders $D^{\dagger}(v_j)D(v_i)$.}
    \label{fig:unary_inner_product}
\end{figure}

We consider two vectors $v_i\in\R^d$ and $v_j\in\R^d$, with respective address $i$ and $j$, stored in a unary memory model. We write $D(v_i)$ and $D(v_j)$ their respective loader circuits, for instance the parallel loader from Fig.\ref{fig:unary_loader_parallel}.

\begin{equation}
    D(v_i)\ket{0} = \ket{v_i} = \frac{1}{\sqrt{d}}\sum_i^d v_i\ket{e_i}
\end{equation}

where $e_i$ is the $i^{th}$ unary vector (e.g. $\ket{0\cdots010\cdots0}$). Note that $\ket{v_i}$ can be projected onto the second vector, and thus expressed in the basis $\{v_j,v_j^\perp\}$ as

\begin{equation}
    \ket{v_i} = \braket{v_i}{v_j} \ket{v_j} + \sqrt{1 - \braket{v_i}{v_j}^2}  \ket{v_j^\perp}
\end{equation}

Recall that the unary data loaders from \cite{johri2020nearest_dataloaders} are using only $RBS(\theta)$ gates from Eq.\ref{eq:RBSgate} and it is simple to show that $RBS(\theta) = RBS^{\dagger}(-\theta)$. The concatenation of the two circuits $D^{\dagger}(v_j)D(v_i)$ will therefore output the state:

\begin{align}\label{eq:unary_inner_product_state}\nonumber
    D^{\dagger}(v_j)D(v_i) =
    D^{\dagger}(v_j)\left(\braket{v_i}{v_j} \ket{v_j} + \sqrt{1 - \braket{v_i}{v_j}^2}  \ket{v_j^\perp}\right) &\\=  
    \braket{v_i}{v_j} \ket{e_1} + \sqrt{1 - \braket{v_i}{v_j}^2}  \ket{e_1^\perp}
\end{align}

Using parallel data loaders as shown in Fig.\ref{fig:unary_inner_product}, the depth of the circuit is $O(2\log(d))$, using O(d) qubits. If the goal is to retrieve a classical approximation of $\braket{v_i}{v_j}$, one can perform several measurements to estimate $|\braket{v_i}{v_j}|^2$, being the probability to the number of times a `1' is seen on the first qubit. 

Notably, we remark that recovering the inner product would be exact only for vectors with the same sign. To generalize and get the sign of the inner product, one can add an extra qubit in superposition to control the whole circuit, which will result in an amplitude of $((1 + \braket{v_i}{v_j})/2)$.

\clearpage

\section{Quantum Circuit for Inner Product and Distance Estimation}\label{sec:inner_product_and_distance_quantum}

In this section, we introduce a quantum algorithm for inner product and distance estimation between amplitude encoded vectors, applicable in superposition. Using the work of \cite{wiebe_nearest_neigbhors}, we adapted the Frobenius distance estimator from \cite{kerenidis2020classification_QSFA} that calculates the average square distance between a single point and all set many other points. It allows us to calculate the square distance or inner product (with its sign) between two vectors. This routine becomes very efficient when having quantum access to the vectors (Definition \ref{def:quantum_access}).

As shown in Fig.\ref{fig:plan_these}, this algorithm will be a core part of many others during this thesis: matrix multiplication, neural network's layer, convolution product, adjacency graph creation, etc.

\subsection{Quantum Circuit}\label{sec:circuit_inner_product_and_distance_quantum}

We first state the Theorem that summarizes the circuit complexity and guarantees, from \cite{qmeans}:

\begin{theorem}{Distance \& Inner Product Estimation}{distance_innpdct_quantum}
Given quantum access in time $T$ to two data matrices $V \in \R^{\n \times d}$ and $C \in \R^{k \times d}$ with rows $v_i$ and $v_j$. For any $\Delta > 0$ and $\epsilon>0$, there exists a quantum algorithm that  computes 
\begin{enumerate}
\item The squared distance: 
$\ket{i}\ket{j}\ket{0} \mapsto \ket{i}\ket{j}\ket{\overline{d^2(v_i,v_j)}}$ where $|\overline{d^{2}(v_i,v_j)}-d^{2}(v_i,v_j)| \leqslant  \epsilon$ 
with probability at least 
$1-2\Delta$, in time 
$O\left(\frac{1}{\epsilon}\norm{v_i}\norm{v_j} T \log(1/\Delta)\right)$. 

\item The unnormalized inner product: 
$\ket{i}\ket{j}\ket{0} \mapsto \ket{i}\ket{j}\ket{\overline{(v_i,v_j)}}$ where 
$|\overline{(v_i,v_j)}-(v_i,v_j)| \leqslant  \epsilon$ 
with probability at least $1-2\Delta$, in time 
$O\left(\frac{1}{\epsilon}\norm{v_i}\norm{v_j} T \log(1/\Delta)\right)$. 

\end{enumerate}

Both of these tasks can be applied in superposition over all vectors of $V$ and $C$, each with a running time upper bounded by $O\left(\frac{1}{\epsilon}\eta T \log(1/\Delta)\right)$ where $\eta = \max_{i,j}(\norm{v_i}\norm{v_j})$ with the assumption $\min_{i}(\norm{v_i}) = \min_{i}(\norm{v_j}) = 1$.
\end{theorem}

Remark that if a QRAM model is used (Theorem \ref{thm:QRAM}), the complexity becomes simply $\widetilde{O}(\eta/\epsilon)$. Note also that the error guarantees can be reformulate for a relative error $|\overline{(v_i,v_j)}-(v_i,v_j)| \leqslant  \epsilon(v_i,v_j)$ which modifies the running time to be $O\left(\frac{1}{\epsilon}\frac{\norm{v_i}\norm{v_j}}{|(v_i,v_j)|} T \log(1/\Delta)\right)$ \\

The parameter $\eta(\cdot)$ will be present in many running times from now on. To this end, we introduce a proper definition:

\begin{definition}{Parameter $\eta(\cdot)$}{eta}
For a matrix $V \in \R^{n\times d}$, its parameter $\eta(V)$ is defined as 
\begin{equation}
\eta(V) = \frac{\max_i(\norm{v_i}^2)}{\min_{i}(\norm{v_i}^2)}
\end{equation}
or as $\max_i(\norm{v_i}^2)$ if we assume $\min_{i}(\norm{v_i})=1$.
\end{definition}

\begin{proof}
Let us start by describing a procedure to estimate the square $\ell_2$ distance between the normalized vectors $\ket{v_i}$ and $\ket{v_j}$. We start with the initial state 
\begin{equation}
\ket{\phi_{ij}} := \ket{i} \ket{j} \frac{1}{\sqrt{2}}(\ket{0}+	\ket{1})\ket{0}
\end{equation}

Then, we query the state preparation oracle controlled on the third register to perform the mappings 
$\ket{i}\ket{j}\ket{0}\ket{0} \mapsto \ket{i}\ket{j}\ket{0}\ket{v_i}$ and $\ket{i}\ket{j}\ket{1}\ket{0} \mapsto \ket{i}\ket{j}\ket{1}\ket{v_j}$. 
The state after this is given by,

\begin{equation}
\frac{1}{\sqrt{2}}\left( \ket{i}\ket{j}\ket{0}\ket{v_i} + \ket{i}\ket{j}\ket{1}\ket{v_j}\right)
\end{equation}
Finally, we apply an Hadamard gate on the the third register to obtain, 
\begin{equation}
\ket{i}\ket{j}\left(
\frac{1}{2}\ket{0}\left(\ket{v_i} + \ket{v_j}\right)
+ \frac{1}{2}\ket{1}\left(\ket{v_i} - \ket{v_j}\right)
\right)
\end{equation}
The probability of obtaining $\ket{1}$ when the third register is measured is,
\begin{equation}
p_{ij} =  \frac{1}{4}(2 - 2\braket{v_i}{v_j}) =  \frac{1}{4} d^2(\ket{v_i}, \ket{v_j}) =  \frac{1 - \langle v_i | v_j\rangle}{2}
\end{equation} 
which is proportional to the square distance between the two normalized vectors.

We can rewrite $\ket{1}\left(\ket{v_i} - \ket{v_j}\right)$ as $\ket{y_{ij},1}$ (by swapping the registers), and hence we have the final mapping
\begin{equation}\label{eq:QDE}
A: \ket{i}\ket{j} \ket{0} \mapsto \ket{i}\ket{j}(\sqrt{p_{ij}}\ket{y_{ij},1}+\sqrt{1-p_{ij}}\ket{G_{ij},0}) 
\end{equation}
where the probability $p_{ij}$ is proportional to the square distance between the normalized vectors and $G_{ij}$ is a garbage state. Note that the running time of $A$ is $T_A=\tilde{O}(T)$.

We then use amplitude estimation (Theorem \ref{thm:amplitude_estimation}) on the unitary $A$ defined in Eq.(\ref{eq:QDE}). This creates an unitary operation that maps
\begin{equation}
\mathcal{U}: \ket{i}\ket{j}  \ket{0} \mapsto \ket{i}\ket{j} \left( \sqrt{\alpha}  \ket{ \overline{p_{ij}}, G, 1} + \sqrt{ (1-\alpha ) }\ket{G', 0}  \right) 
\end{equation} 
where $G, G'$ are garbage registers, $|\overline{p_{ij}} - p_{ij}  |  \leq \epsilon$ and $\alpha \geq 8/\pi^2$. 
The unitary $\mathcal{U}$ requires $P$ iterations of $A$ with $P=O(1/\epsilon)$. Amplitude estimation thus takes time $T_{\mathcal{U}} = \widetilde{O}(T/\epsilon)$.

We then make use of a tool developed in \cite{wiebe_nearest_neigbhors} to boost the probability and precision for the estimation of the distances or inner products. At high level, it takes multiple copies of the estimator from the amplitude estimation procedure, and uses them to compute the median value. It finally reverses the circuit to get rid of the garbage states. We provide more details after the end of the current proof. Here we provide a theorem with respect to time and not query complexity.

\begin{theorem}{Median Evaluation}{median_evaluation}
	Let $U$ be a unitary operation that maps 
	\begin{equation}
	U:\ket{0}\mapsto \sqrt{a}\ket{x,1}+\sqrt{1-a} \ket{G,0}
	\end{equation}
	for some $1/2 < a \le 1$ in time $T$. Then there exists a quantum algorithm that uses $L$ copies of the above state, and for any $\Delta>0$ and for any $1/2<a_0 \le a$, produces a state $\ket{\Psi}$ such that $\norm{\ket{\Psi}-\ket{0}^{\otimes L}\ket{x}} \leq \sqrt{2\Delta}$, in time:  
	\begin{equation}
	2T\left\lceil\frac{\ln(1/\Delta)}{2\left(|a_0|-\frac{1}{2} \right)^2}\right\rceil.
	\end{equation}
	\label{lem:median}
\end{theorem}

We  apply Theorem \ref{thm:median_evaluation} for the unitary $\mathcal{U}$ to obtain a quantum state $\ket{\Psi_{ij}}$ such that, 

\begin{equation}
\norm{\ket{\Psi_{ij}}-\ket{0}^{\otimes L}\ket{\overline{p_{ij}}, G}}_2\le \sqrt{2\Delta}
\end{equation}

The running time of the procedure is 
$O\left( T_{\mathcal{U}} \ln(1/\Delta)\right) = \widetilde{O}\left(\frac{T }{\epsilon}\log (1/\Delta)\right)$.

Note that we can easily multiply the value $\overline{p_{ij}}$ by 4 in order to have the estimator of the square distance of the normalized vectors or compute $1-2\overline{p_{ij}}$ for the normalized inner product. Last, the garbage state does not cause any problem in calculating the minimum in the next step, after which this step is uncomputed. 

%The running time of the procedure is thus
%$O( T_{\mathcal{U}} \ln(1/\Delta)) = O( \frac{T }{\epsilon}\log %(1/\Delta))$. 
%
The last step is to show how to estimate the square distance or the inner product of the unnormalized vectors. Since we know the norms of the vectors, we can simply multiply the estimator of the normalized inner product with the product of the two norms to get an estimate for the inner product of the unnormalized vectors and a similar calculation works for the distance. Note that the absolute error $\epsilon$ now becomes $\epsilon \norm{v_i}\norm{v_j}$ and hence if we want to have in the end an absolute error $\epsilon$ this will incur a factor of $\norm{v_i}\norm{v_j}$ in the running time. This concludes the proof of the Theorem \ref{thm:distance_innpdct_quantum}. 
\end{proof}%\Hsquare

\subsubsection{Median Evaluation Details}

As stated in Theorem \ref{thm:median_evaluation} from \cite{wiebe_nearest_neigbhors}, we can boost the probability of measuring the right state during inner product estimation. The improved precision can be arbitrary with only a logarithmic dependence in the runtime. It also requires $L$ new copies of the current state, which fortunately is also logarithmic in the precision. \\

As before, if we can obtain the state $\sqrt{p}\ket{y} + \sqrt{1-p}\ket{y^{\perp}}$, with $p$ encoding the meaningful information such as a distance or else, we can apply the amplitude estimation and get in time $T$:
\begin{equation}\label{eq:median_evaluation_initial}
    \alpha\ket{y..}\ket{\overline{p}} + \alpha'\ket{y^{\perp}..}\ket{\overline{p}^{\perp}}
\end{equation}
with the informal notation $\ket{y..}$ to denote the potential presence of garbage states, and with $\overline{p}$ being an $\epsilon$-close estimate of $p$ using $m$-qubits for precision. We know from Theorem \ref{thm:amplitude_estimation} that $|\alpha|^2\geq 8/\pi^2$. 

Then, to increase the probability of measuring $\ket{\overline{p}}$ and not $\ket{\overline{p}^{\perp}}$, we will use $L$ copies of the state in Eq.(\ref{eq:median_evaluation_initial}), $(\alpha\ket{y..}\ket{\overline{p}} + \alpha'\ket{y^{\perp}..}\ket{\overline{p}^{\perp}})^{\otimes L}$. We will apply a median evaluation unitary on these $L$ states. This operations write in an extra register the median value $\ket{\tilde{P}}$ observed. 

\begin{equation}
    (\alpha\ket{y..}\ket{\overline{p}} + \alpha'\ket{y^{\perp}..}\ket{\overline{p}^{\perp}})^{\otimes L}\ket{0}^{\otimes m} 
    \mapsto
    (\alpha\ket{y..}\ket{\overline{p}} + \alpha'\ket{y^{\perp}..}\ket{\overline{p}^{\perp}})^{\otimes L}
    \ket{\tilde{P}}
\end{equation}

To understand what is $\ket{\tilde{P}}$, we index each of the $L$ copies gives, such that:

\begin{equation}
    \left(\alpha\ket{..\overline{p_1}} + \alpha'\ket{..\overline{p_1}^{\perp}}\right)
    \otimes
    \left(\alpha\ket{..\overline{p_2}} + \alpha'\ket{..\overline{p_2}^{\perp}}\right)
    \otimes
    \cdots
    \otimes
    \left(\alpha\ket{..\overline{p_L}} + \alpha'\ket{..\overline{p_L}^{\perp}}\right)
\end{equation}

This expression, once developed, is informally equivalent to:

\begin{equation}
    \left(\ket{\overline{p_1}}\ket{\overline{p_2}}\cdots\ket{\overline{p_L}}\right)
    +
    \left(\ket{\overline{p_1}^{\perp}}\ket{\overline{p_2}}\cdots\ket{\overline{p_L}}\right)
    +
    \cdots
    +
    \left(\ket{\overline{p_1}^{\perp}}\ket{\overline{p_2}^{\perp}}\cdots\ket{\overline{p_L}^{\perp}}\right)
\end{equation}

For each term were half of the values are $\ket{\overline{p_i}}$ and not $\ket{\overline{p_i}^{\perp}}$, the median evaluation will gives the right result $\ket{\overline{p}}$. The proof of Theorem \ref{thm:median_evaluation} ensures that the final state $\ket{\tilde{P}}$ is sufficiently close to $\overline{p}$, and that if we reverse the whole computation, we obtain a state $\ket{\Psi}$ such that:

\begin{equation}\label{eq:median_eval_precision}
    \norm{\ket{\Psi} - \ket{0}^{\otimes L}\ket{\overline{p}}}_2 \leq \sqrt{2\Delta} 
\end{equation}

Where $\Delta >0$ is an arbitrary precision parameter linked to the number $L$ of copies requires:

\begin{equation}
    L = \left\lceil\frac{\ln(\Delta^{-1})}{2(|\alpha|-\frac{1}{2})^2}\right\rceil
\end{equation}

We see that the number $L$ of copies is logarithmic in the precision $\Delta$, which is efficient since the median evaluation unitary has a complexity of $O(L)$. Finally, the overall time is given by $O(2TL)$, where the factor $2$ is for reversing the whole computation:

\begin{equation}
    O\left(2T\left\lceil\frac{\ln(\Delta^{-1})}{2(|\alpha|-\frac{1}{2})^2}\right\rceil \right)
\end{equation}

Note that Eq.(\ref{eq:median_eval_precision}) can be interpreted as a result on the probability of observing the right state as well. Indeed, if $\norm{\ket{a} - \ket{b}}_2 \leq \sqrt{\epsilon}$, then we can show that $\ket{a} = \beta\ket{b} + \beta'\ket{G}$ with $\beta \geq \sqrt{1-\epsilon}$ and $\beta' \leq \sqrt{\epsilon}$. $\ket{G}$ is the undesired state. This implies a probability at least $1-2\Delta$ of observing the good state $\ket{0}^{\otimes L}\ket{\overline{p}}$ from $\ket{\Psi}$.

Finally, we can extend this demonstration to the case of $k$ tensor products as it will be used in Chapter \ref{chap:qmeans}. The logic stays the same, but the error between $\ket{\Psi}$ is now upper bounded by $\sqrt{2\Delta k}$, by a Union bound argument. indeed, following Eq.(\ref{eq:median_eval_precision}), the probability to have one bad results out $k$ is $2\Delta k$.

%\Hsquare

\subsection{Error Analysis}

Since the quantum inner product or distance estimation algorithm (Theorem \ref{thm:distance_innpdct_quantum}) will be used frequently, it is important to understand exactly the error it implies for the rest of the computations. Since the algorithm relies on amplitude estimation (Theorem \ref{thm:amplitude_estimation}), itself relying on phase estimation (Theorem \ref{thm:phase_estimation}), we can cascade back the error.\\

At first, the phase estimation algorithm \cite{brassard2002quantum} allows us to compute in superposition the phases $\omega$ of the eigenvectors $e^{i\pi\omega}$ of a unitary. For one value $\omega$, the phase estimation circuit with $n$ qubits will return the state $\ket{\overline{\omega}} = \sum_x \alpha(k)\ket{k}$ where $k$ are series of integers between $0$ and $2^n$. We define the error $\delta = |\omega - k/2^n|$. The probability of obtaining a specific integer $k$ when measuring the output is given by:

\begin{equation}\label{eq:phase_estimation_probability_distribution}
    p(k) = |\alpha(k)|^2 = \frac{1}{2^{2^n}}
    \frac
    {\sin^2(\pi(2^n\delta - k))}
    {\sin^2(\pi(\delta - k/2^n))}
\end{equation}

The most probable $k$ we would measure is the one such that $k/2^n$ is the closest to $\omega$. Unless $\omega=k/2^n$, in which case the computation is perfect. Note that the error made on the estimation of $\omega$ strongly depends on the value of $\omega$ itself. See Fig.\ref{fig:phase_estimation_probability_simulations} for numerical simulations of the distributions. 

\begin{figure}[h]
    \centering
    \includegraphics[width=0.6\textwidth]{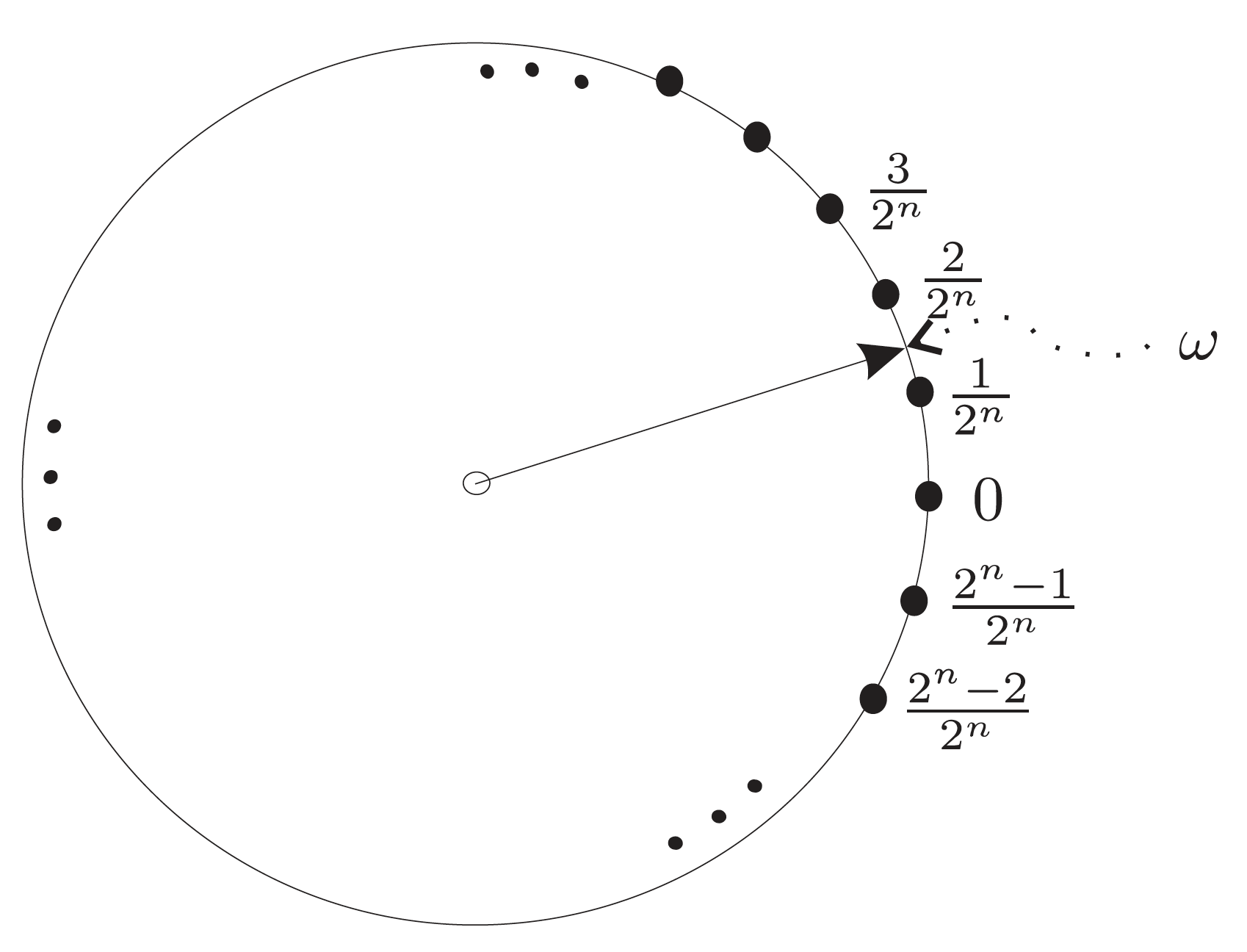}
    \caption{We estimate $\omega$ with the closest $k/2^n$ for any integer $k \in [0,2^n]$ . Source: \cite{kaye2007introduction}.}
    \label{fig:phase_estimation_error_mosca}
\end{figure}

Later on, this error is modified during amplitude estimation and the quantum circuit from \ref{sec:circuit_inner_product_and_distance_quantum}. It is worth noticing that to obtain the unnormalized inner product, or the distance, the error is multiplied by the product of the norm of the two vectors. Therefore, the magnitude of the vectors will increase the absolute error (but not the relative one) as we can see in Fig.\ref{fig:inner_product_error_simulations}.

More precisely, for two vectors $v_i$ and $v_j$, the amplitude we want to estimate is: 
\begin{equation}
a = \frac{1}{2}\left(\frac{\braket{v_i}{v_j}}{\norm{v_i}\norm{v_j}}+1\right)
\end{equation}
Since phase estimation allows to obtain the phase $\omega$ of an amplitude of the form $sin(\pi\omega)$, we further want to obtain the actual normalized inner product:
\begin{equation}
    \omega = \frac{1}{\pi}\arcsin(a)
\end{equation}
As explained above, phase estimation will output, with some probability (see Eq.(\ref{eq:phase_estimation_probability_distribution})), the closest value to $\omega$ of the form $k/2^n$. We write $b = \text{argmin}_{k \in [2^n]}(|k/2^n - \omega|)$. The actual amplitude obtained is an element of the set $\{\overline{\omega}_{\ell}\}_{\ell\in[2^n]}$ such that:
\begin{equation}
    \overline{\omega}_{\ell} = \frac{1}{2^n}(b+\ell\mod(2^n))
\end{equation}
Each $\overline{\omega}_{\ell}$ yields an approximated amplitude $\overline{a}_{\ell} = \sin(\pi\overline{\omega}_{\ell})$
which finally gives the result:
\begin{equation}
\overline{x}_{\ell} = \norm{v_i}\norm{v_j}(2\overline{a}_{\ell}-1)
\end{equation}
With probability: 
\begin{equation}
    p(\ell) = \frac{1}{2^{2^n}}
    \frac
    {\sin^2(\pi(2^n\delta - \ell))}
    {\sin^2(\pi(\delta - \ell/2^n))}
\end{equation}

Note that we don't include median evaluation (Theorem \ref{thm:median_evaluation}) in this analysis, which could have improved the precision arbitrarily.

\subsection{Numerical Simulations}

To gain more intuition on the numerical errors made by the quantum algorithms, we perform extensive classical simulations\footnote{The simulation program was made by Noah Berner} (see Section \ref{sec:classically_simulating_an_algo} for the methodology).

\begin{figure}[h] % "[t!]" placement specifier just for this example
\begin{subfigure}{0.31\textwidth}
\includegraphics[width=\linewidth]{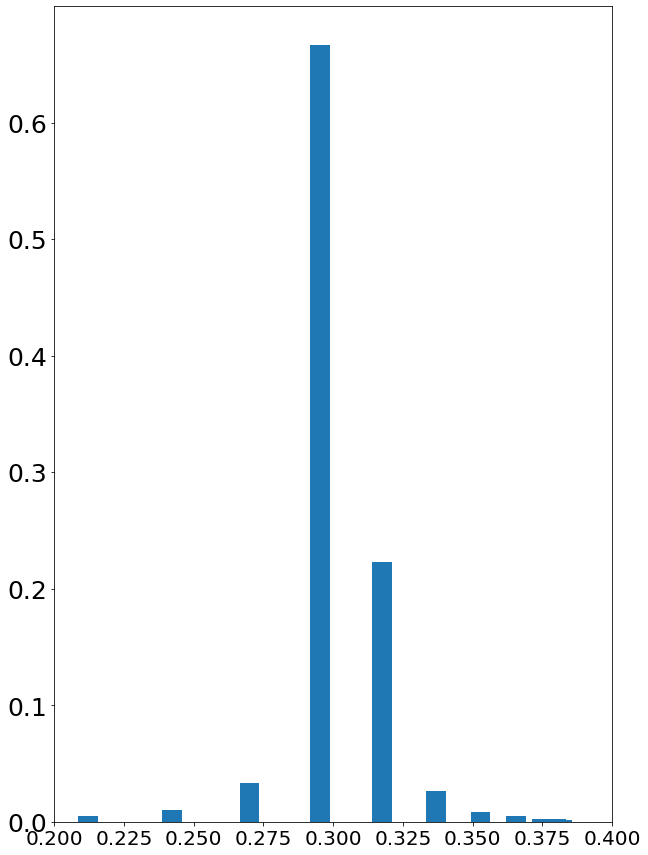}
\caption{$n$=6} \label{fig:phase_estimation_simulations_6}
\end{subfigure}\hspace*{\fill}
\begin{subfigure}{0.31\textwidth}
\includegraphics[width=\linewidth]{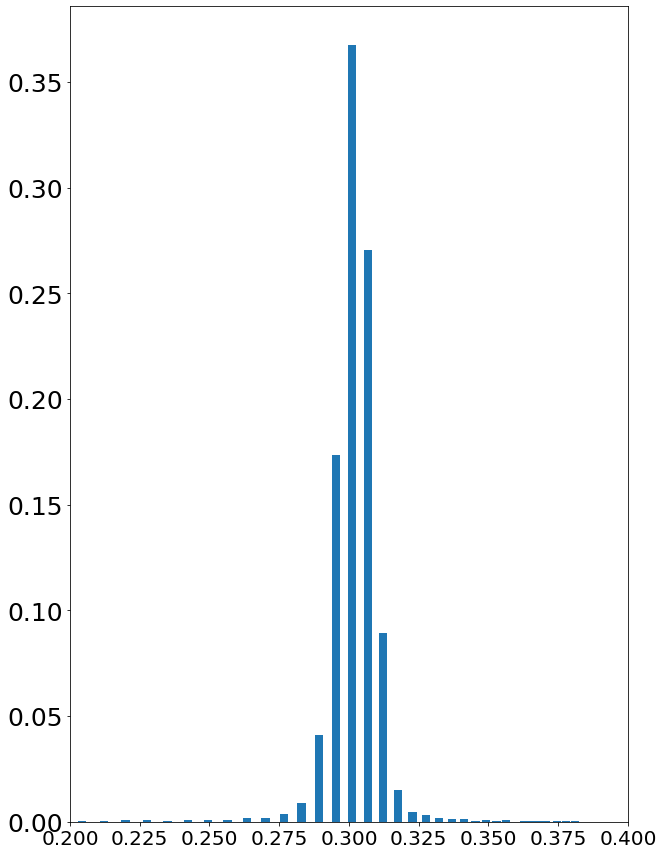}
\caption{$n$=8} \label{fig:phase_estimation_simulations_8}
\end{subfigure}
\begin{subfigure}{0.31\textwidth}
\includegraphics[width=\linewidth]{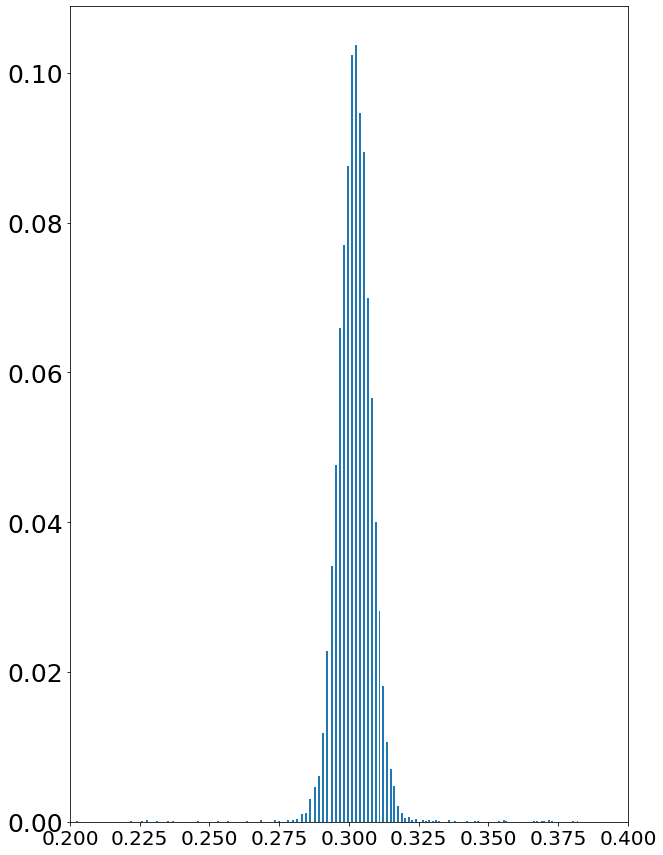}
\caption{$n$=10} \label{fig:phase_estimation_simulations_10}
\end{subfigure}
\caption{Probability distribution for the output of an inner product estimation with $x=(0.25,0.3,-0.5)$ and $y=(-0.15,+0.3,-0.5)$, $\braket{x}{y}=0.302$,  for different number of qubits $n$. The shape of the distribution depends on the value to estimate.}\label{fig:phase_estimation_probability_simulations}
\end{figure}

First, we plotted the distribution from Eq.(\ref{eq:phase_estimation_probability_distribution}) along with the inner product estimation that follows, for two particular vectors. The results are given in Fig.\ref{fig:phase_estimation_probability_simulations}. We can see that the more qubits, the precise is the distribution around the exact value of the inner product. Recall from the previous section that the shape of the distribution depends on the value itself. With more qubits, more values with high probabilities are close to the exact value, and therefore the estimation is more and more precise. 

To gain one dimension, we performed the same simulations by varying one of the two input vectors. We chose two vectors of two dimensions but with only one element of one vector is shown in the x-axis, so that the inner product is shown in the y-axis. The first vector was $(0.544,0.6)$ and the variable vector $(x,1)$.

\begin{figure}[H] % "[t!]" placement specifier just for this example
\begin{subfigure}{0.48\textwidth}
\includegraphics[width=\linewidth]{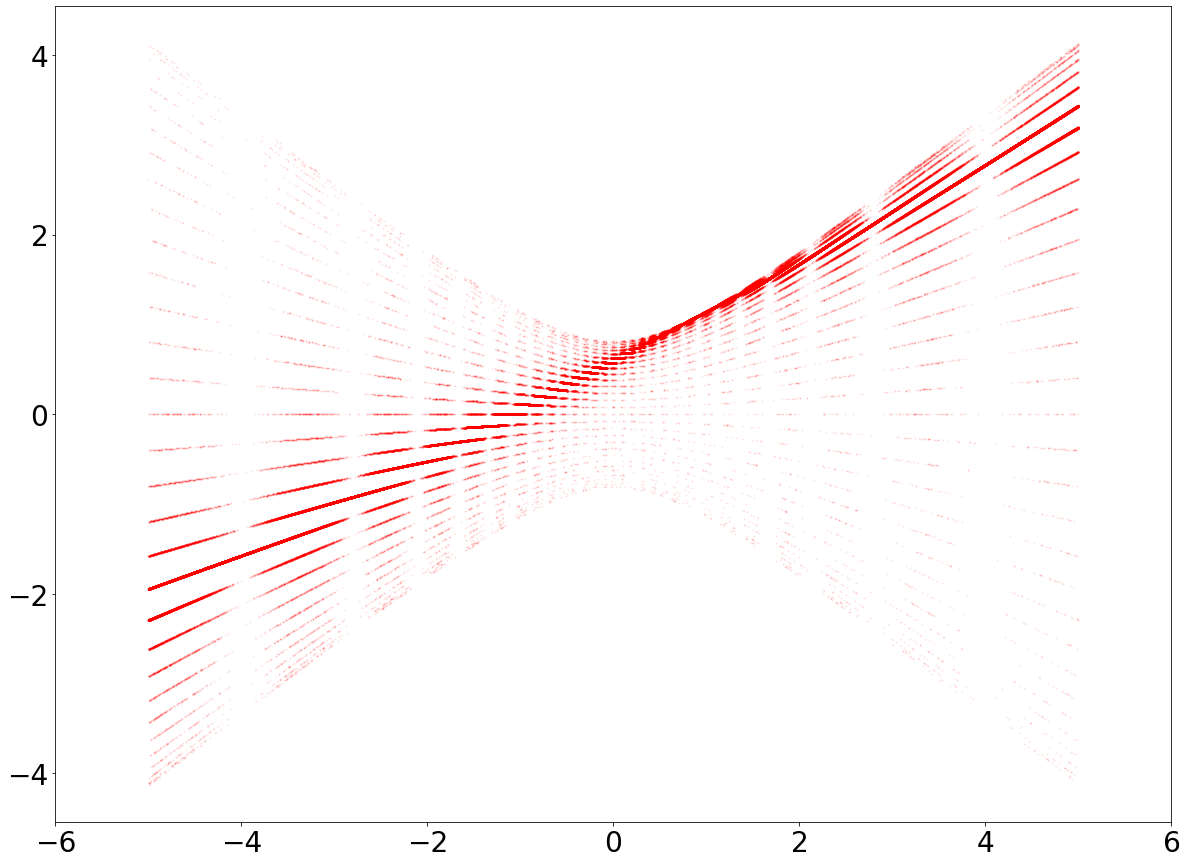}
\caption{$n$=6} \label{fig:error_IPE_500k_6}
\end{subfigure}\hspace*{\fill}
\begin{subfigure}{0.48\textwidth}
\includegraphics[width=\linewidth]{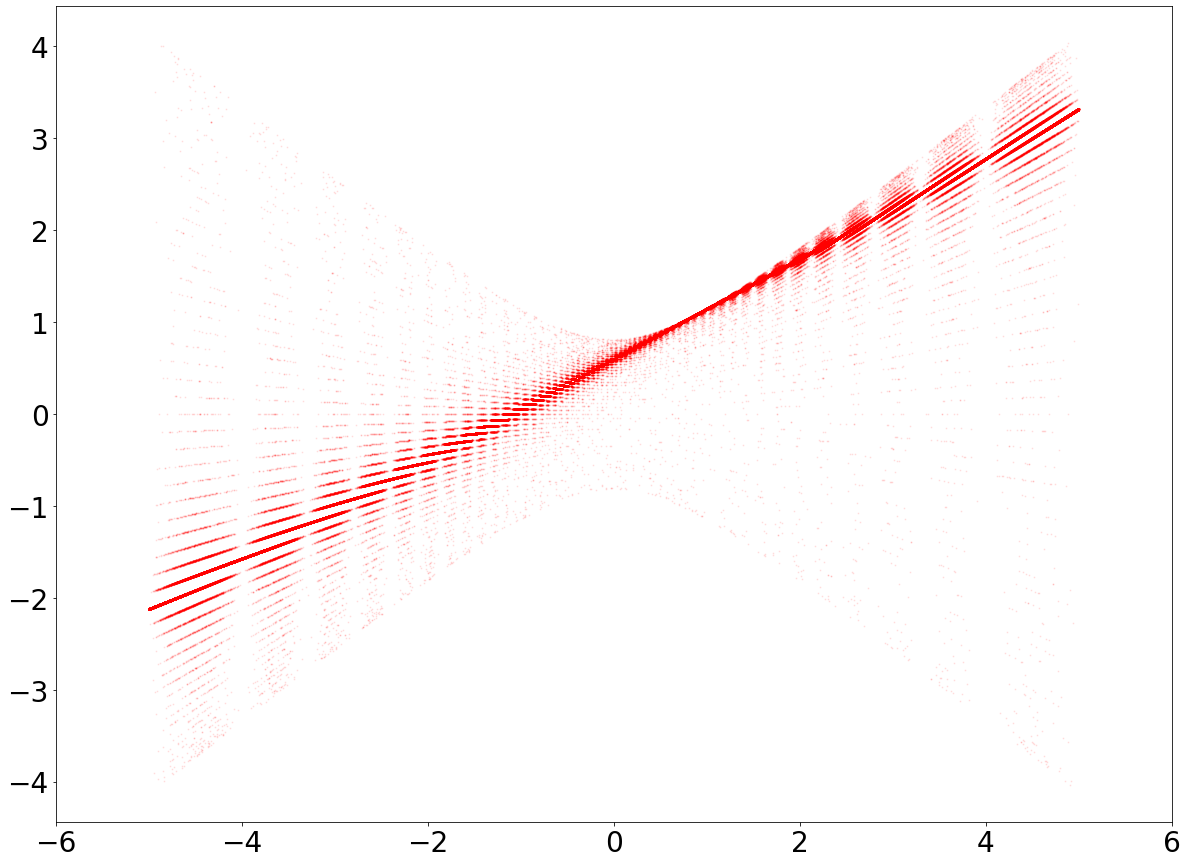}
\caption{$n$=7} \label{fig:error_IPE_500k_7}
\end{subfigure}
\medskip
\begin{subfigure}{0.48\textwidth}
\includegraphics[width=\linewidth]{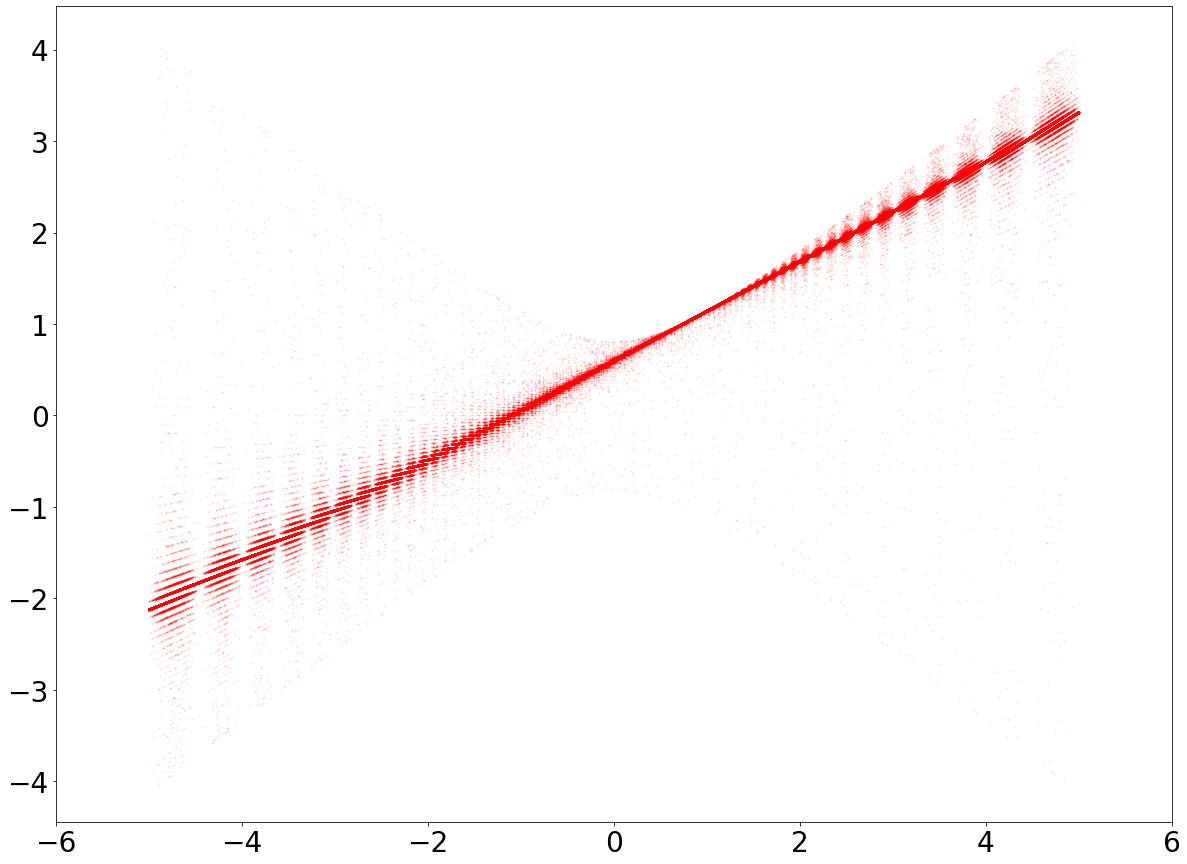}
\caption{$n$=8} \label{fig:error_IPE_500k_8}
\end{subfigure}\hspace*{\fill}
\begin{subfigure}{0.48\textwidth}
\includegraphics[width=\linewidth]{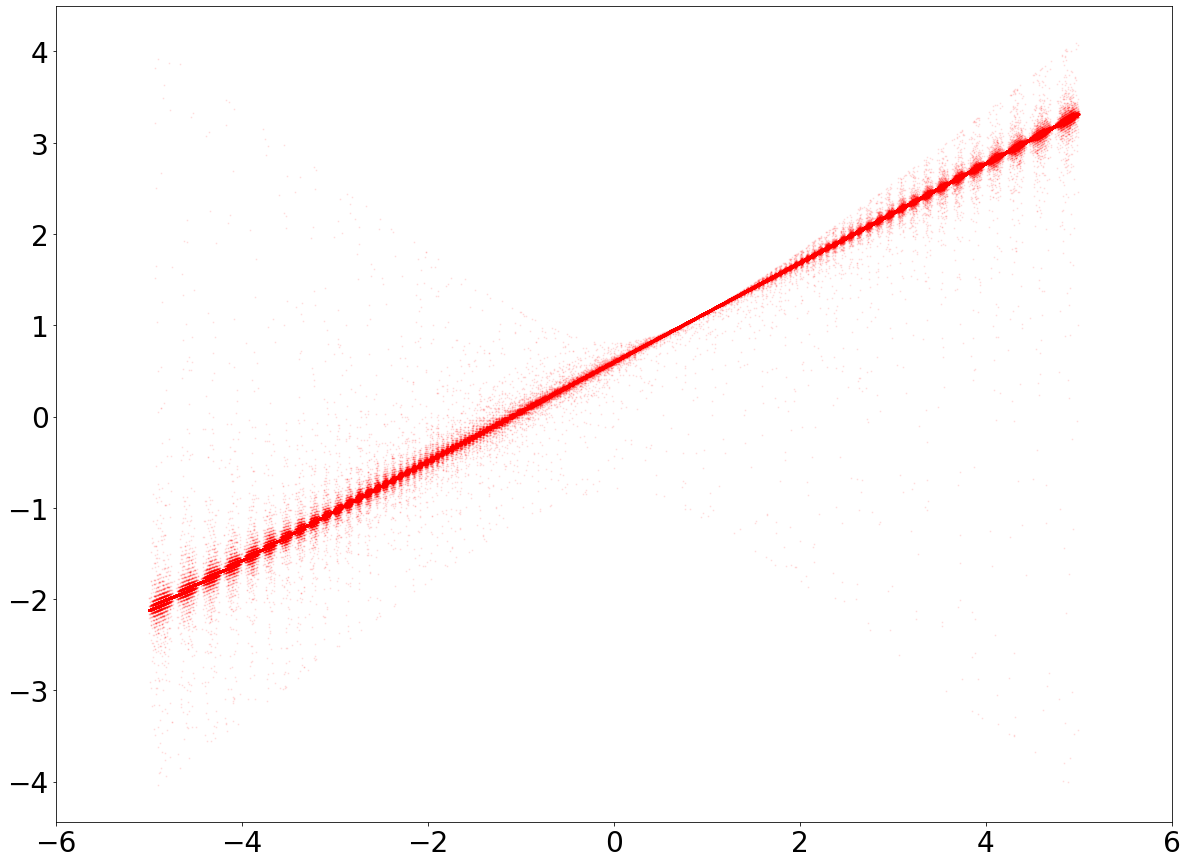}
\caption{$n$=9} \label{fig:error_IPE_500k_9}
\end{subfigure}
\caption{Classical simulation results of the Quantum Inner Product Estimation algorithm. The y-axis is the result of the quantum inner product between a fixed vector $(0.544,0.6)$ and a variable vector $(x,1)$, where $x$ is the x-axis. The estimation is repeated over 500000 points, simulated for different number $n$ of qubits.} 
\label{fig:inner_product_error_simulations}
\end{figure}

\begin{figure}[H] % "[t!]" placement specifier just for this example
\begin{subfigure}{0.48\textwidth}
\includegraphics[width=\linewidth]{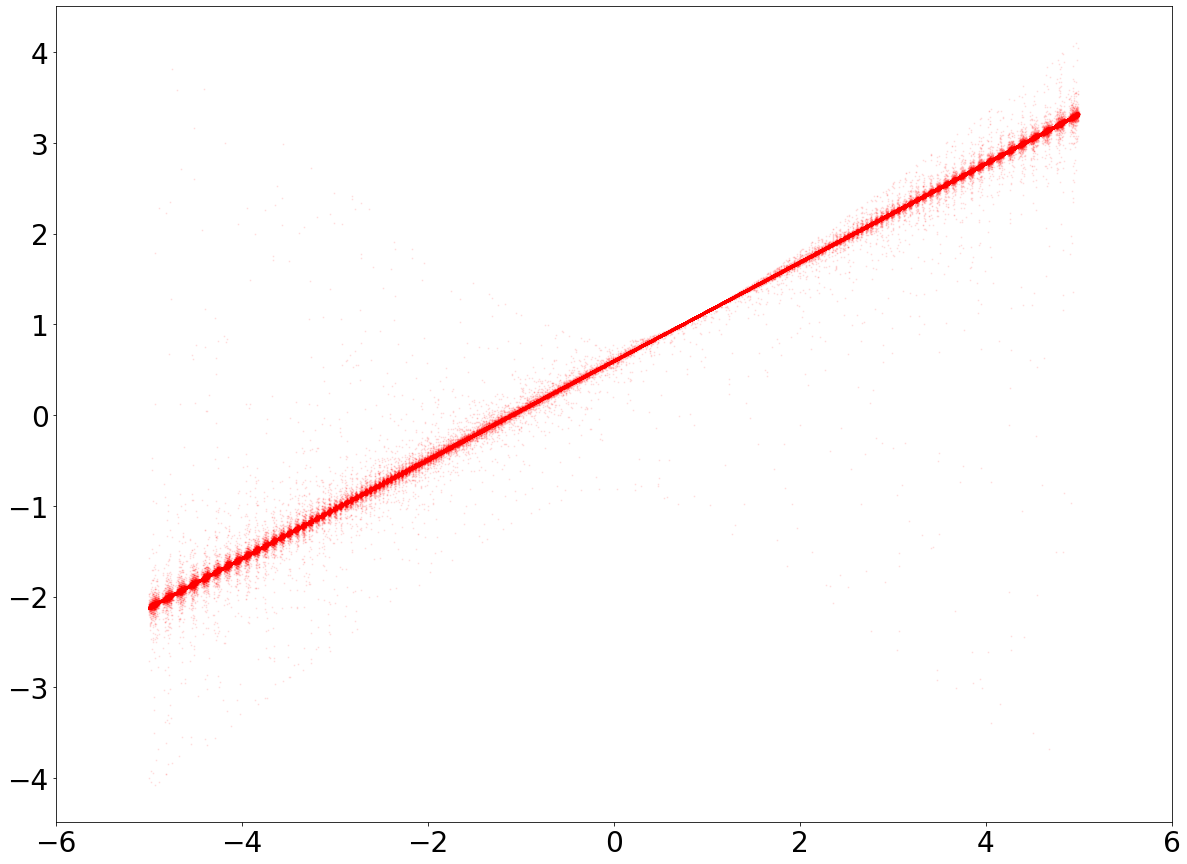}
\caption{$n$=10} \label{fig:error_IPE_500k_10}
\end{subfigure}\hspace*{\fill}
\begin{subfigure}{0.48\textwidth}
\includegraphics[width=\linewidth]{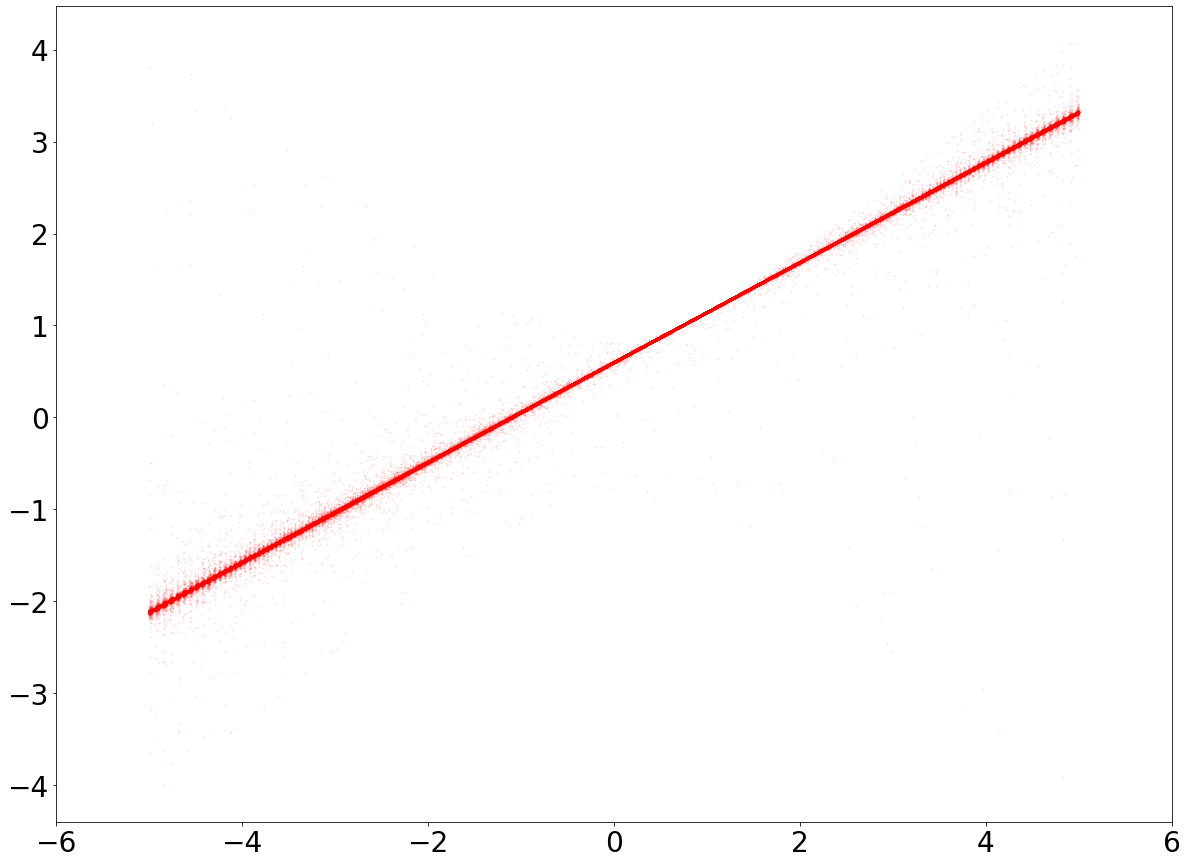}
\caption{$n$=11} \label{fig:error_IPE_500k_11}
\end{subfigure}
\caption{Classical simulation results of the Quantum Inner Product Estimation algorithm. The y-axis is the result of the quantum inner product between a fixed vector $(0.544,0.6)$ and a variable vector $(x,1)$, where $x$ is the x-axis. The estimation is repeated over 500000 points, simulated for different number $n$ of qubits.} 
\label{fig:inner_product_error_simulations_2}
\end{figure}

The results are given in Figures \ref{fig:inner_product_error_simulations} and \ref{fig:inner_product_error_simulations_2} for increasing number of qubits $n$. For each, we sampled randomly 500000 points $x$ and generate an estimation of the inner product using the simulation. In the ideal case, the result of such a simulation should be a straight line $f(x) = \lambda x$ with $\lambda = 0.6 + 0.544 x$. For each graph, we observe a general tendency to follow this line, accompanied by oscillations around it, depending on the $x$ value and the number of qubits. 

The parts of the oscillations with high amplitude show where the error is expected to be the highest. As expected, the error increases on average when the norm of the vector increases. 

Notably, with a small number of qubits such as $n=10$, we are already almost reproducing the ideal line with low noise around.

%\begin{figure}[h] % "[t!]" placement specifier just for this example
%\begin{subfigure}{0.48\textwidth}
%\includegraphics[width=\linewidth]{images/error_IPE_500k_6.png}
%\caption{$n$=6} \label{fig:error_IPE_500k_6}
%\end{subfigure}\hspace*{\fill}
%\begin{subfigure}{0.48\textwidth}
%\includegraphics[width=\linewidth]{images/error_IPE_500k_7.png}
%\caption{$n$=7} \label{fig:error_IPE_500k_7}
%\end{subfigure}
%\medskip
%\begin{subfigure}{0.48\textwidth}
%\includegraphics[width=\linewidth]{images/error_IPE_500k_8.png}
%\caption{$n$=8} \label{fig:error_IPE_500k_8}
%\end{subfigure}\hspace*{\fill}
%\begin{subfigure}{0.48\textwidth}
%\includegraphics[width=\linewidth]{images/error_IPE_500k_9.png}
%\caption{$n$=9} \label{fig:error_IPE_500k_9}
%\end{subfigure}
%\medskip
%\begin{subfigure}{0.48\textwidth}
%\includegraphics[width=\linewidth]{images/error_IPE_500k_10.png}
%\caption{$n$=10} \label{fig:error_IPE_500k_10}
%\end{subfigure}\hspace*{\fill}
%\begin{subfigure}{0.48\textwidth}
%\includegraphics[width=\linewidth]{images/error_IPE_500k_11.png}
%\caption{$n$=11} \label{fig:error_IPE_500k_11}
%\end{subfigure}
%\caption{Classical simulation results of the Quantum Inner Product Estimation algorithm. The y-axis is the result of the quantum inner product between a fixed vector $(0.544,0.6)$ and a variable vector $(x,1)$, where $x$ is the x-axis. The estimation is repeated over 500000 points, simulated for different number $n$ of qubits.} \label{fig:inner_product_error_simulations}
%\end{figure}

Finally, we couldn't help but think of the famous interference pattern one can see in the diffraction of a laser beam passing through a small slit or obstacle. This pattern is the signature of a wave-like phenomenon such as light, but also matter at its smallest scale, as quantum physics tell us.\\   

\begin{figure}[H]
    \centering
    \includegraphics[width=0.7\textwidth]{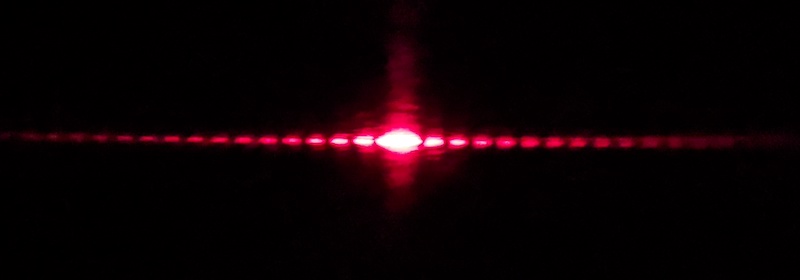}
    \caption{A quantum interference pattern produced by the diffraction of a laser beam. Source: Institute of Quantum Computing, University of Waterloo.}
    \label{fig:laser_inteference}
\end{figure}

Note that in the rest of this thesis, when trying to simulate quantum algorithms classically, we will approximate the quantum noise of inner product or distance estimation by Gaussian noise with mean 0 and variance $\epsilon$, multiplied by the norm of the input vectors. We have seen that this is not exactly what is occurring in theory. The noise at a specific value may seem Gaussian (see Fig.\ref{fig:phase_estimation_probability_simulations}), but in fact it depends on the value itself, which makes it non-isotropic and periodic. That being said, Gaussian noise remains a simpler and fairly accurate approximation of the above effect.

%%%%---------------PART III-------------%%%%
\part{Quantum Unsupervised Learning}\label{part:Unsupervised_QML}
\chapter{Introduction}
\epigraph{\textit{"La science consiste à passer d'un étonnement à un autre."}}{Aristote \\ \emph{La Métaphysique}}

In this Part, we are interested in unsupervised learning and in particular in the canonical problem of clustering: given a dataset represented as $\n$ vectors, we want to find an assignment of the vectors to one of $k$ labels (for a given $k$ that we assume to know) such that similar vectors are assigned to the same cluster. Often, the Euclidean distance is used to measure the similarity of vectors, but other metrics might be used, according to the problem under consideration. Details of the classical algorithms are given in Chapter \ref{chap:intro_classical_ml}. \\

In Chapter \ref{chap:qmeans}, we propose $q$-means \cite{qmeans}, a quantum algorithm for clustering, which can be viewed as a quantum alternative to the classical $k$-means algorithm. More precisely, $q$-means is the equivalent of the $\delta$-$k$-means algorithm, a robust version of $k$-means that will be defined later.  We provide a detailed analysis to show that $q$-means has an output consistent with the classical $\delta$-$k$-means algorithm and further has a running time that depends polylogarithmically on $N$, the number of elements in the dataset. We introduce the notion of \emph{well-clusterable} dataset, allowing us to further refine the final running time. The last part of this work includes simulations which asserts the performance and running time of the $q$-means algorithm. \\

In Chapter \ref{chap:Q_spectral_clustering}, we propose the quantum spectral clustering algorithm \cite{quantumspectralclustering}, a quantum analog of the spectral clustering algorithm. With roots in graph theory, it uses the spectral properties of the Laplacian matrix to project the data in a low dimensional space where clustering is more efficient. Despite its success in clustering tasks, spectral clustering suffers in practice from a fast-growing running time of $O(N^3)$, where $N$ is the number of points in the dataset. In this work we propose an end-to-end quantum algorithm performing spectral clustering, extending several works in quantum machine learning. 

The quantum algorithm is composed of two parts: the first is the efficient creation of the quantum state corresponding to the projected Laplacian matrix, and the second consists of applying $q$-means algorithm presented before. 
Both steps depend polynomially on the number of clusters, as well as precision and data parameters arising from quantum procedures, and polylogarithmically on the dimension of the input vectors. Our numerical simulations show an asymptotic linear growth with $N$ when all terms are taken into account, significantly better than the classical cubic growth.

This work opens the path to other graph-based quantum machine learning algorithms, as it provides routines for efficient computation and quantum access to the Incidence, Adjacency, and projected Laplacian matrices of a graph.

\subsubsection{Related Work}

In this section, we discuss previous work on quantum unsupervised learning and clustering. 
Aimeur, Brassard and Gambs \cite{aimeur2013quantum} gave two quantum algorithms for unsupervised learning using the amplification techniques from \cite{durr1996quantum}. Specifically, they proposed an algorithm for clustering based on minimum spanning trees that runs in time $\Theta(\n^{3/2})$ and a quantum algorithm for $k$-median (a problem related to k-means)  algorithm with complexity time $O(N^{3/2}/\sqrt{k})$.

Lloyd, Mohseni and Rebentrost \cite{LMR13} proposed quantum $k$-means and nearest centroid algorithms using an efficient subroutine for quantum distance estimation assuming as we do quantum access to the data. Given a dataset of $\n$ vectors in a feature space of dimension $d$, the running time of each iteration of the clustering algorithm (using a distance estimation procedure with error $\epsilon$) is  $O(\frac{k\n\log d}{\epsilon})$ to produce the quantum state corresponding to the clusters. Note that the time is linear in the number of data points and it will be linear in the dimension of the vectors if the algorithm needs to output the classical description of the clusters. 
% and of course. 
% the running time has to be multiplied by a factor of $d$ if a classical solution needs to be returned, which is necessary when a new iteration of the $k$-means algorithm needs to be executed. \iordanis{check above}

In the same work, they also proposed an adiabatic algorithm for the assignment step of the $k$-means algorithm, that can potentially provide an exponential speedup in the number of data points as well, in the case the adiabatic algorithm performs exponentially better than the classical algorithm. The adiabatic algorithm is used in two places for this algorithm, the first to select the initial centroids, and the second to assign data points to the closest cluster. However, while arguments are given for its efficiency, it is left as an open problem to determine how well the adiabatic algorithm performs on average, both in terms of the quality of the solution and the running time.  

\cite{wiebe_nearest_neigbhors} applied the minimum finding algorithm \cite{durr1996quantum} to obtain nearest-neighbor methods for supervised and unsupervised learning. At a high level, they recovered a Grover-type quadratic speedup with respect to the number of elements in the dataset in finding the $k$ nearest neighbors of a vector. 
Otterbach et al.  \cite{Otterbach17} performed clustering by exploiting a well-known reduction from clustering to the Maximum-Cut (MAXCUT) problem; the MAXCUT is then solved using QAOA, a quantum algorithm for performing approximate combinatorial optimization \cite{farhi2014quantum}.\\

There is extensive work in quantum computing involving graph problems such as min-cut, max-flow, or the traveling salesman problem \cite{moylett2017quantum,maxcut_cui16}, but only a few are about graph-based machine learning \cite{Otterbach17, drineas2004clustering}. Spectral clustering has been studied in \cite{daskin2017quantum} but no proven speedups were given.
More recently, \cite{apers2019quantum} introduced quantum algorithms using the graph Laplacian for optimization and machine learning applications, including spectral clustering. In that paper, the starting point was the assumption of having superposition-access to the classically-stored weights of the similarity graph, corresponding to the Laplacian directly, from which the authors performed tasks like sparsification of the graph faster than classical algorithms. In our work, we propose an efficient quantum algorithm to construct the projected Laplacian matrix itself from access to the classical input, before proceeding to the clustering algorithm. Note that one could eventually combine the methods from \cite{apers2019quantum} and our procedure, or any other procedure that provides access to the projected Laplacian matrix, and thus construct different quantum spectral clustering algorithms.\\

\chapter{Q-means}\label{chap:qmeans}
\epigraph{\textit{"Mathematics may be defined as the subject in which we never know what we are talking about, nor whether what we are saying is true."}}{Bertrand Russell \\ \emph{Mysticism and Logic and Other Essays} (1910)}

\section{Preliminaries}

Preliminaries on the $k$-means algorithm , along with all notations are given in Section \ref{sec:classical_kmeans_preliminaries}.

\subsection{Main Results}
We define and analyze a new quantum algorithm for clustering, the $q$-means algorithm, whose performance is similar to that of the classical $k$-means algorithm defined in Section \ref{sec:classical_kmeans_preliminaries} and whose running time provides substantial savings, especially for the case of large data sets. To be more precise and to take into account all quantum randomness and noise, we define the classical $\delta$-$k$-means algorithm in Section \ref{sec:delta_k_means}, a noisy version of the $k$-means that is equivalent to our quantum algorithm.

The $q$-means algorithm combines most of the advantages that quantum machine learning algorithms can offer for clustering. First, the running time is polylogarithmic in the number of elements of the dataset and depends only linearly on the dimension of the feature space. Second, $q$-means returns explicit classical descriptions of the cluster centroids that are obtained by the $\delta$-$k$-means algorithm. As the algorithm outputs a classical description of the centroids, it is possible to use them in further (classical or quantum) classification algorithms.

Our $q$-means algorithm requires that the dataset is stored in a QRAM (see Theorem \ref{thm:QRAM}), which allows the algorithm to use efficient linear algebra routines (Part \ref{part:preliminaries_QML}) that have been developed using QRAM data structures. Of course, our algorithm can also be used for clustering datasets for which the data points can be efficiently prepared even without a QRAM, for example if the data points are the outputs of quantum circuits. \\

We start by providing a worst case analysis of the running time of our algorithm, which depends on parameters of the data matrix, for example the condition number and the parameter $\mu$ that appears in the quantum linear algebra procedures. Note that with $\widetilde{O}$ we hide polylogarithmic factors.

\begin{result}{$q$-means - General Case}{}
Given dataset $V \in \mathbb{R}^{\n \times d} $ stored in QRAM, the q-means algorithm outputs with high probability centroids $c_1, \cdots,  c_k$ that are consistent with an output of the $\delta$-$k$-means algorithm in time 
\begin{equation}
\widetilde{O}\left(    k d \frac{\eta}{\delta^2}\kappa(V)(\mu(V) + k \frac{\eta}{\delta}) + k^2 \frac{\eta^{1.5}}{\delta^2} \kappa(V)\mu(V)
\right)
\end{equation}
per iteration,  where $\kappa(V)$ is the condition number, $\mu(V)$ is a parameter that appears in quantum linear algebra procedures and $1\leq \norm{v_i}^2 \leq \eta$. 
\end{result}

When we say that the $q$-means output is consistent with the $\delta$-$k$-means, we mean that with high probability the clusters that the $q$-means algorithm outputs are also possible outputs of the $\delta$-$k$-means.

We go further in our analysis and study a well-motivated model for datasets that allows for good clustering. We call these datasets {\em well-clusterable}. One possible way to think of such datasets is the following: a dataset is well-clusterable when the $k$ clusters arise from picking $k$ well-separated vectors as their centroids, and then each point in the cluster is sampled from a Gaussian distribution with small variance centered on the centroid of the cluster. We provide a rigorous definition in the following sections. For such well-clusterable datasets, we can provide a tighter analysis of the running time and have the following result, whose formal version appears as Theorem \ref{thm:qmeans_main}. 

\begin{result}{$q$-means - Well-Clusterable Case}{}
Given a well-clusterable dataset $V \in \mathbb{R}^{\n \times d} $ stored in QRAM, the q-means algorithm outputs with high probability $k$ centroids $c_1, \cdots ,c_k$ that are consistent with the output of the $\delta$-$k$-means algorithm in time 
\begin{equation}
\widetilde{O}\left( k^2 d \frac{\eta^{2.5}}{\delta^3} + k^{2.5} \frac{\eta^2}{\delta^3} \right)
\end{equation}
per iteration, where $1\leq \norm{v_i}^2 \leq \eta$. 
\end{result}

\noindent 
In order to assess the running time and performance of our algorithm, we performed extensive simulations for different datasets. The running time of the $q$-means algorithm is linear in the dimension $d$, which is necessary when outputting a classical description of the centroids, and polynomial in the number of clusters $k$ which is typically a small constant. The main advantage of the $q$-means algorithm is that it provably  depends logarithmically on the number of points, which can in many cases provide a substantial speedup. The parameter $\delta$ (which plays the same role as in the $\delta$-$k$-means) is expected to be a large enough constant that depends on the data and the parameter $\eta$ is again expected to be a small constant for datasets whose data points have roughly the same norm. For example, for the MNIST dataset, $\eta$ can be less than $8$ and $\delta$ can be taken to be equal to $0.5$. In Section \ref{sec:qmeans_numerical_simulations} we present the results of our simulations. For different datasets we find parameters $\delta$ such that the number of iterations is practically the same as in the $k$-means, and the $\delta$-$k$-means algorithm converges to a clustering that achieves an accuracy similar to the $k$-means algorithm or in times better. We obtained these simulation results by simulating the operations executed by the quantum algorithm adding the appropriate errors in the procedures.\\

As a side note, a generalization of the $q$-means algorithms has been proposed to solve the Gaussian mixture model in \cite{kerenidis2020quantum_guassianmixture}. Instead of looking for centroids only, it also finds the covariance matrix for each cluster that approximates the multi dimensional Gaussian distribution of data points around the centroids. This ensures a more precise clustering and allows for clusters to be of different sizes and shapes.

\subsection{$\delta$-$k$-means}\label{sec:delta_k_means}

We now consider a $\delta$-robust version of the $k$-means defined in Section \ref{sec:classical_kmeans_preliminaries}, in which we introduce some noise. The noise affects the algorithms in both of the steps of k-means: label assignment and centroid estimation. 

Let us describe the rules for the assignment step of $\delta$-$k$-means more precisely. Let $c^{*}_i$ be the closest centroid to the data point $v_i$. Then, the set of possible labels $L_{\delta}(v_i)$ for $v_i$ is 
defined as follows:
\begin{equation}
L_{\delta}(v_i)  =  \{c_p  : | d^2(c^*_i, v_i ) - d^2(c_p, v_i) | \leq \delta \: \}
\end{equation}
The assignment rule selects arbitrarily a cluster label from the set $L_{\delta}(v_i)$.  

Second, we add $\delta/2$ noise during the calculation of the centroid. Let $\mathcal{C}_j^{t+1}$ be the set of points which has been labeled by $j$ in the previous step. For $\delta$-k-means we pick a centroid $c^{t+1}_j $ with the property

\begin{equation}\norm{ c^{t+1}_j - \frac{1}{|\mathcal{C}^{t+1}_j|}\sum_{v_i \in \mathcal{C}^{t+1}_j} v_i} < \frac{\delta}{2}
\end{equation}

One way to do this is to calculate the centroid exactly and then add some small Gaussian noise to the vector to obtain the robust version of the centroid. 

Let us add two remarks on the $\delta$-$k$-means. First, for a well-clusterable data set and for a small $\delta$, the number of vectors on the boundary that risk to be misclassified in each step, that is the vectors for which $|L_{\delta}(v_i)|>1$ is typically much smaller compared to the vectors that are close to a unique centroid. Second, we also increase by $\delta/2$  the convergence threshold from the $k$-means algorithm.
All in all, $\delta$-$k$-means is able to find a clustering that is robust when the data points and the centroids are perturbed with some noise of magnitude $O(\delta)$. 
As we will see in this work, $q$-means is the quantum equivalent of $\delta$-$k$-means.

\subsection{Well-Clusterable Datasets}\label{sec:well_clusterable_datasets}

In this section, we define a model for the dataset in order to provide a tight analysis on the running time of our clustering algorithm. 
Note that we do not need this assumption for our general $q$-means algorithm, but in this model we can provide tighter bounds for its running time. 
Without loss of generality we consider in the following that the dataset $V$ is normalized so that for all $i \in [N]$, we have $1 \leq \norm{v_{i}}$, and we define the parameter $\eta = \max_i{\norm{v_i}^2}$. We will also assume that the number $k$ is the ``right'' number of clusters, meaning that we assume each cluster has at least some $\Omega(N/k)$ data points.

We now introduce the notion of a \emph{well-clusterable} dataset. The definition aims to capture some properties that we can expect from datasets that can be clustered efficiently using a k-means algorithm. Our notion of a well-clusterable dataset shares some similarity with the assumptions made in \cite{drineas2002competitive}, but there are also some differences specific to the clustering problem.

\begin{definition}{Well-clusterable dataset}{wcdataset}
A data matrix $V \in \R^{\n\times d}$ with rows $v_{i} \in \R^{d}, i \in [N]$ is said to be well-clusterable if there exist constants $\xi, \beta>0$, $\lambda \in [0,1]$, $\eta \leq 1$, and cluster centroids $c_i$ for $i\in [k]$ such that:
\begin{enumerate}
    \item (separation of cluster centroids): $d(c_i, c_j) \geq \xi \quad \forall i,j \in[k]  $
    \item (proximity to cluster centroid): At least $\lambda \n$ points $v_i$ in the dataset satisfy $d(v_i, c_{l(v_i)}) \leq \beta$ where $c_{l(v_i)}$ is the centroid 
    nearest to $v_{i}$. 
 \item (Intra-cluster smaller than inter-cluster square distances): 
 The following inequality is satisfied 
 \begin{equation}
 4\sqrt{\eta} \sqrt{ \lambda \beta^{2} + (1-\lambda) 4\eta} \leq \xi^{2} - 2\sqrt{\eta} \beta
 \end{equation}
 
\end{enumerate}
\end{definition}

Intuitively, the assumptions guarantee that most of the data can be easily assigned to one of $k$ clusters, since these points are close to the centroids, and the centroids are sufficiently far from each other. The exact inequality comes from the error analysis, but in spirit it says that $\xi^2$ should be bigger than a quantity that depends on $\beta$ and the maximum norm $\eta$.

 We now show that a well-clusterable dataset has a good rank-$k$ approximation where $k$ is the number of clusters. This result will later be used for giving tight upper bounds on the running time of the quantum algorithm for well-clusterable datasets. As we said, one can easily construct such datasets by picking $k$ well separated vectors to serve as cluster centers and then each point in the cluster is sampled from a Gaussian distribution with small variance centered on the centroid of the cluster. 

\begin{claim}{}{low-rank}
 Let $V_{k}$ be the optimal $k$-rank approximation for a well-clusterable data matrix $V$, then $\norm{V-V_{k}}_{F}^2  \leq (  \lambda \beta^2 +(1-\lambda)4\eta)\norm{V}^2_F$.
\end{claim}

\begin{proof}
Let $W \in \R^{\n\times d}$ be the matrix with row $w_{i} = c_{l(v_i)}$, where $c_{l(v_i)}$ is the centroid closest to $v_i$. 
The matrix $W$ has rank at most $k$ as it has exactly $k$ distinct rows. As $V_k$ is the optimal 
rank-$k$ approximation to $V$, we have $\norm{V-V_{k}}_{F}^{2} \leq \norm{V-W}_F^2$. It therefore suffices to upper bound $\norm{V-W}_F^2$. 
Using the fact that $V$ is  well-clusterable, we have
\begin{equation}
\norm{V-W}_F^2 = \sum_{ij} (v_{ij} - w_{ij})^2 = \sum_{i} d(v_i, c_{l(v_i)})^2 \leq \lambda \n \beta^2 + (1-\lambda)\n4\eta
\end{equation}
where we used Definition \ref{def:wcdataset} to say that for a $\lambda N$ fraction of the 
points $d(v_i, c_{l(v_i)})^2 \leq \beta^{2}$ and for the remaining points $d(v_i, c_{l(v_i)})^2 \leq 4\eta$.     
Also, as all $v_{i}$ have norm at least $1$ we have $N \leq \norm{V}_F$, implying that 
  $\norm{V-V_k}^{2} \leq  \norm{V-W}_F^2 \leq ( \lambda \beta^2 +(1-\lambda)4\eta)\norm{V}_F^2$.

\end{proof}%\Hsquare

\noindent The running time of the quantum linear algebra routines for the data matrix $V$ in Theorem \ref{thm:quantum_matrix_multiplication_inversion} depend on the parameters $\mu(V)$ 
and $\kappa(V)$. We establish bounds on both of these parameters using the fact that $V$ is well-clusterable
\begin{claim}{}{mu}
Let $V$ be a well-clusterable data matrix, then $\mu(V):= \frac{\norm{V}_{F}}{\norm{V}} = O(\sqrt{k})$. 
\end{claim}
\begin{proof} 
We show that when we rescale $V$ so that $\norm{V} = 1$, then we have $\norm{V}_{F}= O(\sqrt{k})$ for the rescaled matrix. 
From the triangle inequality we have that
$\norm{V}_F \leq \norm{V-V_k}_F + \norm{V_k}_F$.
Using the fact that $\norm{V_k}_F^2 = \sum_{i \in [k]} \sigma_i^2 \leq k$ and Claim \ref{thm:low-rank}, we have, 
\begin{equation}
\norm{V}_F \leq \sqrt{ (\lambda\beta^2+(1-\lambda)4\eta)} \norm{V}_F + \sqrt{k}
\end{equation}
Rearranging, we have that
$\norm{V}_F \leq \frac{\sqrt{k}}{1-\sqrt{(\lambda\beta^2+(1-\lambda)4\eta)}} = O(\sqrt{k})$.  
\end{proof}%\Hsquare 
We next show that if we use a condition threshold $\kappa_\tau(V)$ instead of the true condition number $\kappa(V)$, that is we consider 
the matrix $V_{\geq \tau} = \sum_{\sigma_{i} \geq \tau} \sigma_{i} u_{i} v_{i}^{T}$ by discarding the smaller singular values $\sigma_{i} < \tau$, the resulting matrix remains close to the original one, i.e. we have that $\norm{V - V_{\geq \tau}}_F$ is bounded.  
 
\begin{claim}{}{kappa}
Let $V$ be a matrix with a rank-$k$ approximation given by $\norm{V- V_{k}}_{F}\leq \epsilon' \norm{V}_{F}$ and let $\tau = \frac{\errkappa}{\sqrt{k}}\norm{V}_F$, then $\norm{V- V_{\geq \tau}}_{F}\leq (\epsilon' + \errkappa) \norm{V}_{F}$. 
\end{claim}
\begin{proof}
Let $l$ be the smallest index such that $\sigma_{l} \geq \tau$, so that we have $\norm{V-V_{\geq \tau}}_F = \norm{V-V_l}_F$. 
We split the argument into two cases depending on whether $l$ is smaller or greater than $k$. 
\begin{itemize}
	\item If $l \geq k$ then $\norm{V-V_l}_F \leq \norm{V-V_k}_F \leq  \epsilon' \norm{V}_F$.
	\item If $l < k$ then, $\norm{V-V_l}_F \leq \norm{V-V_k}_F + \norm{V_k-V_l}_F \leq  \epsilon' \norm{V}_F + \sqrt{\sum_{i=l+1}^k \sigma_i^2}$. 
	
	As each $\sigma_{i} < \tau$ and the sum is over at most $k$ indices, we have the upper bound  $(\epsilon' + \errkappa) \norm{V}_F$. 	
\end{itemize}
\end{proof}%\Hsquare

\noindent
The reason we defined the notion of well-clusterable dataset is to be able to provide some strong guarantees for the clustering of most points in the dataset. Note that the clustering problem in the worst case is NP-hard and we only expect to have good results for datasets that have some good property. Intuitively, we should only expect $k$-means to work when the dataset can actually be clustered in $k$ clusters. We next show that for a well-clusterable dataset $V$, there is a constant $\delta$ that can be computed in terms of the parameters in Definition \ref{def:wcdataset} such that the $\delta$-$k$-means clusters correctly most of the data points. 

\begin{claim}{}{distcentroid} 
Let $V$ be a well-clusterable data matrix. Then, for at least $\lambda \n$ data points $v_i$, we have 
\begin{equation}
\min_{j\neq \ell(i)}(d^2(v_i,c_j)-d^2(v_i,c_{\ell(i)}))\geq \xi^2 - 2\sqrt{\eta}\beta
\end{equation}
which implies that a $\delta$-$k$-means algorithm with any $\delta < \xi^2 - 2\sqrt{\eta}\beta$ will cluster these points correctly.
\end{claim}
\begin{proof}
    By Definition \ref{def:wcdataset}, we know that for a well-clusterable dataset $V$, 
    we have that $d(v_i, c_{l(v_i)}) \leq \beta$ for at least $\lambda \n$ data points and where $c_{l(v_{i})}$ is the centroid closest to $v_{i}$. 
    Further, the distance between each pair of the $k$ centroids satisfies the bounds $2\sqrt{\eta} \geq d(c_i, c_j) \geq \xi$. By the triangle inequality, we have $d(v_i,c_j) \geq d(c_j,c_{\ell(i)})-d(v_i,c_{\ell(i)})$. Squaring both sides of the inequality and rearranging, 
  
\begin{equation}
d^2(v_i,c_j) - d^2(v_i,c_{\ell(i)})\geq d^2(c_j,c_{\ell(i)})  - 2d(c_j,c_{\ell(i)})d(v_i,c_{\ell(i)}))
\end{equation}
Substituting the bounds on the distances implied by the well-clusterability assumption, we obtain $d^2(v_i,c_j)-d^2(v_i,c_{\ell(i)}) \geq \xi^2 - 2\sqrt{\eta} \beta$. This implies that as long as we pick $\delta <  \xi^2 - 2\sqrt{\eta}\beta$, these points are assigned to the correct cluster, since all other centroids are more than $\delta$ further away than the correct centroid. 
    
\end{proof}%\Hsquare

\section{The $q$-means Algorithm}\label{sec:qmeans}

\subsection{Quantum Circuit}

The $q$-means algorithm is given as Algorithm \ref{$q$-means-algo}. At a high level, it follows the same steps as the classical $k$-means algorithm described in Section \ref{sec:classical_kmeans_preliminaries}. Compared to $k$-means, $q$-means uses quantum subroutines for distance estimation, finding the minimum value among a set of elements, matrix multiplication for obtaining the new centroids as quantum states, and efficient tomography. First, we pick some random initial points, using some classical techniques, for example $k$-means$++$ \cite{arthur2007k}. Then, in Steps 1 and 2 all data points are assigned to clusters, and in Steps 3 and 4 we update the centroids of the clusters. The process is repeated until convergence. 

\begin{algorithm} 
\caption{$q$-means.} \label{$q$-means-algo}
\begin{algorithmic}[1]
 
\REQUIRE  Data matrix $V \in \R^{\n \times d}$ stored in QRAM data structure. Precision parameters $\delta$ for $k$-means, error parameters
$\errdist$ for distance estimation, $\errmult$ and $\errnorms$ for matrix multiplication and $\errtom$ for tomography. 
\ENSURE Outputs vectors $c_{1}, c_{2}, \cdots, c_{k} \in \R^{d}$ that correspond to the centroids at the final step of the $\delta$-$k$-means algorithm.\\
\vspace{10pt} 
\STATE Select $k$ initial centroids $c_{1}^{0}, \cdots, c_{k}^{0}$ and store them in QRAM data structure. 
\STATE t=0
\REPEAT 
\STATE {\bf Step 1: Centroid Distance Estimation}\\
Perform the mapping (Theorem \ref{thm:dist}) 
\begin{equation}\label{initialstate}
\frac{1}{\sqrt{N}}\sum_{i=1}^{\n}   \ket{i} \otimes_{j \in [k]} \ket{j}\ket{0} \mapsto \frac{1}{\sqrt{N}}\sum_{i=1}^{\n}  \ket{i} \otimes_{j \in [k]} \ket{j}\ket{\overline{d^2(v_{i}, c_{j}^{t})}}
\end{equation}
where $|\overline{d^2(v_{i}, c_{j}^{t})} -  d^2(v_{i}, c_{j}^{t}) | \leq \epsilon_{1}. $\\
\STATE {\bf Step 2: Cluster Assignment}\\
Find the minimum distance among $\{d^2(v_{i}, c_{j}^{t})\}_{j\in[k]}$ (Lemma $\ref{thm:minimum}$), then uncompute Step 1 to create the superposition of all points and their labels
\begin{equation}
\frac{1}{\sqrt{N}}\sum_{i=1}^{\n}  \ket{i} \otimes_{j \in [k]} \ket{j}\ket{\overline{d^2(v_{i}, c_{j}^{t})}}
 \mapsto \frac{1}{\sqrt{N}}\sum_{i=1}^{\n} \ket{i} \ket{ \ell^t(v_{i})}
\end{equation} 

\STATE {\bf Step 3: Centroid states creation} \\
{\bf 3.1} Measure the label register to obtain a state $\ket{\chi_{j}^t} = \frac{1}{ \sqrt{ |\mathcal{C}^t_{j}|} }  \sum_{i\in \mathcal{C}^t_j}\ket{i}$, with prob. $\frac{|\mathcal{C}^{t}_j|}{N} $ \\ %= O(\frac{1}{k})$.\\
{\bf 3.2} Perform matrix multiplication with matrix $V^T$ and vector  $\ket{\chi_{j}^t}$  to obtain the state $\ket{c_{j}^{t+1}}$ with error $\errmult$, along with an estimation of $\norm{c_{j}^{t+1}}$ with relative error $\errnorms$ (Theorem \ref{thm:quantum_matrix_multiplication_inversion}). \\

\STATE {\bf Step 4: Centroid Update} \\
{\bf 4.1} Perform tomography for the states $\ket{c_{j}^{t+1}}$ 
with precision $\errtom$ using the operation from Steps 1-3 (Theorem \ref{thm:tomography_l2}) and get a classical estimate $\overline{c}_j^{t+1}$ for the new centroids such that $|c_j^{t+1} - \overline{c}_j^{t+1}| \leq \sqrt{\eta}(\errnorms+\errtom) = \epsilon_{centroids}$\\ %$\norm{\ket{\overline{c_j^{t+1}}}-\ket{c_j^{t+1}}} \leqslant \errtom$ and $| \norm{c_j} - \overline{\norm{c_j}} | \leq \errnorms\norm{c_j}$
{\bf 4.2} Update the QRAM data structure for the centroids with the new vectors $\overline{c}^{t+1}_0 \cdots \overline{c}^{t+1}_k$. 

\STATE t=t+1
\UNTIL convergence condition is satisfied. 

\end{algorithmic}
\end{algorithm}

\subsubsection{Step 1: Centroid Distance Estimation} 
The first step of the algorithm estimates the square distance between data points and clusters using a quantum procedure. For this we used the method presented in Section \ref{sec:inner_product_and_distance_quantum}, assuming quantum access to the vectors $\{v_i\}_{i\in [\n}$ and centroids $\{c_j\}_{j\in [k}$. 

For $q$-means, we need to estimate distances or inner products between vectors which have different norms. At a high level, if we first estimate the inner between the quantum states $\ket{v_i}$ and $\ket{c_j}$ corresponding to the normalized vectors and then multiply our estimator by the product of the vector norms we will get an estimator for the inner product of the unnormalized vectors. A similar calculation works for the square distance instead of the inner product.
If we have an absolute error $\epsilon$ for the square distance estimation of the normalized vectors, then the final error is of the order of $\epsilon \norm{v_i} \norm{c_j}$.  

We now rewrite the Theorem \ref{thm:distance_innpdct_quantum}, suited for the setup and notations of $q$-means. The proof of the new theorem follows rather straightforwardly. In fact one just needs to apply the distance estimation procedure from Theorem \ref{thm:distance_innpdct_quantum} $k$ times in parallel. Note also that the norms of the centroids are always smaller than the maximum norm of a data point which gives us the factor $\eta$.

\begin{theorem}{Centroid Distance estimation}{dist}
	Let a data matrix $V \in \mathbb{R}^{\n \times d}$ and a centroid matrix $C \in \mathbb{R}^{k \times d}$ be stored in QRAM, such that the following unitaries $
\ket{i}\ket{0} \mapsto \ket{i}\ket{v_i}, $ and $\ket{j}\ket{0} \mapsto \ket{j}\ket{c_j}
$ can be performed in time $O(\log (Nd))$ and the norms of the vectors are known.
For any $\Delta > 0$ and $\errdist>0$, there exists a quantum algorithm that performs the mapping
\begin{equation}\label{eq:qmeans_step1}
\frac{1}{\sqrt{N}}\sum_{i=1}^{\n}  \ket{i} \otimes_{j \in [k]} ( \ket{j}\ket{0}) \mapsto\ \frac{1}{\sqrt{N}}\sum_{i=1}^{\n}  \ket{i} \otimes_{j \in [k]}(\ket{j}\ket{\overline{d^2(v_i,c_j)}})
\end{equation}
where $|\overline{d^{2}(v_i,c_j)}-d^{2}(v_i,c_j)| \leqslant  \errdist$ with probability at least $1-2\Delta$, in time $\widetilde{O}\left(k \frac{ \eta}{ \errdist} \log(1/\Delta) \right)$ where $\eta=\max_{i}(\norm{v_i}^2)$. 
\end{theorem}

\subsubsection{Step 2: Cluster Assignment}\label{labels1} 
At the end of step 1, we have coherently estimated the square distance between each point in the dataset and the $k$ centroids in separate registers, as written in Eq.(\ref{eq:qmeans_step1}):
\begin{equation}\label{eq:qmeans_step1_bis}
\frac{1}{\sqrt{N}}\sum_{i=1}^{\n}  \ket{i} \otimes_{j \in [k]}(\ket{j}\ket{\overline{d^2(v_i,c_j)}})
\end{equation}
We can now select the index $j$ that corresponds to the centroid closest to the given data point, written as $\ell(v_{i}) = \text{argmin}_{j \in [k]}(d(v_{i}, c_{j}))$. 
As the square is a monotone function, we do not need to compute the square root of the distance in order to find $\ell(v_{i})$. 

\begin{claim}{}{minimum}%[Circuit for finding the minimum]
	Given $k$ different $\log(p)$-bit registers $\otimes_{j \in [k]} \ket{a_j}$, there is a quantum circuit $U_{min}$ that maps
	$(\otimes_{j \in [p]} \ket{a_j})\ket{0} \to (\otimes_{j \in [k]}\ket{a_j})\ket{\text{\em argmin}(a_j)}$ in time ${O}(k \log p)$. 
\end{claim}

\begin{proof} 
We append an additional register for the result that is initialized to $\ket{1}$. We then repeat the following operation for $2\leq j \leq k$, we compare registers $1$ and $j$, if the value in register $j$ is smaller we swap registers $1$ and $j$ and update the result register to $j$. The cost of the procedure is ${O}(k \log p)$.   	
\end{proof}%\Hsquare
\noindent The cost of finding the minimum is $\widetilde{O}(k)$ in step 2 of the $q$-means algorithm, while we also need to uncompute the distances by repeating Step 1. 
Once we apply the minimum finding Claim $\ref{thm:minimum}$ and undo the computation we obtain the state
\begin{equation}\label{labels} 
\ket{\psi^t} := \frac{1}{\sqrt{N}}\sum_{i=1}^{\n} \ket{i} \ket{ \ell^t(v_{i})}.
\end{equation}

Remark that if, instead of using the above minimum finding classical circuit, we used the algorithm from \cite{durr1996quantum} as in \cite{wiebe_nearest_neigbhors}, we could certainly save a factor $\widetilde{O}(\sqrt{k})$ in some part of the running time.

\subsubsection{Step 3: Centroid State Creation} \label{create_centroids}
The previous step gave us the state $\ket{\psi^t}= \frac{1}{\sqrt{N}}\sum_{i=1}^{\n} \ket{i} \ket{ \ell^t(v_{i})}$.
The first register of this state stores the index of the data points while the second register stores the label for the data point in the current iteration. 
Given these states, we need to find the new centroids  $\ket{c_j^{t+1}}$, which are the average of the data points having the same label. 

Let $\chi_{j}^{t} \in \R^{N}$ be the characteristic vector for cluster $j \in [k]$ at iteration $t$ scaled to unit $\ell_{1}$ norm, that is $(\chi_{j}^{t})_{i} = \frac{1}{  |C_{j}^{t}|} $ if $i\in \mathcal{C}_{j}$ and $0$ if $i \not \in \mathcal{C}_{j}$. The creation of the quantum states corresponding to the centroids is based on the 
following simple claim. 
\begin{claim}{}{simple} 
Let $\chi_{j}^{t} \in \R^{N}$ be the scaled characteristic vector for $\mathcal{C}_{j}$ at iteration $t$ and $V \in\R^{\n\times d}$ be the data matrix, then $c_{j}^{t+1} = V^{T} \chi_{j}^{t}$. 
\end{claim} 
\begin{proof} 
The $k$-means update rule for the centroids is given by $c_{j}^{t+1} = \frac{1}{ |C_{j}^{t}|} \sum_{i \in C_{j} } v_{i}$. As the columns of $V^{T}$ are the vectors $v_{i}$, this can be rewritten as 
$c_{j}^{t+1} = V^{T} \chi_{j}^{t}$. 
\end{proof}%\Hsquare 
\noindent The above claim allows us to compute the updated centroids $c_{j}^{t+1}$ using quantum linear algebra operations.
In fact, the state $\ket{\psi^t}$ can be written as a weighted superposition of the characteristic vectors of the clusters. 
%\als{ 
%\ket{\psi^t} = \sum_{j=1}^{k}\sqrt{\frac{|C_{j}|}{N}} \left( \frac{1}{ %\sqrt{ |C_{j}|} }  \sum_{i\in \mathcal{C}_j}\ket{i}\right)\ket{j} = %\sum_{j=1}^{k}\sqrt{\frac{|C_{j}|}{N}}
% \ket{\chi_{j}^{t} } \ket{j} 
% } 
 
 \begin{equation}
     \ket{\psi^t} = \sum_{j=1}^{k}\sqrt{\frac{|C_{j}|}{N}} \left( \frac{1}{ \sqrt{ |C_{j}|} }  \sum_{i\in \mathcal{C}_j}\ket{i}\right)\ket{j} = \sum_{j=1}^{k}\sqrt{\frac{|C_{j}|}{N}}
 \ket{\chi_{j}^{t} } \ket{j} 
 \end{equation}
 
By measuring the last register, we can sample from the states $\ket{\chi_{j}^{t} }$ for $j \in [k]$, with probability proportional to the size of the cluster. We assume here that all $k$ clusters are non-vanishing, in other words they have size $\Omega(N/k)$. 
Given the ability to create the states $\ket{\chi_{j}^{t} }$ and given that the matrix $V$ is stored in QRAM, we can now perform quantum matrix multiplication by $V^T$ to recover an approximation of the state $\ket{V^T\chi_{j}}=\ket{c_{j}^{t+1}}$ with error $\errmult$, as stated in Theorem \ref{thm:quantum_matrix_multiplication_inversion}. Note that the error $\errmult$ only appears inside a logarithm. The same Theorem allows us to get an estimate of the norm $\norm{V^T\chi_{j}^{t}}=\norm{c_{j}^{t+1}}$ with relative error $\errnorms$. For this, we also need an estimate of the size of each cluster, namely the norms $\norm{\chi_{j}}$. We already have this, since the measurements of the last register give us this estimate, and since the number of measurements made is large compared to $k$ (they depend on $d$), the error from this source is negligible compared to other errors. 

 The running time of this step is derived from Theorem  
\ref{thm:quantum_matrix_multiplication_inversion} where the time to prepare the state $\ket{\chi_{j}^{t}}$ is the time of Steps 1 and 2. Note that we do not have to add an extra $k$ factor due to the sampling, since we can run the matrix multiplication procedures in parallel for all $j$ so that every time we measure a random $\ket{\chi_{j}^{t}}$ we perform one more step of the corresponding matrix multiplication. Assuming that all clusters have size $\Omega(N/k)$ we will have an extra factor of $O(\log k)$ in the running time by a standard coupon collector argument.

\subsubsection{Step 4: Centroids Update} \label{update_centroids}

In Step 4, we need to go from quantum states corresponding to the centroids, to a classical description of the centroids in order to perform the update step. For this, we will apply the vector state tomography algorithm, stated in Theorem \ref{thm:tomography_l2}, on the states $\ket{c_{j}^{t+1}}$ that we create in Step 3. 
Note that for each $j \in [k]$ we will need to invoke the unitary that creates the states $\ket{c_{j}^{t+1}}$ a total of $O(\frac{d \log d}{\errtom^{2}})$ times for achieving $\norm{\ket{c_j} - \ket{\overline{c_j}}} < \errtom$. Hence, for performing the tomography of all clusters, we will invoke the unitary $O(\frac{k (\log k) d(\log d)}{\errtom^{2}})$ times where the $O(k\log k)$ term is the time to get a copy of each centroid state. 

The vector state tomography gives us a classical estimate of the unit norm centroids within error $\errtom$, that is $\norm{\ket{c_j} - \ket{\overline{c_j}}} < \errtom$. Using the approximation of the norms $\norm{c_{j}}$ with relative error $\errnorms$ from Step 3, we can combine these estimates to recover the centroids as vectors. The analysis is described in the following claim:

\begin{claim}{}{epsiloncentroid}
Let $\errtom$ be the error we commit in estimating $\ket{c_j}$ such that $\norm{ \ket{c_j} - \ket{\overline{c_j}}} < \errtom$, and $\errnorms$ the error we commit in the estimating the norms,  $|\norm{c_j} - \overline{\norm{c_j}}| \leq \errnorms \norm{c_{j}} $. Then $\norm{\overline{c_j} - c_j} \leq \sqrt{\eta} (\errnorms +  \errtom) = \epsilon_{centroid}$. 
\end{claim}

\begin{proof}
We can rewrite $\norm{c_j - \overline{c_j}}$ as $\norm{ \norm{c_j}\ket{c_j} - \overline{\norm{c_j}}\ket{\overline{c_j}}}$. It follows from triangle inequality that:
\begin{equation}
\norm{\overline{\norm{c_j}}\ket{\overline{c_j}} - \norm{c_j}\ket{c_j}}  \leq \norm{\overline{\norm{c_j}}\ket{\overline{c_j}} -  \norm{c_j}\ket{\overline{c_j}}} + \norm{\norm{c_j}\ket{\overline{c_j}} -  \norm{c_j}\ket{c_j}}
\end{equation}
We have the upper bound $\norm{c_{j}} \leq \sqrt{\eta}$. 
Using the bounds for the error we have from tomography and norm estimation, we can upper bound the first term by $\sqrt{\eta} \errnorms$ and the second term by $\sqrt{\eta} \errtom$. The claim follows.  	
\end{proof}%\Hsquare

Let us make a remark about the ability to use Theorem \ref{thm:tomography_l2} to perform tomography in our case.
The updated centroids will be recovered in step 4 using the vector state tomography algorithm in Theorem \ref{thm:tomography_l2} on the composition of the unitary that prepares $\ket{\psi^{t}}$ and the unitary that multiplies the first register of $\ket{\psi^t}$ by the matrix $V^{T}$. The input of the tomography algorithm requires a unitary $U$ 
such that $U\ket{0} = \ket{x}$ for a fixed quantum state $\ket{x}$. However, the labels $\ell(v_{i})$ are not deterministic due to errors in distance estimation, 
hence the composed unitary $U$ as defined above therefore does not produce a fixed pure state $\ket{x}$. 

We therefore need a procedure that finds labels $\ell(v_{i})$ that are a deterministic function of $v_{i}$ and the centroids $c_{j}$ for $j \in [k]$. One solution is to 
change the update rule of the $\delta$-$k$-means algorithm to the following: Let $\ell(v_{i})= j$ if $\overline{d(v_{i}, c_{j})}   < \overline{d(v_{i}, c_{j'})} - 2\delta$ for $j' \neq j$ where 
we discard the points to which no label can be assigned. This assignment rule ensures that if the second register is measured and found to be in state 
$\ket{j}$, then the first register contains a uniform superposition of points from cluster $j$ that are $\delta$ far from the cluster boundary (and possibly a few points 
that are $\delta$ close to the cluster boundary). Note that this simulates exactly the $\delta$-$k$-means update rule while discarding some of the data points 
close to the cluster boundary. The $k$-means centroids are robust under such perturbations, so we expect this assignment rule to produce good results in practice. 

A better solution is to use consistent phase estimation instead of the usual phase estimation for the distance estimation step , which can be found in \cite{T13, A12}. 
The distance estimates are generated by the phase estimation algorithm applied to a certain unitary in the amplitude estimation step. The usual phase estimation algorithm does not produce a deterministic answer and 
instead for each eigenvalue $\lambda$ outputs with high probability one of two possible estimates $\overline{\lambda}$ such that $|\lambda - \overline{\lambda}|\leq \epsilon$. 
Instead, here as in some other applications we need the consistent phase estimation algorithm that with high probability outputs a deterministic estimate such that $|\lambda - \overline{\lambda}|\leq \epsilon$. 

We also describe another simple method of getting such consistent phase estimation, which is to combine phase estimation estimates that are obtained for two different precision values. Let us assume that the 
eigenvalues for the unitary $U$ are $e^{2\pi i \theta_{i}}$ for $\theta_{i} \in [0, 1]$. First, we perform phase estimation with precision $\frac{1}{N_{1}}$ where $N_{1}=2^{l}$ is a power 
of $2$. We repeat this procedure $O(\log N/\theta^{2})$ times and output the median estimate. If the value being estimated 
is $\frac{\lambda + \alpha }{2^{l}}$ for $\lambda \in \Z$ and $\alpha \in [0,1]$ and $|\alpha - 1/2 | \geq \theta'$ for an explicit constant $\theta'$ (depending on $\theta$) then 
with probability at least $1-1/\text{poly}(N)$ the median estimate will be unique and will equal to $1/2^{l}$ times the closest integer to $(\lambda+ \alpha)$. 
In order to also produce a consistent estimate for the eigenvalues for the cases where the above procedure fails, we perform a second phase estimation with precision $2/3N_{1}$. 
We repeat this procedure as above for $O(\log N/\theta^{2})$ iterations and taking the median estimate. The second procedure fails to produce a consistent estimate only 
for eigenvalues $\frac{\lambda + \alpha }{2^{l}}$ for $\lambda \in \Z$ and $\alpha \in [0,1]$ and $|\alpha - 1/3 | \leq \theta'$ or $|\alpha - 2/3 | \leq \theta'$ for a suitable constant 
$\theta'$. Since the cases where the two procedures fail are mutually exclusive, one of them succeeds with probability $1-1/\text{poly}(N)$. The estimate produced 
by the phase estimation procedure is therefore deterministic with very high probability. In order to complete this proof sketch, we would have to give explicit values of the constants $\theta$ and $\theta'$ 
and the success probability, using the known distribution of outcomes for phase estimation. 

For what follows, we assume that indeed the state in Eq.(\ref{labels}) is almost a  deterministic state, meaning that when we repeat the procedure we get the same state with very high probability.

We set the error on the matrix multiplication to be $\errmult \ll \frac{\errtom^2}{d\log d}$ as we need to call the unitary that builds $c^{t+1}_j$ for $O(\frac{d\log d}{\errtom^2})$ times. We will see that this does not increase the runtime of the algorithm, as the dependence of the runtime for matrix multiplication is logarithmic in the error.

\subsection{Analysis}

We provide our general theorem about the running time and accuracy of the $q$-means algorithm.

\begin{theorem}{$q$-means - General Case}{q-meansgeneral}
For a data matrix $V \in \mathbb{R}^{\n \times d}$ stored in an appropriate QRAM data structure and parameter $\delta >0$, the q-means algorithm with high probability outputs centroids consistent with the classical $\delta$-$k$-means algorithm, in time 
\begin{equation}
\widetilde{O}\left(    k d \frac{\eta}{\delta^2}\kappa(V)(\mu(V) + k \frac{\eta}{\delta}) + k^2 \frac{\eta^{1.5}}{\delta^2} \kappa(V)\mu(V)
\right)
\end{equation}
per iteration,  where $\kappa(V)$ is the condition number, $1\leq \norm{v_i}^2 \leq \eta$ and $\mu(M)=\min_{p\in [0,1]} (\norm{M}_{F}, \sqrt{s_{2p}(M)s_{(1-2p)}(M^{T})})$. 
\end{theorem}

We prove the theorem in the following Sections and then provide the running time of the algorithm for well-clusterable datasets as Theorem \ref{thm:qmeans_main}.

\subsubsection{Error analysis}\label{erroranalysis}
In this section we determine the error parameters in the different steps of the quantum algorithm so that the quantum algorithm behaves the same as the classical $\delta$-$k$-means. More precisely, we will determine the values of the errors $\errdist, \errmult, \errnorms,\errtom$ in terms of $\delta$ so that firstly, the cluster assignment of all data points made by the $q$-means algorithm is consistent with a classical run of the $\delta$-$k$-means algorithm, and also that the centroids computed by the $q$-means after each iteration are again consistent with centroids that can be returned by the $\delta$-$k$-means algorithm. 

The cluster assignment in $q$-means happens in two steps. The first step estimates the square distances between all points and all centroids. The error in this procedure is of the form
\begin{equation}
|\overline{d^2(c_j,v_i)} - d^2(c_j,v_i) | < \errdist
\end{equation}
for a point $v_i$ and a centroid $c_j$.
The second step finds the minimum of these distances without adding any error.

For the $q$-means to output a cluster assignment consistent with the $\delta$-$k$-means algorithm, we require that: 

\begin{equation}
\forall j \in [k], \quad | \overline{d^2(c_j,v_i) } - d^2(c_j,v_i)  | \leq \frac{\delta}{2}
\end{equation}
which implies that no centroid with distance more than $\delta$ above the minimum distance can be chosen by the $q$-means algorithm as the label. Thus we need to take 
$\errdist < \delta/2$. 

After the cluster assignment of the $q$-means (which happens in superposition), we update the clusters, by first performing a matrix multiplication to create the centroid states and estimate their norms, and then a tomography to get a classical description of the centroids.
The error in this part is $\epsilon_{centroids}$, as defined in Claim \ref{thm:epsiloncentroid}, namely 

\begin{equation}
\norm{\overline{c}_{j} - c_j} \leq \epsilon_{centroid}  = \sqrt{\eta} (\errnorms + \errtom)
\end{equation}

Again, for ensuring that the $q$-means is consistent with the classical $\delta$-$k$-means algorithm we take
$\errnorms < \frac{\delta}{4\sqrt{\eta}}$ and $\errtom < \frac{\delta}{4\sqrt{\eta}}$. Note also that we have ignored the error $\errmult$ that we can easily deal with since it only appears in a logarithmic factor.

\subsubsection{Runtime analysis}\label{runtimeanalysis}
As the classical algorithm, the runtime of $q$-means depends linearly on the number of iterations, so here we analyze the cost of a single step.

The cost of tomography for the $k$ centroid vectors is $O(\frac{kd \log k \log d}{{\errtom}^{2}})$ times the cost of preparation of a single centroid state $\ket{c_{j}^{t}}$. 
A single copy of $\ket{c_{j}^{t}}$ is prepared applying the matrix multiplication by $V^{T}$ procedure on the state $\ket{\chi_{j}^{t}}$ obtained using square distance estimation. The time required for preparing a single copy of $\ket{c_{j}^{t}}$ is $O( \kappa(V) (\mu(V) + T_{\chi}) \log (1/\epsilon_{2}))$ by  Theorem \ref{thm:quantum_matrix_multiplication_inversion} where $T_{\chi}$ is the time for preparing $\ket{\chi_{j}^{t}}$.  The time $T_{\chi}$  is $\widetilde{O}\left(\frac{k\eta\log(\Delta^{-1})\log(\n d)}{ \errdist}\right)= \widetilde{O}(\frac{k\eta} { \errdist} )$ by Theorem \ref{thm:dist}. 

The cost of norm estimation for $k$ different centroids is independent of the tomography cost and is $\widetilde{O}( \frac{k T_{\chi} \kappa(V) \mu(V) }{\epsilon_{3}} )$. 
Combining together all these costs and suppressing all the logarithmic factors we have a total running time of, 

\begin{equation}
\widetilde{O} \left( kd \frac{1 }{ \errtom^{2}}  \kappa(V) \left(  \mu(V) +  k \frac{\eta} { \errdist}\right)  + k^{2} \frac{ \eta }{\epsilon_{3} \epsilon_{1} } \kappa(V) \mu(V) \right) 
\end{equation}
The analysis in section \ref{erroranalysis} shows that we can take $\errdist = \delta/2$, $\errnorms = \frac{\delta}{4\sqrt{\eta}}$ and $\errtom = \frac{\delta}{4\sqrt{\eta}}$. 
Substituting these values in the above running time, it follows that the running time of the $q$-means algorithm is 
\begin{equation}
\widetilde{O} \left(  k d  \frac{\eta}{\delta^2} \kappa(V) \left( \mu(V) + k \frac{ \eta}{\delta} \right) + k^{2}\frac{ \eta^{1.5} }{ \delta^2} \kappa(V) \mu(V) \right)
\end{equation}

%= \widetilde{O}\left( \frac{k\eta \kappa(V)} {\delta^{2}} \en{  \mu(V)(d+ k\sqrt{\eta}) + \frac{dk\eta}{ \delta}  }  \right) $.

\noindent This completes the proof of Theorem \ref{thm:q-meansgeneral}. 
We next state our main result when applied to a well-clusterable dataset, as in Definition \ref{def:wcdataset}.

\begin{theorem}{$q$-means - Well-Clusterable Data}{qmeans_main}
For a well-clusterable dataset $V \in \mathbb{R}^{\n \times d}$ stored in appropriate QRAM, the q-means algorithm returns with high probability the $k$ centroids consistently with the classical $\delta$-$k$-means algorithm for a constant $\delta$ in time 
$\widetilde{O}\left( k^2 d \frac{\eta^{2.5}}{\delta^3} + k^{2.5} \frac{\eta^2}{\delta^3} \right)$ per iteration, for $1\leq \norm{v_i}^2 \leq \eta$.  
\end{theorem}

\begin{proof}
Let $V \in \mathbb{R}^{\n \times d}$ be a well-clusterable dataset as in Definition \ref{def:wcdataset}. In this case, we know by Claim \ref{thm:kappa} that $\kappa(V)=\frac{1}{\sigma_{min}}$ can be replaced by a thresholded condition number $\kappa_\tau(V)=\frac{1}{\tau}$. In practice, this is done by discarding the singular values smaller than a certain threshold during quantum matrix multiplication. Remember that by Claim \ref{thm:mu}  we know that $\norm{V}_F = O(\sqrt{k})$. Therefore we need to pick $\epsilon_\tau$ for a threshold $\tau = \frac{\errkappa}{\sqrt{k}}\norm{V}_F$ such that $\kappa_\tau(V) = O(\frac{1}{\epsilon_{\tau}})$.   

Thresholding the singular values in the matrix multiplication step introduces an additional additive error in $\epsilon_{centroid}$. By Claim \ref{thm:kappa} and Claim \ref{thm:epsiloncentroid} , we have that the error $\epsilon_{centroid}$ in approximating the true centroids becomes $\sqrt{\eta} (\errnorms +\errtom + \epsilon'+ \errkappa)$ where $\epsilon'= \sqrt{ \lambda \beta^{2} + (1-\lambda) 4\eta}$ is a dataset dependent parameter computed in Claim \ref{thm:low-rank}. We can set $\errkappa = \errnorms = \errtom = \epsilon'/3$ to obtain 
$\epsilon_{centroid} = 2\sqrt{\eta} \epsilon'$. 

The definition of the $\delta$-$k$-means update rule requires that $\epsilon_{centroid} \leq \delta/2$. Further, Claim \ref{thm:distcentroid} shows that if the 
error $\delta$ in the assignment step satsifies $\delta \leq \xi^{2} - 2\sqrt{\eta} \beta$, then the $\delta$-$k$-means algorithm finds the corrects clusters. 
By Definition  \ref{def:wcdataset} of a well-clusterable dataset, we can find a suitable constant $\delta$
 satisfying both these constraints, namely satisfying
\begin{equation}
4\sqrt{\eta} \sqrt{ \lambda \beta^{2} + (1-\lambda) 4\eta} < \delta <  \xi^{2} - 2\sqrt{\eta} \beta
\end{equation}

Substituting the values $\mu(V) = O(\sqrt{k})$ from Claim \ref{thm:mu}, $\kappa(V) = O(\frac{1}{\epsilon_{\tau}})$ and $\errkappa = \errnorms = \errtom = \epsilon'/3 = O(\sqrt{\eta}/\delta)$ in the running time for the general $q$-means algorithm, we obtain that the running time for the $q$-means algorithm on a well-clusterable dataset is 
$\widetilde{O}\left( k^2 d \frac{\eta^{2.5}}{\delta^3} + k^{2.5} \frac{\eta^2}{\delta^3} \right)$ per iteration.

\end{proof}%\Hsquare
%\noindent If a dataset is well-clusterable not only we have a good clustering assignment, but we can also choose a constant value for $\delta$ for which the $\delta$-$k$-means algorithm is able to find the clusters. In the case of a dataset satisfying the condition that the minimum norm is one of Definition \ref{datasetassumption}, the parameter $\eta/\delta$ can be interpreted as a relative error. In the next Section, we will try to get an estimate of these parameters, and conclude that for a simple dataset, are small enough.

Let us make some concluding remarks regarding the running time of $q$-means. For dataset where the number of points is much bigger compared to the other parameters, the running time for the $q$-means algorithm is an improvement compared to the classical $k$-means algorithm. For instance, for most problems in data analysis, $k$ is eventually small ($<100$). The number of features $d\leq N$ in most situations, and it can eventually be reduced by applying a quantum dimensionality reduction algorithm first (which has running time polylogarithmic in $d$). To sum up, $q$-means has the same output as the classical $\delta$-$k$-means algorithm (which approximates k-means), it conserves the same number of iterations, but has a running time only polylogarithmic in $\n$, giving an exponential speedup with respect to the size of the dataset.

\subsection{Initialization: $q$-means++}\label{q-means++proof}
We now show that the quantum analogue of the initialization procedure of $k$-means++ can be implemented efficiently using the square distance subroutine estimation for the $q$-means algorithm given in Theorem \ref{thm:dist}. Starting with a random index $j$ we compute the state $\frac{1}{\sqrt{N}}\sum_{i=0}^{N-1}\ket{i}\ket{j}\ket{d^2(v_i,v_j)}$ in time $\tilde{O}(\frac{\eta}{\errdist})$, where $v_j$ is the initial centroid, using our quantum procedure for distance estimation. By applying some arithmetic preprocessing and a controlled rotation we can transfer the distance information as an amplitude to obtain the following state:

\begin{equation}
\frac{1}{\sqrt{N}}\sum_{i=0}^{N-1}\ket{i}\ket{j}\ket{d^2(v_i,v_j)}\left(\frac{d(v_i,v_j)}{2\sqrt{\eta}}\ket{0} + \beta\ket{1}\right)
\end{equation}

Each square distance has been normalized by $2\sqrt{\eta} \geq max_{i,j}(d(v_i,v_j))$ to be a valid amplitude. Note that postselecting on $\ket{0}$ and measuring the register $\ket{i}$ samples exactly from the probability distribution in the $k$-means++ algorithm as the probability of measuring $(i,0)$ on second and fourth registers is $\frac{d^2(v_i,v_j)}{4\eta N}$. 

We can perform amplitude amplification to boost the probability of measuring $\ket{0}$. For this we need to repeat $O(1/\sqrt{P(0)})$ times the previous steps, with $P(0)$ being the probability of measuring $\ket{0}$. Since $P(0)=\frac{1}{N}\left(\sum \frac{d(v_i,v_j)}{2\sqrt{\eta}} \right)^2$, it is simple to show that $\frac{1}{\sqrt{P(0)}} \leq  \frac{2\sqrt{\eta}}{\sqrt{\mathbb{E}(d^2(v_i,v_j))}}$, where $\mathbb{E}(d^2(v_i,v_j))$ is the mean squared distance. 
Note that for the next steps we can use a tensor product of the squared distance from previous centroids to compute the minimum distance among them, using Lemma \ref{thm:minimum}.
In the end we repeat $k-1$ times this circuit, for a total time of $\tilde{O}(k^2\frac{2\eta^{1.5}}{\errdist \sqrt{\mathbb{E}(d^2(v_i,v_j))}})$. The running time for the $q$-means++ initialization is 
smaller than that for the $q$-means algorithm, showing than $q$-means++ initialization doesn't cancel any benefit of the $q$-means algorithm. Thus, we can use the $q$-means++ algorithm to provide a speedup compared to the classical $k$-means++.

\section{Numerical Simulations}\label{sec:qmeans_numerical_simulations}

We would like to assert the capability of the quantum algorithm to provide accurate classification results, by simulating on several datasets. However, since neither quantum simulators nor quantum computers large enough to test $q$-means are available currently, we tested the equivalent classical implementation of $\delta$-$k$-means. 
For implementing the $\delta$-$k$-means, we changed the assignment step of the $k$-means algorithm to select a random centroid among those that are $\delta$-close to the closest centroid and added $\delta/2$ error to the updated clusters. 

We benchmarked our $q$-means algorithm on two datasets: a synthetic dataset of Gaussian clusters, and the well known MNIST dataset of handwritten digits. To measure and compare the accuracy of our clustering algorithm, we ran the $k$-means and the $\delta$-$k$-means algorithms for different values of $\delta$ on a training dataset and then we compared the accuracy of the classification on a test set, containing data points on which the algorithms have not been trained, using a number of widely-used performance measures.

\subsubsection{Gaussian clusters dataset}
\begin{figure} 
\centering
\includegraphics[width=0.8\textwidth]{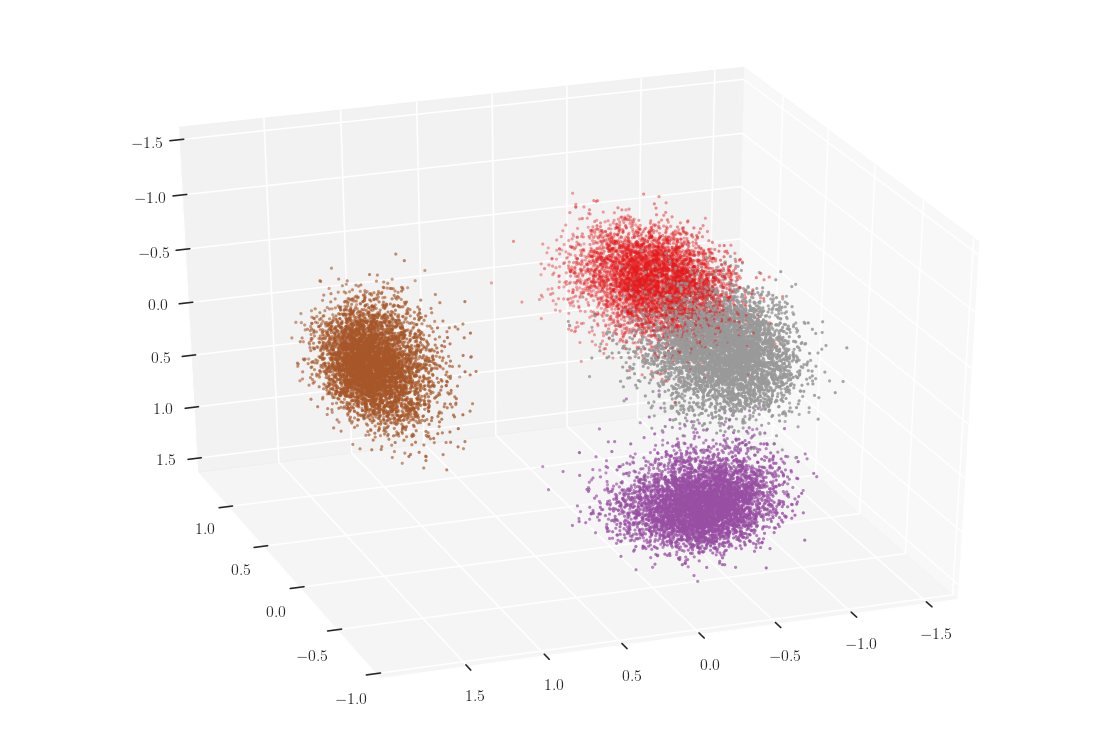} 
%\captionsetup{justification=raggedright, margin=1cm}
\caption{Representation of $4$ Gaussian clusters of $10$ dimensions in a 3D space spanned by the first three PCA dimensions.}\label{gaussian-cluster-1}
\end{figure}

We describe numerical simulations of the $\delta$-$k$-means algorithm on a synthetic dataset made of several clusters formed by random Gaussian distributions. These clusters are naturally well suited for clustering by construction, close to what we defined to be a well-clusterable dataset in Definition \ref{def:wcdataset} of Section \ref{sec:well_clusterable_datasets}. Doing so, we can start by comparing $k$-means and $\delta$-$k$-means algorithms on high accuracy results, even though this may not be the case on real-world datasets. Without loss of generality, we preprocessed the data so that the minimum norm in the dataset is $1$, in which case $\eta = 4$. This is why we defined $\eta$ as a maximum instead of the ratio of the maximum over the minimum which is really the interesting quantity. Note that the running time basically depends on the ratio $\eta/\delta$.
We present a simulation where $20.000$ points in a feature space of dimension $10$ form $4$ Gaussian clusters with standard deviation $2.5$, that we can see in Fig.\ref{gaussian-cluster-1}.  The condition number of dataset is calculated to be $5.06$. We ran $k$-means and $\delta$-$k$-means for $7$ different values of $\delta$ to understand when the $\delta$-$k$-means becomes less accurate.

\begin{figure}  [H] 
\centering
\includegraphics[width=0.8\textwidth]{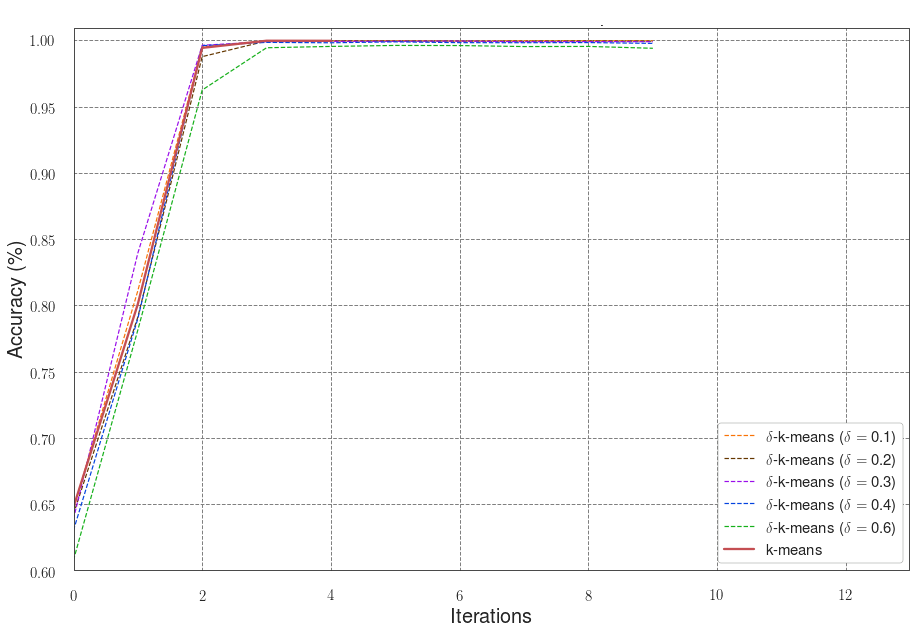} 
%\captionsetup{justification=raggedright, margin=1cm}
\caption{Accuracy evolution during $k$-means and $\delta$-$k$-means on well-clusterable Gaussians for $5$ values of $\delta$. All versions converged to a 100\% accuracy in few steps.}\label{results-gaussian-cluster-1}
\end{figure}

In Fig.\ref{results-gaussian-cluster-1} we can see that until $\eta/\delta = 3$ (for $\delta=1.2$), the $\delta$-$k$-means algorithm converges on this dataset. We can now make some remarks about the impact of $\delta$ on the efficiency. It seems natural that for small values of $\delta$ both algorithms are equivalent. For higher values of $\delta$, we observed a late start in the evolution of the accuracy, witnessing random assignments for points on the clusters' boundaries. However, the accuracy still reaches $100$\% in a few more steps. The increase in the number of steps is a tradeoff with the parameter $\eta/\delta$. 

\subsubsection{MNIST dataset}

The MNIST dataset is composed of $60.000$ handwritten digits as images of 28x28 pixels (784 dimensions). From this raw data we first performed some dimensionality reduction processing, then we normalized the data such that the minimum norm is one. Note that, if we were doing $q$-means with a quantum computer, we could use efficient quantum procedures equivalent to Linear Discriminant Analysis, such as \cite{kerenidis2020classification_QSFA}, or other quantum dimensionality reduction algorithms like \cite{Lloyd_PCA_quantum, cong2015quantum}. 

\begin{figure} [H]
\centering
\includegraphics[width=0.8\textwidth]{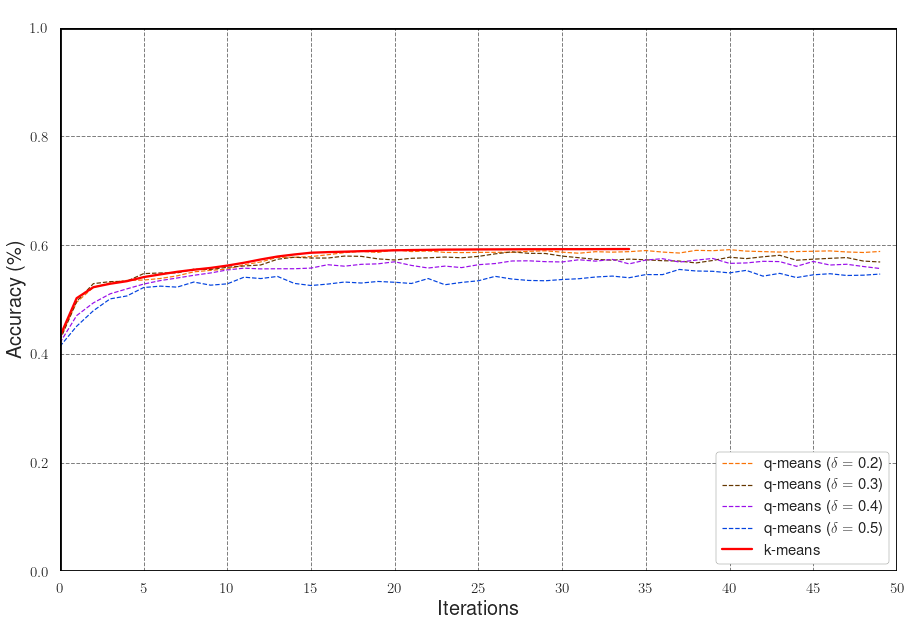}
%\captionsetup{justification=raggedright, margin=0.5cm}
\caption{Accuracy evolution on the MNIST dataset under $k$-means and $\delta$-$k$-means for $4$ different values of $\delta$. Data has been preprocessed by a PCA to 40 dimensions. All versions converge in the same number of steps, with a drop in the accuracy while $\delta$ increases. The apparent difference in the number of steps until convergence is just due to the stopping condition for $k$-means and $\delta$-$k$-means.} \label{mnist-results-1} 
\end{figure}

As preprocessing of the data, we first performed a Principal Component Analysis (PCA), retaining data projected in a subspace of dimension 40. After normalization, the value of $\eta$ was 8.25 (maximum norm of 2.87), and the condition number was 4.53. Fig.\ref{mnist-results-1} represents the evolution of the accuracy during the $k$-means and $\delta$-$k$-means for $4$ different values of $\delta$. In this numerical experiment, we can see that for values of the parameter $\eta/\delta$ of order 20, both $k$-means and $\delta$-$k$-means reached a similar, yet low accuracy in the classification in the same number of steps. It is important to notice that the MNIST dataset, without other preprocessing than dimensionality reduction, is known not to be well-clusterable under the $k$-means algorithm.

On top of the accuracy measure (ACC), we also evaluated the performance of $q$-means against many other metrics, reported in Table \ref{tablecomparison}. More detailed information about these metrics can be found in \cite{friedman2001elements}. We introduce a specific measure of error, the Root Mean Square Error of Centroids (RMSEC), which a direct comparison between the centroids predicted by the k-means algorithm and the ones predicted by the $\delta$-$k$-means. It is a way to know how far the centroids are predicted. Note that this metric can only be applied to the training set. For all these measures, except RMSEC, a bigger value is better. Our simulations show that $\delta$-$k$-means, and thus the $q$-means, even for values of $\delta$ (between $0.2-0.5$) achieves similar performance to $k$-means, and in most cases the difference is on the third decimal point.

\begin{table}[H]
\centering
\begin{tabular}{|c|c|c|c|c|c|c|c|c|}
\hline
Algo                                  & Dataset & ACC   & HOM   & COMP  & V-M   & AMI   & ARI   & RMSEC \\ \hline
\multirow{2}{*}{k-means}              & Train   & 0.582 & 0.488 & 0.523 & 0.505 & 0.389 & 0.488 & 0     \\ \cline{2-9} 
                                      & Test    & 0.592 & 0.500 & 0.535 & 0.517 & 0.404 & 0.499 & -     \\ \hline
$\delta$-$k$-means 					  & Train   & 0.580 & 0.488 & 0.523 & 0.505 & 0.387 & 0.488 & 0.009 \\ \cline{2-9} 
($\delta=0.2$)                        & Test    & 0.591 & 0.499 & 0.535 & 0.516 & 0.404 & 0.498 & -     \\ \hline
$\delta$-$k$-means 					  & Train   & 0.577 & 0.481 & 0.517 & 0.498 & 0.379 & 0.481 & 0.019 \\ \cline{2-9} 
($\delta=0.3$)                        & Test    & 0.589 & 0.494 & 0.530 & 0.511 & 0.396 & 0.493 & -     \\ \hline
$\delta$-$k$-means 					  & Train   & 0.573 & 0.464 & 0.526 & 0.493 & 0.377 & 0.464 & 0.020 \\ \cline{2-9} 
($\delta=0.4$)                        & Test    & 0.585 & 0.492 & 0.527 & 0.509 & 0.394 & 0.491 & -     \\ \hline
$\delta$-$k$-means 					  & Train   & 0.573 & 0.459 & 0.522 & 0.488 & 0.371 & 0.459 & 0.034 \\ \cline{2-9} 
($\delta=0.5$)                        & Test    & 0.584 & 0.487 & 0.523 & 0.505 & 0.389 & 0.487 & -     \\ \hline
\end{tabular}
\caption{A sample of results collected from the same experiments as in Fig.\ref{mnist-results-1}. Different metrics are presented for the train set and the test set. ACC: accuracy. HOM: homogeneity. COMP: completeness. V-M: v-measure. AMI: adjusted mutual information. ARI: adjusted rand index. RMSEC: Root Mean Square Error of Centroids.\label{tablecomparison}} 
\end{table}

These experiments have been repeated several times and each of them presented a similar behavior despite the random initialization of the centroids. 

\begin{figure}[h]
\centering
\begin{minipage}{0.33\textwidth}
  \centering
\includegraphics[width=\linewidth]{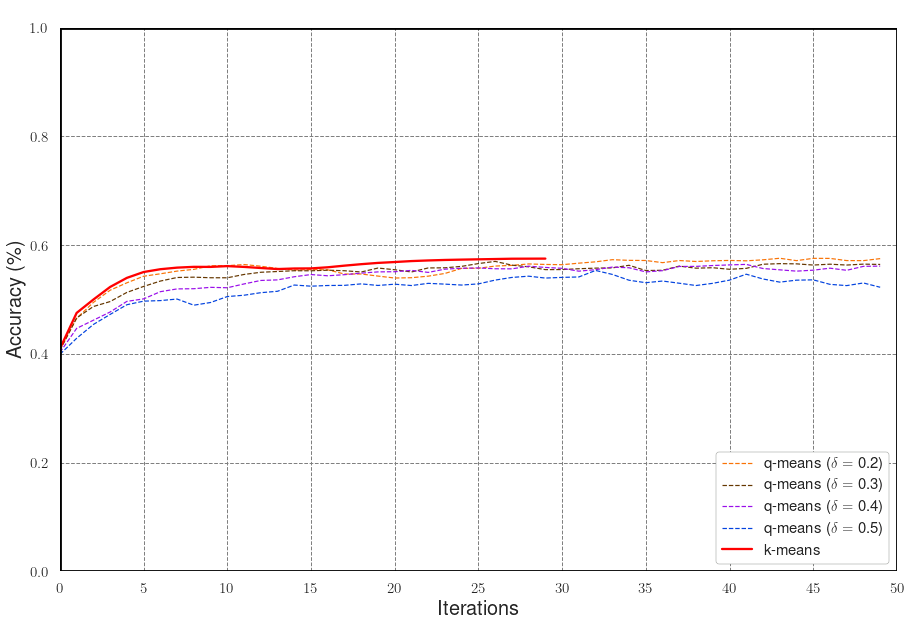}
\subcaption[second caption.]{}\label{fig:1a}
\end{minipage}%
\begin{minipage}{0.33\textwidth}
  \centering
\includegraphics[width=\linewidth]{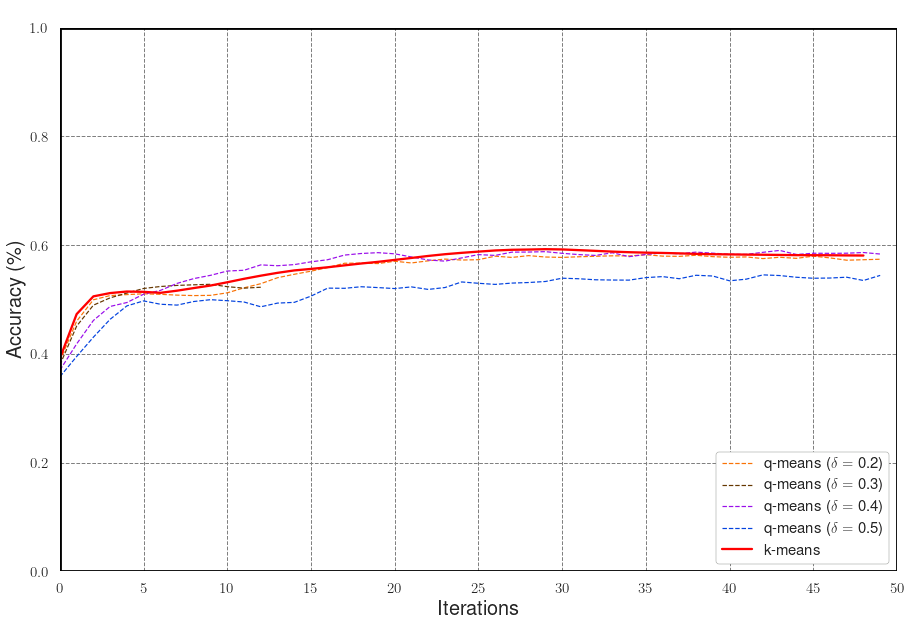}
\subcaption[third caption.]{}\label{fig:1b}
\end{minipage}
\begin{minipage}{0.33\textwidth}
  \centering
\includegraphics[width=\linewidth]{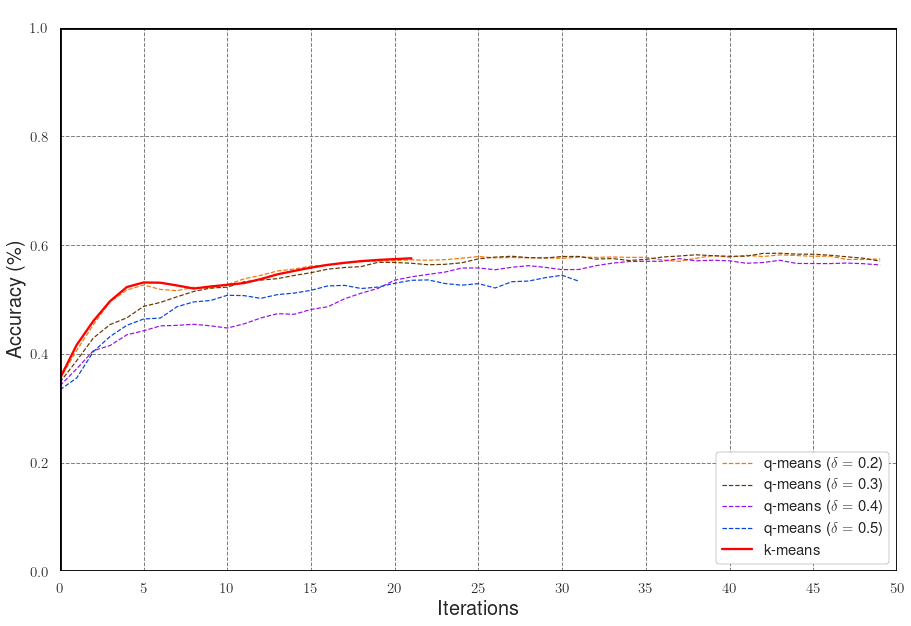}
\subcaption[third caption.]{}\label{fig:1c}
\end{minipage}
%\captionsetup{justification=raggedright, margin=1cm}
\caption{Three accuracy evolution on the MNIST dataset under $k$-means and $\delta$-$k$-means for $4$ different values of $\delta$. Each different behavior is due to the random initialization of the centroids} \label{fig:1}
\end{figure}

\begin{figure} [H]
\centering
\includegraphics[width=0.8\textwidth] {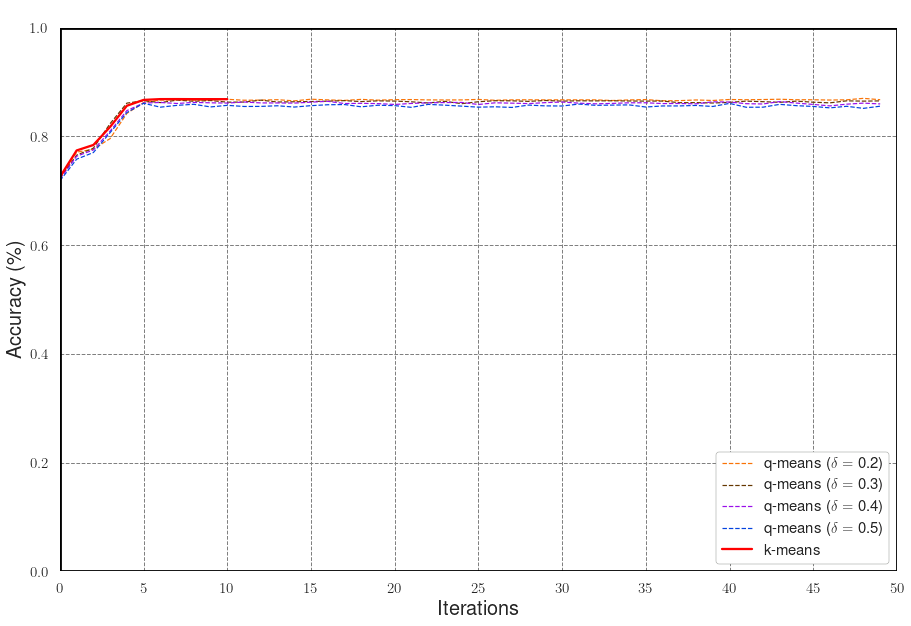} 
%\captionsetup{justification=raggedright, margin=0.5cm}
\caption{Accuracy evolution on the MNIST dataset under $k$-means and $\delta$-$k$-means for $4$ different values of $\delta$. Data has been preprocessed to 9 dimensions with a LDA reduction. All versions of  $\delta$-$k$-means converge to the same accuracy than $k$-means in the same number of steps.} \label{mnist-results-2} 
\end{figure}

Finally, we present a last experiment with the MNIST dataset with a different data preprocessing. In order to reach higher accuracy in the clustering, we replace the previous dimensionality reduction by a Linear Discriminant Analysis (LDA). Note that a LDA is a supervised process that uses the labels (here, the digits) to project points in a well chosen lower dimensional subspace. Thus this preprocessing cannot be applied in practice in unsupervised machine learning. However, for the sake of benchmarking, by doing so $k$-means is able to reach a 87\% accuracy, therefore it allows us to compare $k$-means and $\delta$-$k$-means on a real and almost well-clusterable dataset. In the following, the MNIST dataset is reduced to 9 dimensions. The results in Fig.\ref{mnist-results-2} and Table \ref{tablecomparison_2} show that $\delta$-$k$-means converges to the same accuracy than $k$-means even for values of $\eta/\delta$ down to $16$. In some other cases, $\delta$-$k$-means shows a faster convergence, due to random fluctuations that can help escape faster from a temporary equilibrium of the clusters.

\begin{table}[H]
\centering
\begin{tabular}{|c|c|c|c|c|c|c|c|c|}
\hline
Algo                                   & Dataset & ACC   & HOM   & COMP  & V-M   & AMI   & ARI   & RMSEC \\ \hline
\multirow{2}{*}{k-means}               & Train   & 0.868 & 0.736 & 0.737 & 0.737 & 0.735 & 0.736 & 0     \\ \cline{2-9} 
                                       & Test    & 0.891 & 0.772 & 0.773 & 0.773 & 0.776 & 0.771 & -     \\ \hline
q-means  							   & Train   & 0.868 & 0.737 & 0.738 & 0.738 & 0.736 & 0.737 & 0.031 \\ \cline{2-9} 
($\delta=0.2$)                           & Test    & 0.891 & 0.774 & 0.775 & 0.775 & 0.777 & 0.774 & -     \\ \hline
q-means  							   & Train   & 0.869 & 0.737 & 0.739 & 0.738 & 0.736 & 0.737 & 0.049 \\ \cline{2-9} 
($\delta=0.3$)                           & Test    & 0.890 & 0.772 & 0.774 & 0.773 & 0.775 & 0.772 & -     \\ \hline
q-means  							   & Train   & 0.865 & 0.733 & 0.735 & 0.734 & 0.730 & 0.733 & 0.064  \\ \cline{2-9} 
($\delta=0.4$)                           & Test    & 0.889 & 0.770 & 0.771 & 0.770 & 0.773 & 0.769 & -     \\ \hline
q-means  							   & Train   & 0.866 & 0.733 & 0.735 & 0.734 & 0.731 & 0.733 & 0.079 \\ \cline{2-9} 
($\delta=0.5$)                           & Test    & 0.884 & 0.764 & 0.766 & 0.765 & 0.764 & 0.764 & -     \\ \hline
\end{tabular}
\caption{A sample of results collected from the same experiments as in Fig.\ref{mnist-results-2}. Different metrics are presented for the train set and the test set. ACC: accuracy. HOM: homogeneity. COMP: completeness. V-M: v-measure. AMI: adjusted mutual information. ARI: adjusted rand index. RMSEC: Root Mean Square Error of Centroids.\label{tablecomparison_2}} 
\end{table}

Let us remark, that the values of $\eta/\delta$ in our experiment remained between 3 and 20. Moreover, the parameter $\eta$, which is the maximum square norm of the points, provides a worst case guarantee for the algorithm, while one can expect that the running time in practice will scale with the average square norm of the points. For the MNIST dataset after PCA, this value is 2.65 whereas $\eta = 8.3$. 

In conclusion, our simulations show that the convergence of $\delta$-$k$-means is almost the same as the regular $k$-means algorithms for large enough values of $\delta$. This provides evidence that the $q$-means algorithm will have as good performance as the classical $k$-means, and its running time will be significantly lower for large datasets.

\chapter{Quantum Spectral Clustering}\label{chap:Q_spectral_clustering}
\epigraph{\textit{"All generalisations - perhaps except this one - are false."}}{Kurt Gödel}

Preliminaries on spectral clustering, along with all notations are given in Section \ref{sec:classical_spectral_clustering}. Variables are summarized in Table \ref{table:spectral_clustering_variable_summary}.

\begin{table}[h]
\centering
\begin{tabular}{|l|c|c|l|}
\hline
\multicolumn{1}{|c|}{Variable}                                  & Dimension                  & \multicolumn{1}{c|}{Remark} \\ \hline
Input Matrix:                   $S$                             & $N\times d$                & \vtop{\hbox{\strut Each input vector is }\hbox{\strut a row $s_i \in \R^d$}} \\ \hline
Adjacency Matrix:               $A$                             & $N\times N$                & \vtop{\hbox{\strut Nodes connectivity. Elements $a_{ij}$ =  }\hbox{\strut  $1$ if $d(s_i,s_j)\leq d_{min}$ and 0 otherwise}} \\ \hline
Incidence Matrix:               $B$                             & $N\times \frac{N(N-1)}{2}$ & \vtop{\hbox{\strut Elements $B_{i,(p,q)} = \pm a_{pq}$ if }\hbox{\strut edge $(p,q)$ is incident to node $i$}}\\ \hline
Normalized Incidence Matrix:    $\mathcal{B}$                   & $N\times \frac{N(N-1)}{2}$ & \vtop{\hbox{\strut Each row is normalized  }\hbox{\strut  Eigenvalues $\lambda_1^{\mathcal{B}}\leq\cdots\leq \lambda_n^{\mathcal{B}}$}}\\ \hline
Normalized Laplacian:           $\mathcal{L}$                   & $N\times N$                & \vtop{\hbox{\strut $\mathcal{L} = \mathcal{B}\mathcal{B}^T$ }\hbox{\strut  Eigenvalues $\lambda_j = (\lambda_j^{\mathcal{B}})^2$}} \\ \hline
Projected Normalized Laplacian: $\widetilde{\mathcal{L}}^{(k)}$ & $N\times k$                & \vtop{\hbox{\strut $\mathcal{L}$ projected on its }\hbox{\strut $k$ lowest eigenvectors}}  \\ \hline
\end{tabular}
\caption{Summary of variables for Classical and Quantum spectral clustering.\label{table:spectral_clustering_variable_summary}}%ex - \label{variable_summary}
\end{table}

\section{Main Results}

In this work, we develop an end-to-end quantum algorithm for spectral clustering \cite{ng2002spectral}. As detailed in Section \ref{sec:classical_spectral_clustering}, this unsupervised learning algorithm shows great accuracy in identifying complex and interlacing clusters, and allows a high level of explainability. However, it suffers from a fast-growing runtime, namely cubic in the number $N$ of vectors in the dataset, that inhibits its use in practice. The goal of the spectral clustering algorithm is to perform the clustering task in a low dimensional space derived from the data. More precisely, one starts with the set of input vectors and constructs a similarity graph, where the edge between two nodes is built from the distance between the two associated vectors. From this graph, one can define the Incidence matrix and the Laplacian matrix. By extracting the eigenvectors of the Laplacian matrix and keeping the subspace spanned by the lowest ones, a lower dimensional space is defined. The initial data can then be projected onto this new space, where one can expect the data to be better organized, as clusters. Therefore, to obtain the clusters, the $k$-means algorithm is applied to the projected vectors. \\

We provide a quantum algorithm for spectral clustering following a similar methodology, while having to carefully extend or alter the specifics of each step of the algorithm. We first adapt and extend recent and efficient quantum subroutines for linear algebra and distance estimation. These methods are used to create the similarity graph, as well as the Incidence and Laplacian matrices, extract their eigenvectors and project the data points onto the right subspace to finally apply a quantum analog of $k$-means. While the steps of the algorithm follow the classical ones, we had to adjust most of the definitions, for example the definitions of the Incidence and Laplacian matrices, in order to make the quantum algorithm efficient. We will detail all the needed changes in the following sections.

In high level, the running time of the quantum spectral clustering algorithm reflects the two stages of spectral clustering and is given by
\begin{equation}
O( T_{\widetilde{\mathcal{L}}^{(k)}} T_{qmeans})
\end{equation} 
The first term $T_{\widetilde{\mathcal{L}}^{(k)}}$ is the time to create a quantum state corresponding to the normalized Laplacian matrix of the graph projected on its lowest eigenvectors. The resulting quantum state is the input state to the quantum clustering algorithm whose overall running time contains another multiplicative term that we denote by  $T_{qmeans}$.

For the first part, we will propose an algorithm in time 
\begin{equation}
T_{\widetilde{\mathcal{L}}^{(k)}}= \widetilde{O}\left( T_S 
    \frac{\eta(S)}{\epsilon_{dist}\epsilon_B} 
    \frac{\mu(\mathcal{B})\kappa(\widetilde{\mathcal{L}}^{(k)})}{\epsilon_{\lambda}} \right).
\end{equation}
Here, the term $T_S$ is the time to load the input vectors in a quantum state, which becomes efficient is we assume fast quantum access (see Definition \ref{def:quantum_access}). The terms $\epsilon_{dist},\epsilon_B$ and $ \epsilon_{\lambda}$ correspond to error or precision parameters that appear in several quantum routines. The matrices $S, \mathcal{B},$ and $\widetilde{\mathcal{L}}^{(k)}$ are respectively the input data matrix, the normalized incidence matrix, and the projection of the normalized Laplacian matrix. For these matrices, the condition number $\kappa(\cdot)$ is the ratio between the largest and the lowest singular values, and the terms $\mu(\cdot)$ and $\eta(\cdot)$ are two specific norm parameters defined respectively in Definitions \ref{def:mu} and \ref{def:eta} (see Chapter \ref{chap:quantum_linalg} for more details).

We will show that the term $\mu(B)$ in the above expression is in fact upper bounded by $O(N)$, in the worst case scenario. In our basic numerical experiments (see Section \ref{NumericalSimulations}), we indeed observed a quantum running time scaling linearly with $N$, when all terms are taken into account. This implies a significant polynomial speedup over the classical algorithm.
    
The second stage, the quantum clustering, adds to the running time the multiplicative term $T_{qmeans}$, a rather complex expression detailed in Chapter \ref{chap:qmeans}. In the case of well-clusterability, namely when the vectors can effectively be organized in clusters that can be detected efficiently (See Section \ref{sec:well_clusterable_datasets} for details), this runtime can be rewritten as follows (see also Theorem \ref{thm:qmeans_main}), where $k$ is the number of clusters and $\delta$ is a precision parameter : 
\begin{equation}
T_{qmeans} = \widetilde{O}\left(\frac{k^3\eta({\widetilde{\mathcal{L}}^{(k)}})^{2.5}}{\delta^3}\right)
\end{equation}

We expect our quantum spectral clustering algorithm to be accurate and efficient when the classical spectral clustering algorithm also works well. In fact, the classical algorithm works well in the case when, once projected onto the reduced spectral space, the vectors follow the well-clusterability assumption, allowing for efficient clustering. In particular, in this case, the term $\kappa(\widetilde{\mathcal{L}}^{(k)})$ is close to $k$, the number of clusters. Note however that our algorithm could work without this well-clusterability assumption, but the theoretical runtime would be bounded differently. \\

In conclusion, our algorithm provides a considerable theoretical speedup that could allow for new applications of spectral clustering on larger, high-dimensional datasets. The quantum subroutines developed in this section could be useful independently, and we hope for substantial improvements in several other graph based machine learning algorithms.

\section{Quantum Graph Based Machine Learning}\label{quantumspectralclusteringsection}

In this part, we will detail the quantum algorithm that performs spectral clustering with similar guarantees and more efficient running time compared to its classical analog. We start by presenting all the theorems that will help us construct the quantum spectral clustering algorithm, then in the following subsections we provide details and proofs.

We first state the theorem that allows to compute the edge's value of the data adjacency graph:

\begin{theorem}{Quantum Algorithm for Data Point Similarity}{quantumaccestosimilarity}
Given quantum access to the data matrix $S \in \R^{N\times d}$ in time $T_S$ and two indices $p,q \in [N]^2$, we can obtain the following mapping: $\ket{p}\ket{q}\ket{0} \mapsto \ket{p}\ket{q}\ket{a_{pq}}$ in time $O(T_S \frac{\eta(S)}{\epsilon_{dist}})$. The elements $a_{pq}$ correspond to the edge's values of the data adjacency graph, using the rule of construction based on a threshold distance. $\eta(S)$ is a data parameter defined in Definition \ref{def:eta}, $\epsilon_{dist}>0$ is the precision parameter in the estimation of the distance between input points. 
\end{theorem}

Using the previous theorem we can have quantum access to the normalized Incidence matrix.

\begin{theorem}{Quantum access to the Normalized Incidence Matrix}{quantumaccesstoB}
Given quantum access to the data matrix $S \in \R^{N\times d}$ in time $T_S$ we can have quantum access to the normalized incidence matrix $\mathcal{B} \in \R^{N\times \frac{N(N-1)}{2}}$ in time 
\begin{equation}
T_{\mathcal{B}} = \widetilde{O}\left(T_S \frac{\eta(S)}{\epsilon_{dist}}  \frac{1}{\epsilon_{B}}\right)
\end{equation}
where $\eta(S)$ is defined in Definition \ref{def:eta}, $\epsilon_{dist} >0$ is the precision of distance estimation between vectors, and $\epsilon_{B}$ is the substitute of the zeros in $B$.
\end{theorem}

Using the previous theorem we can finally have quantum access to the projected Laplacian matrix.

\begin{theorem}{Quantum access to projected Laplacian matrix}{quantumaccesstoL}
Given quantum access to the normalized incidence matrix $\mathcal{B} \in R^{N\times \frac{N(N-1)}{2}}$ in time $T_{\mathcal{B}}$, we can have quantum access to $\widetilde{\mathcal{L}}^{(k)} \in R^{N\times k}$, the Laplacian matrix projected onto its $k$ lowest eigenvalues, in time 
%\begin{equation}
%T_{\widetilde{\mathcal{L}}^{(k)}} = \widetilde{O}\left(T_{\mathcal{B}}  %\frac{\mu(\mathcal{B\widetilde{\mathcal{L}}^{(k)}})}{\epsilon_{\lambda}} %\kappa(\widetilde{\mathcal{L}}^{(k)})\right)
%\end{equation}
\begin{equation}
T_{\widetilde{\mathcal{L}}^{(k)}} = \widetilde{O}\left(T_{\mathcal{B}}  \frac{\mu(\mathcal{B})\kappa(\widetilde{\mathcal{L}}^{(k)})}{\epsilon_{\lambda}} \right)
\end{equation}
where $\epsilon_\lambda$ is the precision parameter for estimating the eigenvalues of $\mathcal{L}$, $\mu(\mathcal{B})$ is a data parameter defined in Definition \ref{def:mu}, and $\kappa(\widetilde{\mathcal{L}}^{(k)})$ is the condition number of $\widetilde{\mathcal{L}}^{(k)}$.
\end{theorem}

The quantum spectral clustering algorithm consists then in applying the quantum $k$-means algorithm (Theorem \ref{thm:qmeans_main}) using the fact that we have quantum access to the projected Laplacian matrix.

The main algorithm can thus be summarized in the following theorem. 

\begin{theorem}{Quantum Spectral Clustering}{Qspectralclustering_main}
Given quantum access to a data matrix $S \in \R^{N\times d}$ in time $T_S$, there is a quantum algorithm that with high probability outputs $k$ centroids in the Laplacian spectral space, in time 
\begin{equation}
\widetilde{O}
    \left( T_S
    \frac{\eta(S)}{\epsilon_{dist}\epsilon_B}
    \frac{\mu(\mathcal{B})\kappa(\widetilde{\mathcal{L}}^{(k)})}{\epsilon_{\lambda}} T_{qmeans}
    \right)
\end{equation}
where $T_{qmeans}$ is the multiplicative factor in the running time of the quantum clustering algorithm (Theorem  \ref{thm:qmeans_main}) on the vectors projected onto the Laplacian spectral space. In the case where the vectors projected onto the spectral space are well-clusterable (see Definition \ref{def:wcdataset}), the running time becomes
\begin{equation}
\widetilde{O}
    \left( T_S
    \frac{\eta(S)}{\epsilon_{dist}\epsilon_B}
    \frac{\mu(\mathcal{B})\kappa(\widetilde{\mathcal{L}}^{(k)})}{\epsilon_{\lambda}}
    \frac{k^3\eta({\widetilde{\mathcal{L}}^{(k)}})^{2.5}}{\delta^3}
    \right)
\end{equation}
In the above formulas, $\mathcal{B}$ refers to the normalized incidence matrix of the data, and $\widetilde{\mathcal{L}}^{(k)}$ to the projection of the Laplacian matrix. $\epsilon_{dist}$, $\epsilon_{B}$, $\epsilon_\lambda$ and $\delta$ are error or precision parameters. $\eta(S)$, $\eta(\widetilde{\mathcal{L}}^{(k)})$, and $\mu(\mathcal{B})$ are data parameters defined in Definition \ref{def:eta} and Definition \ref{def:mu}, and $\kappa(\widetilde{\mathcal{L}}^{(k)})$ is the condition number of $\widetilde{\mathcal{L}}^{(k)}$.
\end{theorem}

\subsection{Quantum Circuits}

\subsubsection{Computing the similarity between two nodes}\label{similarity_section}

We now detail the algorithm for Theorem \ref{thm:quantumaccestosimilarity}, which allows to compute the elements $a_{pq}$ of the Adjacency Matrix, corresponding to a pair of input vectors $s_p$ and $s_q$. This will be used as a subroutine in the algorithm that gives quantum access to the normalized incidence matrix.

For any two quantum states corresponding to the indices $\ket{p}$ and $\ket{q}$, we can use Theorem \ref{thm:distance_innpdct_quantum} to obtain $\ket{p}\ket{q}\ket{\overline{d(s_p,s_q)^2}}$. The square distance obtained is approximated with a precision $\epsilon_{dist} > 0$ such that $|\overline{d(s_p,s_q)^2} - d(s_p,s_q)^2|\leq \epsilon_{dist}$. 

These distances are then converted into the edge values $a_{pq} \in \{0,1\}$ using our modified graph construction rule from Section \ref{notationsanddefinition}. For doing so, we can use the comparison operator (see Claim \ref{thm:boolean_circuits}) to check if $\overline{d(s_p,s_q)^2}$ is smaller than the desired threshold below which we consider that two nodes are connected. Therefore we can easily obtain $\ket{p}\ket{q}\ket{a_{pq}}$.

Note that it was necessary to use this particular definition of the Adjacency Matrix in order to be able to perform this operation efficiently quantumly. Other definitions include the direct use of the distance, or sometimes a mapping of the distance using diverse kernels.

Finally, note that using this theorem, it is possible to have quantum access to a normalized vesrion of the Adjacency Matrix, which could be useful in many applications. In our case, however, it is not clear how to use it to obtain quantum access to the normalized Laplacian, because it would require the Degree matrix and the norms $\norm{a_i}$, which are not accessible efficiently. Therefore, we chose to work with the \emph{normalized} Incidence matrix.

\subsubsection{Quantum Access to the Normalized Incidence Matrix $\mathcal{B}$}\label{NormalizedB_section}

To obtain quantum access to $\mathcal{B}$ (see Definition \ref{def:quantum_access}), we start with a quantum state encoding one node's index $i \in [N]$, along with the superposition of all $\frac{N(N-1)}{2}$ possible edges encoded as pairs $(p,q) \in [N]^2$ with $p<q$. We use the following quantum state, whose preparation requires $O{(\log(N))}$ qubits and a simple quantum circuit.
\begin{equation}
\frac{\sqrt{2}}{\sqrt{N(N-1)}}
\ket{i}
\sum_{p<q}\ket{p}\ket{q}
\end{equation}

From this state, we first determine which edges are incident using two ancillary qubits as flags, using the equality operator (see Section \ref{thm:boolean_circuits}). This allows us to separate the cases that appear in Eq.(\ref{Bdefinition}). For simplicity, we will use the notation $\ket{p}\ket{q} = \ket{p,q}$.

\begin{equation}\label{afterdist}
    \frac{\sqrt{2}}{\sqrt{N(N-1)}}
    \ket{i}
    \left(
    \sum_{\substack{p<q \\ i=p}}\ket{p,q}\ket{11}
    +
    \sum_{\substack{p<q \\ i=q}}\ket{p,q}\ket{10}
    +
    \sum_{\substack{p<q \\ i\notin \{p,q\}}}\ket{p,q}\ket{00}
    \right)
\end{equation}

We then use another register to write the values of $B$, the unormalized incidence matrix, given by Eq.(\ref{Bdefinition}). For the flagged edges ($i=p$ or $i=q$), using a controlled version of Theorem \ref{thm:quantumaccestosimilarity}, we obtain the superposition of the similarity between all input points and point $i$, in a quantum register. For the other edges ($i\notin \{p,q\}$), instead of simply writing 0 as in Eq.(\ref{Bdefinition}) we modify the zero elements of the matrix to a value $\epsilon_B > 0$ in order to retain the efficiency of the quantum algorithm in the next step. The running time for this step is $O(T_S  \eta(S)/\epsilon_{dist})$, where $\eta(S)$ is a data parameter defined in Definition \ref{def:eta}, and $\epsilon_{dist}>0$ is the precision parameter in the estimation of $d^2(s_i,s_j)$. Finally $T_S$ is the time to have quantum access to the input points $s_1,\cdots,s_N$, which becomes $T_S = O(\log(Nd))$ if we assume QRAM access.

We obtain the following state:

\begin{equation}
    \frac{\sqrt{2}}{\sqrt{N(N-1)}}
    \ket{i}
    \left(
    \sum_{\substack{p<q \\ i=p}}\ket{p,q}\ket{11}\ket{a_{pq}}
    +
    \sum_{\substack{p<q \\ i=q}}\ket{p,q}\ket{10}\ket{a_{pq}}
    +
    \sum_{\substack{p<q \\ i\notin \{p,q\}}}\ket{p,q}\ket{00}\ket{\epsilon_B}
    \right)
\end{equation}

Which, by Eq.(\ref{Bdefinition}) and after uncomputing the flags would be equal to
\begin{equation}
   \frac{\sqrt{2}}{\sqrt{N(N-1)}}
    \ket{i}
    \sum_{p<q}\ket{p,q}\ket{B_{i,(p,q)}}
\end{equation}
From this state, we use a Conditional Rotation (see Theorem \ref{thm:conditionrotation}) to encode, in superposition, the values of the $B$ into the amplitude of a new qubit, and after uncomputing the values of the matrix $B$ from the registers, we have the state:  

\begin{equation}
    \frac{\sqrt{2}}{\sqrt{N(N-1)}}
    \ket{i}
    \left(
    \sum_{p<q}B_{i,(p,q)} \ket{p,q} \ket{0}+ \sum_{p<q}\sqrt{1-B_{i,(p,q)}^2} \ket{p,q} \ket{1}
    \right)
\end{equation}

Finally, an amplitude amplification (see Theorem \ref{thm:amplitude_amplification}) is performed to select the $\ket{0}$ part of the state, and obtain a valid quantum encoding of the vector $B_i$. This requires to repeat the previous steps $O(1/\sqrt{P_i(0)})$ times where $P_i(0)$ is the probability of reading $``0"$ in the last register. Since all elements of the matrix $B$ have norm at least $\epsilon_B$ we therefore have $O(1/\sqrt{P_i(0)})$ is at least $O(1/\epsilon_B)$. We finally obtain with high probability the state
$
    \ket{i}
    \frac{1}{\norm{B_i}}\sum_{p<q}B_{i,(p,q)}\ket{p,q}
    = \ket{i}\ket{\mathcal{B}_i}
$.

Recall from Section \ref{notationsanddefinition} that $\mathcal{B}_{i,(p,q)} = \frac{B_{i,(p,q)}}{\norm{B_i}}$ and that $\norm{B_i} = \norm{a_i}$. In addition, we have access to the norm of each row of $\mathcal{B}$ since by definition they are all equal to $1$. We can therefore conclude that we have quantum access to $\mathcal{B}$, the normalized incidence matrix, according to Definition (\ref{def:quantum_access}). The global running time to have quantum access to $\mathcal{B}$ is given by $O(T_S \frac{\eta(S)}{\epsilon_{dist}} \frac{1}{\epsilon_B}))$. This proves Theorem \ref{thm:quantumaccesstoB}.

\subsubsection{Quantum Access to the Projected Normalized Laplacian Matrix $\widetilde{\mathcal{L}}^{(k)}$}\label{SVEsection}

With the quantum access to the normalized incidence matrix $\mathcal{B}$ in time $T_{\mathcal{B}}$, we will use the fact that the $i^{th}$ row of $\mathcal{L}=\mathcal{B}\mathcal{B}^T$ can be written as $\mathcal{L}_i = \mathcal{L}\cdot e_i$ where $e_i$ represents the $i^{th}$ vector of the standard basis, for which the corresponding quantum state is simply $\ket{i}$. We also use the fact that this state can be naturally expressed as $\ket{i} = \sum_j\sigma_{ij}\ket{u_j}$ in the basis made of the left singular vectors $u_j$ of $\mathcal{B}$, with unknown coefficients $\sigma_{ij}$, such that $\sum_j\sigma_{ij}^2=1$.

On this initial state $\ket{i}$ we apply the SVE algorithm\footnote{Since SVE must be applied to a square matrix, we use in fact $\begin{pmatrix}0 & \mathcal{B}\\\mathcal{B}^T & 0\end{pmatrix}$ see \cite{kerenidis_recommendation_system}.} 
(Theorem \ref{thm:SVE}) to estimate the singular values of $\mathcal{B}$ in superposition and obtain $\sum_j\sigma_{ij}\ket{u_j}\ket{\lambda^{\mathcal{B}}_j}$. The running time for this step is $O(T_\mathcal{B}  \mu(\mathcal{B})/\epsilon_{\lambda})$, where $\mu(\mathcal{B})$ is a data parameter defined in Definition \ref{def:mu}, and $\epsilon_{\lambda}>0$ is the desired precision in the estimation of the singular values. We then square these values to obtain the state $\sum_j\sigma_{ij}\ket{u_j}\ket{\lambda_j}$, with the eigenvalues of $\mathcal{L}$. 

Note that $\mu(\mathcal{B})$ is upper bounded by $N$. Indeed, from Definition \ref{def:mu} we have $\mu(\mathcal{B}) \leq \norm{\mathcal{B}}_F$. Recall that $\mathcal{B}\in\R^{N \times N(N-1)/2}$. Since by construction each $\norm{B_i}_2 = 1$, and since $N(N-1)/2 < N^2$, we have finally:
\begin{equation}
\mu(\mathcal{B}) 
\leq \norm{\mathcal{B}}_F = \sqrt{\sum_{i}^{\frac{N(N-1)}{2}}\norm{B_i}_2^2} 
= \sqrt{\frac{N(N-1)}{2}}
\leq N
\end{equation}

At this point, we can prepare the projection on the $k$ lowest eigenvectors of $\mathcal{L}$, as in Section \ref{sec:quantum_projection_SVE}. We first separate the eigenvalues lower that a threshold $\nu > 0$ with an ancillary qubit, such that the $k$ lowest eigenvalues are flagged by $``0"$ :
\begin{equation}
    \sum_{\substack{j | \lambda_j \leq \nu}}\sigma_{ij}\ket{u_j}\ket{\lambda_j}\ket{0}
    + 
    \sum_{\substack{j | \lambda_j > \nu}}\sigma_{ij}\ket{u_j}\ket{\lambda_j}\ket{1}
\end{equation}

If the flag qubit is $``0"$, we perform a conditional rotation on the eigenvalue register using Theorem \ref{thm:conditionrotation} in a controlled fashion. For this we introduce again an ancillary qubit. Since the amplitude of the rotation must be inferior to 1, we first divide each eigenvalue by $\nu$, which is higher than the largest eigenvalue of $\widetilde{\mathcal{L}}^{(k)}$:  
\begin{equation}\label{LaplacianCR}
    \sum_{\substack{j | \lambda_j \leq \nu}}\sigma_{ij}\ket{u_j}\ket{\lambda_j}\ket{0}
    \left(\frac{\lambda_{j}}{\nu}\ket{0} + \sqrt{1-\frac{\lambda^2_j}{\nu^2}}\ket{1}\right)
    +
    \sum_{\substack{j | \lambda_j > \nu}}\sigma_{ij}\ket{u_j}\ket{\lambda_j}\ket{1}\ket{0}
\end{equation}

From this quantum state, we will show how to obtain quantum access to $\ket{\widetilde{\mathcal{L}}^{(k)}}$ using Amplitude Estimation and amplitude amplification on the $``00"$ value on the last two registers. We recall that quantum access is guaranteed if we can recover, for each row $\widetilde{\mathcal{L}}^{(k)}_i$, its norm and the corresponding quantum state.\\
Let  $P_i(00)$ be the probability of reading $``00"$ on the last two registers. It is easy to show that $ P_i(00) = \frac{1}{\nu^2}\sum_{\substack{j | \lambda_j \leq \nu}}\sigma_{ij}^2\lambda_{j}^2$. 
We will prove that $\norm{\widetilde{\mathcal{L}}^{(k)}_i}= \nu\sqrt{P_i(00)}$, and that amplifying the state will yield to $\ket{\widetilde{\mathcal{L}}^{(k)}_i}$. \\
The normalized Laplacian matrix can be written in its eigenbasis using the outer product $\mathcal{L} = \sum_j \lambda_j \ket{u_j}\bra{u_j}$. 
Similarly, recall that we used $\sum_j \sigma_{ij} \ket{u_j}$ to encode $e_i$, the $i^{th}$ vector in the standard basis. Therefore, each row $\mathcal{L}_i = \mathcal{L} \cdot e_i$ can be written as $\ket{\mathcal{L}_i} = \sum_j \lambda_j \ket{u_j}\bra{u_j} \cdot \sum_j \sigma_{ij} \ket{u_j} = \sum_j \sigma_{ij}\lambda_j\ket{u_j}$. We thus have a formula for the norm of the rows $\norm{\mathcal{L}_i}=\sqrt{\sum_j \sigma_{ij}^2\lambda_j^2}$. The same idea holds for $\widetilde{\mathcal{L}}^{(k)}$, the projection of the normalized Laplacian, and we obtain $\norm{\widetilde{\mathcal{L}}^{(k)}_i}=\sqrt{\sum_{j=1}^{k}\sigma_{ij}^2\lambda_j^2} = \nu\sqrt{P_i(00)}$.\\
Therefore, using Amplitude Estimation (Theorem \ref{thm:amplitude_amplification}), we can estimate $P(00)$ and have access to the norms of $\widetilde{\mathcal{L}}^{(k)}$ to a multiplicative constant. Similarly, using amplitude amplification (Theorem \ref{thm:amplitude_amplification} also), we can amplify the $\ket{00}$ state and obtain the state $\sum_{\lambda_j \leq \nu}\sigma_{ij}\lambda_j\ket{u_j} = \ket{\widetilde{\mathcal{L}}^{(k)}_i}$.\\
The number of iterations for Amplitude Estimation and Amplification is $\widetilde{O}(1/\sqrt{P_i(00)})$. Let $\lambda_k$ be the $k^{th}$ eigenvalue of $\mathcal{L}$ and therefore the largest eigenvalue of $\widetilde{\mathcal{L}}^{(k)}$. By correctly choosing the threshold $\nu$ close to the largest eigenvalue, for instance $\nu \leq \gamma \lambda_k$ with $\gamma>1$ a small positive constance, we can write:
\begin{equation}
P_i(00) =\sum_{j | \lambda_j \leq \nu}\sigma_{ij}^2 \frac{\lambda_{j}^2}{\nu^2} 
\geq 
\sum_{j | \lambda_j \leq \lambda_k}\sigma_{ij}^2 \frac{\lambda_{min}^2}{\gamma^2 \lambda_k^2}
= \frac{1}{\gamma^2} \frac{\lambda_{min}^2}{\lambda_k^2} 
= \frac{1}{\gamma^2} \frac{1}{\kappa(\widetilde{\mathcal{L}}^{(k)})^2} 
= O\left(1 /\kappa(\widetilde{\mathcal{L}}^{(k)})^2\right)
\end{equation}

Therefore the number of iterations for this step can be upper bounded by $\widetilde{O}(\kappa(\widetilde{\mathcal{L}}^{(k)}))$. Overall, the running time is given by $O(T_{\mathcal{B}}  \frac{\mu(\mathcal{B})}{\epsilon_{\lambda}} 
\kappa(\widetilde{\mathcal{L}}^{(k)}))$. This concludes the algorithm that gives quantum access to the projected Laplacian of a graph and proves Theorem \ref{thm:quantumaccesstoL}.

\begin{figure}[t!]
\centering
   \includegraphics[width=\textwidth]{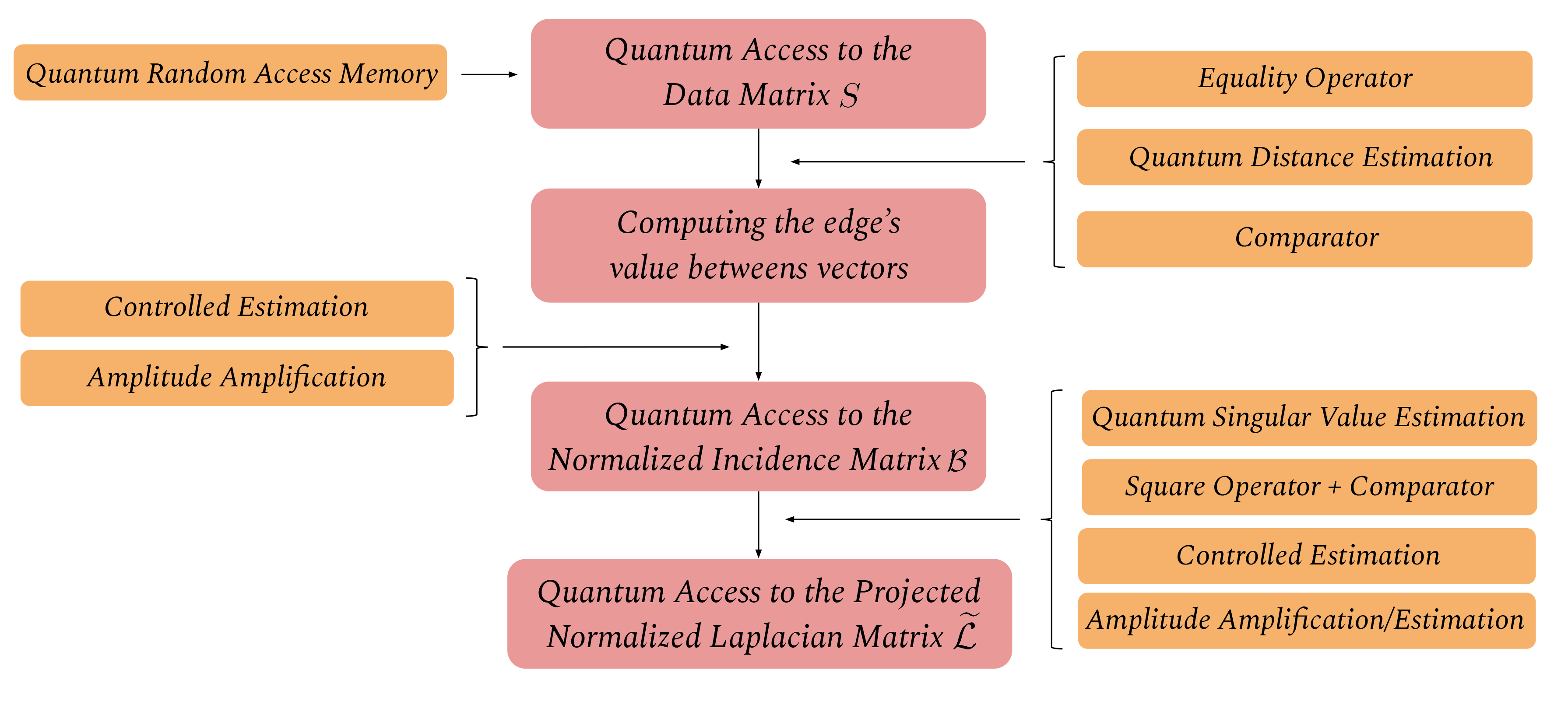}
   \label{fig_diagram} 
\caption{Diagram for Sections \ref{similarity_section} to \ref{SVEsection}}
\label{fig:diagram_q_spectral_clust}
\end{figure}

\subsection{Quantum Clustering in the Spectral Space}
Having quantum access to the projected normalized Laplacian $\widetilde{\mathcal{L}}^{(k)}$, we possess all requirements to apply the quantum $k$-means algorithm, or $q$-means (Theorem \ref{thm:qmeans_main}). We initialize $k$ centroids at random or using $q$-means++ (see Section \ref{q-means++proof}), equivalent to the $k$-means++ initialization. The $q$-means algorithm first consists in constructing a state where all distances between rows $\widetilde{\mathcal{L}}^{(k)}_i$ and the current centroids are computed in parallel. The rest of the algorithm consists in finding the label for each row, and updating the centroids as the average of the vectors composing the cluster. After several iterations, the centroids should have converged and can be retrieved. 

Several approximations are necessary during the steps, hence the presence of precision parameter $\delta>0$. It expresses the approximation error committed during the distance estimation, and during the classical description of the new centroids at each step. Therefore, to be more specific, $q$-means is a noisy version of $k$-means. 

In high level, denoting with $T_{\widetilde{\mathcal{L}}^{(k)}}$ the time to have quantum access to $\widetilde{\mathcal{L}}^{(k)} \in \R^{N\times k}$, which plays the role of the initial state for the clustering algorithm,  and $T_{qmeans}$ the remaining multiplicative factor in the running time of the quantum clustering algorithm (see Theorem \ref{thm:qmeans_main}), the overall running time of our algorithm is given by 
\begin{equation}
    T_{\widetilde{\mathcal{L}}^{(k)}} T_{qmeans}
\end{equation}
To go further, we can assume that the vectors are effectively made of well separated clusters. Indeed, it should be the case in the spectral clustering method, once projected onto the spectral space (see Section \ref{NumericalSimulations} for numerical simulations). This well-clusterability assumption (See Definition \ref{def:wcdataset}) ensures the classical spectral clustering to classify the data accurately, and the quantum algorithm to work efficiently. Indeed, the running time of the $q$-means algorithm is bounded with better guarantees in this case (Chapter \ref{chap:qmeans}). With this assumption, and with input dimension $k$ for the spectral space of $\widetilde{\mathcal{L}}^{(k)}$, the running time to update the $k$ centroids in the case of well-clusterable data is given by
\begin{equation}\label{runningtimeqmeansonL}
\widetilde{O}\left(T_{\widetilde{\mathcal{L}}^{(k)}}  \frac{k^3\eta({\widetilde{\mathcal{L}}^{(k)}})^{2.5}}{\delta^3}\right)
\end{equation}
Again, it is important to note that our algorithm could work without this well-clusterability assumption, the only difference would be a different bound on the theoretical runtime (See Theorem \ref{thm:qmeans_main} for the general formulation).

\subsection{Running Time}\label{conclusionrunningtime}
Using the running times obtained in Theorems \ref{thm:quantumaccesstoB}
and \ref{thm:quantumaccesstoL}, as well as result (\ref{runningtimeqmeansonL}) for the well-clusterable case, we can conclude that our quantum algorithm for spectral clustering has the following running time :

\begin{equation}\label{runtime}
    \widetilde{O}
    \left( T_S
    \frac{\eta(S)}{\epsilon_{dist}\epsilon_B}
    \frac{\mu(\mathcal{B})\kappa(\widetilde{\mathcal{L}}^{(k)})}{\epsilon_{\lambda}}
    \frac{k^3\eta({\widetilde{\mathcal{L}}^{(k)}})^{2.5}}{\delta^3}
    \right)
\end{equation}

\noindent
$T_S$ is the time to have \emph{quantum access} (see Definition \ref{def:quantum_access}) to the input vectors $S \in \R^{N\times d}$ which becomes $O(polylog(N,d))$ if we assume access to the QRAM. $k$ is the number of clusters. $\kappa(\widetilde{\mathcal{L}}^{(k)})$ is the condition number of $\widetilde{\mathcal{L}}^{(k)}$. $\mu(\mathcal{B})$, $\eta(S)$, and $\eta({\widetilde{\mathcal{L}}^{(k)}})$ are data parameters defined in Definitions \ref{def:mu} and \ref{def:eta}. $\epsilon_B$ is the chosen minimum value in the incidence matrix. $\epsilon_{dist}$ is the precision in the distance calculation between input points. $\delta$ is the precision of the $q$-means algorithm.

\section{Numerical Simulations}

The quantum algorithm performs the same steps as the classical one while introducing noise or randomness along the way. We present numerical simulations on simple synthetic datasets made of two concentric circles, as in the original work on spectral clustering \cite{ng2002spectral}, in order to benchmark the quality of the quantum algorithm. In this way, we see how the quantum effects are impacting the graph and the spectral space, and we obtain numerical estimates on the quantum running time. 

\begin{figure}[h]
\begin{subfigure}[b]{\linewidth}
\centering
   \includegraphics[width=450px]{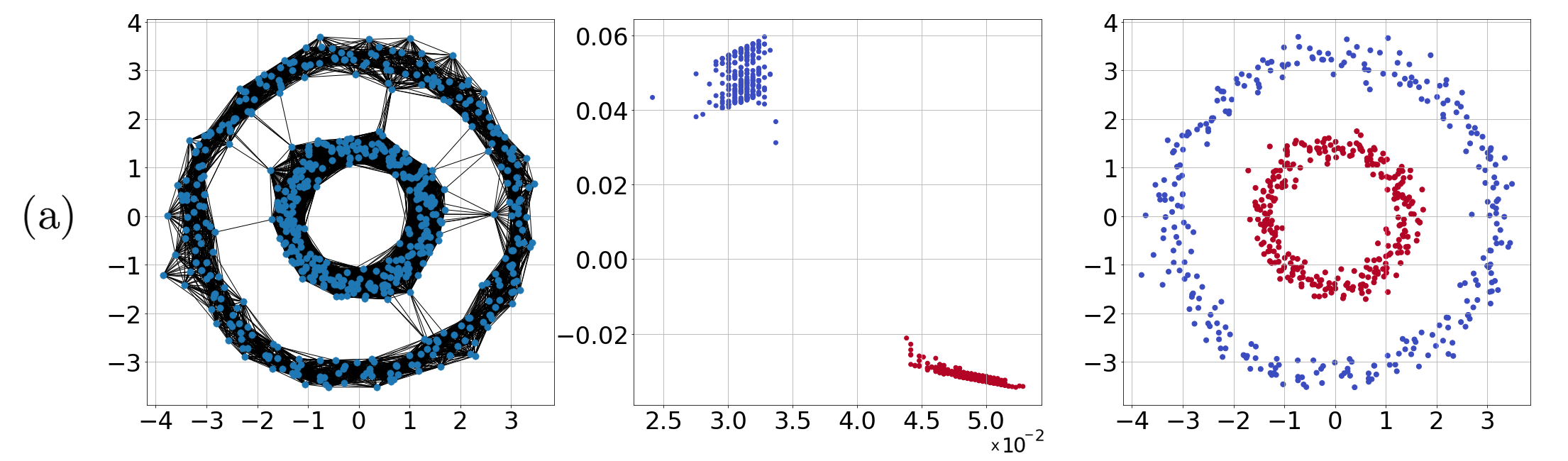}
%   \caption{Classical spectral clustering}
%   \label{fig:Ng1} 
\end{subfigure}
\begin{subfigure}[b]{\linewidth}
\centering
   \includegraphics[width=450px]{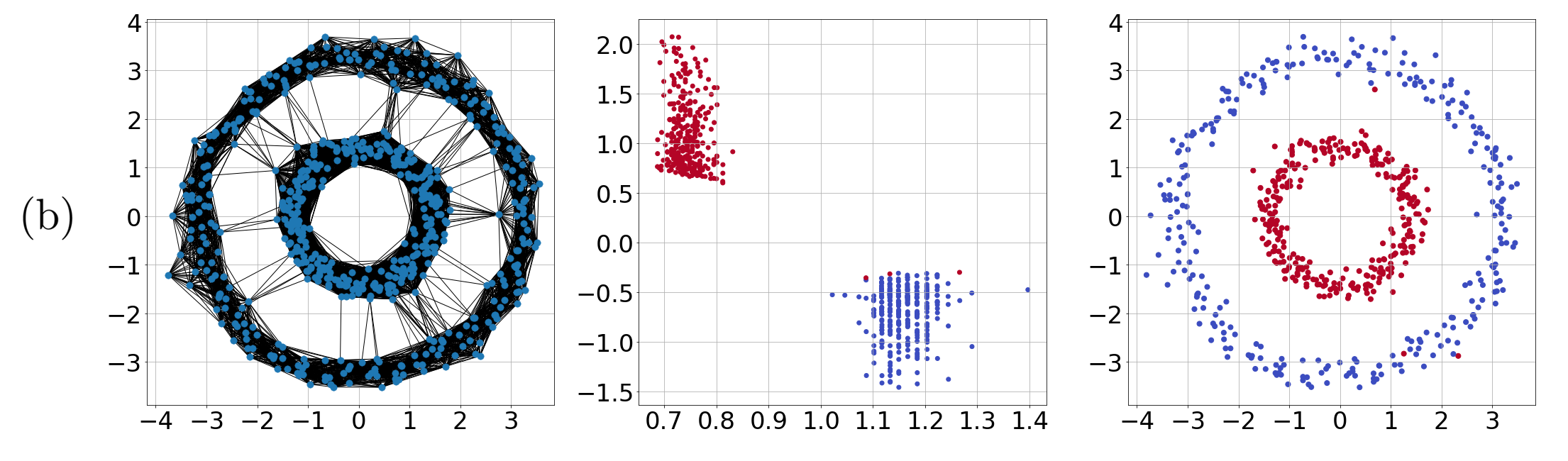}
%   \caption{Quantum spectral clustering}
%   \label{fig:Ng2}
\end{subfigure}
\caption{(a) Classical and (b) Quantum spectral clustering algorithms on two non linearly separable sets of 600 input vectors. The Laplacian is derived from the adjacency graph \emph{(left)}, itself constructed from the data points. The clustering is shown in both spectral \emph{(center)} and input \emph{(right)} space domains. Three points were misclassified in the quantum case.}
\label{figurecircles}
\end{figure}

These simulations are made with a classical computer that simulates the quantum steps and introduces equivalent noise and randomness. It would be very interesting to perform the same experiments using a real quantum computer, alas such computers are not yet available. While simulating the quantum steps, the computation becomes very soon impractical, in fact the simulations we present already take several hours to execute, and thus we leave numerical simulations on larger datasets as future work. Our goal was to design a quantum spectral clustering algorithm with a rigorous theoretical analysis of its running time and provide initial evidence of its practical efficiency and accuracy on a canonical dataset. 

\begin{figure}[h!]
\begin{floatrow}
\ffigbox{%
 \includegraphics[width=0.5\textwidth]{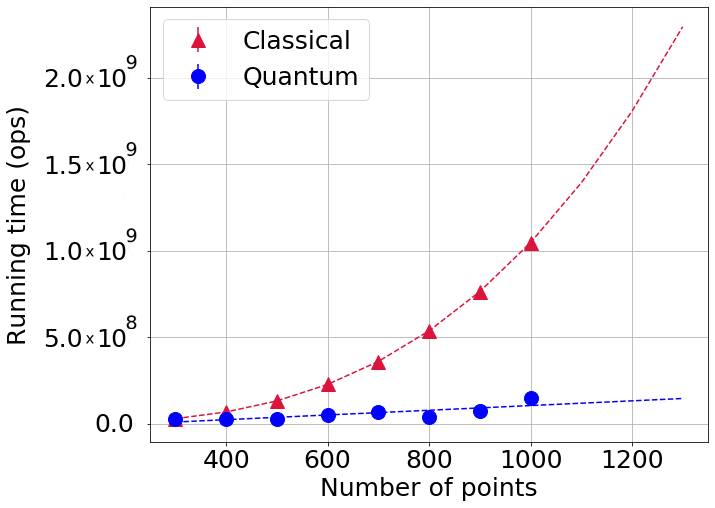} 
}{%
\caption{Running times for quantum and classical spectral clustering. Error bars are present.} 
\label{runningtimes}
}
\capbtabbox{%
  \begin{tabular}{ccc}
    \toprule
    \multicolumn{1}{c}{\multirow{2}{*}{\shortstack[l]{Numbers \\ of points}}} & \multicolumn{2}{c}{Accuracy}                   \\
    \cmidrule(r){2-3}
         &  Classical    & Quantum \\
    \midrule
300 & $100\%$ & $99.6 \% \pm 1.05 \%$  \\
400 & $100\%$ & $99.5 \% \pm 0.83 \%$  \\
500 & $100\%$ & $99.3 \% \pm 1.52 \%$  \\
600 & $100\%$ & $99.9 \% \pm 0.31 \%$  \\
700 & $100\%$ & $100.0 \% \pm 0.0 \%$  \\
800 & $100\%$ & $99.6 \% \pm 0.62 \%$  \\
900 & $100\%$ & $99.6 \% \pm 0.68 \%$  \\
1000 & $100\%$ & $99.9 \% \pm 0.19 \%$  \\
    \bottomrule
  \end{tabular}
}{%
  \caption{Our quantum algorithm finds the clusters with a similar high accuracy.}%
  \label{accuracytable}
}
\end{floatrow}
\end{figure}

The precision parameters used in the quantum case were: $\epsilon_{dist} = 0.1$ for the creation of edges, $\epsilon_{B} = 0.1$ for the creation of the incidence matrix, $\epsilon_{\lambda} = 0.9$ during the estimation of the eigenvalues of $\mathcal{L}$ and finally the precision parameter in $q$-means $\delta = 0.9$. The quantum algorithm was able to classify the two sets with high accuracy (Table \ref{accuracytable}). The clustering was simulated for 300 to 1000 points, repeated 10 times each. In Fig.\ref{figurecircles} we observe clearly the impact of the quantum effects in the graph, where the edges are different, and in the spectral space, where the clusters are more spread out. It is surprising that the quantum algorithm is already faster for small values of $N$, below 1000 points (Fig.\ref{runningtimes}). This difference should substantially increase as $N$ grows. Indeed, compared to the classical algorithm used in practice which scales as $O(N^3)$, the quantum running time is advantageous as its scaling appears to be linear in $N$. In fact, the scaling comes from the factor $\mu(B)$ which is upper bounded by $N$ (see Section \ref{SVEsection}). Note, of course, that both for the classical and quantum running time we used as a proxy the order of steps in the theoretical analysis, disregarding questions of clock time or error correction. Our results show more than anything that it is certainly worth pursuing quantum algorithms for spectral clustering and other graph based machine learning algorithms, since at least at a first level of comparison they can offer considerable advantages compared to the classical algorithms. It remains an open question to see when and if quantum hardware can become good enough to offer such advantages in practice.

\subsubsection{Conclusions}\label{conclusions}
In this work, we described a new quantum machine learning algorithm, inspired by the classical spectral clustering algorithm. While introducing a number of modifications, approximations, and randomness in the process, the quantum algorithm can still perform clustering tasks with similar very good accuracy, and with a more efficient running time thanks to a weaker dependence on the number of input points: at most linear in our preliminary experiments. This could allow quantum spectral clustering to be applied on datasets that are now considered infeasible in practice. Our quantum algorithm is end-to-end, from classical input to classical output, and could pave the way to other graph based methods in machine learning and optimization, for example using our methods for obtaining access to the normalized Adjacency, Incident, and Laplacian matrices.

%%%%---------------PART IV-------------%%%%
\part{Quantum Neural Networks}\label{part:Q_NeuralNet}
\chapter{Introduction}
\epigraph{\textit{“Les machines un jour pourront résoudre tous les problèmes, mais jamais aucune d'entre elles ne pourra en poser un !”}}{Albert Einstein \\ \emph{Comment je vois le monde} (1934)}

We refer the reader to Section \ref{sec:preliminaries_classical_nn} for a basic introduction to classical neural network theory and notations. 

\section{Why is it Hard to Implement a Neural Network on a Quantum Computer?}\label{sec:difficult_qnn}

In recent years, many attempts have been made to provide quantum algorithms for neural networks. this proliferation is due to the significance that it would represent but also to the fact that it is a problem difficult to solve \cite{schuld2014quest}. 

The main issues in creating quantum neural networks are non-linearities, modularity and diversity:

\paragraph{Non-linearities} As we have seen in Section \ref{sec:classical_fcnn_intro}, non-linear functions are systematically present at the end of each neural network layer. They allow the model to reach universality by approximating any functions \cite{cybenko1989approximation,leshno1993multilayer}. However, quantum computing is the realm of unitary, therefore linear, transformations. In particular, when the amplitudes of a quantum state encode the meaningful data, one cannot apply an arbitrary non-linear function on it. This represents a strong limit on the ability of quantum computers to enhance deep learning by directly converting them as quantum circuits and harnessing quantum superposition across layers. 

To deal with this paradox, several attempts have been made \cite{tacchino2019artificial, beer2020training,wiebe2014quantum_deeplearning, cao2017quantum}. Most use measurements to trigger non-linear behaviors, others transfer the data in binary encoding for classical applications of non-linearities \cite{allcock2020quantum}. Finally, it is often the case that quantum neural networks are proposed without any non-linearity \cite{Lloyd_hopfield_nn}. In this thesis, we will pursue the approach of \cite{allcock2020quantum}, which comes at the cost of performing amplitude estimation and thus limits the speedup to a quadratic one at the end. 

\paragraph{Modular approach} The success of deep learning relies on the infinite number of architectures that are possible to produce in order to solve one particular task. It allows us to adapt to any input or output size, or to test different numbers and sizes of hidden layers (often called \emph{hyper parameter optimization}). Recent architectures such as ResNet \cite{tai2017image} or LSTM \cite{gers2000learning} even require connections between input layer non adjacent hidden layers. In deep learning for recommendation systems \cite{zhang2019deep} several inputs are processed on their own neural network, then branched later on. 

The way we define quantum neural networks should include this modularity and offer similar handling. This would require a certain architecture for the quantum circuit. First, the quantum circuit should correspond to one layer, and potentially be end-to-end (from classical input to classical output). 

\paragraph{Diversity of modules} Even though quantum deep learning should start at the basic, namely fully connected neural networks (see Section \ref{sec:classical_fcnn_intro}), there is a large variety of layers that exist in the literature. Different layer types are combined in a single network, including convolutional layer, but also pooling (see Section \ref{sec:classical_cnn_preliminaries}), averaging, dropout. A zoo of non-linearities are also used at different steps (ReLu, Sigmoid, tanh, etc.). More recently, additional constraints have been introduced on the layer such as orthogonality, isometry, hyperbolic geometry \cite{ganea2018hyperbolic} and more. 

Quantum neural networks don't have to mimic them all at first, but we should bear in mind that this is the level of diversity we could aim for. Quantum computing may allow for certain properties easily (see Chapter \ref{chap:OrthoNN_nisq}).

\paragraph{Training} Finally, we must not overlook the difficulty of training the weights of the neural networks, which is most commonly done using backpropagation (see Section \ref{sec:bacpropagation_fcnn}). This also represents a challenge for quantum computing since it imposes to access the state of each layer. It is then a strong argument for modularity and the classical inputs and outputs. Having only a quantum state for each layer and pursuing the computation would present to do the backpropagation unless we destroy the quantum state by measuring it. for instance, faced with this problem, variational quantum circuits have chosen to abandon backpropagation for less efficient techniques (see Section \ref{sec:NISQ}).\\

Before going into the details of quantum neural networks, it is useful to remind that there exist different approaches for short term or NISQ applications. As detailed in Section \ref{sec:NISQ}, variational quantum circuits don't properly have weights but gate parameters that are tunable as well. As in \cite{cong2019quantum, verdon2017quantum}, the appellation \emph{quantum neural networks} is used for their conceptual similarities. In Chapter \ref{chap:OrthoNN_nisq}, we will propose a similar NISQ neural network that has a specific and fast training paradigm, and that is equivalent to its classical counterpart. There also exist recent proposals for fault tolerant quantum computers using the \emph{Neural Tangent Kernel} approach and the HHL algorithm \cite{zlokapa2021quantum}.

\section{Quantum Algorithm for Fully Connected Neural Networks}\label{sec:quantum_fcnn}

In \cite{allcock2020quantum}, the authors introduced a quantum algorithm for fully connected neural network defined in Section \ref{sec:classical_fcnn_intro}, based on the quantum inner product estimation (QIPE) from \cite{qmeans} (Theorem \ref{thm:distance_innpdct_quantum}). 

As stated in Eq.(\ref{eq:nn_basic_equation_1}) and Eq.(\ref{eq:nn_basic_equation_2}), the $\ell^{th}$ layer of a fully connected neural network's with input vector $a^\ell$, weight matrix $W^\ell$ and bias vector $ b^\ell$ can be written as

\begin{equation}\label{eq:nn_basic_equation_1_bis}
    z^{\ell+1} = W^\ell a^\ell + b^\ell 
\end{equation}
    
\begin{equation}\label{eq:nn_basic_equation_2_bis}
    a^{\ell+1} = \sigma(z^{\ell+1})
\end{equation}

We see that the matrix product between $W^\ell$ and $a^\ell$ is the core of this operation, followed by a non-linear function $\sigma$. This matrix multiplication can be replaced by a series of inner product between the input $a^\ell$ and each row of the weight matrix $W^\ell_j$ for $j \in [n_{\ell}]$, where $n_{\ell}$ is the number of input nodes or the dimension of $a^\ell$.

\begin{equation}
    z^{\ell+1}_j = (W^\ell_j, a^\ell) + b^\ell_j 
\end{equation}

where $(x, y)$ is the inner product between vectors $x$ and $y$.\\

Using the QIPE algorithm from Theorem \ref{thm:distance_innpdct_quantum}, with relative error parameters $\epsilon$ and probability parameter $\Delta$, the authors defined a $(\epsilon,\Delta)$-feedforward neural network. They achieve a running time of:

\begin{equation}
    \widetilde{O}\left(N\frac{\log(1/\Delta)}{\epsilon}R\right)
\end{equation}

where $N$ is the number of neurons in the whole neural network, and $R$ is an aggregate of data dependant parameters (based on the norms of the weights and inputs).

To this end, they apply the QIPE in superposition over all the rows of the weights matrix, assuming quantum access to $W^\ell$ and $a^\ell$. 

\begin{figure}[h]
    \centering
    \includegraphics[width=0.7\textwidth]{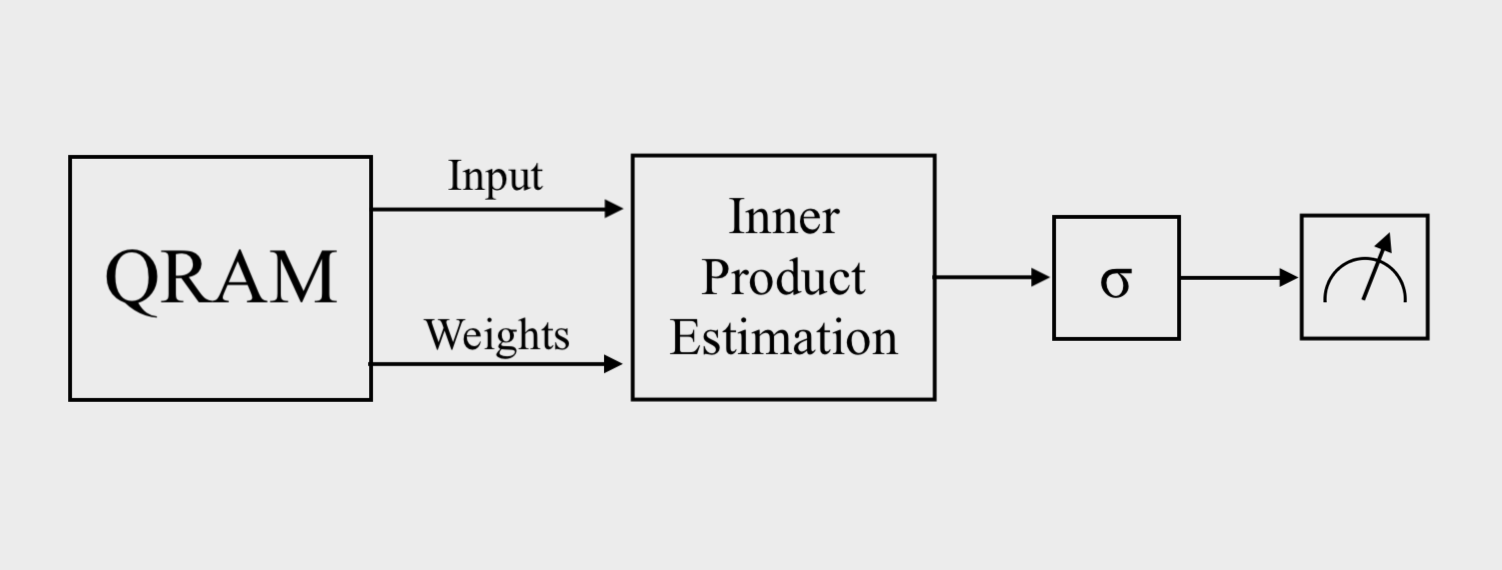}
    \caption{Diagram representation of a fully connected neural network's layer using the quantum inner product estimation from Theorem \ref{thm:distance_innpdct_quantum}. It performs the matrix multiplication and the following non-linearity $\sigma$.}
    \label{fig:FCNN_schema}
\end{figure}

Further, a quantum backpropagation algorithm was introduced. Indeed, as explained in Section \ref{sec:bacpropagation_fcnn}, backpropagation is also a series of matrix multiplication performed layer by layer backward. Using again the QIPE, the authors proposed a quantum algorithm for training the networks in time

\begin{equation}
    \widetilde{O}\left((TM)^{1.5}N\frac{\log(1/\Delta)}{\epsilon}R'\right)
\end{equation}

where $T$ is the number of update iterations, $M$ the
size of the mini batches, and $R'$ another aggregate of data dependant parameters. Classical simulations were conducted and showed that the numerical parameters $R$ and $R'$ were small in practice. \\

In addition, the difficulties presented in Section \ref{sec:difficult_qnn} forced the authors to be subtle with the weight matrix. They first imagined a low rank initialization, to avoid having too many values stored in the QRAM. Then, at each iteration, instead of updating the whole matrix, which would have ruined the benefit of previous quantum computations, they store instead the layers and a derivative of the error vectors, which allow for a fast recomputation of the weight matrix.

The authors suggest that the noise and errors caused by the quantum nature of this algorithm could be beneficial for neural networks. They argue that training a neural network while performing noisy computations will help it generalize correctly, preventing the weights to adjust too well to the training set also known as \emph{overfitting}. This \emph{robust} training using quantum effects is interesting, and could be further developed in terms of privacy preserving, data enhancement, and more. \\

Note finally that it is also possible to perform fully connected neural networks in a different setting, using inner products from unary data loaders \cite{QMedicalImagingRoche} (see Section \ref{sec:inner_product_and_distance_quantum}).

\chapter{Quantum Convolutional Neural Networks}\label{chap:QCNN}
\epigraph{\textit{"L’ignorance rend plus sûr de soi que la connaissance."}}{Etienne Klein \\ \emph{Le goût du vrai} (2020)}

Preliminaries on convolutional neural networks, along with all notations are given in Section \ref{sec:classical_cnn_preliminaries}. Variables are summarized in Section \ref{variable_summary_section}.

\section{Main Results}

We design a quantum algorithm for a complete CNN, with a modular architecture that allows any number of layers, any number and size of kernels,  and that includes a large variety of non-linearity and pooling methods. We introduce a quantum convolution product and a specific quantum sampling technique well suited for information recovery in the context of CNN. We also propose a quantum algorithm for backpropagation that allows an efficient training of our quantum CNN.

As explained in Section \ref{sec:classical_cnn_preliminaries}, a single layer $\ell$ of the classical CNN does the following operations: from an input image $X^{\ell}$ seen as a 3D tensor, and a kernel $K^{\ell}$ seen as a 4D tensor, it performs a convolution $X^{\ell+1}=X^{\ell} * K^{\ell}$, followed by a non-linear function, and followed by pooling. In the quantum case, we will obtain a quantum state corresponding to this output, approximated with error $\epsilon > 0$. To retrieve a classical description from this quantum state, we will apply an $\ell_{\infty}$ tomography (see Theorem \ref{thm:tomography_linfinity}) 
and sample the elements with high values, in order to reduce the computation while still getting the information that matters (see Section \ref{QCNNForward}). 

The QCNN can be directly compared to the classical CNN as it has the same inputs and outputs. 
We show that it offers a speedup compared to the classical CNN for both the forward pass and 
for training using backpropagation in certain cases.  For each layer, on the forward pass (Algorithm \ref{QCNNLayer}), the speedup is exponential in the size of the layer (number of kernels) and almost quadratic on the spatial dimension of the input. We next state informally the speedup for the forward pass, the formal version appears as Theorem \ref{thm:convolution_layer}. 

\begin{result}{Quantum Convolution Layer}{qcnn1}
Let $X^{\ell}$ be the input and $K^{\ell}$ be the kernel for layer $\ell$ of a convolutional neural network, and $f : \mathbb{R} \mapsto [0,C]$ with $C > 0$ be a non-linear function so that 
$f(X^{\ell+1}) := f(X^{\ell}*K^{\ell})$ is the output for layer $\ell$. 

Given $X^{\ell}$ and $K^{\ell}$ stored in QRAM, there is a quantum algorithm that, for any precision parameters $\epsilon > 0$ and $\nu > 0$, creates quantum state $\ket{f(\overline{X}^{\ell+1})}$ 
such that $ \norm{ f(\overline{X}^{\ell+1}) - f(X^{\ell+1}) }_{\infty}  \leq 2M\epsilon$ and retrieves classical tensor $\mathcal{X}^{\ell+1}$ such that for each pixel $j$,

\begin{equation}
	\begin{cases}
	|\mathcal{X}^{\ell+1}_{j} - f(X^{\ell+1}_j)| \leq 2\epsilon \quad \text{if} \quad f(\overline{X}^{\ell+1}_{j}) \geq \nu\\
	\mathcal{X}^{\ell+1}_{j} = 0 \qquad\qquad\qquad \text{if} \quad f(\overline{X}^{\ell+1}_{j}) < \nu\\
	\end{cases}
\end{equation}
This algorithm runs in time 
 \begin{equation}
 \widetilde{O}\left(\frac{1}{\epsilon \nu^2 } \frac{M \sqrt{C}}{\sqrt{\mathbb{E}(f(\overline{X}^{\ell+1}))}}\right)
 \end{equation}
 where $\mathbb{E}(f(\overline{X}^{\ell+1}))$ represents the average value of $f(\overline{X}^{\ell+1})$, and $\widetilde{O}$ hides factors polylogarithmic in the size of $X^{\ell}$ and $K^{\ell}$ and the parameter $M$ (defined in Eq.(\ref{Mdefinition})) is the maximum product of norms from subregions of $X^{\ell}$ and $K^{\ell}$.

\end{result} 

We see that the number of kernels contributes only polylogarithmically to the running time, allowing the QCNN to work with larger and in particular exponentially deeper kernels. The contribution from the input size is hidden in the precision parameter $\nu$ such that (see also Eq.(\ref{def:nu_qcnn})):
\begin{equation}
    \nu = \frac{1}{\sqrt{\sigma \cdot H^{\ell +1}W^{\ell +1}D^{\ell+1}}}
\end{equation}
Therefore the running time has a factor $O(\sigma \cdot H^{\ell +1}W^{\ell +1}D^{\ell+1})$, which is the number of pixels classically retrieved from the quantum state. Indeed, a sufficiently large fraction of pixels must be sampled from the output of the quantum convolution to retrieve meaningful information. In the Numerical Simulations (Section \ref{NumericalSimulations}) we give estimates for $\nu$. The cost of generating the output $\mathcal{X}^{\ell+1}$ of the quantum convolutional layer 
can thus be much smaller than that for the classical CNN in certain cases, Section \ref{runningtime} provides a detailed comparison to the classical running time.

Following the forward pass, a loss function $\mathcal{L}$ is computed for the output of a classical CNN. The backpropagation algorithm is then used to calculate, layer by layer, the gradient of this loss with respect to the elements of the kernels $K^{\ell}$, in order to update them through gradient descent. The formal version of our quantum backpropagation algorithm is given as Theorem \ref{thm:qbackprop}

\begin{result}{Quantum Backpropagation for Quantum CNN}{qcnn2}
Given the forward pass quantum algorithm in Result \ref{res:qcnn1}, and given the kernel matrix $F^{\ell}$, the input matrices $A^{\ell}$ and $Y^{\ell}$, stored in the QRAM for each layer $\ell$, and a loss function $\mathcal{L}$, there is a quantum backpropagation algorithm that estimates each element of the gradient tensor $\frac{\partial \mathcal{L}}{\partial F^{\ell}}$ within additive error $\delta \norm{ \frac{\partial \mathcal{L}}{\partial F^{\ell}} } $, and updates them to perform a gradient descent. The running time of a single layer $\ell$ for quantum backpropagation is given by
\begin{equation}
O\left(\left(\left(\mu(A^{\ell})+\mu(\frac{\partial \mathcal{L}}{\partial Y^{\ell+1}})\right)\kappa(\frac{\partial \mathcal{L}}{\partial F^{\ell}})+\left(\mu(\frac{\partial \mathcal{L}}{\partial Y^{\ell+1}})+\mu(F^{\ell})\right)\kappa(\frac{\partial \mathcal{L}}{\partial Y^{\ell}})\right) \frac{\log(1/\delta)}{\delta^{2}}\right)
\end{equation}
where for a matrix $V \in \R^{n\times n}$, $\kappa(V)$ is the condition number and $\mu(V)\leq \sqrt{n}$ is a matrix dependent parameter defined in Definition \ref{def:mu}.
\end{result}

Details concerning the tensors and their matrix expansion or reshaping are given in Section \ref{tensors}, and a summary of all variables with their meaning and dimension is given in Section \ref{variable_summary_section}. Note that $X^{\ell}$, $Y^{\ell}$ and $A^{\ell}$ are different forms of the same input. Similarly $K^{\ell}$ and $F^{\ell}$ both refer to the kernels.

For the quantum backpropagation algorithm, we introduce a quantum tomography algorithm with $\ell_{\infty}$ norm guarantees, that could be of independent interest. It is exponentially faster than the tomography with $\ell_2$ norm guarantees and is given as Theorem \ref{thm:tomography_l2}. Numerical simulations on classifying the MNIST dataset show that our quantum CNN achieves a similar classification accuracy as the classical CNN. 

The rest of the chapter is organized as follows: we explain our quantum algorithm in two parts: the forward quantum convolution layer (Section \ref{QCNNForward}) and the quantum backpropagation (Section \ref{quantumalgobackprop}). The final part presents the results of our numerical simulations (Section \ref{NumericalSimulations}) and our conclusions (Section \ref{conclusions_qcnn}). A summary of the variables is given as Section \ref{variable_summary_section}, and the two algorithms for the forward and backpropagation phase of the QCNN are given as Algorithm \ref{QCNNLayer} and Algorithm \ref{QBackpropagation}.\\

\section{Quantum Algorithm}

\subsection{Quantum Feedforward Algorithm}\label{QCNNForward}

%\section{Forward pass for the QCNN}\label{QCNNForward}
 
In this section we will design quantum procedures for the usual operations in a CNN layer. We start by describing the main ideas before providing the details. The forward pass algorithm for the QCNN is given as Algorithm \ref{QCNNLayer}. 

First, to perform a convolution product between an input and a kernel, we use the mapping between convolution of tensors and matrix multiplication from Section \ref{tensors}, which can further be reduced to inner product estimation between vectors. The output will be a quantum state representing the result of the convolution product, from which we can sample to retrieve classical information to feed the next layer. This is stated in the following Theorem:

\begin{theorem}{Quantum Convolution Layer}{convolution_layer}
Given 3D tensor input $X^{\ell} \in \mathbb{R}^{H^{\ell}\times W^{\ell}\times D^{\ell}}$ and 4D tensor kernel $K^{\ell} \in \mathbb{R}^{H\times W\times D^{\ell}\times D^{\ell+1}}$ stored in QRAM, and precision parameters $\epsilon, \Delta >0$, there is a quantum algorithm that computes a quantum states $\Delta$-close to $\ket{f(\overline{X}^{\ell+1})}$ where $X^{\ell+1} = X^{\ell}*K^{\ell}$ and $f : \mathbb{R} \mapsto [0,C]$ is a non-linear function. 
A classical approximation such that 
\begin{equation}
\norm{f(\overline{X}^{\ell+1}) - f(X^{\ell+1})}_{\infty} \leq \epsilon
\end{equation}
The time complexity of this procedure is given by
$\widetilde{O}\left( M/\epsilon\right)$, where $M$ is the maximum norm of a product between one of the $D^{\ell+1}$ kernels, and one of the regions of $X^{\ell}$ of size $HWD^{\ell}$ 
and $\widetilde{O}$ hides factors polylogarithmic in $\Delta$ and in the size of $X^{\ell}$ and $K^{\ell}$.
\end{theorem} 
 Recall that a convolution product can be seen as a pattern detection on the input image, where the pattern is the kernel. The output values correspond to ``how much" the pattern was present in the corresponding region of the input. Low value pixels in the output indicate the absence of the pattern in the input at the corresponding regions. Therefore, by sampling according to these output values, where the high value pixels are sampled with more probability, we could retrieve less but only meaningful information for the neural network to learn. It is a singular use case where amplitudes of a quantum state are proportional to the importance of the information they carry, giving a new utility to the probabilistic nature of quantum sampling. Numerical simulations in Section \ref{NumericalSimulations} provide an empirical estimate of the 
 sampling rate to achieve good classification accuracy.

\begin{algorithm} 
\caption{QCNN Layer} \label{QCNNLayer}
\begin{algorithmic}[1]

\REQUIRE  Data input matrix $A^{\ell}$ and kernel matrix $F^{\ell}$ stored in QRAM. Precision parameters $\epsilon$ and $\nu$, a non-linearity function $f : \mathbb{R} \mapsto [0,C]$. 
\ENSURE Outputs the data matrix $A^{\ell+1}$ for the next layer, result of the convolution between the input and the kernel, followed by a non-linearity and pooling.\\
\vspace{10pt} 

\STATE {\bf Step 1: Quantum Convolution}\\

{\bf 1.1: Inner product estimation}\\
Perform the following mapping, using QRAM queries on rows $A^{\ell}_{p}$ and columns $F^{\ell}_{q}$, along with Theorems \ref{thm:amplitude_amplification} and \ref{thm:median_evaluation} to obtain
\begin{equation}
\frac{1}{K} \sum_{p,q} \ket{p}\ket{q}
\mapsto 
\frac{1}{K} \sum_{p,q} \ket{p}\ket{q}\ket{\overline{P}_{pq}}\ket{g_{p q}}, 
\end{equation}
where $\overline{P}_{p q}$ is $\epsilon$-close to $P_{p q} = \frac{1+\braket{A^{\ell}_{p}}{F^{\ell}_{q}}}{2}$ and $K = \sqrt{H^{\ell+1}W^{\ell+1}D^{\ell+1}}$ is a normalisation factor. $\ket{g_{p q}}$ is some garbage quantum state. 

{\bf 1.2: Non-linearity}\\
Use an arithmetic circuit and two QRAM queries to obtain $\overline{Y}^{\ell+1}$, an $\epsilon$-approximation of the convolution output $Y^{\ell+1}_{p, q} = (A^{\ell}_{p},F^{\ell}_{q})$ and apply the non-linear function $f$ as a boolean circuit to obatin 
\begin{equation}
\frac{1}{K} \sum_{p,q} \ket{p}\ket{q}\ket{f(\overline{Y}^{\ell+1}_{p, q})}\ket{g_{p q}}. 
\end{equation}

\STATE {\bf Step 2: Quantum Sampling} \\
Use Conditional Rotation and Amplitude Amplification to obtain the state 
\begin{equation}
\frac{1}{K} \sum_{p,q} \alpha'_{pq} \ket{p}\ket{q}\ket{f(\overline{Y}^{\ell+1}_{pq})}\ket{g_{p q}}.
\end{equation}
Perform $\ell_{\infty}$ tomography from Theorem \ref{thm:tomography_linfinity} with precision $\nu$, and obtain classically all positions and values $(p,q,f(\overline{Y}^{\ell+1}_{pq}))$ such that, with high probability, values above $\nu$ are known exactly, while others are set to 0.

\STATE {\bf Step 3: QRAM Update and Pooling} \\
Update the QRAM for the next layer $A^{\ell+1}$ while sampling. The implementation of pooling (Max, Average, etc.) can be done by a specific update or the QRAM data structure described in Section \ref{pooling}.

\end{algorithmic}
\end{algorithm}

\subsection{Single Quantum Convolution Layer}\label{singlequantumlayer}
In order to develop a quantum algorithm to perform the convolution as described above, we will make use of quantum linear algebra procedures. We will use quantum states proportional to the rows of $A^{\ell}$, denoted $\ket{A_{p}}$, and the columns of $F^{\ell}$, denoted $\ket{F_{q}}$ (we omit the $\ell$ exponent in the quantum states to simplify the notation). These states are given by
\begin{equation}
\ket{A_{p}} = \frac{1}{\norm{A_{p}}}\sum_{r=0}^{HWD^{\ell}-1}A_{pr}\ket{r}
\end{equation}
\begin{equation}
\ket{F_{q}} = \frac{1}{\norm{F_{q}}}\sum_{s=0}^{D^{\ell+1}-1}F_{sq}\ket{s}
\end{equation}

\noindent We suppose we can load these vectors in quantum states by performing the following queries:
\begin{equation}
\begin{cases}
\ket{p}\ket{0} \mapsto \ket{p}\ket{A_{p}}\\
\ket{q}\ket{0} \mapsto \ket{q}\ket{F_{q}}
\end{cases}
\end{equation}
Such queries, in time polylogarithmic in the dimension of the vector, can be implemented with a Quantum Random Access Memory (QRAM). See Section \ref{qramupdate} for more details on the QRAM update rules and its integration layer by layer.\\

\subsubsection{Inner Product Estimation}\label{innerproduct}
The following method to estimate inner products is derived from previous work \cite{qmeans}. With the initial state $\ket{p}\ket{q}\frac{1}{\sqrt{2}}(\ket{0}+\ket{1})\ket{0}$ we apply the queries detailed above in a controlled fashion, followed simply by a Hadamard gate to extract the inner product $\braket{A_{p}}{F_{q}}$ in an amplitude.  
\begin{equation}
\frac{1}{\sqrt{2}}\left(\ket{p}\ket{q}\ket{0}\ket{0}+\ket{p}\ket{q}\ket{1}\ket{0}\right) \mapsto \frac{1}{\sqrt{2}}\left(\ket{p}\ket{q}\ket{0}\ket{A_{p}}+\ket{p}\ket{q}\ket{1}\ket{F_{q}}\right)
\end{equation}
By applying a Hadamard gate on the third register we obtain the following state,
\begin{equation}
\frac{1}{2}\ket{p}\ket{q}\Big(\ket{0}(\ket{A_{p}}+\ket{F_{q}}) + \ket{1}(\ket{A_{p}}-\ket{F_{q}})\Big)
\end{equation}
The probability of measuring $0$ on the third register is given by $P_{p q} = \frac{1+\braket{A_{p}}{F_{q}}}{2}$. Thus we can rewrite the previous state as
\begin{equation}
\ket{p}\ket{q}\Big( \sqrt{P_{p q}}\ket{0, y_{p q}} + \sqrt{1-P_{p q}}\ket{1, y^{'}_{pq}} \Big)
\end{equation}
where $\ket{y_{pq}}$ and $\ket{y'_{pq}}$ are some garbage states. \\
We can perform the previous circuit in superposition. Since $A^{\ell}$ has $H^{\ell+1}W^{\ell+1}$ rows, and $F^{\ell}$ has $D^{\ell+1}$ columns, we obtain the state:
\begin{equation}
\ket{u} =  \frac{1}{\sqrt{H^{\ell+1}W^{\ell+1}D^{\ell+1}}} \sum_{p}\sum_{q} \ket{p}\ket{q}\Big( \sqrt{P_{pq}}\ket{0, y_{pq}} + \sqrt{1-P_{pq}}\ket{1, y^{'}_{pq}} \Big)
\end{equation}
Therefore the probability of measuring the triplet $(p,q,0)$ in the first three registers is given by 
\begin{equation}
P_0(p,q) = \frac{P_{pq}}{H^{\ell+1}W^{\ell+1}D^{\ell+1}} = \frac{1+\braket{A_{p}}{F_{q}}}{2H^{\ell+1}W^{\ell+1}D^{\ell+1}}
\end{equation}
Now we can relate to the Convolution product. Indeed, the triplets $(p,q,0)$ that are the most probable to be measured are the ones for which the value $\braket{A_{p}}{F_{q}}$ is the highest. Recall that each element of $Y^{\ell+1}$ is given by $Y^{\ell+1}_{pq}=(A_{p},F_{q})$, where $``(\cdot,\cdot)"$ denotes the inner product. We see here that we will sample most probably the positions $(p,q)$ for the highest values of $Y^{\ell+1}$, that corresponds to the most important points of $X^{\ell+1}$, by the Eq.(\ref{YtoX}). Note that the values of $Y^{\ell+1}$ can be either positive or negative, which is not an issue thanks to the positiveness of $P_0(p,q)$.

A first approach could be to measure indices $(p,q)$ and rely on the fact that pixels with high values, hence a high amplitude, would have a higher probability to be measured. However we have not exactly the final result, since $\braket{A_{p}}{F_{q}} \neq (A_{p},F_{q}) = \norm{A_{p}}\norm{F_{q}}\braket{A_{p}}{F_{q}}$. Most importantly we then want to apply a non-linearity $f(Y^{\ell+1}_{pq})$ to each pixel, for instance the ReLu function, which seems not possible with unitary quantum gates if the data is encoded in the amplitudes only. Moreover, due to the normalization of the quantum amplitudes and the high dimension of the Hilbert space of the input, the probability of measuring each pixel is roughly the same, making the sampling inefficient. Given these facts, we have added steps to the circuit, in order to measure $(p,q,f(Y^{\ell+1}_{pq}))$, therefore know the value of a pixel when measuring it, while still measuring the most important points in priority. 

\subsubsection{Encoding the amplitude in a register}\label{registerencoding}
Let $\mathcal{U}$ be the unitary that map $\ket{0}$ to $\ket{u}$ 
\begin{equation}
\ket{u} =  \frac{1}{\sqrt{H^{\ell+1}W^{\ell+1}D^{\ell+1}}} \sum_{p,q} \ket{p}\ket{q}\Big( \sqrt{P_{pq}}\ket{0, y_{pq}} + \sqrt{1-P_{pq}}\ket{1, y^{'}_{pq}} \Big)
\end{equation}
The amplitude $\sqrt{P_{pq}}$ can be encoded in an ancillary register by using Amplitude Estimation (Theorem \ref{thm:amplitude_amplification}) followed by a Median Evaluation (Theorem \ref{thm:median_evaluation}). 

For any $\Delta>0$ and $\epsilon>0$, we can have a state $\Delta$-close to 
\begin{equation}
\ket{u^{'}} =  \frac{1}{\sqrt{H^{\ell+1}W^{\ell+1}D^{\ell+1}}} \sum_{p,q} \ket{p}\ket{q}\ket{0}\ket{\overline{P}_{pq}}\ket{g_{p q}}
\end{equation}
with probability at least $1-2\Delta$, where $|P_{pq} - {\overline{P}_{pq}}| \leq \epsilon$ and $\ket{g_{p q}}$ is a garbage state. This requires $O(\frac{\ln(1/\Delta)}{\epsilon})$ queries of $\mathcal{U}$. In the following we discard the third register $\ket{0}$ for simplicity.

The benefit of having $\overline{P}_{pq}$ in a register is to be able to perform operations on it (arithmetic or even non-linear). Therefore we can simply obtain a state corresponding to the exact value of the convolution product. Since we've built a circuit such that  $P_{pq} = \frac{1+\braket{A_{p}}{F_{q}}}{2}$, with two QRAM calls, we can retrieve the norm of the vectors by applying the following unitary:

\begin{equation}
\ket{p}\ket{q}\ket{\overline{P}_{pq}}\ket{g_{p q}}\ket{0}\ket{0}
\mapsto 
\ket{p}\ket{q}\ket{\overline{P}_{pq}}\ket{g_{p q}}\ket{\norm{A_{p}}}\ket{\norm{F_{q}}}
\end{equation}

On the fourth register, we can then write $Y^{\ell+1}_{pq} = \norm{A_{p}}\norm{F_{q}} \braket{A_{p}}{F_{q}}$ using some arithmetic circuits (addition, multiplication by a scalar, multiplication between registers). We then apply a boolean circuit that implements the ReLu function on the same register, to obtain an estimate of $f(Y^{\ell+1}_{pq})$ in the fourth register. We finish by inverting the previous computations and obtain the final state

\begin{equation}\label{afternonlinearity}
\ket{f(\overline{Y}^{\ell+1})} = \frac{1}{\sqrt{H^{\ell+1}W^{\ell+1}D^{\ell+1}}} \sum_{p,q} \ket{p}\ket{q}\ket{f(\overline{Y}^{\ell+1}_{pq})}\ket{g_{p q}}
\end{equation}

Because of the precision $\epsilon$ on $\ket{\overline{P}_{pq}}$, our estimation $\overline{Y}^{\ell+1}_{pq} = (2\overline{P}_{pq}-1)\norm{A_{p}}\norm{F_{q}}$, is obtained with error such that
\begin{equation}\label{errorAE}
|\overline{Y}^{\ell+1}_{pq} - Y^{\ell+1}_{pq}| \leq 2\epsilon \norm{A_{p}}\norm{F_{q}}
\end{equation}

In superposition, we can bound this error by $ |\overline{Y}^{\ell+1}_{pq} - Y^{\ell+1}_{pq}| \leq 2M\epsilon$ where we define 
\begin{equation}\label{Mdefinition}
M = \max_{p,q}{\norm{A_{p}}\norm{F_{q}}}
\end{equation}
$M$ is the maximum product between norms of one of the $D^{\ell+1}$ kernels, and one of the regions of $X^{\ell}$ of size $HWD^{\ell}$. Finally, since Eq.(\ref{errorAE}) is valid for all pairs $(p,q)$, the overall error committed on the convolution product can be bounded by $\norm{\overline{Y}^{\ell+1}-Y^{\ell+1}}_{\infty} \leq 2M\epsilon$, where $\norm{.}_{\infty}$ denotes the $\ell_{\infty}$ norm. Recall that $Y^{\ell+1}$ is just a reshaped version of $X^{\ell+1}$. Since the non-linearity adds no approximation, we can conclude on the final error committed for a layer of our QCNN
\begin{equation}\label{errorAEfinal}
\norm{f(\overline{X}^{\ell+1})-f(X^{\ell+1})}_{\infty} \leq 2M\epsilon
\end{equation}

At this point, we have established Theorem \ref{thm:convolution_layer} as we have created the quantum state (\ref{afternonlinearity}), with given precision guarantees, in time polylogarithmic in $\Delta$ and in the size of $X^{\ell}$ and $K^{\ell}$.

We know aim to retrieve classical information from this quantum state. Note that $\ket{Y^{\ell+1}_{pq}}$ is representing a scalar encoded in as many qubits as needed for the precision, whereas $\ket{A_{p}}$ was representing a vector as a quantum state in superposition, where each element $A_{p,r}$ is encoded in one amplitude (See Section \ref{sec:amplitude_encoding}). The next step can be seen as a way to retrieve both encodings at the same time, which will allow an efficient tomography focus on the values of high magnitude.

\subsubsection{Conditional rotation}
In the following sections, we omit the $\ell+1$ exponent for simplicity.
Garbage states are removed as they will not perturb the final measurement. We now aim to modify the amplitudes, such that the highest values of $\ket{f(\overline{Y})}$ are measured with higher probability. As shown in Section \ref{sec:amplitude_encoding}, a way to do so consists in applying a conditional rotation on an ancillary qubit, proportionally to $f(\overline{Y}_{pq})$. We will detail the calculation since in the general case $f(\overline{Y}_{pq})$ can be greater than 1. To simplify the notation, we denote it by $x=f(\overline{Y}_{pq})$. 

This step consists of applying the following rotation on an ancillary qubit:
\begin{equation}
\ket{x}\ket{0} \mapsto \ket{x}\left(\sqrt{\frac{x}{\max{x}}}\ket{0}+\beta \ket{1}\right)
\end{equation}
Where $\max{x} = \max_{p,q}{f(\overline{Y}_{pq})}$ and $\beta = \sqrt{1-(\frac{x}{\max{x}})^2}$. Note that in practice it is not possible to have access to $\ket{\max{x}}$ from the state (\ref{afternonlinearity}), but we will present a method to know \emph{a priori} this value or an upper bound in section \ref{capReLu}.

Let's denote $\alpha_{pq} = \sqrt{\frac{f(\overline{Y}_{pq})}{\\max_{p,q}(f(\overline{Y}_{pq}))}}$. The ouput of this conditional rotation in superposition on state (\ref{afternonlinearity}) is then 
\begin{equation}
\frac{1}{\sqrt{HWD}} \sum_{p,q} \ket{p}\ket{q}\ket{f(\overline{Y}_{pq})}(\alpha_{pq}\ket{0}+\sqrt{1-\alpha_{pq}^2}\ket{1})
\end{equation}

\subsubsection{Amplitude Amplification}\label{amplitudeamplification}
In order to measure $(p,q,f(\overline{Y}_{pq}))$ with higher probability where $f(\overline{Y}_{pq})$ has high value, we could post select on the measurement of $\ket{0}$ on the last register. Otherwise, we can perform an amplitude amplification on this ancillary qubit. Let's rewrite the previous state as
\begin{equation}
\frac{1}{\sqrt{HWD}} \sum_{p,q} \alpha_{pq}\ket{p}\ket{q}\ket{f(\overline{Y}_{pq})}\ket{0}+\sqrt{1-\alpha_{pq}^2}\ket{g'_{p q}}\ket{1}
\end{equation}
Where $\ket{g'_{p q}}$ is another garbage state. The overall probability of measuring $\ket{0}$ on the last register is $P(0) = \frac{1}{HWD}\sum_{pq}|\alpha_{pq}|^2$. The number of queries required to amplify the state $\ket{0}$ is $O(\frac{1}{\sqrt{P(0)}})$ (Theorem \ref{thm:amplitude_amplification}). Since $f(\overline{Y}_{pq}) \in \mathbb{R}^{+}$, we have $\alpha_{pq}^2 = \frac{f(\overline{Y}_{pq})}{\max_{p,q}(f(\overline{Y}_{pq}))}$. Therefore the number of queries is 
\begin{equation}
O\left(\sqrt{\max_{p,q}(f(\overline{Y}_{pq}))}\frac{1}{\sqrt{\frac{1}{HWD}\sum_{p,q}f(\overline{Y}_{pq})}}\right) = 
O\left(\frac{\sqrt{\max_{p,q}(f(\overline{Y}_{pq}))}}{\sqrt{\mathbb{E}_{p,q}(f(\overline{Y}_{pq}))}}\right)
\end{equation}
Where the notation $\mathbb{E}_{p,q}(f(\overline{Y}_{pq}))$ represents the average value of the matrix $f(\overline{Y})$. It can also be written $\mathbb{E}(f(\overline{X}))$ as in Result \ref{res:qcnn1}:
\begin{equation}\label{averagevalue}
\mathbb{E}_{p,q}(f(\overline{Y}_{pq})) = \frac{1}{HWD}\sum_{p,q}f(\overline{Y}_{pq})
\end{equation}
At the end of these iterations, we have modified the state to the following: 

\begin{equation}\label{afteramplitudeencoding}
\ket{f(\overline{Y})} = \frac{1}{\sqrt{HWD}} \sum_{p,q} \alpha'_{pq}\ket{p}\ket{q}\ket{f(\overline{Y}_{pq})}
\end{equation}

Where, to respect the normalization of the quantum state, $\alpha'_{pq} = \frac{\alpha_{pq}}{\sqrt{\sum_{p,q}\frac{\alpha^2_{pq}}{HWD}}}$. Eventually, the probability of measuring $(p,q,f(\overline{Y}_{pq}))$ is given by
\begin{equation}
p(p,q,f(\overline{Y}_{pq})) = \frac{(\alpha'_{pq})^2}{HWD} = \frac{(\alpha_{pq})^2}{\sum_{p,q}(\alpha_{pq})^2}=\frac{f(\overline{Y}_{pq})}{\sum_{p,q}{f(\overline{Y}_{pq})}}
\end{equation}

Note that we have used the same type of name $\ket{f(\overline{Y})}$ for both state (\ref{afternonlinearity}) and state (\ref{afteramplitudeencoding}). For now on, this state name will refer only to the latter (\ref{afteramplitudeencoding}). \\

\subsubsection{$\ell_{\infty}$ tomography and probabilistic sampling}\label{tomographyconvolution}
We can rewrite the final quantum state obtained in (\ref{afteramplitudeencoding}) as 
\begin{equation}\label{finalstate}
\ket{f(\overline{Y}^{\ell+1})} = \frac{1}{\sqrt{\sum_{p,q}{f(\overline{Y}^{\ell+1}_{pq})}}} \sum_{p,q} \sqrt{f(\overline{Y}^{\ell+1}_{pq})} \ket{p}\ket{q}\ket{f(\overline{Y}^{\ell+1}_{pq})}
\end{equation}

We see here that $f(\overline{Y}^{\ell+1}_{pq})$, the values of each pixel, are encoded in both the last register and in the amplitude. We will use this property to extract efficiently the exact values of high magnitude pixels. 
For simplicity, we will use instead the notation $f(\overline{X}^{\ell+1}_{n})$ to denote a pixel's value, with $n \in [H^{\ell+1}W^{\ell+1}D^{\ell+1}]$. Recall that $Y^{\ell+1}$ and $X^{\ell+1}$ are reshaped version of the same object.  

The pixels with high values will have more probability of being sampled. Specifically, we perform a tomography with $\ell_{\infty}$ guarantee and precision parameter $\nu > 0$. See Theorem \ref{thm:tomography_linfinity} and Section \ref{sec:l_2_and_l_infinite_tomography} for details. The $\ell_{\infty}$ guarantee allows to obtain each pixel with error at most $\nu$, and require $\widetilde{O}(1/\nu^2)$ samples from the state (\ref{finalstate}). 
Pixels with low values $f(\overline{X}^{\ell+1}_{n}) < \nu$ will probably not be sampled due to their low amplitude. Therefore the error committed will be significative and we adopt the rule of setting them to 0.
Pixels with higher values $f(\overline{X}^{\ell+1}_{n}) \geq \nu$, will be sample with high probability, and only one appearance is enough to get the exact register value $f(\overline{X}^{\ell+1}_{n})$ of the pixel, as is it also written in the last register. 

To conclude, let's denote $\mathcal{X}^{\ell+1}_{n}$ the resulting pixel values after the tomography, and compare it to the real classical outputs $f(X^{\ell+1}_{n})$. Recall that the measured values $f(\overline{X}^{\ell+1}_{n})$ are approximated with error at most $2M\epsilon$ with $M = \max_{p,q}{\norm{A_{p}}\norm{F_{q}}}$. The algorithm described above implements the following rules:

\begin{equation}
	\begin{cases}
	|\mathcal{X}^{\ell+1}_{n} - f(X^{\ell+1}_n)| \leq 2M\epsilon \quad \text{if} \quad f(\overline{X}^{\ell+1}_{n}) \geq \nu\\
	\mathcal{X}^{\ell+1}_{n} = 0 \qquad\qquad\qquad\qquad \text{if} \quad f(\overline{X}^{\ell+1}_{n}) < \nu\\
	\end{cases}
\end{equation}

Concerning the running time, one could ask what values of $\nu$ are sufficient to obtain enough meaningful pixels. Certainly, this highly depends on the output's size $H^{\ell +1}W^{\ell +1}D^{\ell+1}$ and on the output's content itself. But we can view this question from another perspective, by considering that we sample a constant fraction of pixels given by $\sigma \cdot (H^{\ell +1}W^{\ell +1}D^{\ell+1})$ where $\sigma \in [0,1]$ is a sampling ratio. Because of the particular amplitudes of state (\ref{finalstate}), the high value pixels will be measured and known with higher probability. The points that are not sampled are being set to 0. We see that this approach is equivalent to the $\ell_{\infty}$ tomography, therefore we have 
\begin{equation}\label{simga_eta_equivalence}
\frac{1}{\nu^2} = \sigma \cdot H^{\ell +1}W^{\ell +1}D^{\ell+1}
\end{equation}
\begin{equation}\label{def:nu_qcnn}
\nu = \frac{1}{\sqrt{\sigma \cdot H^{\ell +1}W^{\ell +1}D^{\ell+1}}}
\end{equation}

We will use this analogy in the numerical simulations (Section \ref{NumericalSimulations}) to estimate, for a particular QCNN architecture and a particular dataset of images, which values of $\sigma$ are enough to allow the neural network to learn.

\subsubsection{Regularization of the Non Linearity}\label{capReLu}
In the previous steps, we see several appearances of the parameter $\max_{p,q}(f(\overline{Y}^{\ell+1}_{pq}))$.  First, for the conditional rotation preprocessing, we need to know this value or an upper bound. Then for the running time, we would like to bound this parameter. Both problems can be solved by replacing the usual ReLu non-linearity with a particular activation function, that we denote by $capReLu$. This function is simply a parametrized ReLu function with an upper threshold, the cap $C$, after which the function remains constant. The choice of $C$ will be tuned for each particular QCNN, as a tradeoff between accuracy and speed. Otherwise, the only other requirement of the QCNN activation function would be not to allow negative values. This is already often the case for most of the classical CNN. In practice, we expect the capReLu to be as good as a usual ReLu, for convenient values of the cap $C$ ($\leq10$). We performed numerical simulations to compare the learning curve of the same CNN with several values of $C$. See the numerical experiments presented in Section \ref{NumericalSimulations} for more details.

\begin{figure}[h]
\minipage{0.5\textwidth}
	\centering
	\includegraphics[width=60mm] {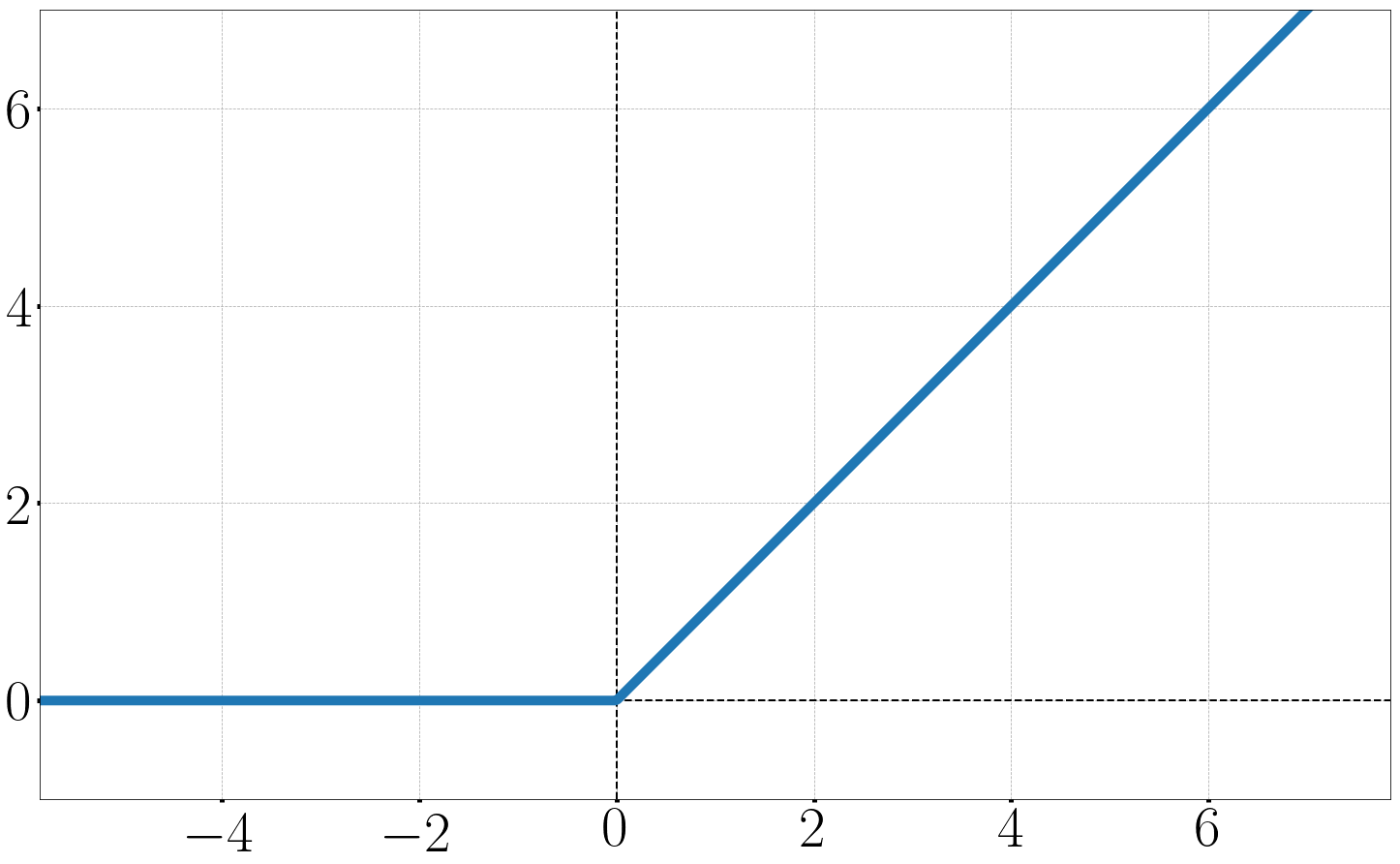} 
\endminipage\hfill
\minipage{0.5\textwidth}
	\centering
	\includegraphics[width=60mm] {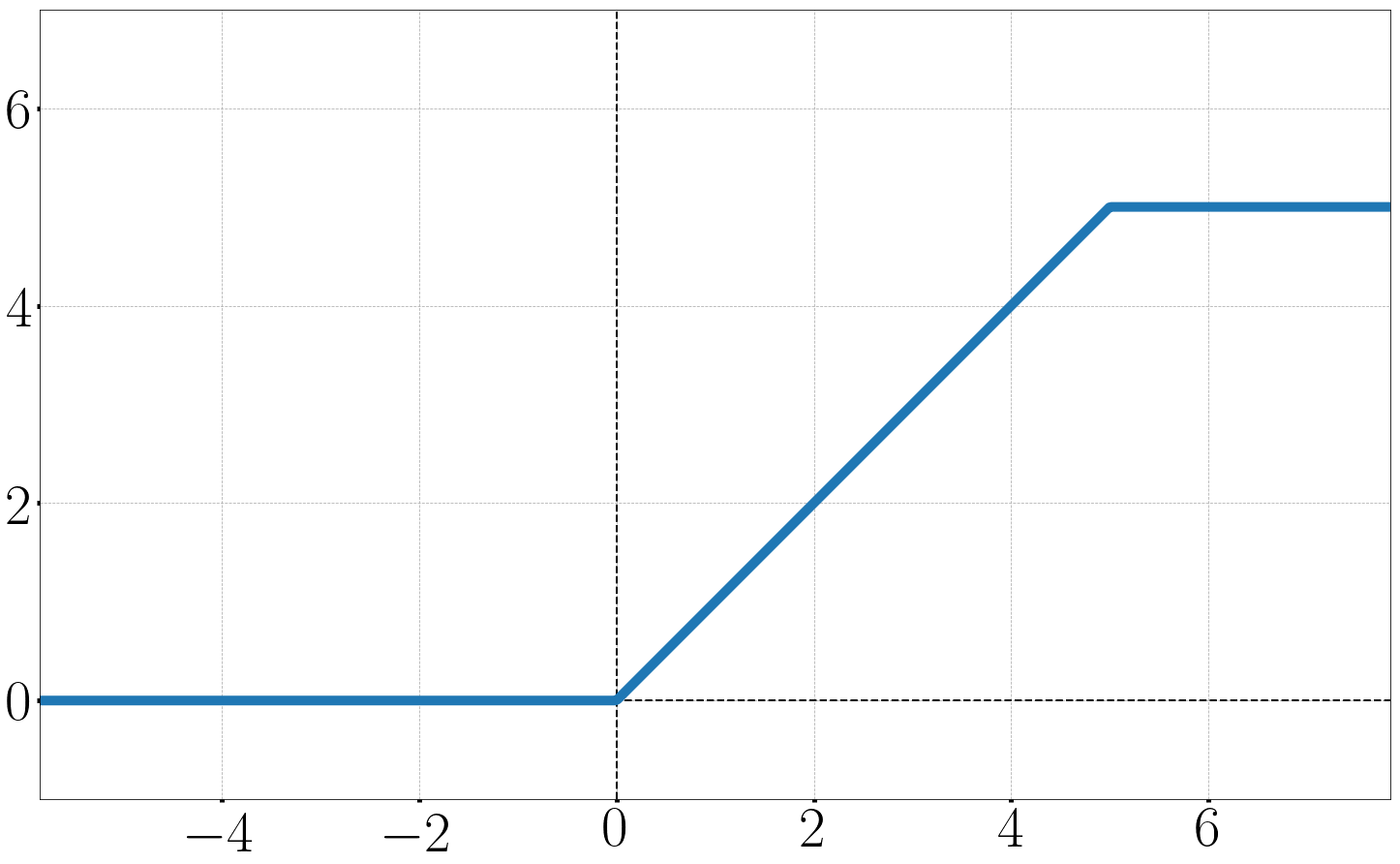} 
\endminipage
\captionsetup{justification=raggedright, margin=1cm}
\caption{Activation functions: ReLu (Left) and capReLu (Right) with a cap $C$ at 5.}
\label{capReLu_5}
\end{figure} 

\subsection{Quantum Memory Update}\label{qramupdate}

We wish to detail the use of the QRAM between each quantum convolution layer and present how the pooling operation can happen during this phase. General results about the QRAM are given as Theorem \ref{thm:QRAM}. Implementation details can be found Section \ref{sec:quantum_memory_models}. In this section, we will show how to store samples from the output of the layer $\ell$, to create the input of layer $\ell+1$.

\subsubsection{Storing the output values during the sampling}

At the beginning of layer $\ell+1$, the QRAM must store $A^{\ell+1}$, a matrix where each elements is indexed by $(p',r')$, and perform $\ket{p'}\ket{0} \mapsto \ket{p'}\ket{A^{\ell+1}_{p'}}$. The data is stored in the QRAM as a tree structure described in Fig.\ref{fig:QRAM_tree}. Each row $A^{\ell+1}_{p'}$ is stored in such a tree $T^{\ell+1}_{p'}$. Each leaf $A^{\ell+1}_{p'r'}$ correspond to a value sampled from the previous quantum state $\ket{f(\overline{Y}^{\ell+1})}$, output of the layer $\ell$. The question is to know where to store a sample from $\ket{f(\overline{Y}^{\ell+1})}$ in the tree $T^{\ell+1}_{p'}$.

When a point is sampled from the final state of the quantum convolution, at layer $\ell$, as described in Section \ref{amplitudeamplification}, we obtain a triplet corresponding to the two positions and the value of a point in the matrix $f(\overline{Y}^{\ell+1})$. 
We can know where this point belongs in the input of layer $\ell+1$, the tensor $X^{\ell+1}$, by Eq.(\ref{YtoX}), since $Y^{\ell}$ is a reshaped version of $X^{\ell}$.

The position in $X^{\ell+1}$, denoted $(i^{\ell+1},j^{\ell+1},d^{\ell+1})$, is then matched to several positions $(p',r')$ in $A^{\ell+1}$. For each $p'$, we write in the tree $T^{\ell+1}_{p'}$ the sampled value at leaf $r'$ and update its parent nodes. Note that leaves that weren't updated will be considered as zeros, corresponding to pixels with too low values, or not selected during pooling (see next section). 

Having stored pixels in this way, we can then query $\ket{p'}\ket{0} \mapsto \ket{p'}\ket{A^{\ell}_{p'}}$, using Theorem \ref{thm:QRAM}, where we correctly have $\ket{A^{\ell+1}_{p'}} = \frac{1}{\norm{A^{\ell+1}_{p'}}}\sum_{r'}A^{\ell+1}_{p'r'}\ket{r'}$. Note that each tree has a logarithmic depth in the number of leaves, hence the running time of writing the output of the quantum convolution layer in the QRAM gives a marginal multiplicative increase, polylogarithmic in the number of points sampled from $\ket{f(\overline{Y}^{\ell+1})}$, namely $O(\log(1/\nu^2))$.

\subsubsection{Quantum Pooling}\label{pooling}
As for the classical CNN, a QCNN should be able to perform pooling operations. We first detail the notations for classical pooling. At the end of layer $\ell$, we wish to apply a pooling operation of size P on the output $f(X^{\ell+1})$. We denote by $\tilde{X}^{\ell+1}$ the tensor after the pooling operation. For a point in $f(X^{\ell+1})$ at position $(i^{\ell+1},j^{\ell+1},d^{\ell+1})$, we know to which \emph{pooling region} it belongs, corresponding to a position $(\tilde{i}^{\ell+1},\tilde{j}^{\ell+1},\tilde{d}^{\ell+1})$ in $\tilde{X}^{\ell+1}$:

\begin{equation}
\begin{cases}
    \tilde{d}^{\ell+1} = d^{\ell+1} \\
    \tilde{j}^{\ell+1} = \floor{\frac{j^{\ell+1}}{P}}\\
    \tilde{i}^{\ell+1} = \floor{\frac{i^{\ell+1}}{P}}
\end{cases}
\end{equation}

\begin{figure}[h]
\centering
\includegraphics[scale=0.8]{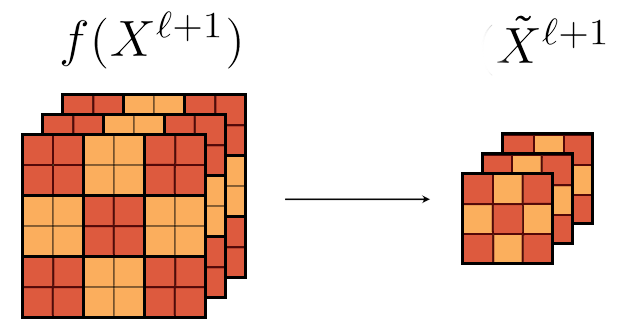} 
\captionsetup{justification=raggedright, margin=1cm}
\caption{A $2\times 2$ tensor pooling. A point in $f(X^{\ell+1})$ (left) is given by its position $(i^{\ell+1},j^{\ell+1},d^{\ell+1})$. A point in $\tilde{X}^{\ell+1}$ (right) is given by its position $(\tilde{i}^{\ell+1},\tilde{j}^{\ell+1},\tilde{d}^{\ell+1})$. Different \emph{pooling regions} in $f(X^{\ell+1})$ have separate colours, and each one corresponds to a unique point in $\tilde{X}^{\ell+1}$.}\label{max-pool}
\end{figure}

We now show how any kind of pooling can be efficiently integrated into our QCNN structure. Indeed the pooling operation will occur during the QRAM update described above, at the end of a convolution layer. At this moment we will store sampled values according to the pooling rules. 

In the quantum setting, the output of layer $\ell$ after tomography is denoted by $\mathcal{X}^{\ell+1}$. After pooling, we will describe it by $\mathcal{\tilde{X}}^{\ell+1}$, which has dimensions $\frac{H^{\ell+1}}{P} \times \frac{W^{\ell+1}}{P} \times D^{\ell+1}$. $\mathcal{\tilde{X}}^{\ell+1}$ will be effectively used as input for layer $\ell+1$ and its values should be stored in the QRAM to form the trees $\tilde{T}^{\ell+1}_{p'}$, related to the matrix expansion $\tilde{A}^{\ell+1}$. 

However $\mathcal{X}^{\ell+1}$ is not known before the tomography is over. Therefore we have to modify the update rule of the QRAM to implement the pooling in an online fashion, each time a sample from $\ket{f(\overline{X}^{\ell+1})}$ is drawn. Since several sampled values of $\ket{f(\overline{X}^{\ell+1})}$ can correspond to the same leaf $\tilde{A}^{\ell+1}_{p'r'}$ (points in the same \emph{pooling region}), we need an overwrite rule, that will depend on the type of pooling. In the case of Maximum Pooling, we simply update the leaf and the parent nodes if the new sampled value is higher than the one already written. In the case of Average Polling, we replace the actual value with the new averaged value. 

In the end, any pooling can be included in the already existing QRAM update. In the worst case, the running time is increased by $\widetilde{O}(P/\nu^2)$, an overhead corresponding to the number of times we need to overwrite existing leaves, with $P$ being a small constant in most cases. 

As we will see in Section \ref{quantumalgobackprop}, the final positions $(p, q)$ that were sampled from $\ket{f(\overline{X}^{\ell+1})}$ and selected after pooling must be stored for further use during the backpropagation phase.

\subsubsection{Running Time}\label{runningtime}
We will now summarize the running time for one forward pass of convolution layer $\ell$. With $\tilde{O}$ we hide the polylogarithmic factors. We first write the running time of the classical CNN layer, which is given by
\begin{equation}
\widetilde{O}\left(H^{\ell +1}W^{\ell +1}D^{\ell+1} \cdot HWD^{\ell}\right)
\end{equation}

\noindent For the QCNN, the previous steps prove Result \ref{res:qcnn1} and can be implemented in time
\begin{equation}
\widetilde{O}\left(\frac{1}{\epsilon \nu^2}\cdot \frac{M\sqrt{C}}{\sqrt{\mathbb{E}(f(\overline{X}^{\ell+1}))}} \right)
\end{equation}

\noindent Note that, as explain in Section \ref{tomographyconvolution}, the quantum running time can also be written 
\begin{equation}\label{final_runtime_2}
\widetilde{O}\left(\sigma H^{\ell +1}W^{\ell +1}D^{\ell+1} \cdot \frac{M\sqrt{C}}{\epsilon \sqrt{\mathbb{E}(f(\overline{X}^{\ell+1}))}} \right)
\end{equation}
 with $\sigma \in [0,1]$ being the fraction of sampled elements among $H^{\ell +1}W^{\ell +1}D^{\ell+1}$ of them.

It is interesting to notice that the one quantum convolution layer can also include the ReLu operation and the Pooling operation in the same circuit, for no significant increase in the running time, whereas in the classical CNN each operation must be done on the whole data again. 

Let's go through all the important parameters that appear in the quantum running time:

- The error $\epsilon$ committed during the inner product estimation, is an empirical parameter
Our simulations tend to show that this error can be high without compromising the learning. Indeed the introduction of noise is sometimes interesting in machine learning applications, providing more robust learning \cite{goodfellow2016deep, bishop1995training}.

- The parameter $M = \max_{p,q}{\norm{A_{p}}\norm{F_{q}}}$ as a worst case upper bound during inner product estimation.

- Precision parameter $\nu$ can be related to the fraction of sampled elements in the quantum output $\ket{f(\overline{X}^{\ell+1})}$ of the convolution layer, during $\ell_{\infty}$ tomography.  

- Amplitude amplification adds a multiplicative term $\sqrt{\max{(f(\overline{X}^{\ell+1}))}}$ to the running time, replaced here by $\sqrt{C}$, a constant parameter of order $O(1)$, corresponding to the \emph{cap}, or upper bound, of the activation function. See Section \ref{capReLu} for details. This parameter appears at the conditional rotation step.

- Similarly, the data related value $\mathbb{E}(f(\overline{X}^{\ell+1}))$, appearing during amplitude amplification, denotes the average value in the tensor $f(\overline{X}^{\ell+1})$, as defined in Eq.(\ref{averagevalue}).

Finally, in most cases, to recognize kernel features in the input tensor, the size $H \times W$ of the kernels is a sufficient constant fraction of the input size $H^{\ell} \times W^{\ell}$. Since $H^{\ell+1} = H^{\ell} - H  + 1$, the classical running time can be seen as quadratic in the input size, whereas the quantum algorithm is almost linear.

\subsubsection{Variable Summary}\label{variable_summary_section}

We recall the most important variables for layer $\ell$. They represent tensors, their approximations, and their reshaped versions.

\begin{table}[H]
\centering
\begin{tabular}{|c|c|c|c|}
\hline
Data                                & Variable             & Dimensions                                                           & Indices                                              \\ \hline
\multirow{3}{*}{Input}       & $X^{\ell}$           & $H^{\ell}\times W^{\ell}\times D^{\ell}$                             & $(i^{\ell},j^{\ell},d^{\ell})$                       \\ \cline{2-4} 
                                       & $Y^{\ell}$           & $(H^{\ell}W^{\ell})\times D^{\ell}$                                  & -                                                    \\ \cline{2-4} 
                                       & $A^{\ell}$           & $(H^{\ell+1}W^{\ell+1})\times(HWD^{\ell})$                           & $(p,r)$                                              \\ \hline
\multirow{2}{*}{Kernel}    & $K^{\ell}$           & $H\times W \times D^{\ell}\times D^{\ell+1}$                         & $(i,j,d,d')$                                         \\ \cline{2-4} 
                                       & $F^{\ell}$           & $(HWD^{\ell})\times D^{\ell+1}$                                      & $(s,q)$                                             \\ \hline
\end{tabular}
\caption{Summary of input variables for the $\ell^{th}$ layer, along with their meaning, dimensions and corresponding notations. These variables are common for both \emph{quantum} and \emph{classical} algorithms. We have omitted indices for $Y^{\ell}$ which don't appear in our work.\label{variable_summary_1}}
\end{table}

\begin{table}[H]
\centering
\begin{tabular}{|c|c|c|c|}
\hline
Data                                   & Variable             & Dimensions                                                           & Indices                                              \\ \hline

\multirow{2}{*}{Output of Quantum Convolution} 
				     & $f(\overline{Y}^{\ell+1})$         & $(H^{\ell+1}W^{\ell+1})\times D^{\ell+1}$                            & $(p,q)$                               \\ \cline{2-4} 
				     & $f(\overline{X}^{\ell+1})$         & $H^{\ell+1}\times W^{\ell+1}\times D^{\ell+1}$                    & $(i^{\ell+1},j^{\ell+1},d^{\ell+1})$                    \\ \hline

Output of Quantum Tomography   
				     & $\mathcal{X}^{\ell+1}$ & $H^{\ell+1}\times W^{\ell+1}\times D^{\ell+1}$ & 				$(i^{\ell+1},j^{\ell+1},d^{\ell+1})$        		     \\ \hline

Output of Quantum Pooling           & $\tilde{\mathcal{X}}^{\ell+1}$ & $\frac{H^{\ell+1}}{P} \times \frac{W^{\ell+1}}{P} \times D^{\ell+1}$ &   $(\tilde{i}^{\ell+1},\tilde{j}^{\ell+1},\tilde{d}^{\ell+1})$ 		      \\ \hline

\end{tabular}
\caption{Summary of variables describing outputs of the layer $\ell$, with the \emph{quantum} algorithm.\label{variable_summary_2}}
\end{table}

\begin{table}[H]
\centering
\begin{tabular}{|c|c|c|c|}
\hline
Data                                   & Variable             & Dimensions                                                           & Indices                                              \\ \hline

\multirow{2}{*}{Output of Classical Convolution} 

				     & $f(Y^{\ell+1})$         & $(H^{\ell+1}W^{\ell+1})\times D^{\ell+1}$                            		& $(p,q)$         	                      \\ \cline{2-4} 
				     & $f(X^{\ell+1})$         & $H^{\ell+1}\times W^{\ell+1}\times D^{\ell+1}$                       		& $(i^{\ell+1},j^{\ell+1},d^{\ell+1})$         	     \\ \hline

Output of Classical Pooling           & $\tilde{X}^{\ell+1}$ & $\frac{H^{\ell+1}}{P} \times \frac{W^{\ell+1}}{P} \times D^{\ell+1}$ &   $(\tilde{i}^{\ell+1},\tilde{j}^{\ell+1},\tilde{d}^{\ell+1})$     \\ \hline

\end{tabular}
\caption{Summary of variables describing outputs of the layer $\ell$, with the \emph{classical} algorithm.\label{variable_summary_3}} 
\end{table}

Classical and quantum algorithms can be compared with these two diagrams:

\begin{equation}
\begin{cases}

\text{Quantum convolution layer : }
X^{\ell} \rightarrow 
\ket{\overline{X}^{\ell+1}} \rightarrow 
\ket{f(\overline{X}^{\ell+1})} \rightarrow 
\mathcal{X}^{\ell+1} \rightarrow 
\tilde{\mathcal{X}}^{\ell+1}\\

\text{Classical convolution layer : }
X^{\ell} \rightarrow 
X^{\ell+1} \rightarrow 
f(X^{\ell+1}) \rightarrow 
\tilde{X}^{\ell+1}

\end{cases}
\end{equation} 

~\\
We finally provide some remarks that could clarify some notations ambiguity:

- Formally, the output of the quantum algorithm is $\tilde{\mathcal{X}}^{\ell+1}$. It is used as input for the next layer $\ell+1$. But we consider that all variables' names are \emph{reset} when starting a new layer: $X^{\ell+1} \leftarrow \tilde{\mathcal{X}}^{\ell+1}$.

- For simplicity, we have sometimes replaced the indices $(i^{\ell+1},j^{\ell+1},d^{\ell+1})$ by $n$ to index the elements of the output.

- In Section \ref{pooling}, the input for layer $\ell+1$ is stored as $A^{\ell+1}$, for which the elements are indexed by $(p',r')$.

\begin{figure}[h]
    \centering
    \includegraphics[width=\textwidth]{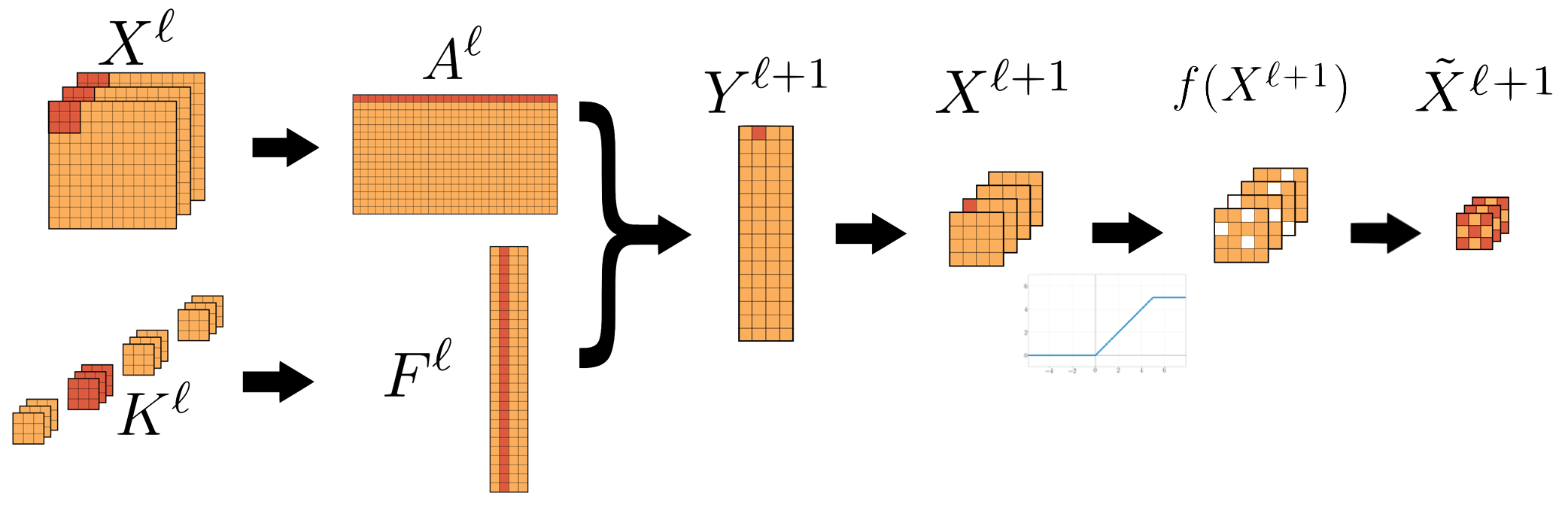}
    \caption{Variable summary and representation of a full QCNN layer. From left to right: matrix expansion for input and kernel, parallel matrix product, non-linearity, tomography and pooling.}
    \label{fig:qcnn_full_layer}
\end{figure}

\subsubsection{Quantum-Inspired Classical Algorithm}

In Section \ref{sec:quantum_inspired_algo}, we detailed recent works that have provided quantum inspired classical algorithms for linear algebra that rely on $\ell_2$ sampling using classical analogs of the binary search tree (BST) data structure to efficiently estimate inner products. Indeed the inner product can be efficiently approximated classically, analogous to quantum inner product estimation. As shown in \cite{allcock2020quantum}, if $x,y \in \R^{n}$ are stored in $\ell_2$-BST then, with probability at least $1 - \Delta$, a value $s$ that approximates the inner product $\braket{x}{y}$ can be computed with the following guarantees, 
\begin{equation}
|s - \braket{x}{y}| \leq 
\begin{cases}
\epsilon \qquad\qquad \text{in time} \quad  \widetilde{O}\left( \frac{\log(1/\Delta)}{\epsilon^2}\frac{\norm{x}^2\norm{y}^2}{|\braket{x}{y}|} \right) \\
\epsilon |\braket{x}{y}| \quad \text{in time} \quad  \widetilde{O}\left( \frac{\log(1/\Delta)}{\epsilon^2}\norm{x}^2\norm{y}^2 \right)
\end{cases}
\end{equation}

The running time is similar to the quantum inner product estimation presented in Section \ref{innerproduct}, but presents a quadratic overhead on the precision $\epsilon$ and the norms of the vectors $x$ and $y$, which in our case would be $A^{\ell}_{p}$ and $F^{\ell}_{q}$, input and kernel vectors. Similarly, the steps corresponding to the amplitude amplification of Section \ref{amplitudeamplification} can be done classically with a quadratically worse dependence on the parameters. 

It is therefore possible to define a classical algorithm inspired by this work if the matrices $A^{\ell}$ and $F^{\ell}$ are stored in classical $\ell_2$ BST. Using the above result, and applying classically non-linearity and pooling, would give a forward pass algorithm with running time,  

\begin{equation}
\widetilde{O}\left(H^{\ell +1}W^{\ell +1}D^{\ell+1} \cdot \frac{M^2 C}{\epsilon^2 \mathbb{E}(f(\overline{X}^{\ell+1}))} \right).
\end{equation}

Similar to the quantum algorithm's running time (\ref{final_runtime_2}), we obtain a polylogarithmic dependence on the size of the kernels. We however see a quadratically worse dependence with respect to $\epsilon$, $M = \max_{p,q}{\norm{A_{p}}\norm{F_{q}}}$, $C$ the upper bound of the non-linearity, and the average value of $f(\overline{X}^{\ell+1})$, too. Recent numerical experiments \cite{arrazola2019quantum, allcock2020quantum} showed that such quantum inspired algorithms were less efficient than the quantum ones, and even than the standard classical algorithms for performing the same tasks. 

Finally, the quantum inspired algorithm doesn't provide the speedup characterized by $\sigma \in [0,1]$, the fraction of sampled elements among $H^{\ell +1}W^{\ell +1}D^{\ell+1}$ of them. Indeed, the probabilistic importance sampling described in Section \ref{tomographyconvolution}, that allows sampling the highest values of the convolution product output does not have a classical analog. The importance sampling does not offer an asymptotic speedup, however it could offer constant factor savings that are relevant in practice.

\subsection{Quantum Backpropagation Algorithm}\label{quantumalgobackprop}

Preliminaries on classical CNN backpropagation, along with all notations are given in Section Section \ref{sec:backpropagaton_cnn}.\\

The entire QCNN is made of multiple layers. For the last layer's output, we expect only one possible outcome, or a few in the case of a classification task, which means that the dimension of the quantum output is very small. A full tomography can be performed on the last layer's output in order to calculate the outcome. The loss $\mathcal{L}$ is then calculated, as a measure of correctness of the predictions compared to the ground truth. As the classical CNN, our QCNN should be able to perform the optimization of its weights (elements of the kernels) to minimize the loss by an iterative method.

\begin{theorem}{Quantum Backpropagation for Quantum CNN}{qbackprop} 
Given the forward pass quantum algorithm in Algorithm \ref{QCNNLayer}, the input matrix $A^{\ell}$ and the kernel matrix $F^{\ell}$ stored in the QRAM for each layer $\ell$, and a loss function $\mathcal{L}$, there is a quantum backpropagation algorithm that 
estimates, for any precision $\delta> 0$, the gradient tensor $\frac{\partial \mathcal{L}}{\partial F^{\ell}}$ and update each element to perform gradient descent such that
\begin{equation}
\forall (s,q), \left|\frac{\partial \mathcal{L}}{\partial F^{\ell}_{s,q}} - \overline{\frac{\partial \mathcal{L}}{\partial F^{\ell}_{s,q}}} \right| 
\leq 2\delta \norm{\frac{\partial \mathcal{L}}{\partial F^{\ell}}}_{2}
\end{equation}
Let $\frac{\partial \mathcal{L}}{\partial Y^{\ell}}$ be the gradient with respect to the $\ell^{th}$ layer. The running time of a single layer $\ell$ for quantum backpropagation is given by
\begin{equation}
O\left(\left(\left(\mu(A^{\ell})+\mu(\frac{\partial \mathcal{L}}{\partial Y^{\ell+1}})\right)\kappa(\frac{\partial \mathcal{L}}{\partial F^{\ell}})+\left(\mu(\frac{\partial \mathcal{L}}{\partial Y^{\ell+1}})+\mu(F^{\ell})\right)\kappa(\frac{\partial \mathcal{L}}{\partial Y^{\ell}})\right) \frac{\log(1/\delta)}{\delta^{2}}\right)
\end{equation}
where for a matrix $V$, $\kappa(V)$ is the condition number and $\mu(V)$ is defined in Definition \ref{def:mu}.
\end{theorem}

%Theorem \ref{thm:qbackprop} is proved in Section \ref{quantumalgobackprop} for the running time and Section \ref{gradientdescenterror} for the error guarantees. 

\begin{algorithm}[h]
\caption{Quantum Backpropagation} \label{QBackpropagation}
\begin{algorithmic}[1]
\REQUIRE   Precision parameter $\delta$. Data matrices $A^{\ell}$ and kernel matrices $F^{\ell}$ stored in QRAM for each layer $\ell$.
\ENSURE Outputs gradient matrices $\frac{\partial \mathcal{L}}{\partial F^{\ell}}$ and $\frac{\partial \mathcal{L}}{\partial Y^{\ell}}$ for each layer $\ell$.\\
\vspace{10pt} 
\STATE Calculate the gradient for the last layer $L$ using the outputs and the true labels: $\frac{\partial \mathcal{L}}{\partial Y^{L}}$ 
\FOR{ $\ell = L-1, \cdots, 0$}

\STATE {\bf Step 1 : Modify the gradient}\\
With $\frac{\partial \mathcal{L}}{\partial Y^{\ell+1}}$ stored in QRAM, set to 0 some of its values to take into account pooling, tomography and non-linearity that occurred in the forward pass of layer $\ell$. These values correspond to positions that haven't been sampled nor pooled, since they have no impact on the final loss.

\STATE {\bf Step 2 : Matrix-matrix multiplications}\\
With the modified values of $\frac{\partial \mathcal{L}}{\partial Y^{\ell+1}}$, use quantum linear algebra (Theorem \ref{thm:quantum_matrix_multiplication_inversion}) to perform the following matrix-matrix multiplications
\begin{equation}
\begin{cases}
(A^{\ell})^{T} \cdot \frac{\partial L}{\partial Y^{\ell+1}}\\
\frac{\partial \mathcal{L}}{\partial Y^{\ell+1}} \cdot (F^{\ell})^T
\end{cases}
\end{equation}
to obtain quantum states corresponding to $\frac{\partial \mathcal{L}}{\partial F^{\ell}}$ and $\frac{\partial \mathcal{L}}{\partial Y^{\ell}}$.

\STATE {\bf Step 3 : $\ell_{\infty}$ tomography}\\
Using the $\ell_{\infty}$ tomography procedure given in Algorithm \ref{alg:tom}, estimate each entry of $\frac{\partial \mathcal{L}}{\partial F^{\ell}}$ and $\frac{\partial \mathcal{L}}{\partial Y^{\ell}}$ with errors $\delta \norm{\frac{\partial \mathcal{L}}{\partial F^{\ell}}}$ and $\delta \norm{\frac{\partial \mathcal{L}}{\partial Y^{\ell}}}$ respectively. 
Store all elements of $\frac{\partial \mathcal{L}}{\partial F^{\ell}}$ in QRAM. 
\STATE {\bf Step 4 : Gradient descent }\\
Perform gradient descent using the estimates from step 3 to update the values of $F^{\ell}$ in QRAM:
\begin{equation}
F^{\ell}_{s,q} \gets F^{\ell}_{s,q} - 
\lambda \left(\frac{\partial \mathcal{L}}{\partial F^{\ell}_{s,q}} 
\pm 2\delta\norm{\frac{\partial \mathcal{L}}{\partial F^{\ell}}}_{2} \right)
\end{equation}
\ENDFOR
\end{algorithmic}
\end{algorithm}

We will detail the quantum algorithm to perform backpropagation on a layer $\ell$, and analyze the impact on the derivatives, given by the following diagram:

\begin{equation}\label{backpropagationdiagram}
\begin{cases}
    \frac{\partial \mathcal{L}}{\partial X^{\ell}}\\
    \frac{\partial \mathcal{L}}{\partial F^{\ell}}\\
\end{cases}
 \leftarrow 
\frac{\partial \mathcal{L}}{\partial \overline{X}^{\ell+1}}  \leftarrow 
\frac{\partial \mathcal{L}}{\partial f(\overline{X}^{\ell+1})}  \leftarrow 
\frac{\partial \mathcal{L}}{\partial \mathcal{X}^{\ell+1}}  \leftarrow 
\frac{\partial \mathcal{L}}{\partial \tilde{\mathcal{X}}^{\ell+1}}  =
\frac{\partial \mathcal{L}}{\partial X^{\ell+1}} 
\end{equation}

We assume that backpropagation has been done on layer $\ell+1$. 
This means in particular that $\frac{\partial \mathcal{L}}{\partial X^{\ell+1}}$ is stored in QRAM. 
However, as shown on Diagram (\ref{backpropagationdiagram}), $\frac{\partial \mathcal{L}}{\partial X^{\ell+1}}$ corresponds formally to $\frac{\partial \mathcal{L}}{\partial \tilde{\mathcal{X}}^{\ell+1}}$, and not $\frac{\partial \mathcal{L}}{\partial \overline{X}^{\ell+1}}$. 
Therefore, we will have to modify the values stored in QRAM to take into account non-linearity, tomography and pooling. 
We will first consider how to implement $\frac{\partial \mathcal{L}}{\partial X^{\ell}}$ and $\frac{\partial \mathcal{L}}{\partial F^{\ell}}$ through backpropagation, considering only convolution product, as if $\frac{\partial \mathcal{L}}{\partial \overline{X}^{\ell+1}}$ and $\frac{\partial \mathcal{L}}{\partial X^{\ell+1}}$ where the same. Then we will detail how to simply modify $\frac{\partial \mathcal{L}}{\partial X^{\ell+1}}$ \emph{a priori}, by setting some of its values to 0.

\subsubsection{Quantum Convolution Product}
In this section we consider only the quantum convolution product without non-linearity, tomography nor pooling, hence writing its output directly as $X^{\ell+1}$. Regarding derivatives, the quantum convolution product is equivalent to the classical one. Gradient relations (\ref{updateF}) and (\ref{updateX}) remain the same. Note that the $\epsilon$-approximation from Section \ref{registerencoding} doesn't participate in gradient considerations.

The gradient relations being the same, we still have to specify the quantum algorithm that implements the backpropagation and outputs classical description of $ \frac{\partial \mathcal{L}}{\partial X^{\ell}}$ and $\frac{\partial \mathcal{L}}{\partial F^{\ell}}$.
 We have seen that the two main calculations (\ref{updateF}) and (\ref{updateX}) are in fact matrix-matrix multiplications both involving $\frac{\partial \mathcal{L}}{\partial Y^{\ell+1}}$, the reshaped form of $\frac{\partial \mathcal{L}}{\partial X^{\ell+1}}$. For each, the classical running time is $O(H^{\ell+1}W^{\ell+1}D^{\ell+1}HWD^{\ell})$. 
We know from Theorem \ref{thm:quantum_matrix_multiplication_inversion} and Theorem \ref{thm:tomography_linfinity} a quantum algorithm to perform efficiently a matrix-vector multiplication and return a classical state with $\ell_{\infty}$ norm guarantees. For a matrix $V$ and a vector $b$, both accessible from the QRAM, the running time to perform this operation is 
\begin{equation}
O\left(\frac{\mu(V) \kappa(V) \log(1/\delta)}{\delta^{2}}\right)
\end{equation}
where $\kappa(V)$ is the condition number of the matrix and $\mu(V)$ is a matrix parameter defined in Definition \ref{def:mu}. Precision parameter $\delta > 0$ is the error committed in the approximation for both Theorems \ref{thm:quantum_matrix_multiplication_inversion} and \ref{thm:tomography_linfinity}.

We can therefore apply theses theorems to perform matrix-matrix multiplications, by simply decomposing them in several matrix-vector multiplications. For instance, in Eq.(\ref{updateF}), the matrix could be $ (A^{\ell})^{T}$ and the different vectors would be each column of $\frac{\partial L}{\partial Y^{\ell+1}}$. We know from Section \ref{sec:matrix_multiplication_quantum} that the global running time to perform quantumly Eq.(\ref{updateF}) is obtained by replacing $\mu(V)$ by $\mu(\frac{\partial \mathcal{L}}{\partial Y^{\ell+1}})+\mu(A^{\ell})$ and $\kappa(V)$ by $\kappa((A^{\ell})^T\cdot \frac{\partial \mathcal{L}}{\partial Y^{\ell+1}})$. Likewise, for Eq.(\ref{updateX}), we have $\mu(\frac{\partial \mathcal{L}}{\partial Y^{\ell+1}})+\mu(F^{\ell})$ and $\kappa(\frac{\partial \mathcal{L}}{\partial Y^{\ell+1}}\cdot (F^{\ell})^T)$.

Note that the dimension of the matrix doesn't appear in the running time since we tolerate a $\ell_{\infty}$ norm guarantee for the error, instead of a $\ell_{2}$ guarantee (see Section \ref{sec:l_infinite_proof} for details). The reason why $\ell_\infty$ tomography is the right approximation here is because the result of these linear algebra operations are rows of the gradient matrices, that are not vectors in an euclidean space, but a series of numbers for which we want to be $\delta$-close to the exact values. See the next section for more details. 

It is an open question to see if one can apply the same sub-sampling technique as in the forward pass (Section \ref{singlequantumlayer}) and sample only the highest derivatives of $\frac{\partial \mathcal{L}}{\partial X^{\ell}}$, to reduce the computation cost while maintaining a good optimization.

We then have to understand which elements of $\frac{\partial \mathcal{L}}{\partial X^{\ell+1}}$ must be set to zero to take into account the effects the non-linearity, tomography and pooling.

\subsubsection{Quantum Non-Linearity and Tomography}
To include the impact of the non-linearity, one could apply the same rule as in (\ref{relubackpropagation}), and simply replace ReLu with capReLu. After the non-linearity, we obtain $f(\overline{X}^{\ell+1})$, and the gradient relation would be given by 

\begin{equation}\label{fakecaprelubackpropagation}
\left[\frac{\partial \mathcal{L}}{\partial \overline{X}^{\ell+1}}\right]_{i^{\ell+1},j^{\ell+1},d^{\ell+1}} = 
\begin{cases}
    \left[\frac{\partial \mathcal{L}}{\partial f(\overline{X}^{\ell+1})}\right]_{i^{\ell+1},j^{\ell+1},d^{\ell+1}} \text{ if }  0\leq \overline{X}^{\ell+1}_{i^{\ell+1},j^{\ell+1},d^{\ell+1}} \leq C\\
    0 \text{ otherwise}\\
\end{cases}
\end{equation}

If an element of $\overline{X}^{\ell+1}$ was negative or bigger than the cap $C$, its derivative should be zero during the backpropagation. However, this operation was performed in quantum superposition. In the quantum algorithm, one cannot record at which positions $(i^{\ell+1},j^{\ell+1},d^{\ell+1})$ the activation function was selective or not. The gradient relation (\ref{fakecaprelubackpropagation}) cannot be implemented \emph{a posteriori}.

We provide a partial solution to this problem, using the fact that quantum tomography must also be taken into account for some derivatives. Indeed, only the points $(i^{\ell+1},j^{\ell+1},d^{\ell+1})$ that have been sampled should have an impact on the gradient of the loss. Therefore we replace the previous relation by

\begin{equation}\label{approximationnonlinearitybackpropagation}
\left[\frac{\partial \mathcal{L}}{\partial \overline{X}^{\ell+1}}\right]_{i^{\ell+1},j^{\ell+1},d^{\ell+1}} = 
\begin{cases}
    \left[\frac{\partial \mathcal{L}}{\partial \mathcal{X}^{\ell+1}}\right]_{i^{\ell+1},j^{\ell+1},d^{\ell+1}} \text{ if $(i^{\ell+1},j^{\ell+1},d^{\ell+1})$ was sampled }\\
    0 \text{ otherwise}\\
\end{cases}
\end{equation}

Nonetheless, we can argue that this approximation will be tolerable:

In the first case where $\overline{X}^{\ell+1}_{i^{\ell+1},j^{\ell+1},d^{\ell+1}} < 0$, the derivatives can not be set to zero as they should. But in practice, their values will be zero after the activation function and such points would not have a chance to be sampled. In conclusion, their derivatives would be zero as required. 

In the other case where $\overline{X}^{\ell+1}_{i^{\ell+1},j^{\ell+1},d^{\ell+1}} > C$, the derivatives can not be set to zero as well but the points have a high probability of being sampled. Therefore their derivative will remain unchanged as if we were using a ReLu instead of a capReLu. However in cases where the cap $C$ is high enough, this shouldn't be a source of disadvantage in practice.

\subsubsection{Quantum Pooling}
From relation (\ref{approximationnonlinearitybackpropagation}), we can take into account the impact of quantum pooling (see Section \ref{pooling}) on the derivatives. This case is easier since one can record the selected positions during the QRAM update. Therefore, applying the backpropagation is similar to the classical setting with Eq.(\ref{poolingbackpropagation}).

\begin{equation}\label{quantumpoolingbackpropagation}
\left[\frac{\partial \mathcal{L}}{\partial \mathcal{X}^{\ell+1}}\right]_{i^{\ell+1},j^{\ell+1},d^{\ell+1}} = 
\begin{cases}
    \left[\frac{\partial \mathcal{L}}{\partial \tilde{\mathcal{X}}^{\ell+1}}\right]_{\tilde{i}^{\ell+1},\tilde{j}^{\ell+1},\tilde{d}^{\ell+1}} \text{ if } (i^{\ell+1},j^{\ell+1},d^{\ell+1}) \in \mathcal{P}\\
    0 \text{ otherwise}\\
\end{cases}
\end{equation}
where $\mathcal{P}$ is the set of indices selected during pooling.
Note that we know $\frac{\partial \mathcal{L}}{\partial \tilde{\mathcal{X}}^{\ell+1}}$ as it is equal to $\frac{\partial \mathcal{L}}{\partial X^{\ell+1}}$, the gradient with respect to the input of layer $\ell+1$, known by assumption and stored in the QRAM.

\subsubsection{Conclusion and Running Time}

In conclusion, given $\frac{\partial \mathcal{L}}{\partial Y^{\ell+1}}$ in the QRAM, the quantum backpropagation first consists in applying the relations (\ref{quantumpoolingbackpropagation}) followed by (\ref{approximationnonlinearitybackpropagation}). The effective gradient now take into account non-linearity, tomography and pooling that occurred during layer $\ell$. We can know use apply the quantum algorithm for matrix-matrix multiplication that implements relations (\ref{updateX}) and (\ref{updateF}).

Note that the steps in Algorithm \ref{QBackpropagation} could also be reversed: during backpropagation of layer $\ell+1$, when storing values for each elements of $\frac{\partial \mathcal{L}}{\partial Y^{\ell+1}}$ in the QRAM, one can already take into account (\ref{quantumpoolingbackpropagation}) and (\ref{approximationnonlinearitybackpropagation}) of layer $\ell$. In this case we directly store $\frac{\partial \mathcal{L}}{\partial \overline{X}^{\ell+1}}$, at no supplementary cost.

Therefore, the running time of the quantum backpropagation for one layer $\ell$, given as Algorithm \ref{QBackpropagation}, corresponds to the sum of the running times of the circuits for implementing relations (\ref{updateF}) and (\ref{updateX}). We finally obtain

\begin{multline}
O(((\mu(A^{\ell})+\mu(\frac{\partial \mathcal{L}}{\partial Y^{\ell+1}}))\kappa((A^{\ell})^T \cdot \frac{\partial \mathcal{L}}{\partial Y^{\ell+1}}) \\ 
 +(\mu(\frac{\partial \mathcal{L}}{\partial Y^{\ell+1}})+\mu(F^{\ell}))\kappa(\frac{\partial \mathcal{L}}{\partial Y^{\ell+1}}\cdot (F^{\ell})^T)) \frac{\log(1/\delta)}{\delta^{2}}) 
\end{multline}

which can be rewritten as
\begin{equation}
O\left(\left(\left(\mu(A^{\ell})+\mu(\frac{\partial \mathcal{L}}{\partial Y^{\ell+1}})\right)\kappa(\frac{\partial \mathcal{L}}{\partial F^{\ell}})+\left(\mu(\frac{\partial \mathcal{L}}{\partial Y^{\ell+1}})+\mu(F^{\ell})\right)\kappa(\frac{\partial \mathcal{L}}{\partial Y^{\ell}})\right) \frac{\log(1/\delta)}{\delta^{2}}\right). 
\end{equation}
Besides storing $\frac{\partial \mathcal{L}}{\partial X^{\ell}}$, the main output is a classical description of $\frac{\partial \mathcal{L}}{\partial F^{\ell}}$, necessary to perform gradient descent of the parameters of $F^{\ell}$.

\subsubsection{Gradient Descent and Classical equivalence}\label{gradientdescenterror}
In this part we will see the impact of the quantum backpropagation compared to the classical case, which can be reduced to a simple noise addition during the gradient descent. Recall that gradient descent, in our case, would consist of applying the following update rule

\begin{equation}
F^{\ell} \leftarrow F^{\ell} -\lambda \frac{\partial \mathcal{L}}{\partial F^{\ell}}
\end{equation}
with the learning rate $\lambda$.

Let's denote $x = \frac{\partial \mathcal{L}}{\partial F^{\ell}}$ and its elements $x_{s,q} = \frac{\partial \mathcal{L}}{\partial F^{\ell}_{s,q}}$. From the first result of Theorem \ref{thm:quantum_matrix_multiplication_inversion} with error $\delta<0$, and the tomography procedure from Theorem \ref{thm:tomography_linfinity}, with same error $\delta$, we can obtain a classical description of $\frac{\overline{x}}{\norm{\overline{x}}_2}$ with $\ell_{\infty}$ norm guarantee, such that:
\begin{equation}
\norm{\frac{\overline{x}}{\norm{\overline{x}}_2} - \frac{x}{\norm{x}_2}}_{\infty} \leq \delta
\end{equation}
in time $\widetilde{O}(\frac{\kappa(V)\mu(V)\log(\delta)}{\delta^2})$, where we denote $V$ the matrix stored in the QRAM that allows to obtain $x$, as explained in Section \ref{quantumalgobackprop}. The $\ell_{\infty}$ norm tomography is used so that the error $\delta$ is at most the same for each component
\begin{equation}
\forall (s,q), \left | \frac{\overline{x_{s,q}}}{\norm{\overline{x}}_2} - \frac{x_{s,q}}{\norm{x}_2} \right | \leq \delta
\end{equation}
From the second result of the Theorem \ref{thm:quantum_matrix_multiplication_inversion} we can also obtain an estimate $\norm{\overline{x}}_2$ of the norm, for the same error $\delta$, such that
\begin{equation}
|\norm{\overline{x}}_2 - \norm{x}_2| \leq \delta \norm{x}_2
\end{equation}
in time $\widetilde{O}(\frac{\kappa(V)\mu(V)}{\delta}\log(\delta))$ (which does not affect the overall asymptotic running time). Using both results we can obtain an unnormalized state close to $x$ such that, by the triangular inequality 
\begin{gather}\nonumber
\norm{\overline{x}-x}_{\infty} = \norm{\frac{\overline{x}}{\norm{\overline{x}}_2}\norm{\overline{x}}_2 - \frac{x}{\norm{x}_2}\norm{x}_2}_{\infty}\\ \nonumber%\end{equation}
%\begin{equation}
\leq 
\norm{\frac{\overline{x}}{\norm{\overline{x}}_2}\norm{\overline{x}}_2 - \frac{\overline{x}}{\norm{\overline{x}}_2}\norm{x}_2}_{\infty} + 
\norm{\frac{\overline{x}}{\norm{\overline{x}}_2}\norm{x}_2 - \frac{x}{\norm{x}_2}\norm{x}_2}_{\infty}\\ \nonumber%\end{equation}
%\begin{equation}
\leq 
1 \cdot  |\norm{\overline{x}}_2 - \norm{x}_2| +
\norm{x}_2 \cdot \norm{\frac{\overline{x}}{\norm{\overline{x}}_2} - \frac{x}{\norm{x}_2}}_{\infty}\\ %\end{equation}
%\begin{equation}
\leq \delta \norm{x}_2  + \norm{\overline{x}}_2 \delta
\leq 2\delta\norm{x}_2
\end{gather}
in time $\widetilde{O}(\frac{\kappa(V)\mu(V)\log(\delta)}{\delta^2})$. In conclusion, with $\ell_{\infty}$ norm guarantee, having also access to the norm of the result is costless.

Finally, the noisy gradient descent update rule, expressed as $F^{\ell}_{s,q} \gets F^{\ell}_{s,q} - \lambda \overline{\frac{\partial \mathcal{L}}{\partial F^{\ell}_{s,q}}}$ can written in the worst case with
\begin{equation}
\overline{\frac{\partial \mathcal{L}}{\partial F^{\ell}_{s,q}}} =  
\frac{\partial \mathcal{L}}{\partial F^{\ell}_{s,q}} \pm 2\delta\norm{\frac{\partial \mathcal{L}}{\partial F^{\ell}}}_{2}
\end{equation}

To summarize, using the quantum linear algebra from Theorem \ref{thm:quantum_matrix_multiplication_inversion} with $\ell_{\infty}$ norm tomography from Theorem \ref{thm:tomography_linfinity}, both with error $\delta$, along with norm estimation with relative error $\delta$ too, we can obtain classically the unnormalized values $\overline{\frac{\partial \mathcal{L}}{\partial F^{\ell}}}$ such that $\norm{\overline{\frac{\partial \mathcal{L}}{\partial F^{\ell}}}-\frac{\partial \mathcal{L}}{\partial F^{\ell}}}_{\infty} \leq 2\delta\norm{\frac{\partial \mathcal{L}}{\partial F^{\ell}}}_{2}$ or equivalently
\begin{equation}
\forall (s,q), \left|\overline{\frac{\partial \mathcal{L}}{\partial F^{\ell}_{s,q}}}-\frac{\partial \mathcal{L}}{\partial F^{\ell}_{s,q}}\right|  \leq  2\delta\norm{\frac{\partial \mathcal{L}}{\partial F^{\ell}}}_{2}
\end{equation}
Therefore the gradient descent update rule in the quantum case becomes $F^{\ell}_{s,q} \gets F^{\ell}_{s,q} - \lambda \overline{\frac{\partial \mathcal{L}}{\partial F^{\ell}_{s,q}}}$, which in the worst case becomes
\begin{equation}\label{gradientdescent}
F^{\ell}_{s,q} \gets F^{\ell}_{s,q} - 
\lambda \left(\frac{\partial \mathcal{L}}{\partial F^{\ell}_{s,q}} 
\pm 2\delta\norm{\frac{\partial \mathcal{L}}{\partial F^{\ell}}}_{2} \right)
\end{equation}

This proves the Theorem \ref{thm:qbackprop}. This update rule can be simulated by the addition of a random relative noise given as a Gaussian centered on 0, with a standard deviation equal to $\delta$. This is how we will simulate quantum backpropagation in the next Section.

Compared to the classical update rule, this corresponds to the addition of noise during the optimization step. This noise decreases as $\norm{\frac{\partial \mathcal{L}}{\partial F^{\ell}}}_{2}$, which is expected to happen while converging. Recall that the gradient descent is already a stochastic process. Therefore, we expect that such noise, with acceptable values of $\delta$, will not disturb the convergence of the gradient, as the following numerical simulations tend to confirm.

\section{Numerical Simulations}\label{NumericalSimulations}
As described above, the adaptation of the CNNs to the quantum setting implies some modifications that could alter the efficiency of the learning or classifying phases. We now present some experiments to show that such modified CNNs can converge correctly, as the original ones. 

The experiment, using the PyTorch library \cite{paszke2017automatic}, consists of training classically a small convolutional neural network for which we have added a ``quantum" sampling after each convolution, as in Section \ref{qramupdate}. Instead of parameterizing it with the precision $\nu$, we have chosen to use the sampling ratio $\sigma$ that represents the number of samples drawn during tomography. These two definitions are equivalent, as shown in Section \ref{tomographyconvolution}, but the second one is more intuitive regarding the running time and the simulations. 

We also add a noise simulating the amplitude estimation (Section \ref{registerencoding}, parameter: $\epsilon$), followed by a capReLu instead of the usual ReLu (Section \ref{capReLu}, parameter: $C$), and a noise during the backpropagation (Section \ref{gradientdescenterror}, parameter: $\delta$). In the following results, we observe that our quantum CNN is able to learn and classify visual data from the widely used MNIST dataset. This dataset is made of 60.000 training images and 10.000 testing images of handwritten digits. Each image is a 28x28 grayscale pixels between 0 and 255 (8 bits encoding), before normalization.

Let's first observe the ``quantum" effects on an image of the dataset. In particular, the effect of the capped non-linearity, the introduction of noise and the quantum sampling.

\begin{figure}[H]
\minipage{0.25\textwidth}
  \includegraphics[width=\linewidth]{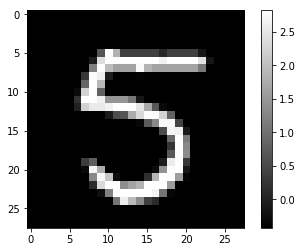}
  \includegraphics[width=\linewidth]{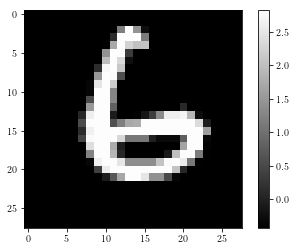}
\endminipage\hfill
\minipage{0.25\textwidth}
  \includegraphics[width=\linewidth]{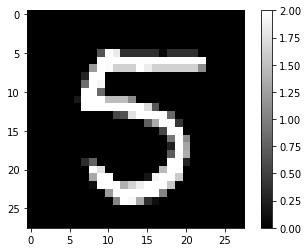}
  \includegraphics[width=\linewidth]{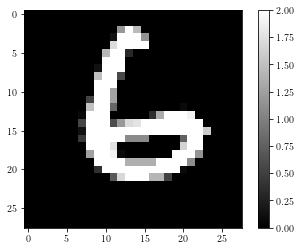}
\endminipage\hfill
\minipage{0.25\textwidth}
  \includegraphics[width=\linewidth]{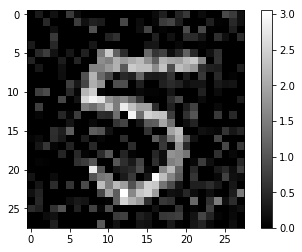}
  \includegraphics[width=\linewidth]{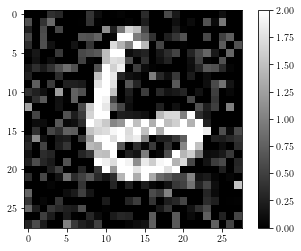}
\endminipage
\minipage{0.25\textwidth}
  \includegraphics[width=\linewidth]{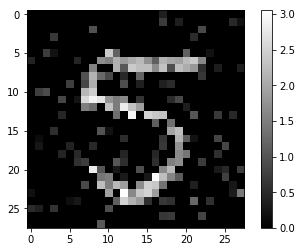}
  \includegraphics[width=\linewidth]{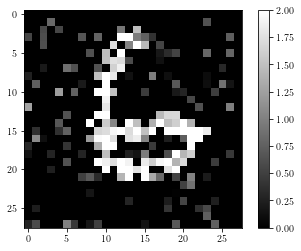}
\endminipage
\caption{Effects of the QCNN on a 28x28 input image. From left to right: original image, image after applying a capReLu activation function with a cap $C$ at $2.0$, introduction of a strong noise during amplitude estimation with $\epsilon=0.5$, quantum sampling with ratio $\sigma=0.4$ that samples the highest values in priority. The useful information tends to be conserved in this example. The side gray scale indicates the value of each pixel. Note that during the QCNN layer, a convolution is supposed to happen before the last image but we chose not to perform it for better visualization.}
\label{fig:quantumeffects}
\end{figure}

We now present the full simulation of our quantum CNN. In the following, we use a simple network made of 2 convolution layers, and compare our quantum CNN to the classical one. The first and second layers are respectively made of 5 and 10 kernels, both of size 7x7. A three-layer fully connected network is applied at the end and a softmax activation function is applied on the last layer to detect the predicted outcome over 10 classes (the ten possible digits). Note that we didn't introduce pooling, being equivalent between quantum and classical algorithms and not improving the results on our CNN. The objective of the learning phase is to minimize the loss function, defined by the negative log-likelihood of the classification on the training set. The optimizer used was a built-in Stochastic Gradient Descent. 

Using PyTorch, we have been able to implement the following quantum effects (the first three points are shown in Fig.\ref{fig:quantumeffects}):
\begin{itemize}
\item The addition of noise, to simulate the approximation of amplitude estimation during the forward quantum convolution layer, by adding Gaussian noise centered on 0 and with standard deviation $2M\epsilon$, with $M = \max_{p,q}{\norm{A_{p}}\norm{F_{q}}}$, as given by Eq.(\ref{errorAEfinal}). 
\item A modification of the non-linearity: a ReLu function that becomes constant above the value $T$ (the cap).
\item A sampling procedure to apply on a tensor with a probability distribution proportional to the tensor itself, reproducing the quantum sampling with ratio $\sigma$.
\item The addition of noise during the gradient descent, to simulate the quantum backpropagation, by adding a Gaussian noise centered on 0 with standard deviation $\delta$, multiplied by the norm of the gradient, as given by Eq.(\ref{gradientdescent}).\\
\end{itemize}

\begin{figure}[H]
\begin{multicols}{2}
    \includegraphics[width=\linewidth]{images/qcnn_training_epsilon001_T10_delta001.png}\par
    \includegraphics[width=\linewidth]{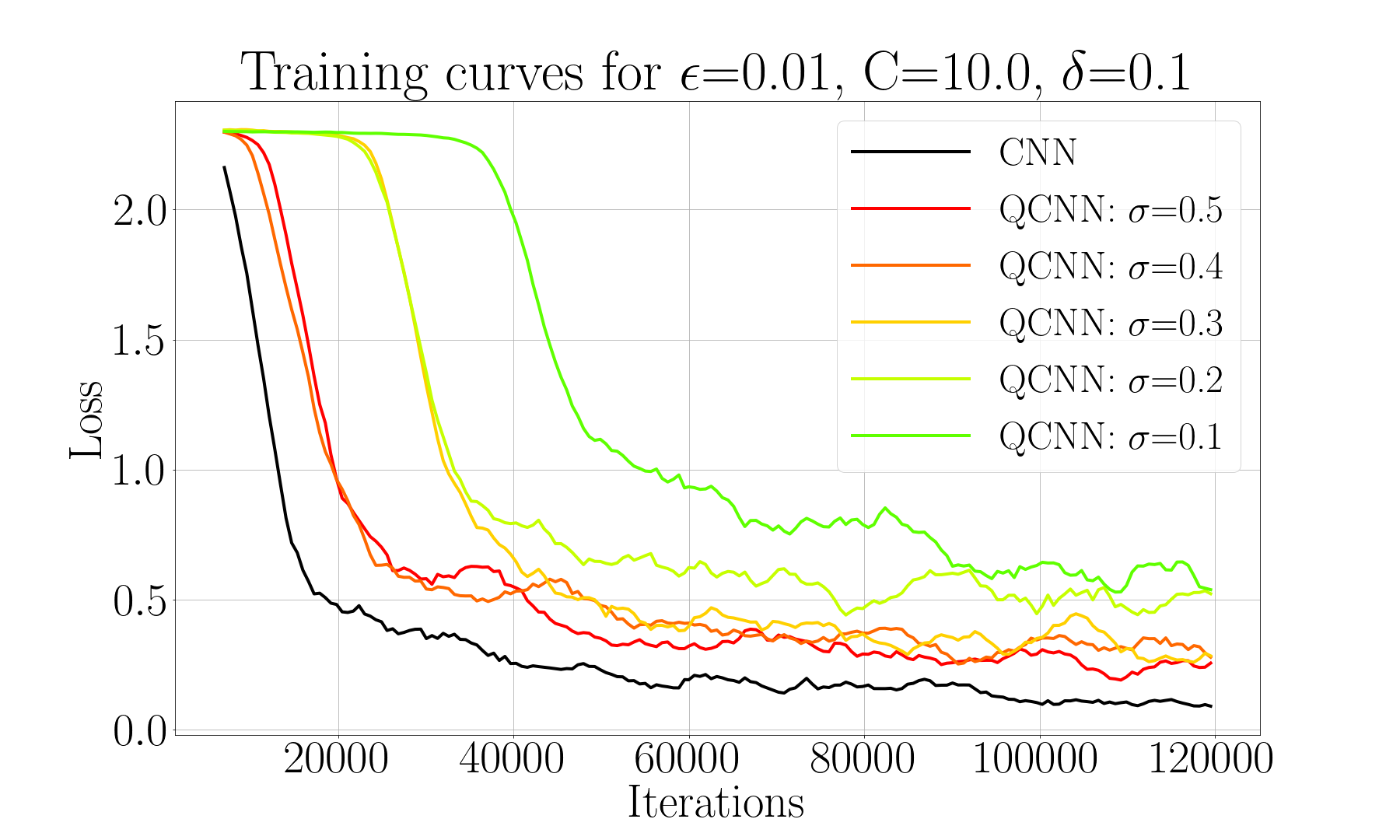}\par 
\end{multicols}
\begin{multicols}{2}
    \includegraphics[width=\linewidth]{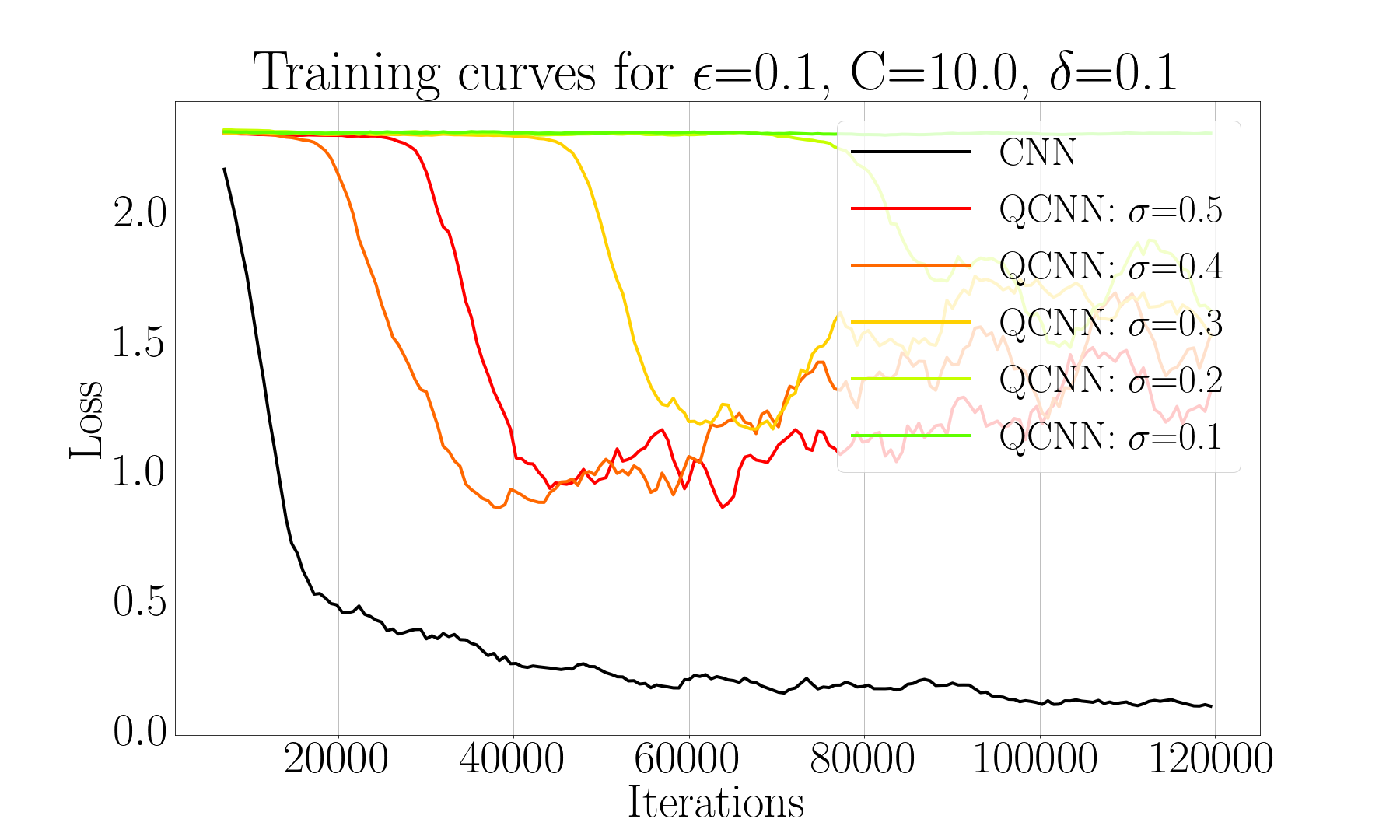}\par
    \includegraphics[width=\linewidth]{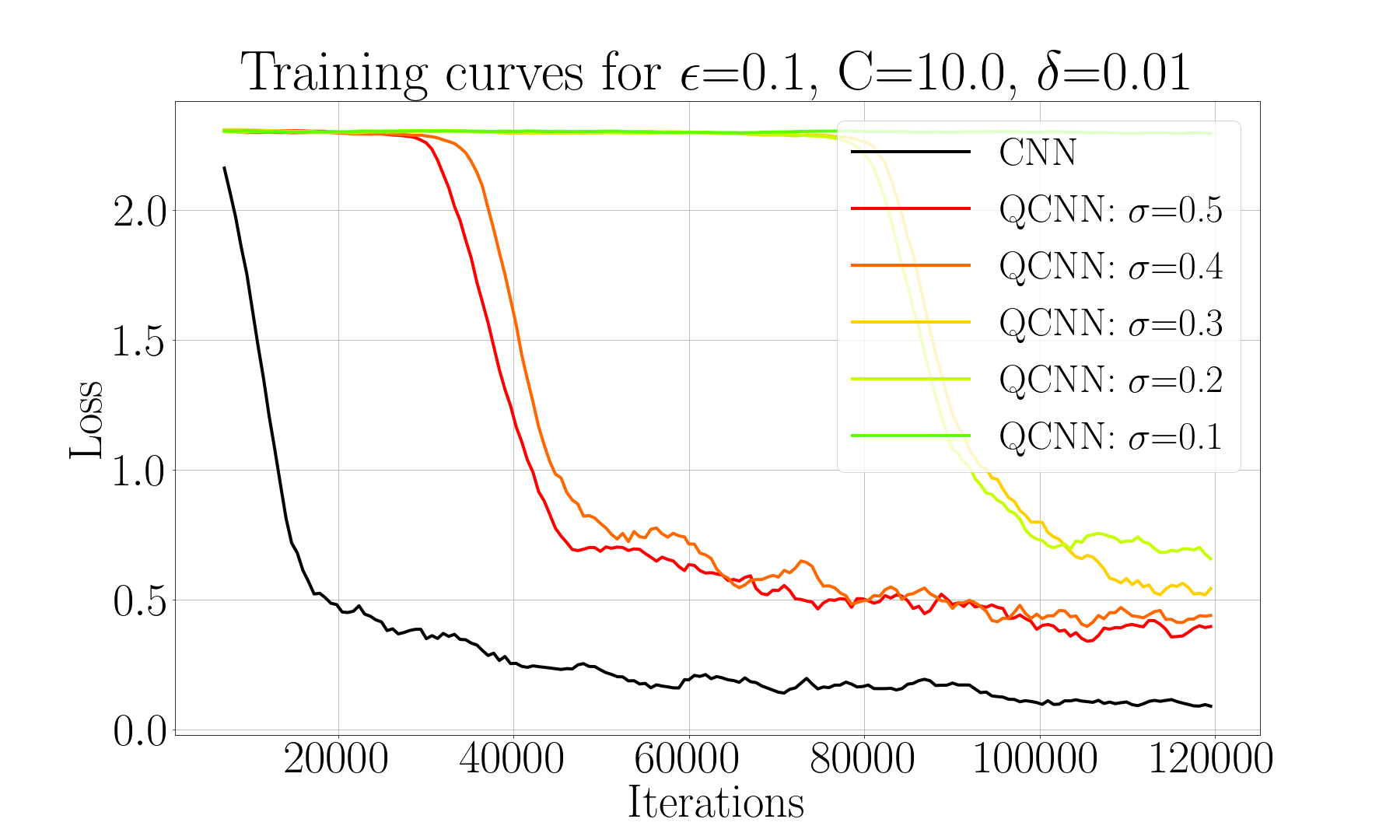}\par 
\end{multicols}
\begin{multicols}{2}
    \includegraphics[width=\linewidth]{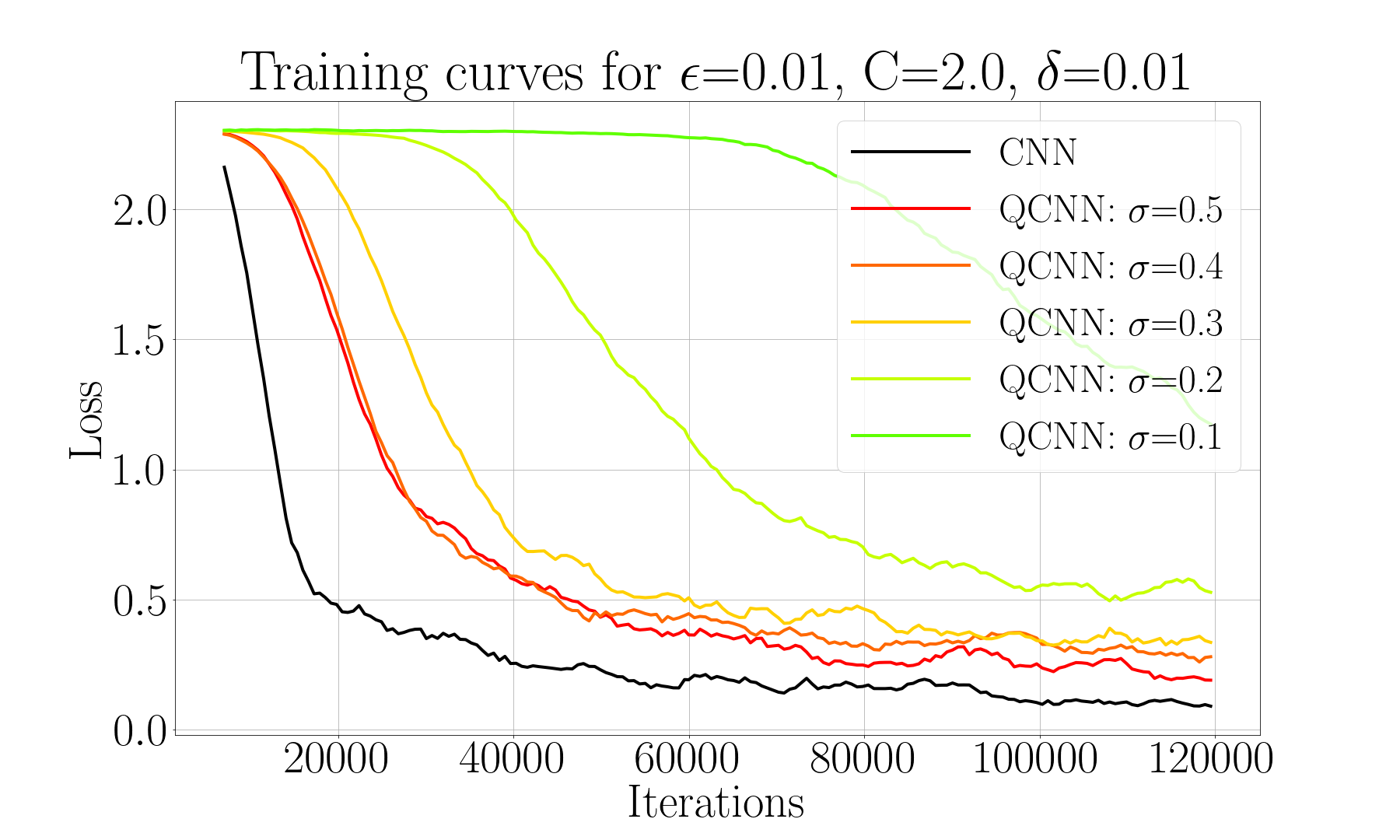}\par 
    \includegraphics[width=\linewidth]{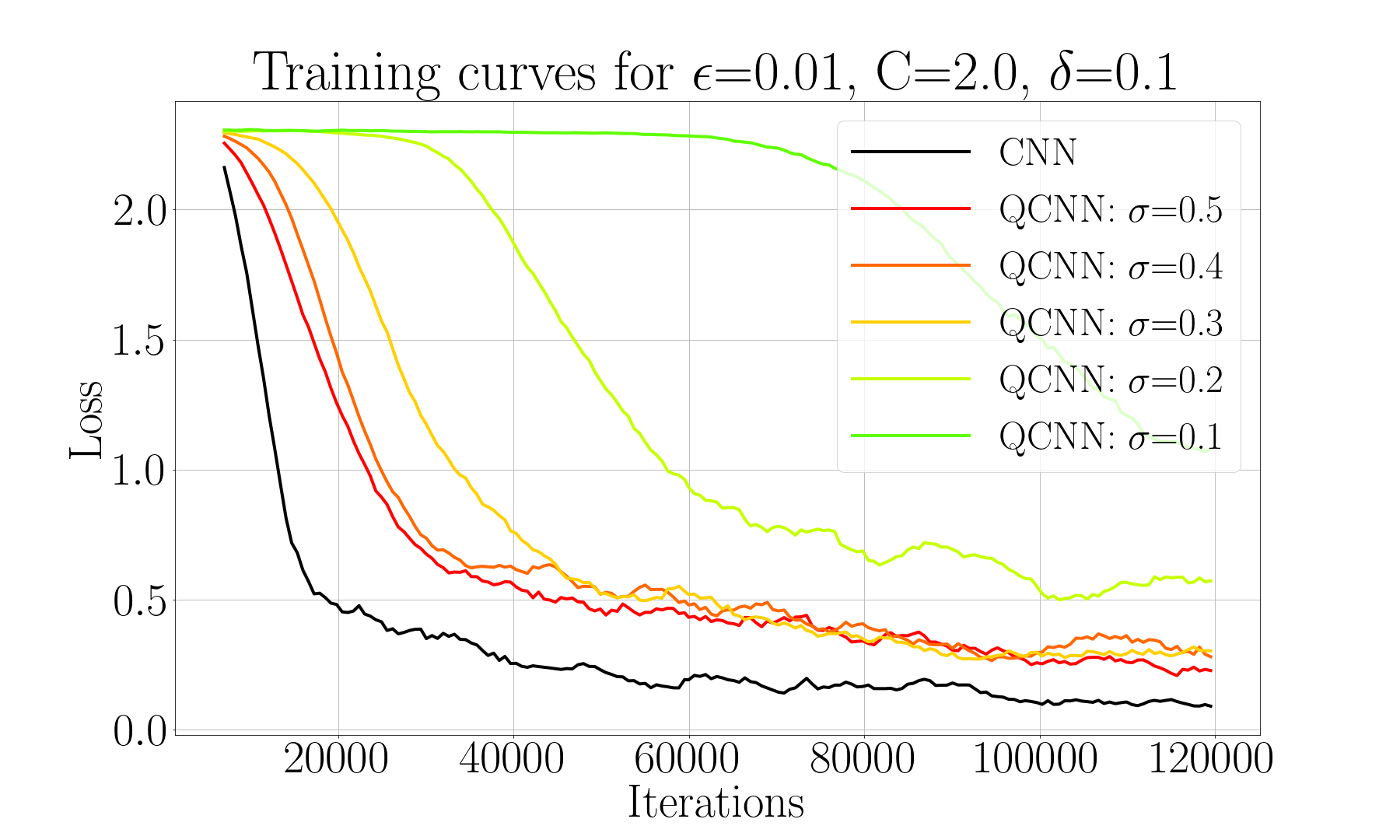}\par 
    \end{multicols}
\caption{Numerical simulations of the training of the QCNN. These training curves represent the evolution of the Loss $\mathcal{L}$ as we iterate through the MNIST dataset. For each graph, the amplitude estimation error $\epsilon$ $(0.1,0.01)$, the non-linearity cap $C$ $(2,10)$, and the backpropagation error $\delta$ $(0.1,0.01)$ are fixed whereas the quantum sampling ratio $\sigma$ varies from 0.1 to 0.5. We can compare each training curve to the classical learning (CNN). Note that these training curves are smoothed, over windows of 12 steps, for readability.}
\label{experiment_1_2}
\end{figure}

The CNN used for this simulation may seem ``small" compared to the standards AlexNet \cite{krizhevsky2012imagenet} or VGG-16 \cite{simonyan2014very}, or those used in industry. However simulating this small QCNN on a classical computer was already very computationally intensive and time consuming, due to the``quantum" sampling task, apparently not optimized for a classical implementation in PyTorch. Every single training curve showed in Fig.\ref{experiment_1_2} could last for 4 to 8 hours. Hence adding more convolutional layers wasn't convenient. Similarly, we didn't compute the loss on the whole testing set (10.000 images) during the training to plot the testing curve. However we have computed the test losses and accuracies once the model trained (see Table \ref{testset_metrics_1}), in order to detect potential overfitting cases.

We now present the result of the training phase for a quantum version of this CNN, where partial quantum sampling is applied, for different sampling ratio (number of samples taken from the resulting convolution). Since the quantum sampling gives more probability to observe high value pixels, we expect to be able to learn correctly even with a small ratio ($\sigma \leq 0.5$). We compare these training curves to the classical one. The learning has been done on two epochs, meaning that the whole dataset is used twice. The following plots show the evolution of the loss $\mathcal{L}$ during the iterations on batches. This is the standard indicator of the good convergence of a neural network learning phase. We can compare the evolution of the loss between a classical CNN and our QCNN for different parameters.

\begin{figure}[H]
\centering
\includegraphics[width=120mm] {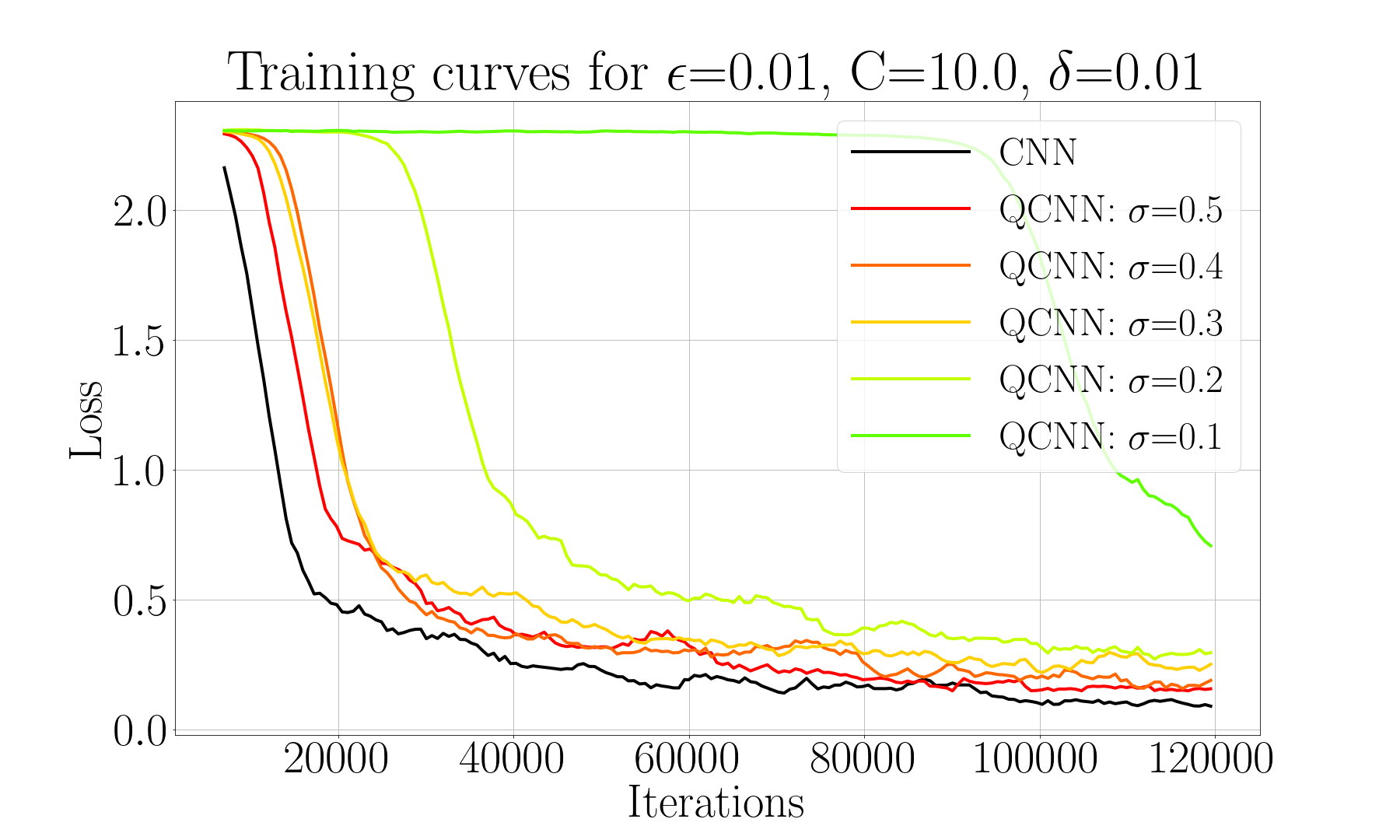} 
\captionsetup{justification=raggedright, margin=1cm}
\caption{Training curves comparison between the classical CNN and the Quantum CNN (QCNN) for $\epsilon=0.01$, $C=10$, $\delta=0.01$ and the sampling ratio $\sigma$ from $0.1$ to $0.5$. We can observe a learning phase similar to the classical one, even for a weak sampling of 20\% or 30\% of each convolution output, which tends to show that the meaningful information is distributed only at certain locations of the images, coherently with the purpose of the convolution layer. Even for a very low sampling ratio of 10\%, we observe a convergence despite a late start.}
\label{Training curves }
\end{figure}

In the following we report the classification results of the QCNN when applied on the test set (10.000 images). We distinguish to use cases: in Table \ref{testset_metrics_1} the QCNN has been trained quantumly as described in this Chapter, whereas in Table \ref{testset_metrics_2} we first have trained the classical CNN, then transferred the weights to the QCNN only for the classification. This second use case has a global running time worst than the first one, but we see it as another concrete application: quantum machine learning could be used only for faster classification from a classically generated model, which could be the case for high rate classification task (e.g. for autonomous systems, classification over many simultaneous inputs). We report the test loss and accuracy for different values of the sampling ratio $\sigma$, the amplitude estimation error $\epsilon$, and for the backpropagation noise $\delta$ in the first case. The cap $C$ is fixed at 10. These values must be compared to the classical CNN classification metrics, for which the loss is 0.129 and the accuracy is 96.1\%. Note that we used a relatively small CNN and hence the accuracy is just over $96\%$, lower than the best possible accuracy with larger CNN. \\

\begin{table}[h]
\centering
\begin{tabular}{|c|c|c|c|c|c|}
\hline
\multicolumn{6}{|c|}{QCNN Test - Classification}                                                          \\ \hline
\multirow{2}{*}{$\sigma$} & $\epsilon$ & \multicolumn{2}{c|}{0.01} & \multicolumn{2}{c|}{0.1} \\ \cline{2-6} 
                                  & $\delta$   & 0.01         & 0.1        & 0.01        & 0.1        \\ \hline
\multirow{2}{*}{0.1}              & Loss       & 0.519            & 0.773          & 2.30           & 2.30          \\
                                  & Accuracy   & 82.8\%          & 74.8\%        & 11.5\%         & 11.7\%        \\ \hline
\multirow{2}{*}{0.2}              & Loss       & 0.334            & 0.348          & 0.439           & 1.367          \\
                                  & Accuracy   & 89.5\%          & 89.0\%        & 86.2\%         & 54.1\%        \\ \hline
\multirow{2}{*}{0.3}              & Loss       & 0.213            & 0.314          & 0.381           & 0.762          \\
                                  & Accuracy   & 93.4\%          & 90.3\%        & 87.9\%         & 76.8\%        \\ \hline
\multirow{2}{*}{0.4}              & Loss       & 0.177            & 0.215          & 0.263           & 1.798          \\
                                  & Accuracy   & 94.7\%          & 93.3\%        & 91.8\%         & 34.9\%        \\ \hline
\multirow{2}{*}{0.5}              & Loss       & 0.142            & 0.211          & 0.337           & 1.457          \\
                                  & Accuracy   & 95.4\%          & 93.5\%        & 89.2\%         & 52.8\%        \\ \hline
\end{tabular}
\caption{QCNN trained with quantum backpropagation on MNIST dataset. With $C=10$ fixed.\label{testset_metrics_1}}
\end{table}

\begin{table}[h]
\centering
\begin{tabular}{|c|c|c|c|}
\hline
\multicolumn{4}{|c|}{QCNN Test - Classification}   \\ \hline
$\sigma$             & $\epsilon$ & 0.01 & 0.1 \\ \hline
\multirow{2}{*}{0.1} & Loss       & 1.07    & 1.33   \\
                     & Accuracy   & 86.1\%  & 78.6\% \\ \hline
\multirow{2}{*}{0.2} & Loss       & 0.552    & 0.840   \\
                     & Accuracy   & 92.8\%  & 86.5\% \\ \hline
\multirow{2}{*}{0.3} & Loss       & 0.391    & 0.706   \\
                     & Accuracy   & 94,3\%  & 85.8\% \\ \hline
\multirow{2}{*}{0.4} & Loss       & 0.327    & 0.670   \\
                     & Accuracy   & 94.4\%  & 84.0\% \\ \hline
\multirow{2}{*}{0.5} & Loss       & 0.163    & 0.292   \\
                     & Accuracy   & 95.9\%  & 93.5\% \\ \hline
\end{tabular}
\caption{QCNN created from a classical CNN trained on MNIST dataset. With $\delta = 0.01$ and $C=10$ fixed.\label{testset_metrics_2}} 
\end{table}

Our simulations show that the QCNN is able to learn despite the introduction of noise, tensor sampling and other modifications. In particular it shows that only a fraction of the information is meaningful for the neural network, and that the quantum algorithm captures this information in priority. This learning can be more or less efficient depending on the choice of the key parameters. For reasonable values of these parameters, the QCNN is able to converge during the training phase. It can then classify correctly on both training and testing set, indicating that it does not overfit the data.  

We notice that the learning curves sometimes present a late start before the convergence initializes, in particular for small sampling ratio. This late start can be due to the random initialization of the kernel weights, which performs a meaningless convolution, a case where the quantum sampling of the output is of no interest. However it is very interesting to see that despite this late start, the kernel can start converging once they have found a good combination.

Overall, the QCNN may present some behaviors that do not have a classical equivalent. Understanding their potential effects, positive or negative, is an open question, all the more so as the effects of the classical CNN's hyperparameters are already a topic an active research \cite{samek2017explainable}. Note also that the size of the neural network used in this simulation is rather small. A following experiment would be to simulate a quantum version of a standard deeper CNN (AlexNet or VGG-16), eventually on more complex datasets, such as CIFAR-10 \cite{krizhevsky2009learning} or Fashion MNIST \cite{xiao2017fashion}.

\section{Conclusions}\label{conclusions_qcnn}
We have presented a quantum algorithm for evaluating and training convolutional neural networks (CNN). At the core of this algorithm, we have developed the first quantum algorithm for computing a convolution product between two tensors, with a substantial speed up. This technique could be reused in other signal processing tasks that would benefit an enhancement by a quantum computer. Layer by layer, convolutional neural networks process and extract meaningful information. Following this idea of learning foremost important features, we have proposed a new approach of quantum tomography where the most meaningful information is sampled with higher probability, hence reducing the complexity of our algorithm. 

Our quantum CNN is complete in the sense that almost all classical architectures can be implemented in a quantum fashion: any (non negative an upper bounded) non-linearity, pooling, number of layers and size of kernels are available. Our circuit is shallow, indeed one could repeat the main loop many times on the same shallow circuit, since performing the convolution product uses shallow linear algebra techniques, and is similar for all layers. The pooling and non-linearity are included in the loop. Our building block approach, layer by layer, allows a high modularity, and can be combined with previous works on quantum feedforward neural network \cite{allcock2020quantum} (see Section \ref{sec:quantum_fcnn}). 

The running time presents a speedup compared to the classical algorithm, due to fast linear algebra when computing the convolution product, and by only sampling the important values from the resulting quantum state. This speedup can be highly significant in cases where the number of channels $D^{\ell}$ in the input tensor is high (high dimensional time series, videos sequences, games play) or when the number of kernels $D^{\ell+1}$ is big, allowing deep architectures for CNN, which was the case in the recent breakthrough of DeepMind AlphaGo algorithm \cite{silver2016mastering}. The quantum CNN also allows larger kernels, that could be used for larger input images, since the size the kernels must be a constant fraction of the input in order to recognize patterns. However, despite our new techniques to reduce the complexity, applying a non-linearity and reusing the result of a layer for the next layer make register encoding and state tomography mandatory, hence preventing from having an exponential speedup on the number of input parameters. 

Finally we have presented a backpropagation algorithm that can also be implemented as a quantum circuit. The numerical simulations on a small CNN show that despite the introduction of noise and sampling, the QCNN can efficiently learn to classify visual data from the MNIST dataset, performing a similar accuracy than the classical CNN.

\chapter{NISQ Algorithm for Orthogonal Neural Networks}\label{chap:OrthoNN_nisq}
\epigraph{\textit{"La vision scientifique et la vision poétique, loin de s'exclure, se rejoignent pour nous faire percevoir le monde dans sa véritable richesse."}}{Hubert Reeves \\ \emph{Malicorne. Réflexions d'un observateur de la nature} (1990)}

In this chapter, we present a new training method for neural networks that preserves perfect orthogonality while having the same running time as usual gradient descent methods without the orthogonality condition, thus achieving the best of both worlds, most efficient training and perfect orthogonality.

The main idea comes from the quantum world, where we know that any quantum circuit corresponds to an operation described by a unitary matrix, which if we only use gates with real amplitudes is an orthogonal matrix. In particular, we propose a novel special-architecture quantum circuit, for which there is an efficient way to map the elements of the orthogonal weight matrix to the parameters of the gates of the quantum circuit and vice versa. In other words, while performing a gradient descent on the elements of the weight matrix individually does not preserve orthogonality, performing a gradient descent on the parameters of the quantum circuit preserves orthogonality (since any quantum circuit with real parameters corresponds to an orthogonal matrix) and is equivalent to updating the weight matrix. We also prove that performing gradient descent on the parameters of the quantum circuit can be done efficiently classically (with constant update cost per parameter) thus concluding that there exists a quantum-inspired, but fully classical way of efficiently training perfectly orthogonal neural networks.

Moreover, the special-architecture quantum circuit we defined has many properties that make it a good candidate for NISQ implementation (see Section \ref{sec:NISQ}): it uses only one type of quantum gates, requires simple connectivity between the qubits, has depth linear in the input and output node sizes, and benefits from powerful error mitigation techniques that make it resilient to noise. This allows us to also propose an inference method running the quantum circuit on data which might offer a faster running time, given the shallow depth of the quantum circuit. 

Our main contributions are summarized in Table \ref{table:runningtimes}, where we have considered the time to perform a feedforward pass, or one gradient descent step. A single neural network layer is considered, with input and output of size $n$.

\begin{table}[]
\begin{tabular}{|c|c|c|}
\hline
Algorithm                                                            & Feedforward Pass              &  Training      \\ \hline
Quantum Pyramidal Circuit (This work)                                & $2n/\delta^2 = O(n/\delta^2)$ & \multirow{2}{*}{$O(n^2)$} \\ \cline{1-2}
Classical Pyramidal Circuit (This work)                              & $2n(n-1) = O(n^2)$            &                           \\ \hline
Classical Approximated OrthoNN (SVB) \cite{jia2019orthogonal}        & $n^2 = O(n^2)$                & $O(n^3)$                  \\ \hline
Classical Strict OrthoNN (Stiefel Manifold) \cite{jia2019orthogonal} & $n^2 = O(n^2)$                & $O(n^3)$                  \\ \hline
Standard Neural Network (non orthogonal)                             & $n^2 = O(n^2)$                & $O(n^2)$                  \\ \hline
\end{tabular}
\caption{Running times summary. $n$ is the size of the input and output vectors, $\delta$ is the error parameter in the quantum implementation. See Section \ref{sec:classical_ortho_nn_preliminaries} for details on related work.\label{table:runningtimes}}
\end{table}

\section{A Parametrized Quantum Circuit for Orthogonal Neural Networks}

Preliminaries on orthogonal neural networks, along with all notations are given in Section \ref{sec:classical_ortho_nn_preliminaries}, as well as Section \ref{sec:unary_data_loaders} for data loaders with unary encoding.

In the following, we will define a special-architecture parametrized quantum circuit that will be useful for performing training and inference on orthogonal neural networks. As we said, the training will be completely classical in the end, but the intuition of the new method comes from this quantum circuit, while the inference can happen both classically or by applying this quantum circuit.

\subsection{The $RBS$ Gate}\label{sec:RBSgate}

\begin{figure}[h]
    \centering
    \includegraphics[width=0.7\textwidth]{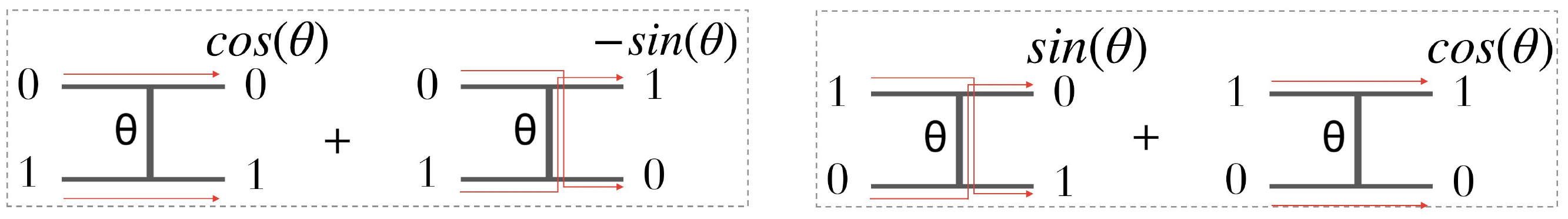}
    \caption{Representation of the quantum mapping from Eq.(\ref{RBSgate}) on two qubits.}
    \label{fig:RBS_representation}
\end{figure}

The quantum circuit proposed in this work (see Fig.\ref{fig:RBS_representation}), which implements a fully connected neural network layer with an orthogonal weight matrix, uses only one type of quantum gate, the Reconfigurable Beam Splitter (\emph{RBS}) gate.
%, or equivalently its complex form the $iRBS$ gate. These two qubits gate are parametrizable with one angle $\theta \in [0,2\pi]$.
This two-qubit gate is parametrizable with one angle $\theta \in [0,2\pi]$. Its matrix representation is given as:
\begin{equation}\label{RBSgate}
RBS(\theta) = \begin{pmatrix}
1 & 0           & 0            & 0 \\
0 & \cos\theta & \sin\theta & 0 \\
0 & -\sin\theta & \cos\theta  & 0 \\
0 & 0           & 0            & 1
\end{pmatrix}
\quad
RBS(\theta) : 
 \begin{cases}
\ket{01} \mapsto \cos\theta\ket{01}-\sin\theta\ket{10}\\
\ket{10} \mapsto \sin\theta\ket{01}+\cos\theta\ket{10}\\
\end{cases}
\end{equation}
We can think of this gate as a rotation in the two-dimensional subspace spanned by the basis $\{\ket{01},\ket{10}\}$, while it acts as the identity in the remaining subspace $\{\ket{00},\ket{11}\}$. Or equivalently, starting with two qubits, one in the $\ket{0}$ state and the other one in the state $\ket{1}$, the qubits can be swapped or not in superposition. The qubit $\ket{1}$ stays on its wire with amplitude $\cos\theta$ or switches with the other qubit with amplitude $+\sin\theta$ if the new wire is below ($\ket{10}\mapsto\ket{01}$) or $-\sin\theta$ if the new wire is above ($\ket{01}\mapsto\ket{10}$). Note that in the two other cases ($\ket{00}$ and $\ket{11}$) the $RBS$ gate acts as identity. 

See Fig.\ref{fig:RBS_decomposition} and Section \ref{sec:real_qc_experiment_rbs} for practical circuit of the $RBS$ gate and implementation with real quantum computers.

\subsection{Quantum Pyramidal Circuit}\label{sec:pyramidal_circuit}
\begin{figure}[h]
\centering
\begin{subfigure}{.5\textwidth}
  \centering
  \includegraphics[width=\linewidth]{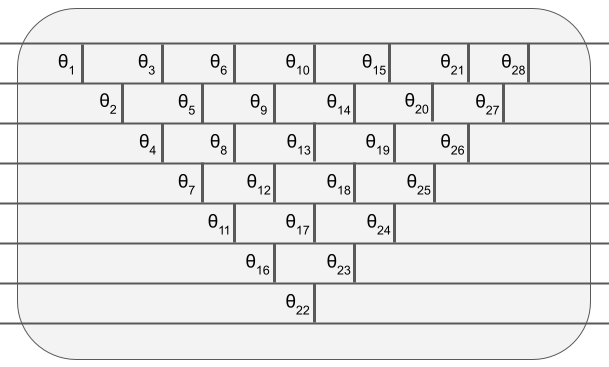}
  \caption{}
  \label{fig:QONNcircuit}
\end{subfigure}%
\begin{subfigure}{.35\textwidth}
  \centering
  \includegraphics[width=\linewidth]{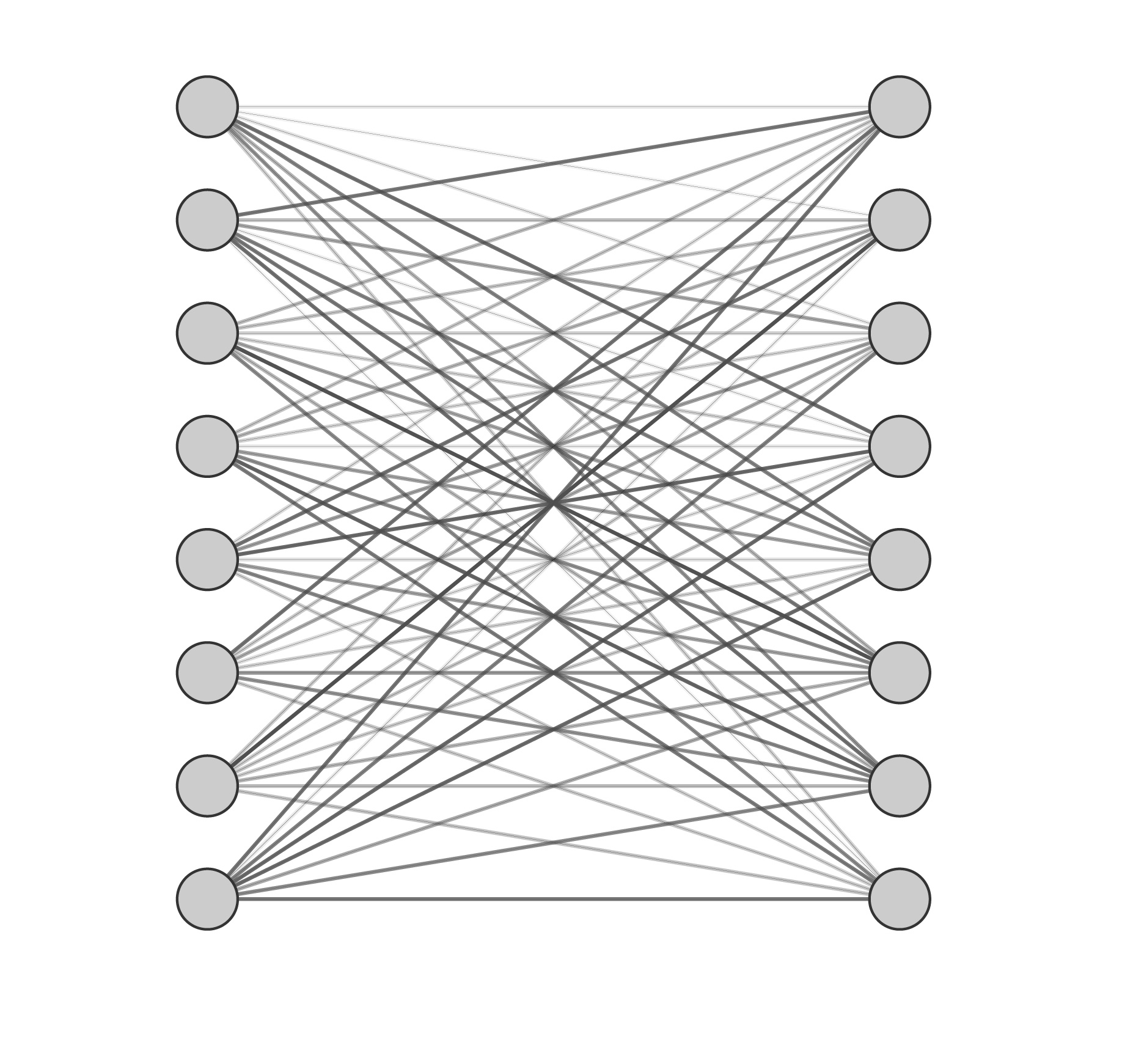}
  \caption{}
  \label{fig:8-8-nn}
\end{subfigure}
\caption{ (a) Quantum circuit for an 8x8 fully connected, orthogonal layer. Each vertical line corresponds to an $RBS$ gate with its angle parameter. And (b), the equivalent classical orthogonal neural network 8x8 layer.}
\label{fig:QONNcircuit_comparison}
\end{figure}

We now propose a quantum circuit that implements an orthogonal layer of a neural network \cite{Quantum_OrthoNN}. The circuit is a pyramidal structure of $RBS$ gates, each with an independent angle, as represented in Fig.\ref{fig:QONNcircuit}. In Section \ref{sec:data_loading} and \ref{QONN_forward}, more details are provided concerning respectively the input loading, and the equivalence with a neural network's orthogonal layer. 

\begin{figure}[h]
\centering
\begin{subfigure}{.4\textwidth}
  \centering
  \includegraphics[width=\linewidth]{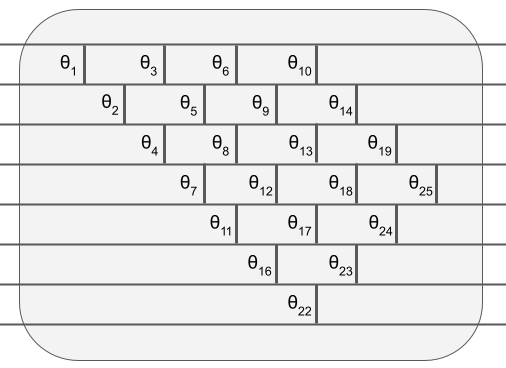}
  \caption{}
  \label{fig:QONNcircuit_rectangular}
\end{subfigure}%
\begin{subfigure}{.28\textwidth}
  \centering
  \includegraphics[width=\linewidth]{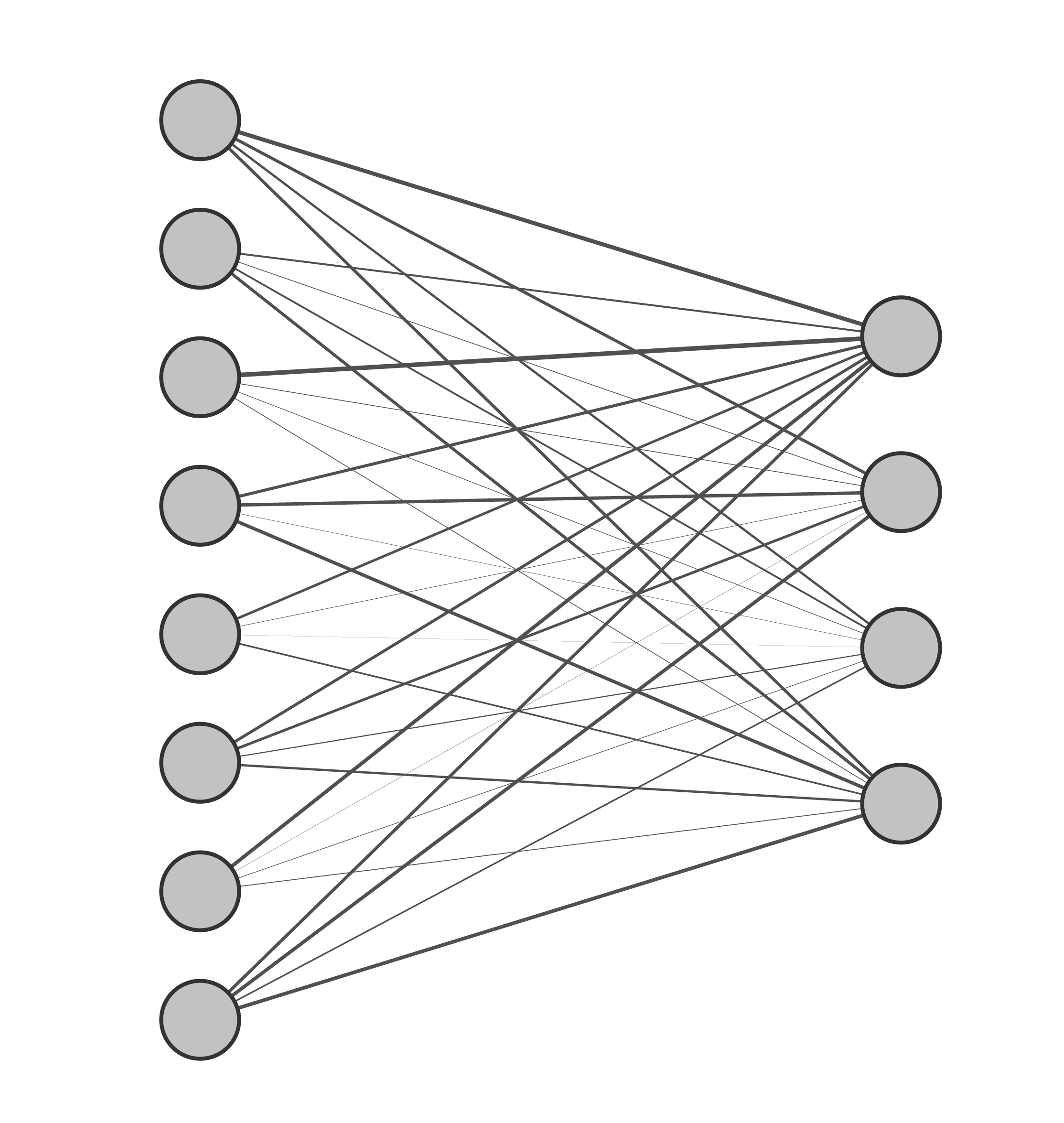}
  \caption{}
  \label{fig:8-8-nn_rectangular}
\end{subfigure}
\caption{ (a) Quantum circuit for a \emph{rectangular} 8x4 fully connected orthogonal layer, and  (b) the equivalent 8x4 classical orthogonal neural network. They both have 22 free parameters.}
\label{fig:QONNcircuit_rectangular_comparison}
\end{figure}

To mimic a given classical layer with a quantum circuit, the number of output qubits should be the size of the classical layer's output. We refer to the \emph{square case} when the input and output sizes are equal, and to the \emph{rectangular case} otherwise (Fig.\ref{fig:QONNcircuit_rectangular}). 

The important property to note is that the number of parameters of the quantum pyramidal circuit corresponding to a neural network layer of size $n \times d$ is $(2n-1-d)*d/2$ exactly the same as the number of degrees of freedom of an orthogonal matrix of dimension $n \times d$.

For simplicity, we pursue our analysis using only the \emph{square case} but everything can be easily extended to the rectangular case. As we said, the full pyramidal structure of the quantum circuit described above imposes the number of free parameters to be $n(n-1)/2$, the exact number of free parameters to specify a $n\times n$ orthogonal matrix.

In Section \ref{QONN_forward} we will show how the parameters of the gates of this pyramidal circuit can be easily related to the elements of the orthogonal matrix of size $n\times n$ that describes it. We note that alternative architectures can be imagined as long as the number of gate parameters is equal to the parameters of the orthogonal weight matrix and a simple mapping between them and the elements of the weight matrix can be found.

Note finally that this circuit has linear depth and is convenient for near term quantum hardware platforms with restricted connectivity. Indeed, the distribution of the $RBS$ gates requires only nearest neighbor connectivity between qubits.

\subsection{Loading the Data}\label{sec:data_loading}

Before applying the quantum pyramidal circuit, we will need to upload the classical data into the quantum circuit. As introduced in Section \ref{sec:unary_encoding}, we will use one qubit per feature of the data. 
For this, we use a unary amplitude encoding of the input data (see Definition \ref{def:unary_encoding}) that we will recall briefly. Let's consider an input sample $x=(x_0,\cdots,x_{n-1}) \in \R^n$, such that $\norm{x}_2=1$. We will encode it in a superposition of unary states:
\begin{equation}\label{eq:x_unary_state}
    \ket{x} = 
    x_0\ket{10\cdots0} + x_1\ket{010\cdots0} + \cdots + x_{n-1}\ket{0\cdots01}
\end{equation}
We can also rewrite the previous state as
$\ket{x} = \sum^{n-1}_{i=0} x_i\ket{e_i}$, where $\ket{e_i}$ represents the i$^{th}$ unary state with a $\ket{1}$ in the i$^{th}$ position $\ket{0\cdots010\cdots0}$.
Recent work \cite{johri2020nearest_dataloaders} proposed a logarithmic depth data loader circuit for loading such states. Here we will use a much simpler circuit. It is a linear depth cascade of $n$-1 $RBS$ gates which, due to the particular structure of our quantum pyramidal circuit, only adds 2 extra steps to our circuit. 

\begin{figure}[!h]
    \centering
    \includegraphics[width=0.6\textwidth]{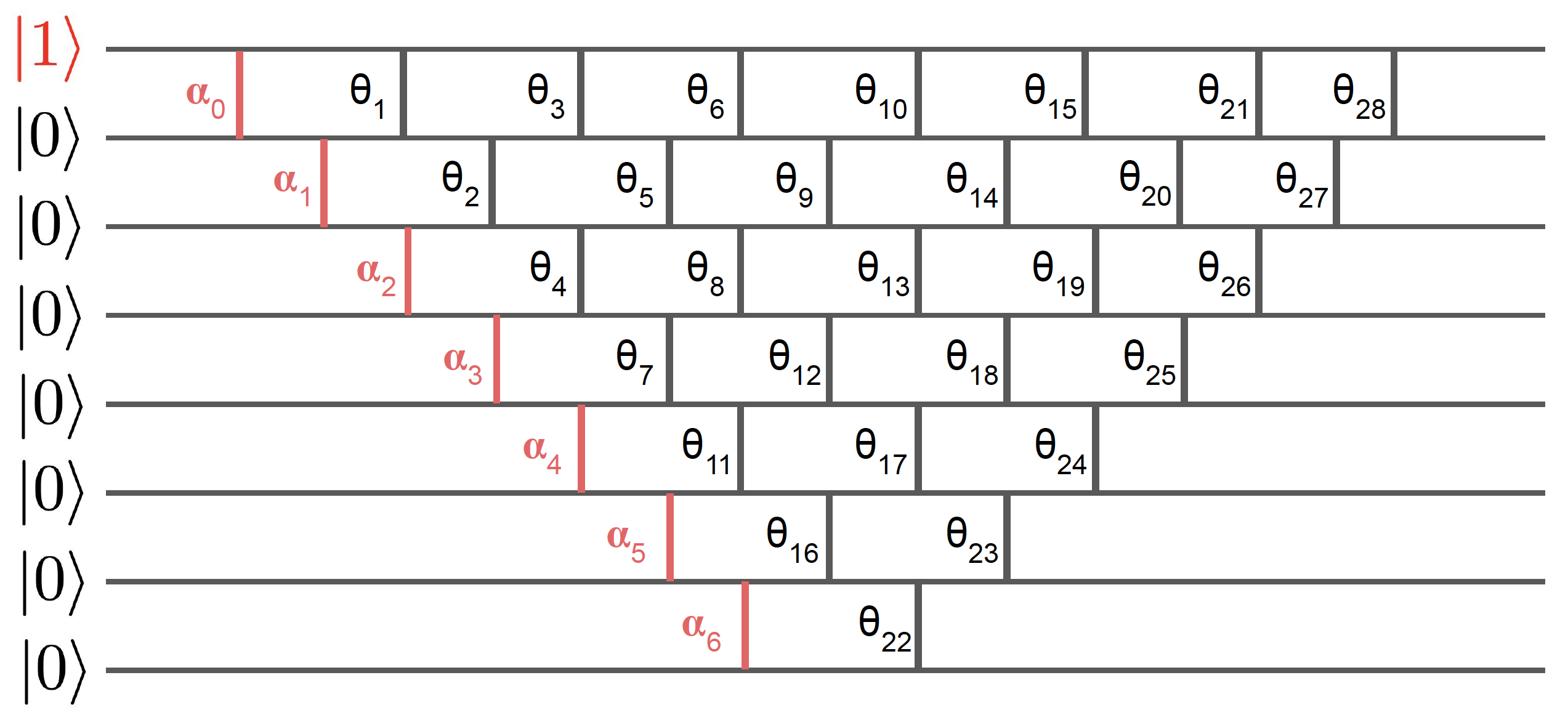}
    \caption{The 8 dimensional linear data loader circuit (in red) is efficiently embedded before the pyramidal circuit. The input state is the first unary state. The angles parameters $\alpha_0,\cdots,\alpha_{n-2}$ are classically pre-computed from the input vector.}
    \label{fig:data_loader}
\end{figure}

The circuit starts in the all $\ket{0}$ state and flips the first qubit using an $X$ gate, in order to obtain the unary state $\ket{10\cdots 0}$ as shown in Fig.\ref{fig:data_loader}. Then a cascade of $RBS$ gates allow to create the state $\ket{x}$ using a set of $n-1$ angles $\alpha_0,\cdots,\alpha_{n-2}$. Using Eq.(\ref{RBSgate}), we will choose the angles such that, after the first $RBS$ gate of the loader, the qubits would be in the state 
$
x_0\ket{100\cdots} 
+ \sin(\alpha_0)\ket{010\cdots}
$
and after the second one in the state 
$
x_0\ket{100\cdots} 
+ x_1\ket{010\cdots}
+ \sin(\alpha_0)\sin(\alpha_1)\ket{001\cdots}
$
and so on, until obtaining $\ket{x}$ as in Eq.(\ref{eq:x_unary_state}). To this end, we simply perform a classical preprocessing to compute recursively the $n$-1 loading angles, in time $O(n)$. We choose $\alpha_0 = \arccos(x_0)$, $\alpha_1 = \arccos(x_1\sin^{-1}(\alpha_0))$, $\alpha_2 = \arccos(x_2\sin^{-1}(\alpha_0)\sin^{-1}(\alpha_1))$ and so on (see Section \ref{sec:unary_data_loaders}). 

%$\alpha_0 = \arccos(x_0)$, $\alpha_1 = \arccos\left(\frac{x_1}{\sin(\alpha_0)}\right)$, $\alpha_2 = \arccos\left(\frac{x_2}{\sin(\alpha_0)\sin(\alpha_1)}\right)$ and so on. 

%\begin{equation}
%\centering
%\begin{cases}
%      \alpha_0 = \arccos(x_0)\\
%      \alpha_1 = %\arccos\left(\frac{x_1}{\sin(\alpha_0)}\right%)\\
%      \alpha_2 = %\arccos\left(\frac{x_2}{\sin(\alpha_0)\sin(\a%lpha_1)}\right) \\
%      \cdots
%    \end{cases} 
%\label{eq:data_loader_angles}
%\end{equation}

The ability of loading data in such a way relies on the assumption that each input vector is normalized, i.e. $\norm{x}_2=1$. This normalization constraint could seem arbitrary and impact the ability to learn from the data. In fact, in the case of an orthogonal neural network, this normalization shouldn't degrade the training because orthogonal weight matrices are in fact orthonormal and thus norm-preserving. Hence, changing the norm of the input vector, by diving each component by $\norm{x}_2$, in both classical and quantum settings is not a problem. The normalization would impose that each input has the same norm, or the same "luminosity" in the context of images, which can be helpful or harmful depending on the case.

\section{Orthogonal Feedforward Pass}\label{QONN_forward}

In this section, we will detail the effect of the quantum pyramidal circuit on an input encoded in a unary basis, as in Eq.(\ref{eq:x_unary_state}). We will also see in the end how to simulate this quantum circuit classically with a small overhead and thus be able to provide a fully classical scheme.

\begin{figure}[h]
    \centering
    \includegraphics[width=0.65\textwidth]{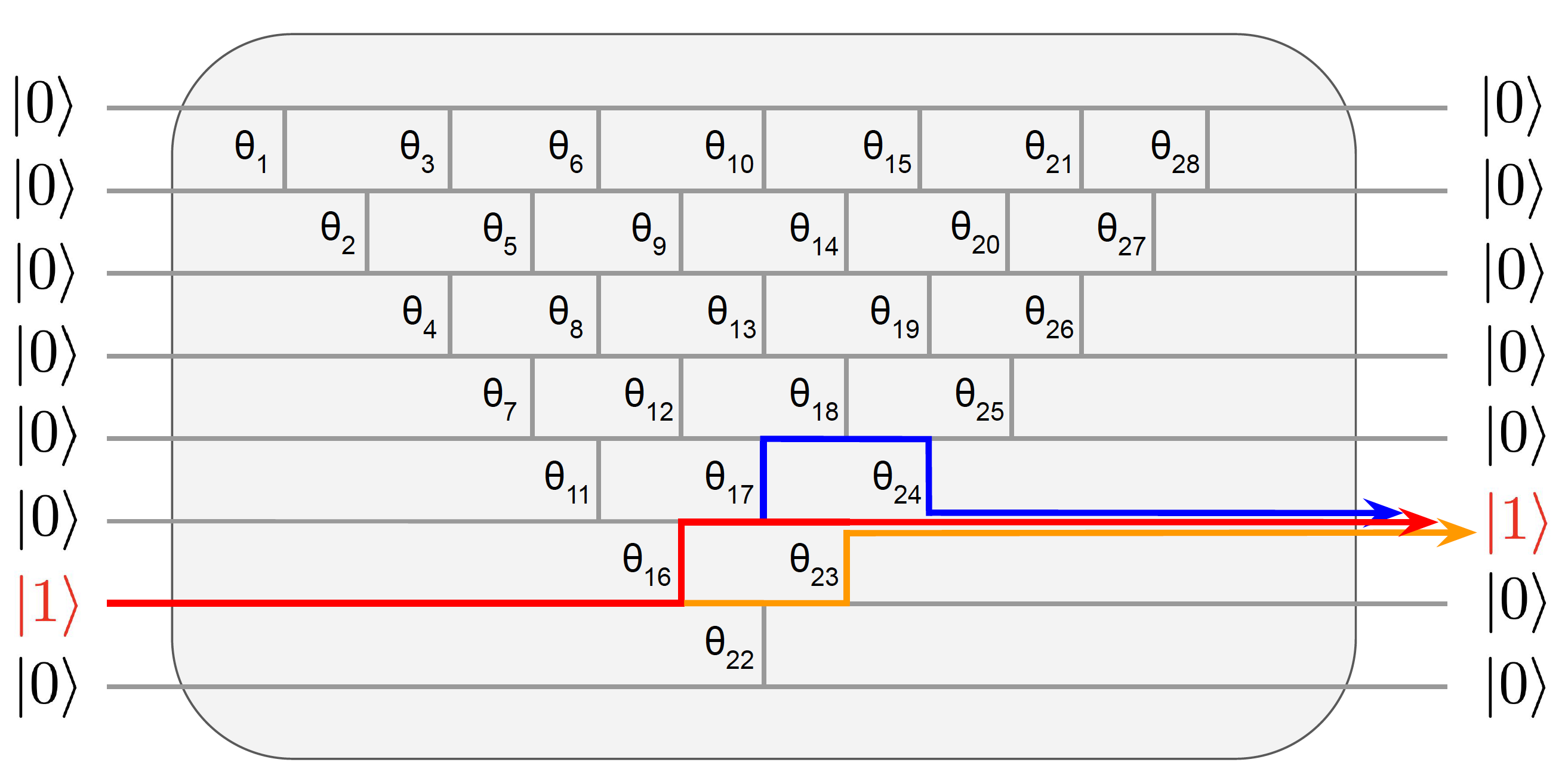}
    \caption{The three possibles paths from the $7^{th}$ unary state to the $6^{th}$ unary state, on an 8x8 quantum pyramidal circuit.}
    \label{fig:QONNcircuit_path}
\end{figure}

Let's first consider one pure unary input, where only the qubit $j$ is in state $\ket{1}$ (e.g. $\ket{00000010}$). This unary input will be transformed into a superposition of unary states, each with an amplitude. If we consider again only one of these possible unary outputs, where only the qubit $i$ is in state $\ket{1}$, its amplitude can be interpreted as a conditional amplitude to transfer the $\ket{1}$ from qubit $j$ to qubit $i$. Intuitively, this value is the sum of the quantum amplitudes associated with each possible path that \emph{connects} the qubit $j$ to qubit $i$, as shown in Fig.\ref{fig:QONNcircuit_path}. 
Using this image of \emph{connectivity} between input and output qubits, we can construct a matrix $W \in \R^{n\times n}$, where each element $W_{ij}$ is the overall conditional amplitude to transfer the $\ket{1}$ from qubit $j$ to qubit $i$. 

Fig.\ref{fig:QONNcircuit_path} shows an example where exactly three paths can be taken to map the input qubit $j=6$ (the 7$^{th}$ unary state) to the qubit $i=5$ (the 6$^{th}$ unary state). Each path comes with a certain amplitude. For instance, one of the paths (the red one in Fig.\ref{fig:QONNcircuit_path}) moves up at the first gate, and then stays put in the next three gates, with a resulting amplitude of $-\sin(\theta_{16}) \cos(\theta_{17}) \cos(\theta_{23}) \cos(\theta_{24})$. The sum of the amplitudes of all possible paths give us the element $W_{56}$ of the matrix $W$ (where, for simplicity, $s(\theta)$ and $c(\theta)$ respectively stand for $\sin(\theta)$ and $\cos(\theta)$):
%\begin{equation}
%\begin{split}
%W_{56} = 
%- \sin(\theta_{16}) \cos(\theta_{22}) \sin(\theta_{23}) %-\sin(\theta_{16}) \cos(\theta_{17}) \cos(\theta_{23}) %\cos(\theta_{24})\\
%\quad +\sin(\theta_{16}) \sin(\theta_{17}) \cos(\theta_{18}) %\sin(\theta_{24}) 
%\end{split}
%\end{equation}
\begin{equation}
W_{56} = 
- c(\theta_{16}) c(\theta_{22}) s(\theta_{23}) c(\theta_{24}) 
-s(\theta_{16}) c(\theta_{17}) c(\theta_{23}) c(\theta_{24}) 
+s(\theta_{16}) s(\theta_{17}) c(\theta_{18}) s(\theta_{24}) 
\end{equation}
%%%% 7 -> 7 subit example
%In the example from Fig.\ref{fig:QONNcircuit_path}, we see that only two possible paths can be taken to map the input qubit n°7 to itself as output. One path remains on the same wire through three consecutive gates, thus with amplitude $\cos(\theta_{15}) \cos(\theta_{22}) \cos(\theta_{23})$. The other path moves up at the first gate, goes through the second gate, and moves down at the last gate, with a resulting amplitude $-\sin(\theta_{16}) \cos(\theta_{17}) \sin(\theta_{23})$. Therefore, we would obtain the element of the weight matrix for $i=j=7$ to be 
%\begin{equation}
%W_{77} = \cos(\theta_{16}) \cos(\theta_{22}) \cos(\theta_{23}) -\sin(\theta_{16}) \cos(\theta_{17}) \sin(\theta_{23}) 
%\end{equation}
In fact, the $n\times n$ matrix $W$ can be seen as the unitary matrix of our quantum circuit if we solely consider the unary basis, which is specified by the parameters of the quantum gates. A unitary is a complex unitary matrix, but in our case, with only real operations, the matrix is simply orthogonal. This proves the correspondence between any matrix $W$ and the pyramidal quantum circuit. 

The full unitary $U_W$ in the Hilbert Space of our $n$-qubit quantum circuit is a $2^n\times 2^n$ matrix with the $n\times n$  matrix $W$ embedded in it as a submatrix on the unary basis. This is achieved by loading the data as unary states and by using only $RBS$ gates that keep the number of 0s and 1s constant.

\begin{figure}[h]
    \centering
    \includegraphics[width=\textwidth]{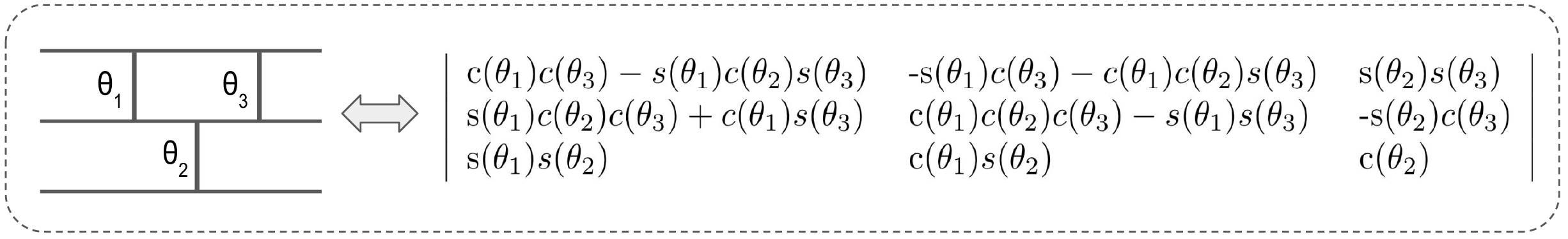}
    \caption{Example of a 3 qubits pyramidal circuit and the equivalent orthogonal matrix. $c(\theta)$ and $s(\theta)$ respectively stand for $\cos(\theta)$ and $\sin(\theta)$.}
    \label{fig:3x3_case}
\end{figure}

For instance, as shown in Fig.\ref{fig:3x3_case}, a 3-qubit pyramidal circuit is described as a unique $3\times 3$ matrix, that can be easily verified to be orthogonal.

In Fig.\ref{fig:QONNcircuit_path}, we considered the case of a single unary for both the input and output. But with actual data, as seen in Section \ref{sec:data_loading}, input and output states are in fact a superposition of unary states. Thanks to the linearity of quantum mechanics in absence of measurements, the previous descriptions remain valid and can be applied on a linear combination of unary states.

\begin{figure}[h!]
    \begin{center}
    \centering
    \includegraphics[width=\textwidth]{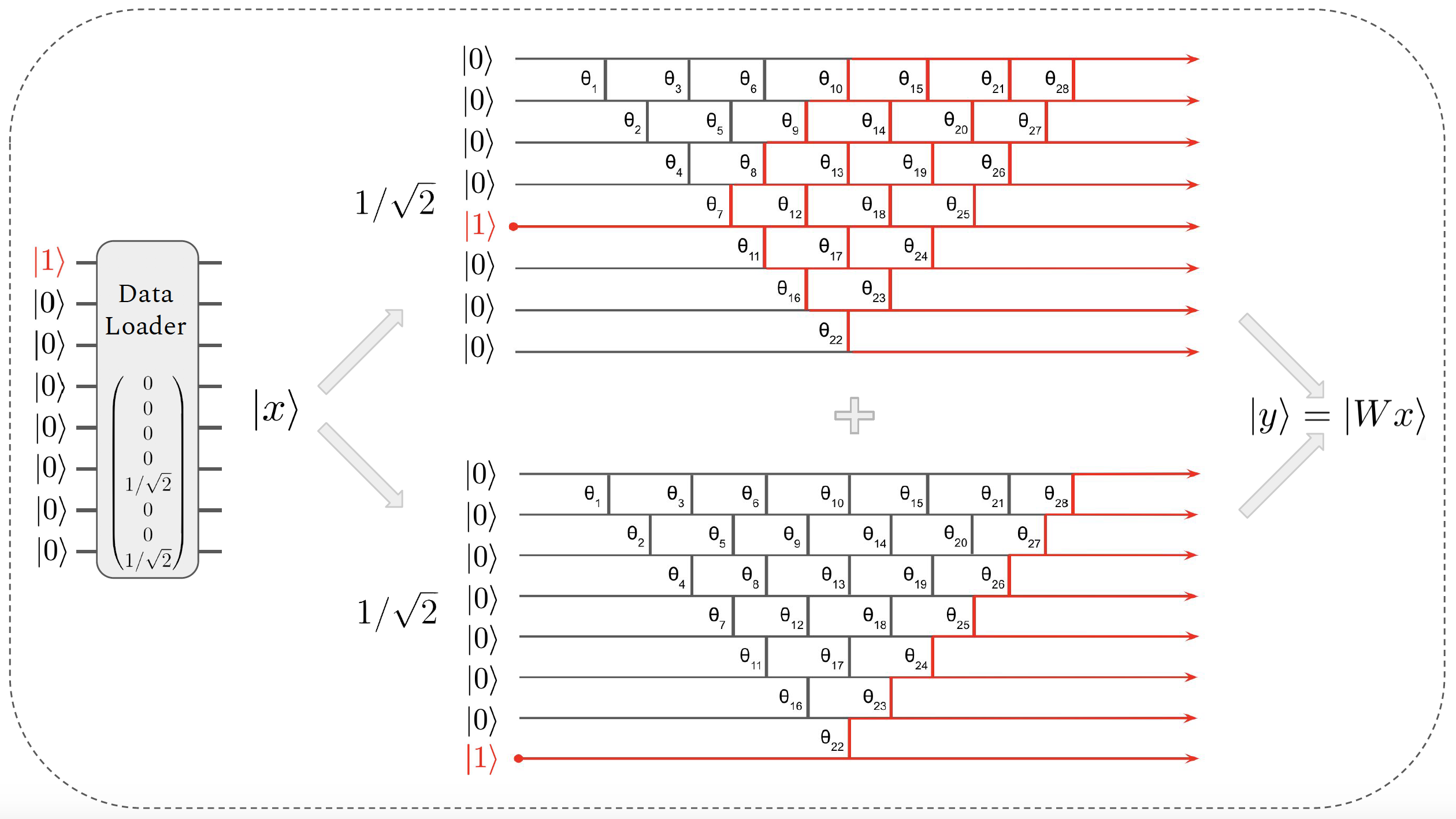}
    \caption{Schematic representation of a pyramidal circuit applied on a loaded vector $x$ with two non-zero values. The output is the unary encoding of $y=Wx$ where $W$ is the corresponding orthogonal matrix associated with the circuit.}
    \label{fig:full_schema}
    \end{center}
\end{figure}

Let's consider an input vector $x \in \R^n$ encoded as a quantum state $\ket{x} = \sum^{n-1}_{i=0} x_i\ket{e_i}$ where $\ket{e_i}$ represents the i$^{th}$ unary state (see Section \ref{sec:data_loading}). By definition of $W$, each unary $\ket{e_i}$ will undergo a proper evolution $\ket{e_i} \mapsto \sum_{j=0}^{n-1} W_{ij}\ket{e_j}$. This yields, by linearity, the following mapping 
\begin{equation}\label{eq:full_map}
\ket{x} \mapsto \sum_{i,j} W_{ij} x_i \ket{e_j}
\end{equation}
As explained above, our quantum circuit is equivalently described by the sparse unitary $U_W\in\R^{2^n\times2^n}$ or on the unary basis by the matrix $W\in\R^{n\times n}$. This can be summarized with 
\begin{equation}\label{eq:equivalence}
    U_W\ket{x} = \ket{Wx}
\end{equation}
We see from Eq.(\ref{eq:full_map}) and Eq.(\ref{eq:equivalence}) that the output is in fact $\ket{y}$, the unary encoding of the vector $y = Wx$, which is the output of a matrix multiplication between the $n\times n$ orthogonal matrix $W$ and the input $x\in\R^n$. As expected, each element of $y$ is given by $y_k = \sum_{i=0}^{n-1}W_{ik}x_i$. See Fig.\ref{fig:full_schema} for a diagram representation of this mapping.

Therefore, for any given neural network's orthogonal layer, there is a quantum pyramidal circuit that reproduces it. On the other hand, any quantum pyramidal circuit is implementing an orthogonal layer of some sort.

As a side note, we can ask if a circuit with only $\log(n)$ qubits could also implement an orthogonal matrix multiplication of size $n\times n$. Indeed, it would be a unitary matrix in $\R^{n\times n}$, but since the circuit should also have $n(n-1)/2$ free parameters to tune, this would come at a cost of large depth, potentially unsuitable for NISQ devices.

\subsubsection{ Error Mitigation}\label{sec:error}

It is important to notice that with our restriction to unary states, strong error mitigation techniques become available. Indeed, as we expect to obtain only quantum superposition of unary states at every layer, we can post process our measurements and discard the ones that present non unary states (i.e. states with more than one qubit in state $\ket{1}$, or the ground state). The most expected error is a bit-flip between $\ket{1}$ and $\ket{0}$. The case where two bit-flips happen at the same time, which would change a unary state to a different unary state and would thus pass through our error mitigation, is even less probable. This error mitigation procedure can be applied efficiently to the results of a hardware demonstration and it has been used in the results presented in this Chapter.

\subsubsection{ Extracting the classical output}\label{sec:tomography}

As shown in Fig.\ref{fig:full_schema}, when using the quantum circuit, the output is a quantum state $\ket{y} = \ket{Wx}$. As often in quantum machine learning \cite{readthefineprint}, it is important to consider the cost of retrieving the classical outputs, using a procedure called tomography. In our case, this is even more crucial since, between each layer, the quantum output will be converted into a classical one in order to apply a non-linear function, and then reloaded for the next layer.

Retrieving the amplitudes of a quantum state comes at cost of multiple measurements, which requires running the circuit multiples times, hence adding a multiplicative overhead in the running time. A finite number of samples is also a source of approximation error in the final result. In this work, we will allow for $\ell_{\infty}$ errors \cite{QCNN}. The $\ell_{\infty}$ tomography on a quantum state $\ket{y}$ with unary encoding on $n$ qubits requires $O(\log(n)/\delta^2)$ measurements, where $\delta>0$ is the error threshold allowed. For each $j\in [n]$, $|y_j|$ will be obtained with an absolute error $\delta$, and if $|y_j| < \delta$, it will most probably not be measured, hence set to 0. In practice, one would perform as many measurements as is convenient during the experiment, and deduce the equivalent precision $\delta$ from the number of measurements made.   

Note that it is important to obtain the amplitudes of the quantum state, which in our case are positive or negative real numbers, and not just the probabilities of the outcomes, which are the squares of the amplitudes. There are different ways of obtaining the sign of the amplitudes and we present two different ways below.

Indeed, a simple measurement in the computational basis will only provide us with estimations of the probabilities that are the squares of the quantum amplitudes (see Section \ref{sec:amplitude_encoding}). In the case of neural networks, it is important to obtain the sign of the layer's components in order to apply certain types of non-linearities. For instance, the ReLu activation function is often used to set all negative components to 0. 

\begin{figure}[h]
    \centering
    \includegraphics[width=\textwidth]{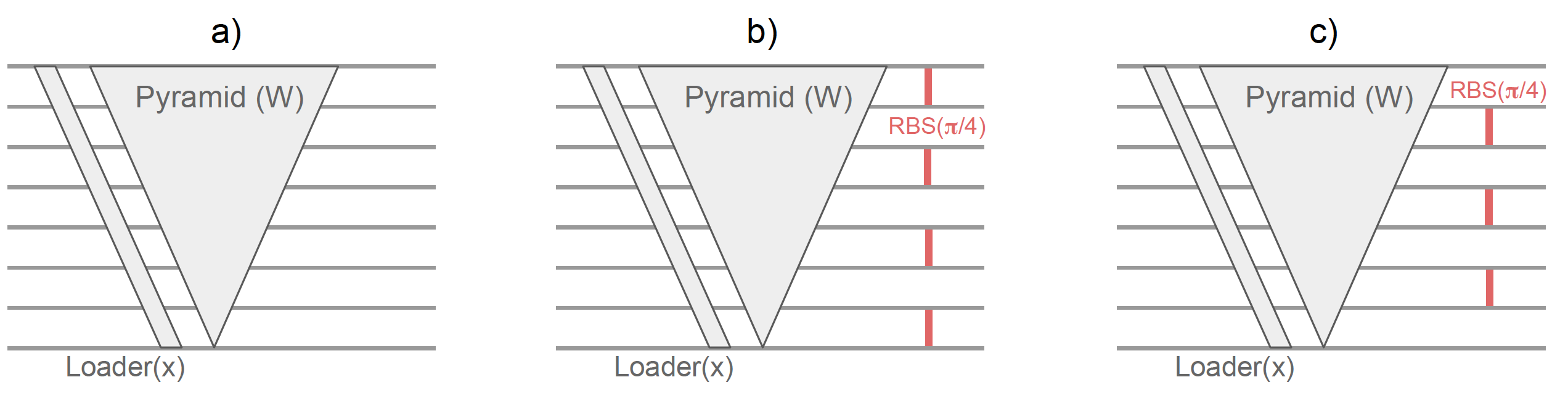}
    \caption{First tomography procedure to retrieve the value and the sign of each component of the resulting vector $\ket{y}=\ket{Wx}$. Circuit a) is the original one while circuits b) and c) have additional \emph{RBS} gates with angle $\pi/4$ at the end to compare the signs between adjacent components. In all three cases an $\ell_{\infty}$ tomography is applied.}
    \label{fig:tomography}
\end{figure}

In Fig.\ref{fig:tomography}, we propose a specific enhancement to our circuit to obtain the signs of the vector's components at low cost. The sign retrieval procedure consists of three parts. 
\begin{enumerate}[a)]
    \item The circuit is first applied as described above, allowing to retrieve each squared amplitude $y_j^2$ with precision $\delta >0$ using the $\ell_{\infty}$ tomography (Section \ref{sec:l_2_and_l_infinite_tomography}). The probability of measuring the unary state $\ket{e_1}$ (i.e. $\ket{100...}$), is $p(e_1) = y_1^2$. 
    \item We apply the same steps a second time on a modified circuit. It has additional \emph{RBS} gates with angle $\pi/4$ at the end, which will mix the amplitudes pair by pair. The probabilities to measure $\ket{e_1}$ and $\ket{e_2}$ are now given by $p(e_1) = (y_1 + y_2)^2$ and $p(e_2) = (y_1 - y_2)^2$. Therefore if $p(e_1)>p(e_2)$, we have $sign(y_1)\neq sign(y_2)$, and if $p(e_1)<p(e_2)$, we have $sign(y_1)= sign(y_2)$. The same holds for the pairs ($y_3$, $y_4$), and so on. 
    \item We finally perform the same where the \emph{RBS} are shifted by one position below. Then we compare the signs of the pairs ($y_2$, $y_3$), ($y_4$, $y_5$) and so on. 
\end{enumerate}

In the end, we are able to recover each value $y_j$ with its sign, assuming that $y_1 > 0$ for instance. This procedure has the benefit of not adding depth to the original circuit but requires 3 times more runs. The overall cost of the tomography procedure with sign retrieval is given by $\widetilde{O}(n/\delta^2)$. 

\begin{figure}[h]
    \centering
    \includegraphics[width=0.7\textwidth]{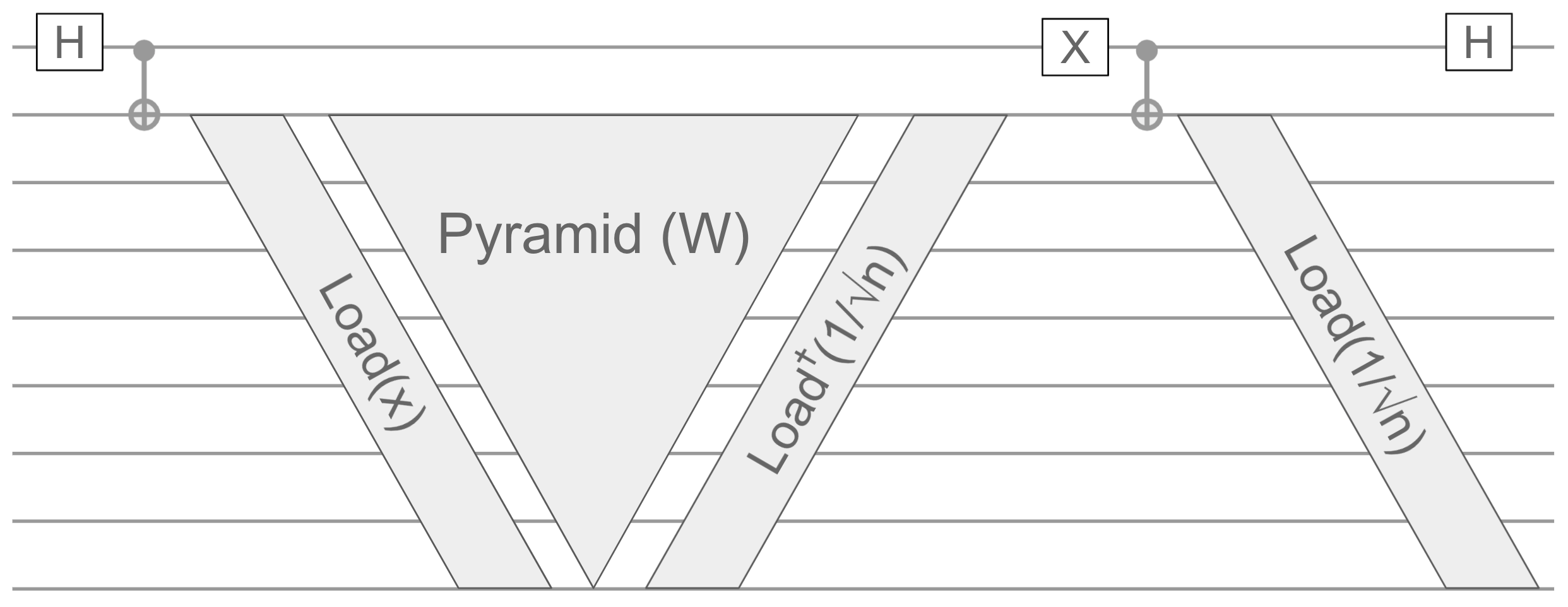}
    \caption{Second tomography procedure to retrieve the value and the sign of each component of the resulting vector $\ket{y}=\ket{Wx}$. For a \emph{rectangular} case with output of size $m$, the two opposite loaders at the end must be on the last $m$ qubits only, and the $CNOT$ gate between them connects the top qubits to the loader's top qubit as well.}
    \label{fig:tomography_2}
\end{figure}

In Fig.\ref{fig:tomography_2} we propose another method to obtain the values of the amplitudes and their signs, which is in fact what we used for the hardware demonstrations. Compared to the above procedure, it relies on one circuit only but requires an extra qubit and a depth of $3n+O(1)$ instead of $2n+O(1)$.

This circuit performs a Hadamard and CNOT gate in order to initialize the qubits in the state $\frac{1}{\sqrt{2}}\ket{0}\ket{0}+\frac{1}{\sqrt{2}}\ket{1}\ket{e_1}$, where the second register corresponds to the $n$ qubits that will be processed by the pyramidal circuit and the loaders. 

Next, applying the data loader for the normalized input vector $x$ (see Section \ref{sec:data_loading}) and the pyramidal circuit will, according to Eq.(\ref{eq:full_map}), map the state to 
\begin{equation}
    \frac{1}{\sqrt{2}}\ket{0}\ket{0}
    +\frac{1}{\sqrt{2}}\ket{1}\sum_{j=1}^n W_j x \ket{e_j}
\end{equation}

In other words, we performed the pyramid circuits controlled on the first qubit being in state $\ket{1}$. Then, we flip the fisrt qubit with an X gate and perform a controlled loading of the uniform norm-1 vector $(\frac{1}{\sqrt{n}},\cdots,\frac{1}{\sqrt{n}})$.
 For this, we add the adjoint data loader for the state, a CNOT gate and the data loader a second time.
 Recall that if a circuit $U$ is followed by $U^{\dagger}$, it is equivalent to the identity.
Therefore, this will load the uniform state only when the first qubit is in state $\ket{1}$:
\begin{equation}
    \frac{1}{\sqrt{2}}\ket{0}\sum_{j=1}^n W_j x \ket{e_j}
    +
    \frac{1}{\sqrt{2}}\ket{1}\sum_{j=1}^n \frac{1}{\sqrt{n}}\ket{e_j}
\end{equation}

Finally, a Hadamard gate will mix both parts of the amplitudes on the extra qubit to give us the desired state:
\begin{equation}
    \frac{1}{2}\ket{0}\sum_{j=1}^n \left(W_j x + \frac{1}{\sqrt{n}} \right)\ket{e_j}
    + \frac{1}{2}\ket{1}\sum_{j=1}^n \left(W_j x - \frac{1}{\sqrt{n}} \right)\ket{e_j}
\end{equation}

On this final state, we can see that the difference in the probabilities of measuring the extra qubit in state $0$ or $1$ and rest in the unary state $e_j$ is given by $\Pr[0,e_j] - \Pr[1,e_j] = \frac{1}{4}\left(W_j x + \frac{1}{\sqrt{n}}\right)^2 - \frac{1}{4}\left(W_j x - \frac{1}{\sqrt{n}}\right)^2 = W_j x /\sqrt{n} $. Therefore, for each $j$, we can deduce the sign of $W_j x$ by looking at the most frequent output of the measurement of the first qubit. To deduce as well the value of $W_j x$, we simply use $\Pr[0,e_j]$ or $\Pr[1,e_j]$ depending on the sign found before. For instance, if $W_j x>0$ we have $W_j x = 2\sqrt{\Pr[0,e_j]} - \frac{1}{\sqrt{n}}$.

Combining with the $\ell_{\infty}$ tomography and the non linearity, the overall cost of this tomography is given by $\widetilde{O}(n/\delta^2)$ as well.

\subsubsection{Multiple Quantum Layers}

\begin{figure}[h]
\centering
\begin{subfigure}[b]{0.85\textwidth}
   \includegraphics[width=1\linewidth]{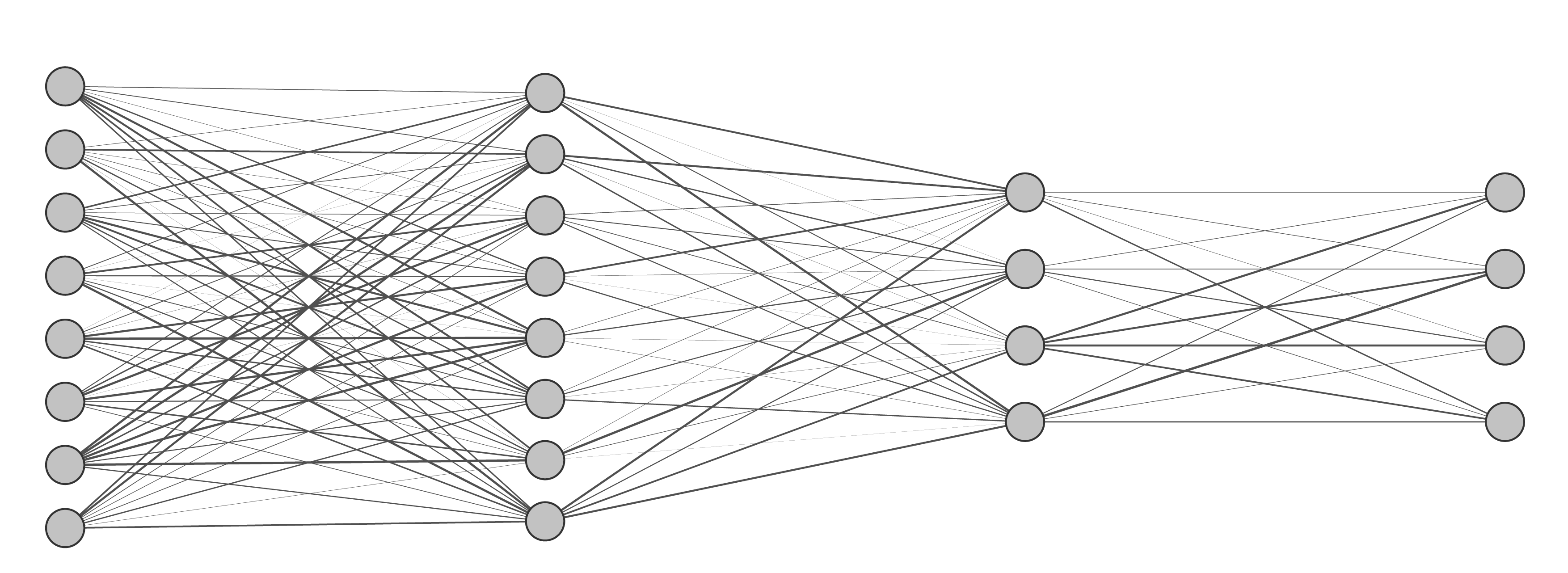}
   \caption{}
   \label{fig:8-8-5-5-nn} 
\end{subfigure}
\begin{subfigure}[b]{\textwidth}
\includegraphics[width=1\linewidth]{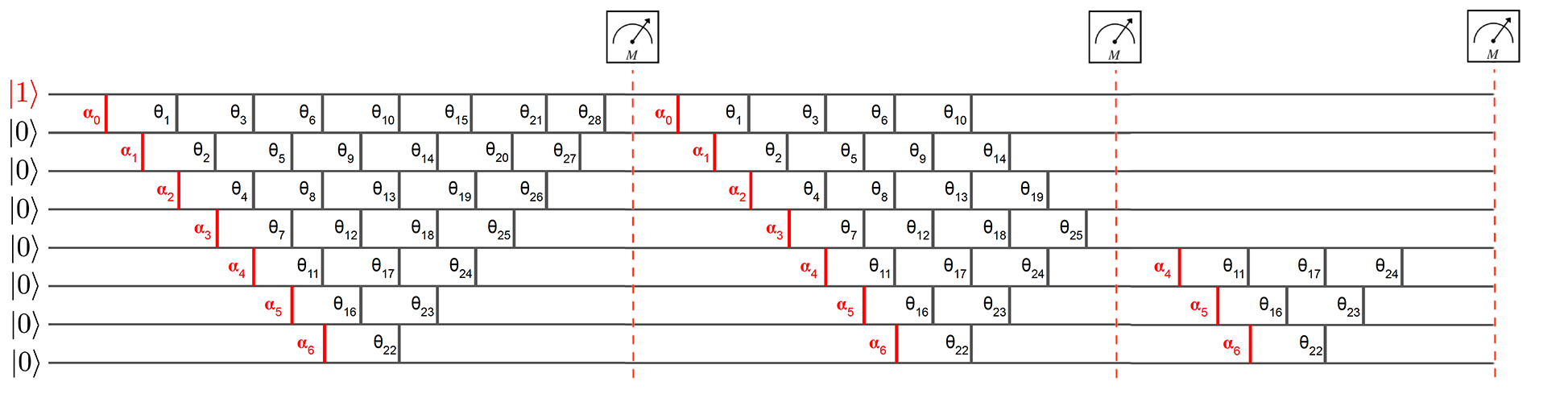}
   \caption{}
   \label{fig:8-8-5-5-Qnn}
\end{subfigure}
\caption{A full neural network with layers [8,8,4,4]. (a) Classical representation. (b) The equivalent quantum circuit is a concatenation of multiple pyramidal circuits. Between each layer one performs a measurement and applies a non-linearity. Each layer starts with a new unary data loader.}
\end{figure}

In the previous sections, we have seen how to implement a quantum circuit to perform the evolution of one orthogonal layer. In classical deep learning, such layers are stacked to gain expressivity and accuracy. Between each layer, a non-linear function is applied to the resulting vector. The presence of these non-linearities is key in the ability of the neural network to learn any function \cite{leshno1993multilayer}. 

The benefit of using our quantum pyramidal circuit is the ability to simply concatenate them to mimic a multi layer neural network. After each layer, a tomography of the output state $\ket{z}$ is performed to retrieve each component, corresponding to its quantum amplitudes (see Section \ref{sec:tomography}). A non-linear function $\sigma$ is then applied classically to obtain $a = \sigma(z)$. The next layer starts with a new unary data loader (See Section \ref{sec:data_loading}). This hybrid scheme allows as well to keep the depth of the quantum circuits reasonable for NISQ devices, by applying the neural network layer by layer.

Note that the quantum neural networks we propose here are close to the behavior of classical neural networks and thus we can control and understand the quantum mapping and implement each layer and its non-linearities in a modular way. They are also different regarding the training strategies which are close to the classical ones but use a different optimization landscape that can provide different models (see Section \ref{sec:trainingquantum} for details). It will be interesting to compare our pyramidal circuit to a quantum variational circuit with $n$ qubits and $n(n-1)/2$ gates of any type, as we usually see in the literature. Using such circuits we would explore among all possible $2^n\times 2^n$ matrices instead of $n\times n$ classical orthogonal matrices, but so far there's no theoretical ground to explain why this should provide an advantage.

As an open outlook, one could imagine incorporating additional entangling gates after each pyramid layer (composed, for instance, of $CNOT$ or $CZ$). This would mark a step out of the unary basis and could effectively allow exploring more interactions in the Hilbert Space.

\paragraph{Classical implementation} 

While we presented the quantum pyramidal circuit as the inspiration of the new methods for orthogonal neural networks, it is not hard to see that these quantum circuits can be simulated classical with a small overhead, thus yielding classical methods for orthogonal neural networks.

This classical algorithm is simply the simulation of the quantum pyramidal circuit, where each $RBS$ gate is replaced by a planar rotation between its two inputs.

\begin{figure}[h!]
    \centering
    \includegraphics[width=\textwidth]{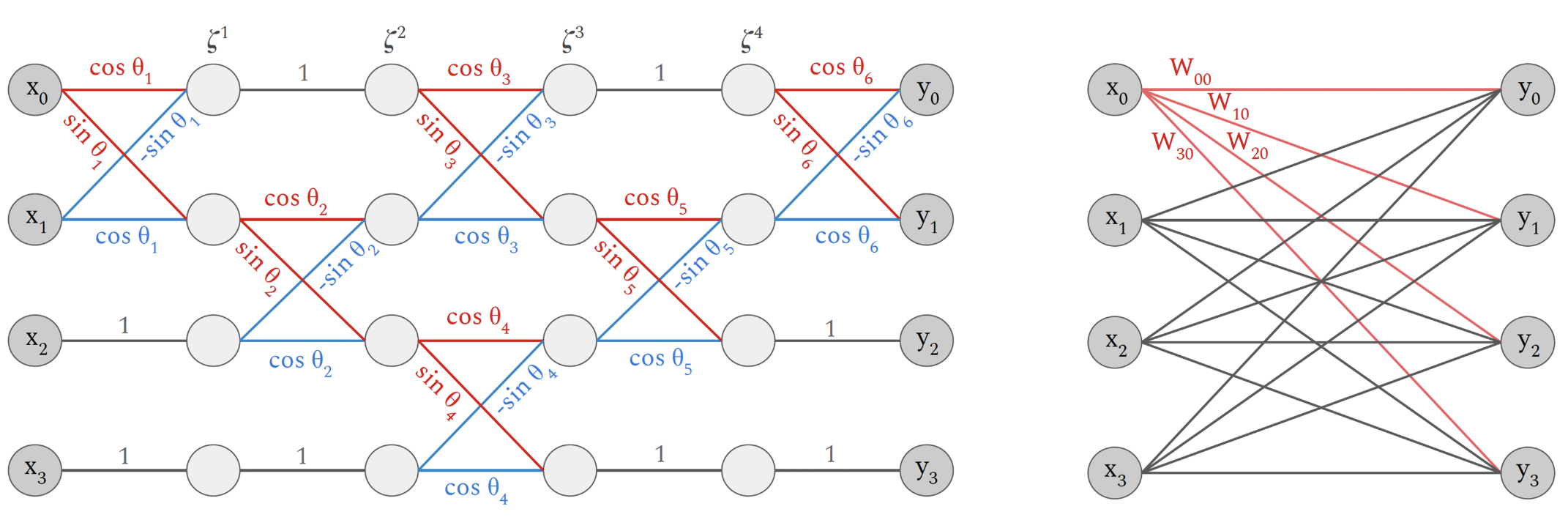}
    \caption{Classical representation of a single orthogonal layer on a 4x4 case ($n$=4) performing $x\mapsto y=Wx$. The angles and the weights can be chosen such that our classical pyramidal circuit (left) and normal classical network (right) are equivalent. Each connecting line represents a scalar multiplication with the value indicated. On the classical pyramidal circuit (left), \emph{inner layers} $\zeta^{\lambda}$ are displayed. A \emph{timestep} corresponds to the lines in between two \emph{inner layers} (see Section \ref{sec:trainingquantum} for definitions).}
    \label{fig:quantum_vs_classical_representation}
\end{figure}

As shown in Fig.\ref{fig:quantum_vs_classical_representation}, we propose a similar classical pyramidal circuit, where each layer is constituted of $\frac{n(n-1)}{2}$ planar rotations, for a total of $4\times\frac{n(n-1)}{2} = O(n^2)$ basic operations. Therefore our single layer feedforward pass has the same complexity $O(n^2)$ as the usual matrix multiplication.

One may still have an advantage in performing the quantum circuit for inference, since the quantum circuit has depth $O(n)$, instead of the $O(n^2)$ classical complexity of the matrix-vector multiplication. In addition, as we will see below, the main advantage of our method is that we can also now train orthogonal weight matrices classically in time $O(n^2)$, instead of the previously best-known $O(n^3)$.

\section{Backpropagation for the Orthogonal Neural Network Circuit}\label{sec:trainingquantum}

\subsubsection{Classical Backpropagation Algorithm}\label{sec:trainingclassical}
Basic introduction and notation to the backpropagation in fully connected neural networks are given in Section \ref{sec:bacpropagation_fcnn}. We recall some facts and notations that will be useful for the following;

The backpropagation in a fully connected neural network is a well known and efficient procedure to update the weight matrix at each layer \cite{hecht1992theory, rojas1996backpropagation}. At layer $\ell$, we denote its weight matrices by $W^{\ell}$ and biases by $b^{\ell}$. Each layer is followed by a non-linear function $\sigma$, and can therefore be written as 
\begin{equation}
a^{\ell} = \sigma(W^{\ell}\cdot a^{\ell-1} + b^\ell) =  \sigma(z^{\ell})
\end{equation}
After the last layer, one can define a cost function $\mathcal{C}$ that compares the output to the ground truth. The goal is to calculate the gradient of $\mathcal{C}$ with respect to each weight and bias, namely 
$\frac{\partial \mathcal{C}}{\partial W^{\ell}}$ and $\frac{\partial \mathcal{C}}{\partial b^{\ell}}$. In the backpropagation, we start by calculating these gradients for the last layer, then propagate back to the first layer. 

We will require to obtain the \emph{error} vector at layer $\ell$ defined by $\Delta^{\ell} = \frac{\partial \mathcal{C}}{\partial z^{\ell}}$. One can show the backward recursive relation $\Delta^{\ell} = (W^{\ell+1})^T\cdot \Delta^{\ell+1}\odot \sigma'(z^{\ell})$, where $\odot$ symbolizes the Hadamard product, or entry-wise multiplication. Note that the previous computation requires simply to apply the layer (i.e. apply matrix multiplication) in reverse. We can then show that each element of the weight gradient matrix at layer $\ell$ is given by $\frac{\partial \mathcal{C}}{\partial W^{\ell}_{jk}} = \Delta^{\ell}_j\cdot a^{\ell-1}_1$. Similarly, the gradient with respect to the biases is easily defined as $\frac{\partial \mathcal{C}}{\partial b^{\ell}_{j}} = \Delta^\ell_j$. 

Once these gradients are computed, we update the parameters using the gradient descent rule, with learning rate $\lambda$:
\begin{equation}\label{gradient_descent}
\centering
W^{\ell}_{jk} \gets W^{\ell}_{jk} -\lambda \frac{\partial \mathcal{C}}{\partial W^{\ell}_{jk}} \quad;\quad
b^{\ell}_{j} \gets b^{\ell}_{j} -\lambda \frac{\partial \mathcal{C}}{\partial b^{\ell}_{j}}
\end{equation}

\subsubsection{Backpropagation for Pyramidal Circuits}

Looking through the prism of our pyramidal quantum circuit, the parameters to update are no longer the individual elements of the weight matrices directly, but the angles of the RBS gates that give rise to these matrices. Thus, we need to adapt the backpropagation method to our setting based on the angles. We will start by introducing some notation for a single layer $\ell$, which will not be explicit in the notation for simplicity. We assume we have as many output bits as input bits, but this can easily be extended to the \emph{rectangular} case. 

\begin{figure}[h]
    \centering
    \includegraphics[width=0.7\textwidth]{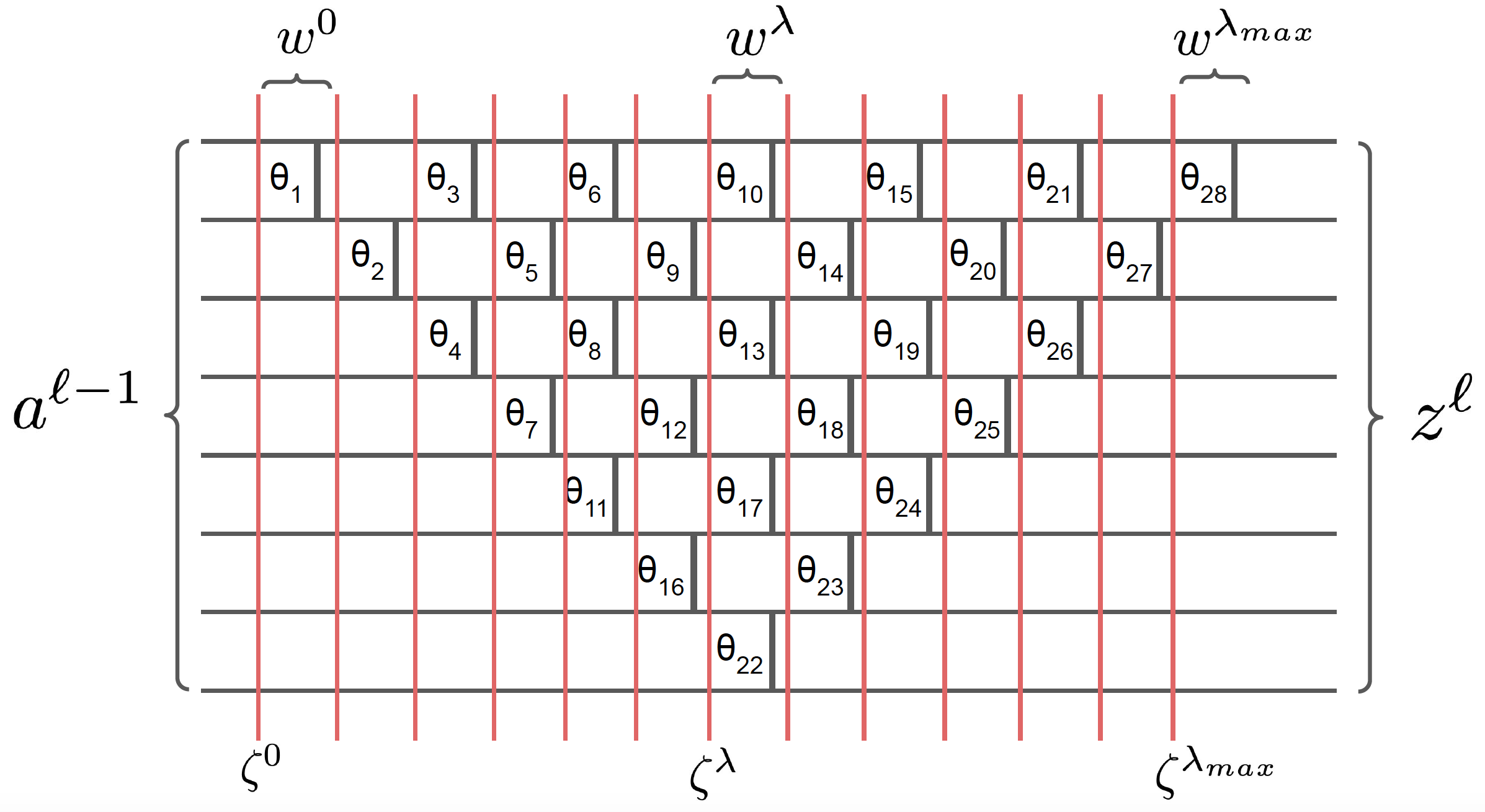}
    \caption{Quantum circuit for one neural network layer divided into \emph{timesteps} (red vertical lines) $\lambda \in [0,\cdots,\lambda_{max}]$. Each timestep corresponds to an \emph{inner layer} $\zeta^{\lambda}$ and an \emph{inner error} $\delta^{\lambda}$. The part of the circuit between two timesteps is an unitary matrix $w^{\lambda}$ in the unary basis.}
    \label{fig:QONNcircuit_timesteps}
\end{figure}

We first introduce the notion of \emph{timesteps} inside each layer, which correspond to the computational steps in the pyramidal structure of the circuit (see Fig.\ref{fig:QONNcircuit_timesteps}). 
It is easy to show that for $n$ inputs, there will be $2n-3$ such \emph{timesteps}, each one indexed by an integer $\lambda \in [0,\cdots,\lambda_{max}]$. 
Applying a timestep consists in applying the matrix $w^{\lambda}$, made of all the RBS gates aligned vertically at this timestep ($w^{\lambda}$ is the unitary in the unary basis, see Section \ref{QONN_forward} for details). Each time a timestep is applied, the resulting state is a vector in the unary basis named \emph{inner layer} and denoted by $\zeta^\lambda$. This evolution can be written as $\zeta^{\lambda+1} = w^{\lambda}\cdot\zeta^{\lambda}$. We use this notation similar to the real layer $\ell$, with the weight matrix $W^{\ell}$ and the resulting vector $z^\ell$ (see Section \ref{sec:trainingclassical}). 

In fact we have the correspondences $\zeta^{0} = a^{\ell-1}$ for the first \emph{inner layer}, which is the input of the actual layer, and $z^\ell = w^{\lambda_{max}}\cdot\zeta^{\lambda_{max}}$ for the last one. We also have $W^\ell = w^{\lambda_{max}}\cdots w^1w^0$.
We use the same kind of notation for the backpropagation errors. At each timestep $\lambda$ we define an \emph{inner error} $\delta^{\lambda} = \frac{\partial \mathcal{C}}{\partial \zeta^{\lambda}}$. This definition is similar to the layer error $\Delta^{\ell} = \frac{\partial \mathcal{C}}{\partial z^{\ell}}$. 
In fact we will use the same backpropagation formulas, without non-linearities, to retrieve each \emph{inner error} vector $\delta^{\lambda} = (w^\lambda)^T\cdot \delta^{\lambda+1}$. In particular, for the last timestep, the first to be calculated, we have $\delta^{\lambda_{max}} = (w^{\lambda_{max}})^T\cdot \Delta^\ell$. Finally, we can retrieve the error at the previous layer $\ell-1$ using the correspondence $\Delta^{\ell-1} = \delta^{0} \odot \sigma'(z^\ell)$.

The reason for this breakdown into timesteps is the ability to efficiently obtain the gradient with respect to each angle. Let's consider the timestep $\lambda$ and one of its gate with angle denoted by $\theta_i$ acting on qubits $i$ and $i+1$ (note that the numbering is different from Fig.\ref{fig:QONNcircuit_timesteps}). We will decompose the gradient $\frac{\partial \mathcal{C}}{\partial \theta_i}$ using each component, indexed by the integer $k$, of the \emph{inner layer} and \emph{inner error} vectors: 
\begin{equation}
\frac{\partial \mathcal{C}}{\partial \theta_i} = 
\sum_k
\frac{\partial \mathcal{C}}{\partial \zeta^{\lambda+1}_k}
\frac{\partial \zeta^{\lambda+1}_k}{\partial \theta_i}
= 
\sum_k
\delta^{\lambda+1}_k
\frac{\partial (w^\lambda_k\cdot\zeta^{\lambda})}{\partial \theta_i}
\end{equation}
Where $w^\lambda_k$ is the k$^{th}$ row of matrix $w^\lambda$. 
Since timestep $\lambda$ is only composed of separated RBS gates, the matrix $w^\lambda$ consists of diagonally arranged 2x2 block submatrices given in Eq.(\ref{RBSgate}). Only one of these submatrices depends on the angle $\theta_i$ considered here, at the position $i$ and $i+1$ in the matrix. We can thus rewrite the above gradient as $
\frac{\partial \mathcal{C}}{\partial \theta_i} = 
\delta^{\lambda+1}_i
\frac{\partial}{\partial \theta_i}\left(w^\lambda_i\cdot\zeta^{\lambda}\right)
+
\delta^{\lambda+1}_{i+1}
\frac{\partial}{\partial \theta_i}\left(w^\lambda_{i+1}\cdot\zeta^{\lambda}\right)
$, or:

\begin{equation}
\frac{\partial \mathcal{C}}{\partial \theta_i} = 
\delta^{\lambda+1}_i
\frac{\partial}{\partial \theta_i}\left(\cos(\theta_i)\zeta^{\lambda}_i+\sin(\theta_i)\zeta^{\lambda}_{i+1}\right)
+
\delta^{\lambda+1}_{i+1}
\frac{\partial}{\partial \theta_i}\left(-\sin(\theta_i)\zeta^{\lambda}_i+\cos(\theta_i)\zeta^{\lambda}_{i+1}\right)
\end{equation}

\begin{equation}\label{eq:gradient_formula_final}
\frac{\partial \mathcal{C}}{\partial \theta_i} = 
\delta^{\lambda+1}_i
(-\sin(\theta_i)\zeta^{\lambda}_i+\cos(\theta_i)\zeta^{\lambda}_{i+1})
+
\delta^{\lambda+1}_{i+1}
(-\cos(\theta_i)\zeta^{\lambda}_i-\sin(\theta_i)\zeta^{\lambda}_{i+1})
\end{equation}

%Therefore we have shown a way to compute each angle gradient: During the feedforward pass, one must apply sequentially each of the $2n-3 = O(n)$ timesteps, and store the resulting vectors, the \emph{inner layers} $\zeta^\lambda$. During the backpropagation, one obtains the \emph{inner errors} $\delta^\lambda$ by applying the timesteps in reverse. One can finally use a gradient descent on each angle $\theta_i$, while preserving the orthogonality of the overall equivalent weight matrix$\theta_i^{\ell} \gets \theta_i^{\ell} -\lambda \frac{\partial \mathcal{C}}{\partial \theta_i^{\ell}}$.

%An interesting aspect of this gradient descent is the fact that the optimization is performed in the angle landscape, and not on the equivalent weight landscape. These landscapes can potentially be different and hence our optimization can produce different models. We leave open the question of finding a theoretical argument to compare the properties of both landscapes.  

Therefore we have shown a way to compute each angle gradient: During the feedforward pass, one must apply sequentially each of the $2n-3 = O(n)$ timesteps, and store the resulting vectors, the \emph{inner layers} $\zeta^\lambda$. During the backpropagation, one obtains the \emph{inner errors} $\delta^\lambda$ by applying the timesteps in reverse. 
One can finally use a gradient descent on each angle $\theta_i$, while preserving the orthogonality of the overall equivalent weight matrix
%\begin{equation}\label{gradient_descent_angle}
%\centering
$
\theta_i^{\ell} \gets \theta_i^{\ell} -\lambda \frac{\partial \mathcal{C}}{\partial \theta_i^{\ell}}
$.
%\end{equation}
Since the optimization is performed in the angle landscape, and not on the equivalent weight landscape, it can potentially be different and produce different models. We leave open the study of the properties of both landscapes.

As one can see from the above description, this is in fact a classical algorithm to obtain the angle's gradients, which allows us to train our OrthoNN efficiently classically while preserving the strict orthogonality. To obtain the angle's gradient, one needs to store the $2n$-3 \emph{inner layers} $\zeta^{\lambda}$ during the feedforward pass. Next, given the error at the following layer, we perform a backward loop on  each \emph{timestep} (see Fig.\ref{fig:quantum_vs_classical_representation}). At each \emph{timestep}, we obtain the gradient for each angle parameter, by simply applying Eq.(\ref{eq:gradient_formula_final}). This requires $O(1)$ operations for each angle. Since there are at most $n/2$ angles per \emph{timesteps}, estimating gradients has a complexity of $O(n^2)$. After each \emph{timestep}, the next \emph{inner error} $\delta^{\lambda-1}$ is computed as well, using at most $4n/2$ operations. 

In the end, our classical algorithm allows us to compute the gradients of the $n(n-1)/2$ angles in time $O(n^2)$, thus performing a gradient descent respecting the strict orthogonality of the weight matrix in the same time. 
This is considerably faster than previous methods based on Singular Value Decomposition methods and provides a training method that is asymptotically as fast as for normal neural networks, while providing the extra property of orthogonality.

\section{Numerical Experiments}\label{sec:numerical_exp}
\begin{figure}[H] % "[t!]" placement specifier just for this example
\begin{subfigure}{0.48\textwidth}
\includegraphics[width=\linewidth]{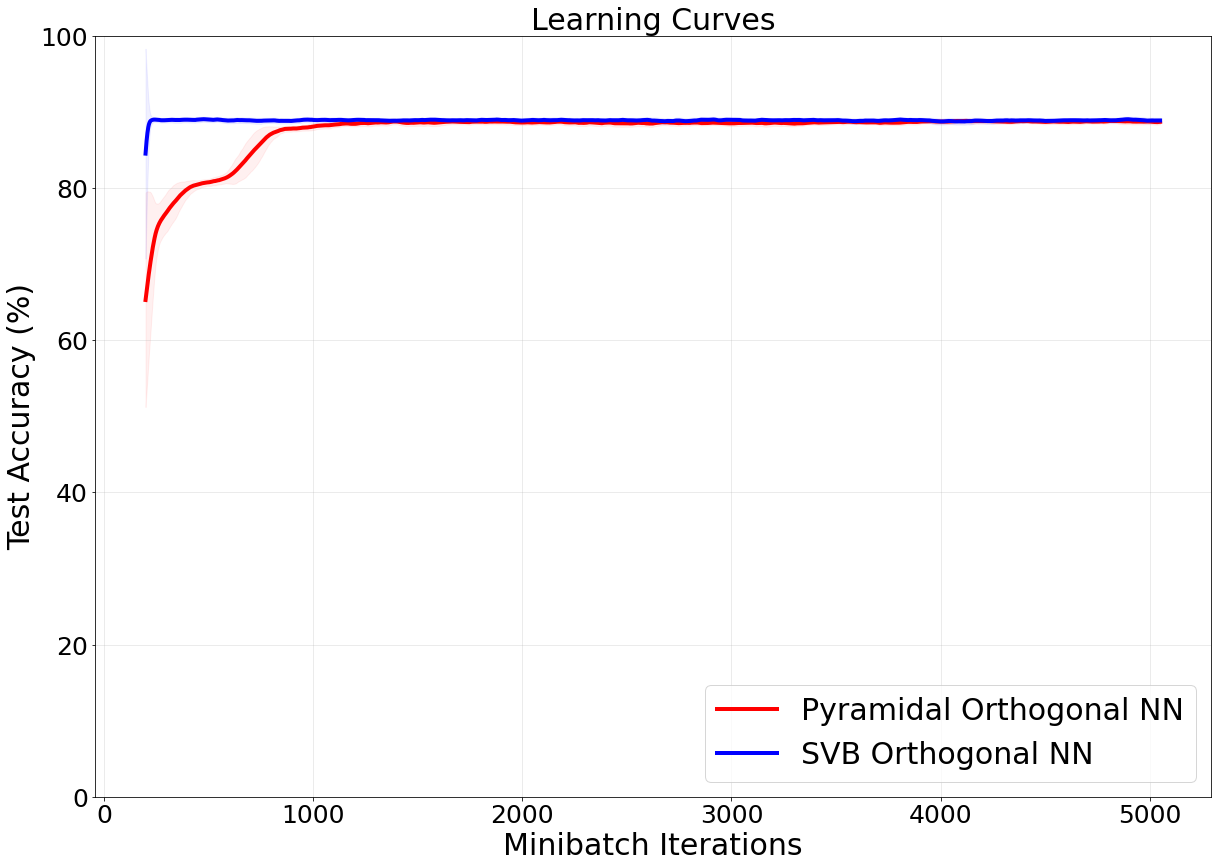}
\caption{[16,4]} 
\end{subfigure}\hspace*{\fill}
\begin{subfigure}{0.48\textwidth}
\includegraphics[width=\linewidth]{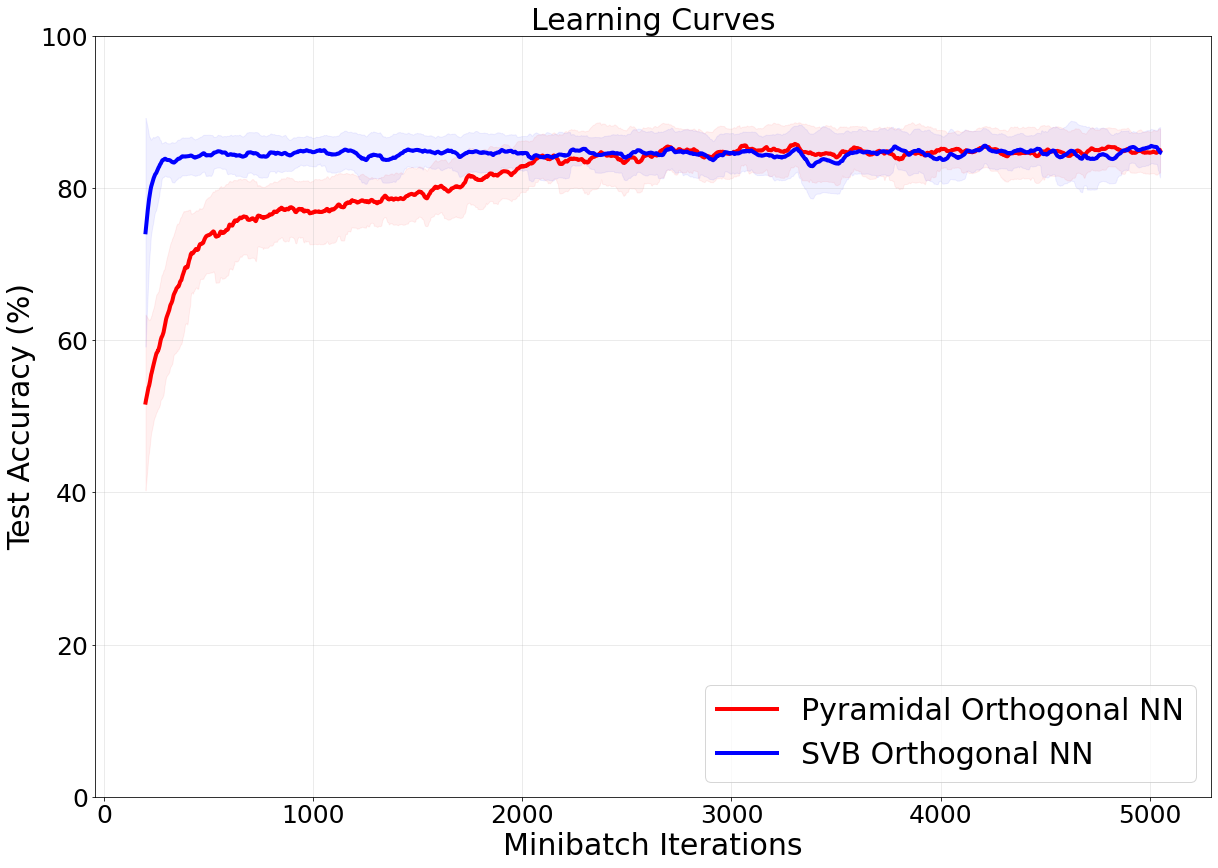}
\caption{[16,8,4]} 
\end{subfigure}
\medskip
\begin{subfigure}{0.48\textwidth}
\includegraphics[width=\linewidth]{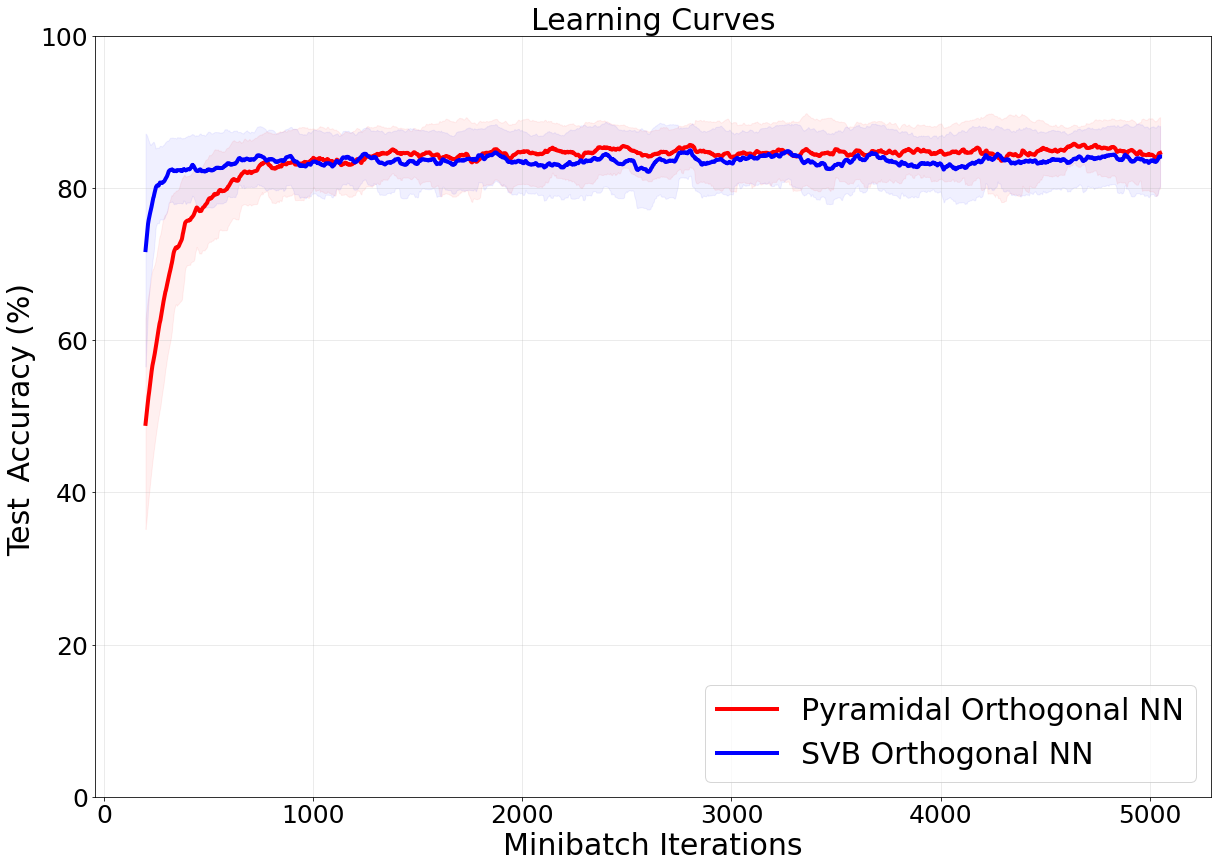}
\caption{[16,16,4]} 
\end{subfigure}\hspace*{\fill}
\begin{subfigure}{0.48\textwidth}
\includegraphics[width=\linewidth]{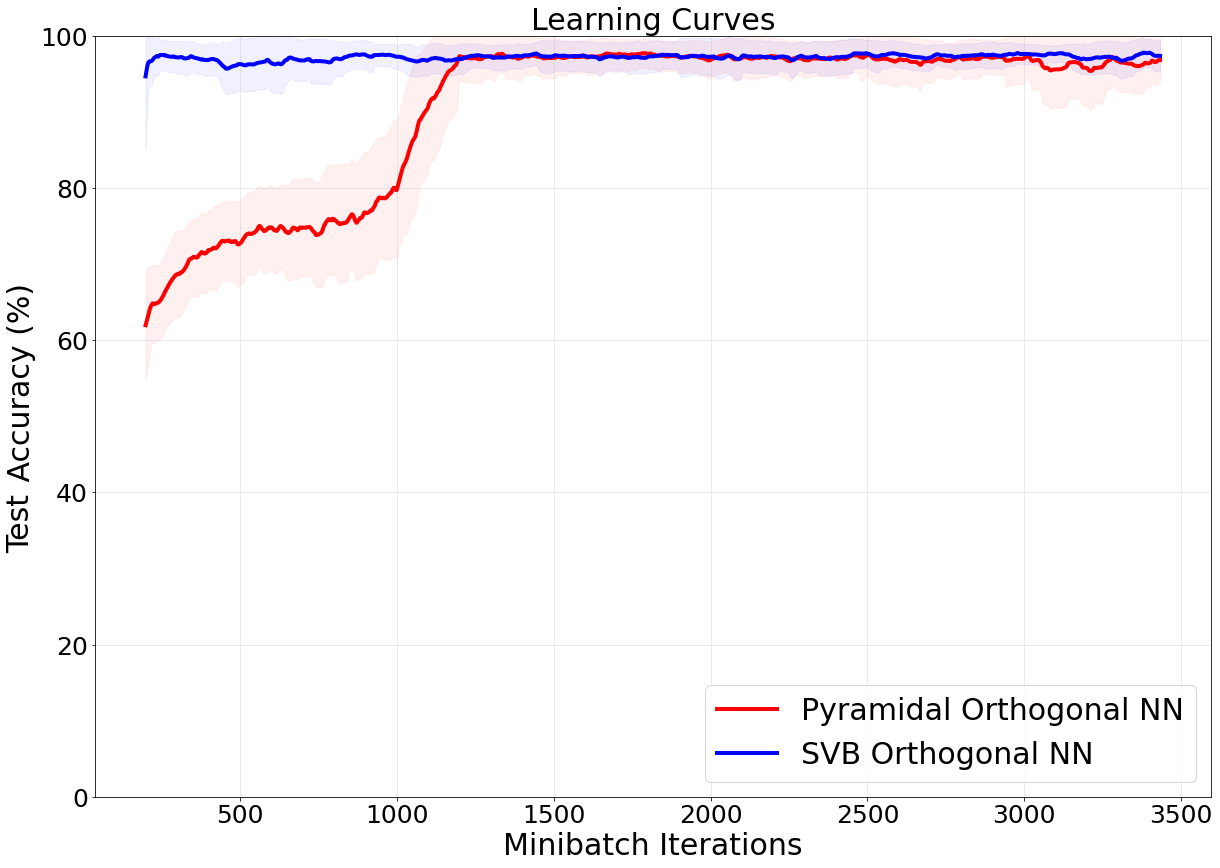}
\caption{[32,8,2]} 
\end{subfigure}
\medskip
\begin{subfigure}{0.48\textwidth}
\includegraphics[width=\linewidth]{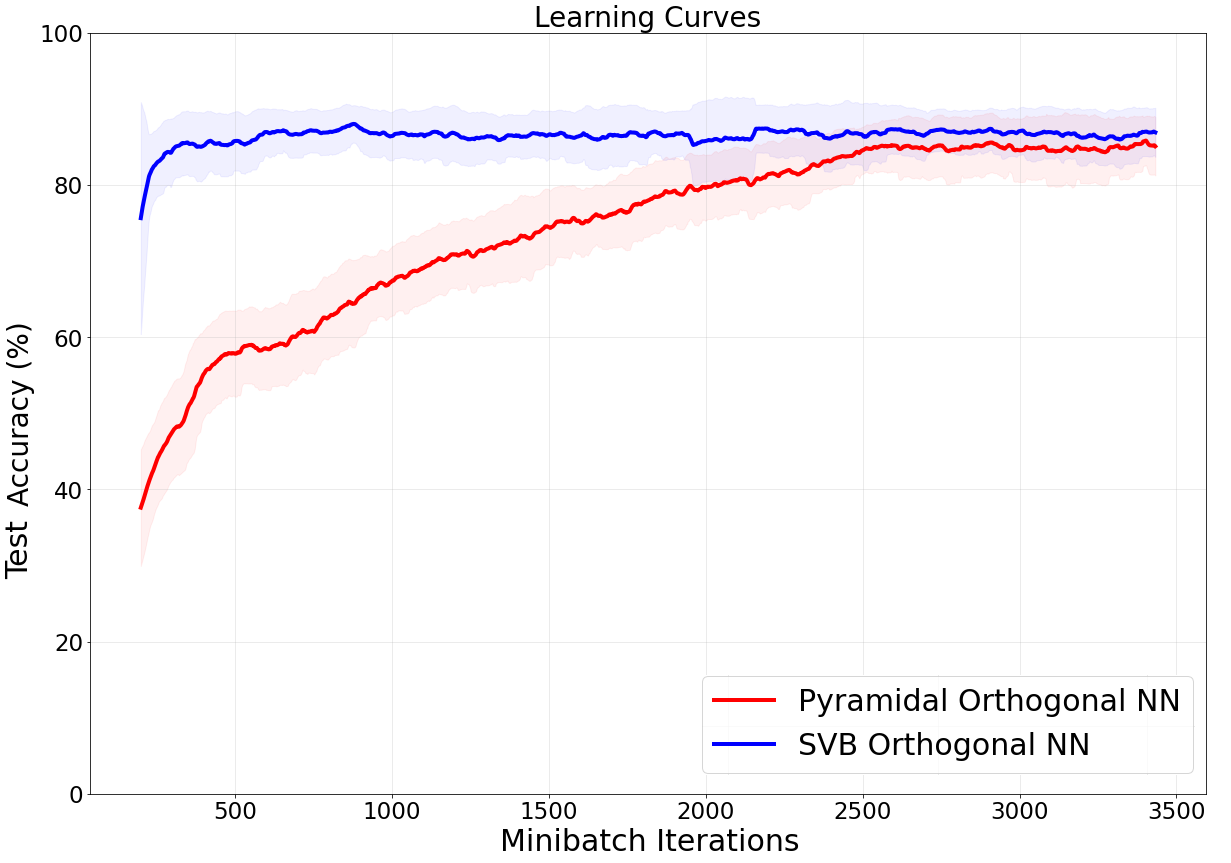}
\caption{[32,16,4]} 
\end{subfigure}\hspace*{\fill}
\begin{subfigure}{0.48\textwidth}
\includegraphics[width=\linewidth]{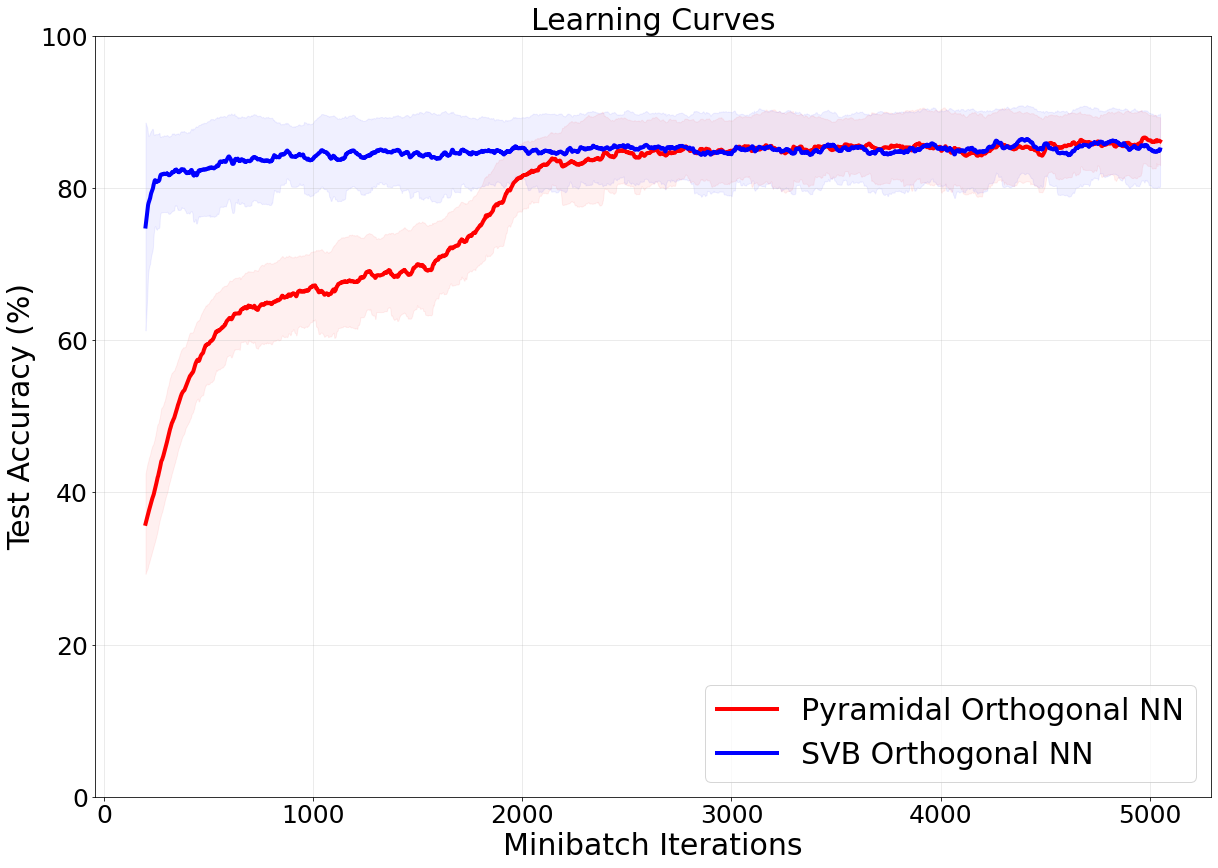}
\caption{[32,32,4]} 
\end{subfigure}
\caption{Training comparison between the SVB OrthoNN from \cite{jia2019orthogonal} and our classical pyramidal OrthoNN. Test accuracy on 1000 samples during several epochs of training on the MNIST dataset on 5000 samples. Initial dimensionality reduction (PCA) was on the samples to fit the input layer of the networks. Shaded areas indicate the accuracy variance during minibatch updates of size 50.} 
%The learning rate was set to 0.2.} 
\label{fig:training_OrthoNN_vs_SVB}
\end{figure}

In \cite{Quantum_OrthoNN}, we performed basic numerical experiments to verify the abilities of our pyramidal circuit, on the standard MNIST dataset. Note that current quantum hardware and software are not yet suited for bigger experiments. We first compared the training of our Classical OrthoNN to the SVB algorithm from \cite{jia2019orthogonal} (see Section \ref{sec:classical_ortho_nn_preliminaries}). Results as reported in Fig.\ref{fig:training_OrthoNN_vs_SVB}. These small scale tests confirmed that the pyramidal circuits and the corresponding gradient descent on the angles were efficient for learning a classification task.   

%\begin{figure}[!h]
%    \centering
%    \includegraphics[width=0.7\textwidth]{images/orthonn_mnist.png}
%    \caption{Training comparison between a [16,16,4] SVB OrthoNN from \cite{jia2019orthogonal} %and our classical pyramidal OrthoNN. Test accuracy on 1000 samples during 50 epochs of %training on the MNIST dataset on 5000 samples. Initial dimensionality reduction (PCA) was %on the samples to fit the input layer of the networks. Shaded areas indicate the variance %during minibatch updates of size 50.}
%    \label{fig:training_OrthoNN_vs_SVB}
%\end{figure}
%
%To complete the results, we provide additional numerical experiments testing our classical %pyramidal circuit for orthogonal neural networks on small use cases. 

~\\
 
Then, we implemented the quantum circuit on a real quantum computer provided by IBM. We used a 16 and 5 qubits device to perform respectively a [8,2] and a [4,2] orthogonal layer. We also branched two layers to perform a [4,4,2] network with non-linearity.
A pyramidal OrthoNN was trained classically, and the resulting angles were transferred to test the quantum circuit on a classification task on classes 6 and 9 of the MNIST dataset, over 500 samples. We compared the real experiment with a simulated one, and the classical pyramidal circuit as well. Results are reported in Table \ref{table:result_quantum_experiment}.

\begin{table}[h]
\begin{tabular}{|c|ccc|}
\hline
\multirow{2}{*}{\begin{tabular}[c]{@{}c@{}}Network\\ Architecture\end{tabular} } & \multicolumn{3}{c|}{Inference Accuracy}                            \\ \cline{2-4} 
                                      & \begin{tabular}[c]{@{}c@{}}Classical Pyramidal\\ Circuit\end{tabular}  & IBM Simulator & IBM Quantum Computer \\ \hline
$[4,2]$                               & 98,4\%                      & 98,4\%        & 98,0\%               \\
$[8,2]$                               & 97,4\%                      & 97,4\%        & 95,0\%               \\
$[4,4,2]$                             & 98,2\%                      & 98,2\%        & 82,8\%               \\ \hline
\end{tabular}
\caption{Results of the Pyramidal OrthoNN on classical simulators and real quantum computers. \emph{ibmq\_bogota v1.4.32} and \emph{ibmq\_guadalupe v1.2.17} are respectively 5 and 16 qubits quantum computers. May 2021.\label{table:result_quantum_experiment}}
\end{table}

% IBM GUADAlUPE 4_4_2 MNIST : 414/500 = 82.8%
% CLASSICAL AND BACKEND SIMULATOR : 491/500

Finally, in \cite{QMedicalImagingRoche}, we performed a benchmark on the MedMNIST dataset for medical imaging. We compared quantum and classical methods for orthogonal neural networks to classify diseases in two datasets: a classification of Chest X-Rays (Pneumonia) and a classification of retina images for retinopathy detection (Retina).

\begin{figure}[H]
    \centering
    \includegraphics[width=\textwidth]{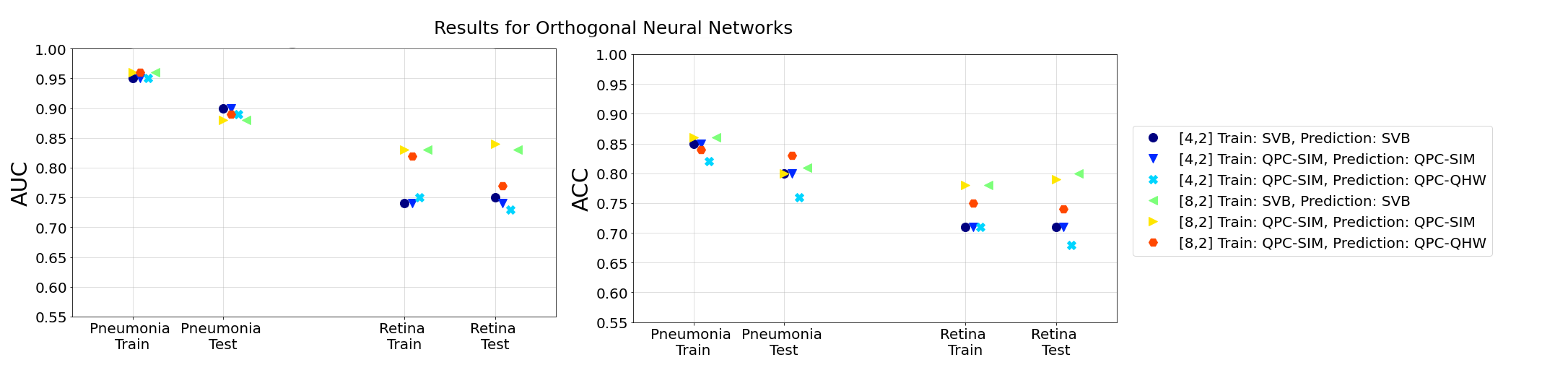}
    \caption{Results of experiments for the Orthogonal Neural Network. QPC stands for Quantum Pyramidal Circuit and is the classical algorithm simulating our quantum circuit. QHW is the quantum circuit on the real quantum hardware. SVB stands for the classical Singular Value Bounded algorithm.}
    \label{fig:Medical_result_graph_OrthoNN}
\end{figure}

\begin{figure}[H]
    \centering
    \includegraphics[width=\textwidth]{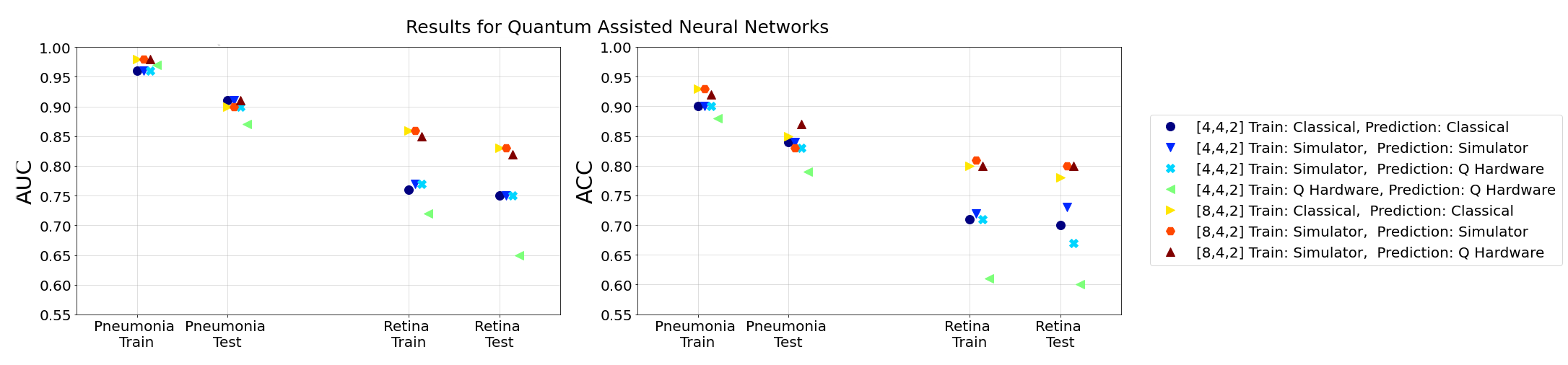}
    \caption{Results of experiments for the Quantum assisted Neural Network.}
    \label{fig:Medical_result_graph_qNN}
\end{figure}

In Fig.\ref{fig:Medical_result_graph_OrthoNN} we show our results, where we provide the AUC (area under curve) and ACC (accuracy) for all different types of neural network experiments, for both the training and test sets, for the Pneumonia and Retina datasets. In Fig.\ref{fig:Medical_result_graph_qNN} we show similar results for non orthogonal quantum neural networks, also called Quantum assisted Neural Network, that are based on simple inner products using data loaders as explained in Section \ref{sec:inner_product_and_distance_quantum}.

\subsubsection{Conclusion and Outlook}

In this chapter, we have proposed for the first time training methods for orthogonal neural networks (OrthoNNs) that run in quadratic time, a significant improvement from previous methods based on Singular Value Decomposition.
The main idea of our method is to replace usual weights and orthogonal matrices with an equivalent pyramidal circuit made of two-dimensional rotations. Each rotation is parametrizable by an angle, and the gradient descent takes place in the angle's optimization landscape. This unique type of gradient backpropagation ensures perfect orthogonality of the weight matrices while substantially improving the running time compared to previous work. 
Moreover, we propose both classical and quantum methods for inference, where the forward pass on a near term quantum computer would provide a provable advantage in the running time. 
This work expands the field of quantum deep learning by introducing new tools, concepts, and equivalences with classical deep learning theory. We have highlighted open questions regarding the construction of such pyramidal circuits for neural networks and their potential new advantages in terms of execution time, accuracy, and learning properties.

%%%%---------------CONCLUSION-------------%%%%
\chapter*{Conclusion}
\addcontentsline{toc}{chapter}{Conclusion}
\epigraph{\textit{"Aux murs de nos laboratoires les cadrans lumineux remplacent les ombres de la caverne."}}{Arthur Koestler \\ Les Somnambules (1959)}

In this thesis, we have proven the existence of new quantum algorithms for machine learning applications in unsupervised learning ($k$-means, spectral clustering) and in deep learning (convolutional and orthogonal neural networks). \\

During the development of these algorithms, we have introduced new quantum subroutines for linear algebra, distance estimation, graph analysis, and tomography of quantum states. These subroutines are fundamental enough to be reused across all machine learning and hopefully in other fields as well. 

Our algorithms are provably faster, using complexity analysis, often with respect to the number of points in the dataset, or in their dimension. These running times should be interpreted with subtlety as they depend on counter intuitive parameters, that often depends on the data values themselves. This behaviour has no classical counterpart and can help us determine where quantum computing can provide an advantage in machine learning. To go further, we have performed extensive classical simulations in order to test the training abilities of our quantum solutions in practice, as well as the scaling of their running times. For quantum orthogonal neural networks, we even implemented real quantum circuits on 8 and 16 qubits quantum computers. \\

In the end, we can expect quantum machine learning to be the most advantageous for hard linear algebra problems, involving spectral analysis (eigenvalue decomposition or projection) for which classical complexity are cubic or more. This should also motivate us to look at new problems that are even harder, in different fields such as topological data analysis and graph problems. On the other hand, it will be interesting to see what quantum can offer for machine learning problems that don't involve data, such as reinforcement learning, approximately solving partial differential equations, and even chemistry or many-body physics. 

Quantum machine learning is definitely an interesting field, and there is still a lot of work to be done to bring these algorithms closer to what can be done in practice. Many efforts are also expected in the physical realization of qubits quality, error correction, and quantum access to data. We are looking forward to the arrival of the first large scale quantum computers to test our theories in practice and hope they will help people solve problems for the general interest.

%%%%---------------BIBLIOGRAPHY------------%%%%
\addcontentsline{toc}{chapter}{Bibliography}
%\printbibliography
\bibliographystyle{alpha}
\bibliography{references}

\newcommand{\etalchar}[1]{$^{#1}$}
\begin{thebibliography}{dPNdSdOL19}

\bibitem[A{\etalchar{+}}19]{Qiskit}
H{\'e}ctor Abraham et~al.
\newblock Qiskit: An open-source framework for quantum computing, 2019.

\bibitem[AAA{\etalchar{+}}19]{M87}
Kazunori Akiyama, Antxon Alberdi, Walter Alef, Keiichi Asada, Rebecca Azulay,
  Anne-Kathrin Baczko, David Ball, Mislav Balokovi{\'c}, John Barrett, Dan
  Bintley, et~al.
\newblock First m87 event horizon telescope results. iv. imaging the central
  supermassive black hole.
\newblock {\em The Astrophysical Journal Letters}, 875(1):L4, 2019.

\bibitem[AAB{\etalchar{+}}19]{googlesupremacy}
Frank Arute, Kunal Arya, Ryan Babbush, Dave Bacon, Joseph~C Bardin, Rami
  Barends, Rupak Biswas, Sergio Boixo, Fernando~GSL Brandao, David~A Buell,
  et~al.
\newblock Quantum supremacy using a programmable superconducting processor.
\newblock {\em Nature}, 574(7779):505--510, 2019.

\bibitem[Aar13]{aaronson2013philosophers}
Scott Aaronson.
\newblock Why philosophers should care about computational complexity.
\newblock {\em Computability: Turing, G{\"o}del, Church, and Beyond}, 261:327,
  2013.

\bibitem[Aar15]{readthefineprint}
Scott Aaronson.
\newblock Read the fine print.
\newblock {\em Nature Physics}, 11(4):291--293, 2015.

\bibitem[ABG13]{aimeur2013quantum}
Esma A{\"\i}meur, Gilles Brassard, and S{\'e}bastien Gambs.
\newblock Quantum speed-up for unsupervised learning.
\newblock {\em Machine Learning}, 90(2):261--287, 2013.

\bibitem[ADBL19]{arrazola2019quantum}
Juan~Miguel Arrazola, Alain Delgado, Bhaskar~Roy Bardhan, and Seth Lloyd.
\newblock Quantum-inspired algorithms in practice.
\newblock {\em arXiv preprint arXiv:1905.10415}, 2019.

\bibitem[ADR82]{aspect1982experimental}
Alain Aspect, Jean Dalibard, and G{\'e}rard Roger.
\newblock Experimental test of bell's inequalities using time-varying
  analyzers.
\newblock {\em Physical review letters}, 49(25):1804, 1982.

\bibitem[AdW20]{apers2019quantum}
Simon Apers and Ronald de~Wolf.
\newblock Quantum speedup for graph sparsification, cut approximation and
  laplacian solving.
\newblock In {\em 2020 IEEE 61st Annual Symposium on Foundations of Computer
  Science (FOCS)}, pages 637--648. IEEE, 2020.

\bibitem[AGJO{\etalchar{+}}15]{arunachalam2015robustness_qram}
Srinivasan Arunachalam, Vlad Gheorghiu, Tomas Jochym-O’Connor, Michele Mosca,
  and Priyaa~Varshinee Srinivasan.
\newblock On the robustness of bucket brigade quantum ram.
\newblock {\em New Journal of Physics}, 17(12):123010, 2015.

\bibitem[AHKZ20]{allcock2020quantum}
Jonathan Allcock, Chang-Yu Hsieh, Iordanis Kerenidis, and Shengyu Zhang.
\newblock Quantum algorithms for feedforward neural networks.
\newblock {\em ACM Transactions on Quantum Computing}, 1(1):1--24, 2020.

\bibitem[AM01]{achlioptas2003fast}
Dimitris Achlioptas and Frank McSherry.
\newblock Fast computation of low rank matrix approximations.
\newblock In {\em Proceedings of the 33rd Annual Symposium on Theory of
  Computing}, pages 611--618, 2001.

\bibitem[Amb12]{A12}
Andris Ambainis.
\newblock Variable time amplitude amplification and quantum algorithms for
  linear algebra problems.
\newblock In {\em STACS'12 (29th Symposium on Theoretical Aspects of Computer
  Science)}, volume~14, pages 636--647. LIPIcs, 2012.

\bibitem[AR20]{aaronson2020quantum}
Scott Aaronson and Patrick Rall.
\newblock Quantum approximate counting, simplified.
\newblock In {\em Symposium on Simplicity in Algorithms}, pages 24--32. SIAM,
  2020.

\bibitem[ASZ{\etalchar{+}}20]{abbas2020power}
Amira Abbas, David Sutter, Christa Zoufal, Aur{\'e}lien Lucchi, Alessio
  Figalli, and Stefan Woerner.
\newblock The power of quantum neural networks.
\newblock {\em arXiv preprint arXiv:2011.00027}, 2020.

\bibitem[AV06]{arthur2006slow}
David Arthur and Sergei Vassilvitskii.
\newblock How slow is the k-means method?
\newblock In {\em Proceedings of the twenty-second annual symposium on
  Computational geometry}, pages 144--153. ACM, 2006.

\bibitem[AV07]{arthur2007k}
David Arthur and Sergei Vassilvitskii.
\newblock k-means++: The advantages of careful seeding.
\newblock In {\em Proceedings of the eighteenth annual ACM-SIAM symposium on
  Discrete algorithms}, pages 1027--1035. Society for Industrial and Applied
  Mathematics, 2007.

\bibitem[BBC{\etalchar{+}}93]{bennett1993teleporting}
Charles~H Bennett, Gilles Brassard, Claude Cr{\'e}peau, Richard Jozsa, Asher
  Peres, and William~K Wootters.
\newblock Teleporting an unknown quantum state via dual classical and
  einstein-podolsky-rosen channels.
\newblock {\em Physical review letters}, 70(13):1895, 1993.

\bibitem[BBF{\etalchar{+}}20]{beer2020training}
Kerstin Beer, Dmytro Bondarenko, Terry Farrelly, Tobias~J Osborne, Robert
  Salzmann, Daniel Scheiermann, and Ramona Wolf.
\newblock Training deep quantum neural networks.
\newblock {\em Nature communications}, 11(1):1--6, 2020.

\bibitem[BCC{\etalchar{+}}16]{bojarski2016visualbackprop}
Mariusz Bojarski, Anna Choromanska, Krzysztof Choromanski, Bernhard Firner,
  Larry Jackel, Urs Muller, and Karol Zieba.
\newblock Visualbackprop: efficient visualization of cnns.
\newblock {\em arXiv preprint arXiv:1611.05418}, 2016.

\bibitem[BCLK{\etalchar{+}}21]{bharti2021noisy}
Kishor Bharti, Alba Cervera-Lierta, Thi~Ha Kyaw, Tobias Haug, Sumner
  Alperin-Lea, Abhinav Anand, Matthias Degroote, Hermanni Heimonen, Jakob~S
  Kottmann, Tim Menke, et~al.
\newblock Noisy intermediate-scale quantum (nisq) algorithms.
\newblock {\em arXiv preprint arXiv:2101.08448}, 2021.

\bibitem[BCW18]{bansal2018can}
Nitin Bansal, Xiaohan Chen, and Zhangyang Wang.
\newblock Can we gain more from orthogonality regularizations in training deep
  cnns?
\newblock {\em arXiv preprint arXiv:1810.09102}, 2018.

\bibitem[BDD{\etalchar{+}}00]{bai2000templates}
Zhaojun Bai, James Demmel, Jack Dongarra, Axel Ruhe, and Henk van~der Vorst.
\newblock {\em Templates for the solution of algebraic eigenvalue problems: a
  practical guide}.
\newblock SIAM, Philadelphia, PA, USA, 2000.

\bibitem[BFL21]{QBNN}
Noah Berner, Vincent Fortuin, and Jonas Landman.
\newblock Quantum bayesian neural networks.
\newblock 2021.

\bibitem[BH97]{brassard1997exact}
Gilles Brassard and Peter Hoyer.
\newblock An exact quantum polynomial-time algorithm for simon's problem.
\newblock In {\em Proceedings of the Fifth Israeli Symposium on Theory of
  Computing and Systems}, pages 12--23. IEEE, 1997.

\bibitem[BHMT02]{brassard2002quantum}
Gilles Brassard, Peter Hoyer, Michele Mosca, and Alain Tapp.
\newblock Quantum amplitude amplification and estimation.
\newblock {\em Contemporary Mathematics}, 305:53--74, 2002.

\bibitem[Bia21]{biamonte2021universal}
Jacob Biamonte.
\newblock Universal variational quantum computation.
\newblock {\em Physical Review A}, 103(3):L030401, 2021.

\bibitem[Bis95]{bishop1995training}
Chris~M Bishop.
\newblock Training with noise is equivalent to tikhonov regularization.
\newblock {\em Neural computation}, 7(1):108--116, 1995.

\bibitem[Bis06]{bishop2006pattern}
Christopher~M Bishop.
\newblock {\em Pattern recognition and machine learning}.
\newblock springer, 2006.

\bibitem[BKG15]{boutsidis2015spectral}
Christos Boutsidis, Prabhanjan Kambadur, and Alex Gittens.
\newblock Spectral clustering via the power method-provably.
\newblock In {\em International conference on machine learning}, pages 40--48,
  2015.

\bibitem[BS17]{brandao2017quantum}
Fernando~GSL Brandao and Krysta~M Svore.
\newblock Quantum speed-ups for solving semidefinite programs.
\newblock In {\em 2017 IEEE 58th Annual Symposium on Foundations of Computer
  Science (FOCS)}, pages 415--426. IEEE, 2017.

\bibitem[BV97]{bernstein1997quantum}
Ethan Bernstein and Umesh Vazirani.
\newblock Quantum complexity theory.
\newblock {\em SIAM Journal on computing}, 26(5):1411--1473, 1997.

\bibitem[BWP{\etalchar{+}}17]{biamonte2017quantum}
Jacob Biamonte, Peter Wittek, Nicola Pancotti, Patrick Rebentrost, Nathan
  Wiebe, and Seth Lloyd.
\newblock Quantum machine learning.
\newblock {\em Nature}, 549(7671):195--202, 2017.

\bibitem[CAB{\etalchar{+}}20]{cerezo2020variational}
Marco Cerezo, Andrew Arrasmith, Ryan Babbush, Simon~C Benjamin, Suguru Endo,
  Keisuke Fujii, Jarrod~R McClean, Kosuke Mitarai, Xiao Yuan, Lukasz Cincio,
  et~al.
\newblock Variational quantum algorithms.
\newblock {\em arXiv preprint arXiv:2012.09265}, 2020.

\bibitem[Cal20]{CASP}
Ewen Callaway.
\newblock 'it will change everything': Deepmind's ai makes gigantic leap in
  solving protein structures.
\newblock {\em Nature}, 2020.

\bibitem[CB18]{cortese2018loading}
John~A Cortese and Timothy~M Braje.
\newblock Loading classical data into a quantum computer.
\newblock {\em arXiv preprint arXiv:1803.01958}, 2018.

\bibitem[CCL19]{cong2019quantum}
Iris Cong, Soonwon Choi, and Mikhail~D Lukin.
\newblock Quantum convolutional neural networks.
\newblock {\em Nature Physics}, 15(12):1273--1278, 2019.

\bibitem[CD15]{cong2015quantum}
Iris Cong and Luming Duan.
\newblock Quantum discriminant analysis for dimensionality reduction and
  classification.
\newblock {\em arXiv preprint arXiv:1510.00113}, 2015.

\bibitem[CDKK20]{coyle2020variational}
Brian Coyle, Mina Doosti, Elham Kashefi, and Niraj Kumar.
\newblock Variational quantum cloning: Improving practicality for quantum
  cryptanalysis.
\newblock {\em arXiv preprint arXiv:2012.11424}, 2020.

\bibitem[CFS{\etalchar{+}}16]{maxcut_cui16}
Shawn~X. Cui, Michael~H. Freedman, Or~Sattath, Richard Stong, and Greg Minton.
\newblock Quantum max-flow/min-cut.
\newblock {\em Journal of Mathematical Physics}, 57(6):062206, 6 2016.

\bibitem[CGAG17]{cao2017quantum}
Yudong Cao, Gian~Giacomo Guerreschi, and Al{\'a}n Aspuru-Guzik.
\newblock Quantum neuron: an elementary building block for machine learning on
  quantum computers.
\newblock {\em arXiv preprint arXiv:1711.11240}, 2017.

\bibitem[CGJ18]{chakraborty2018power}
Shantanav Chakraborty, Andr{\'a}s Gily{\'e}n, and Stacey Jeffery.
\newblock The power of block-encoded matrix powers: improved regression
  techniques via faster hamiltonian simulation.
\newblock {\em 46th International Colloquium on Automata, Languages, and
  Programming}, pages Art. No. 33, 14., 2018.

\bibitem[Chi10]{childs2010relationship}
Andrew~M Childs.
\newblock On the relationship between continuous-and discrete-time quantum
  walk.
\newblock {\em Communications in Mathematical Physics}, 294(2):581--603, 2010.

\bibitem[Chi17]{childs2017lecture}
Andrew~M Childs.
\newblock Lecture notes on quantum algorithms.
\newblock {\em Lecture notes at University of Maryland}, 2017.

\bibitem[CJK{\etalchar{+}}13]{choromanska2013fast}
Anna Choromanska, Tony Jebara, Hyungtae Kim, Mahesh Mohan, and Claire
  Monteleoni.
\newblock Fast spectral clustering via the nystr{\"o}m method.
\newblock In {\em International Conference on Algorithmic Learning Theory},
  pages 367--381. Springer, New York, NY, USA, 2013.

\bibitem[CKS17]{childs2017quantum}
Andrew~M Childs, Robin Kothari, and Rolando~D Somma.
\newblock Quantum algorithm for systems of linear equations with exponentially
  improved dependence on precision.
\newblock {\em SIAM Journal on Computing}, 46(6):1920--1950, 2017.

\bibitem[CMDK20]{coyle2020born}
Brian Coyle, Daniel Mills, Vincent Danos, and Elham Kashefi.
\newblock The born supremacy: quantum advantage and training of an ising born
  machine.
\newblock {\em npj Quantum Information}, 6(1):1--11, 2020.

\bibitem[Cyb89]{cybenko1989approximation}
George Cybenko.
\newblock Approximation by superpositions of a sigmoidal function.
\newblock {\em Mathematics of control, signals and systems}, 2(4):303--314,
  1989.

\bibitem[Das17]{daskin2017quantum}
Ammar Daskin.
\newblock Quantum spectral clustering through a biased phase estimation
  algorithm.
\newblock {\em TWMS Journal of Applied and Engineering Mathematics},
  10(1):24--33, 2017.

\bibitem[DB18]{dunjko2018machine}
Vedran Dunjko and Hans~J Briegel.
\newblock Machine learning \& artificial intelligence in the quantum domain: a
  review of recent progress.
\newblock {\em Reports on Progress in Physics}, 81(7):074001, 2018.

\bibitem[DDK18]{dallaire2018quantum}
Pierre-Luc Dallaire-Demers and Nathan Killoran.
\newblock Quantum generative adversarial networks.
\newblock {\em Physical Review A}, 98(1):012324, 2018.

\bibitem[DFK{\etalchar{+}}04]{drineas2004clustering}
Petros Drineas, Alan Frieze, Ravi Kannan, Santosh Vempala, and V~Vinay.
\newblock Clustering large graphs via the singular value decomposition.
\newblock {\em Machine learning}, 56(1-3):9--33, 2004.

\bibitem[DGV12]{dorner2012towards}
Ross Dorner, John Goold, and Vlatko Vedral.
\newblock Towards quantum simulations of biological information flow.
\newblock {\em Interface focus}, 2(4):522--528, 2012.

\bibitem[DH96]{durr1996quantum}
Christoph Durr and Peter Hoyer.
\newblock A quantum algorithm for finding the minimum.
\newblock {\em arXiv preprint quant-ph/9607014}, 1996.

\bibitem[DJ92]{deutsch1992rapid}
David Deutsch and Richard Jozsa.
\newblock Rapid solution of problems by quantum computation.
\newblock {\em Proceedings of the Royal Society of London. Series A:
  Mathematical and Physical Sciences}, 439(1907):553--558, 1992.

\bibitem[DKR02]{drineas2002competitive}
Petros Drineas, Iordanis Kerenidis, and Prabhakar Raghavan.
\newblock Competitive recommendation systems.
\newblock In {\em Proceedings of the thiry-fourth annual ACM symposium on
  Theory of computing}, pages 82--90. ACM, 2002.

\bibitem[DMGM20]{di2020fault}
Olivia Di~Matteo, Vlad Gheorghiu, and Michele Mosca.
\newblock Fault-tolerant resource estimation of quantum random-access memories.
\newblock {\em IEEE Transactions on Quantum Engineering}, 1:1--13, 2020.

\bibitem[dPNdSdOL19]{de2019quantum_associativememory}
Fernando~M de~Paula~Neto, Adenilton~J da~Silva, Wilson~R de~Oliveira, and
  Teresa~B Ludermir.
\newblock Quantum probabilistic associative memory architecture.
\newblock {\em Neurocomputing}, 351:101--110, 2019.

\bibitem[DW19]{de2019quantum}
Ronald De~Wolf.
\newblock Quantum computing: Lecture notes.
\newblock {\em arXiv preprint arXiv:1907.09415}, 2019.

\bibitem[Fey82]{feynmanquote1}
Richard~P Feynman.
\newblock Simulating physics with computers.
\newblock {\em Int. J. Theor. Phys}, 21(6/7), 1982.

\bibitem[Fey87]{feynmanquote2}
Richard~P Feynman.
\newblock Tiny computers obeying quantum mechanical laws.
\newblock {\em N. Metropolis, DM Kerr, and G. Rota, editors, New Directions in
  Physics: The Los Alamos 40th Anniversary Volume}, pages 7--25, 1987.

\bibitem[FGG14]{farhi2014quantum}
Edward Farhi, Jeffrey Goldstone, and Sam Gutmann.
\newblock A quantum approximate optimization algorithm.
\newblock {\em arXiv preprint arXiv:1411.4028}, 2014.

\bibitem[FHT01]{friedman2001elements}
Jerome Friedman, Trevor Hastie, and Robert Tibshirani.
\newblock {\em The elements of statistical learning}, volume~1.
\newblock Springer series in statistics New York, NY, USA:, 2001.

\bibitem[FKV04]{frieze2004fast}
Alan Frieze, Ravi Kannan, and Santosh Vempala.
\newblock Fast monte-carlo algorithms for finding low-rank approximations.
\newblock {\em Journal of the ACM (JACM)}, 51(6):1025--1041, 2004.

\bibitem[FN18]{farhi2018classification}
Edward Farhi and Hartmut Neven.
\newblock Classification with quantum neural networks on near term processors.
\newblock {\em arXiv preprint arXiv:1802.06002}, 2018.

\bibitem[GBCB16]{goodfellow2016deep}
Ian Goodfellow, Yoshua Bengio, Aaron Courville, and Yoshua Bengio.
\newblock {\em Deep learning}, volume~1.
\newblock MIT press Cambridge, 2016.

\bibitem[GBH18]{ganea2018hyperbolic}
Octavian-Eugen Ganea, Gary B{\'e}cigneul, and Thomas Hofmann.
\newblock Hyperbolic neural networks.
\newblock {\em arXiv preprint arXiv:1805.09112}, 2018.

\bibitem[GGZW21]{grinko2021iterative}
Dmitry Grinko, Julien Gacon, Christa Zoufal, and Stefan Woerner.
\newblock Iterative quantum amplitude estimation.
\newblock {\em npj Quantum Information}, 7(1):1--6, 2021.

\bibitem[GH18]{george2018deep}
Daniel George and EA~Huerta.
\newblock Deep learning for real-time gravitational wave detection and
  parameter estimation: Results with advanced ligo data.
\newblock {\em Physics Letters B}, 778:64--70, 2018.

\bibitem[GLM08a]{giovannetti2008architectures}
Vittorio Giovannetti, Seth Lloyd, and Lorenzo Maccone.
\newblock Architectures for a quantum random access memory.
\newblock {\em Physical Review A}, 78(5):052310, 2008.

\bibitem[GLM08b]{giovannetti2008quantum}
Vittorio Giovannetti, Seth Lloyd, and Lorenzo Maccone.
\newblock Quantum random access memory.
\newblock {\em Physical review letters}, 100(16):160501, 2008.

\bibitem[GLT18]{gilyen2018quantum}
Andr{\'a}s Gily{\'e}n, Seth Lloyd, and Ewin Tang.
\newblock Quantum-inspired low-rank stochastic regression with logarithmic
  dependence on the dimension.
\newblock {\em arXiv preprint arXiv:1811.04909}, 2018.

\bibitem[GPAM{\etalchar{+}}14]{goodfellow2014generative}
Ian~J Goodfellow, Jean Pouget-Abadie, Mehdi Mirza, Bing Xu, David Warde-Farley,
  Sherjil Ozair, Aaron Courville, and Yoshua Bengio.
\newblock Generative adversarial networks.
\newblock {\em arXiv preprint arXiv:1406.2661}, 2014.

\bibitem[Gro96]{groveralgo}
Lov~K Grover.
\newblock A fast quantum mechanical algorithm for database search.
\newblock In {\em Proceedings of the twenty-eighth annual ACM symposium on
  Theory of computing}, pages 212--219, 1996.

\bibitem[Gro05]{grover2005fixed}
Lov~K Grover.
\newblock Fixed-point quantum search.
\newblock {\em Physical Review Letters}, 95(15):150501, 2005.

\bibitem[GSC00]{gers2000learning}
Felix~A Gers, J{\"u}rgen Schmidhuber, and Fred Cummins.
\newblock Learning to forget: Continual prediction with lstm.
\newblock {\em Neural computation}, 12(10):2451--2471, 2000.

\bibitem[GSLW19]{gilyen2019quantum}
Andr{\'a}s Gily{\'e}n, Yuan Su, Guang~Hao Low, and Nathan Wiebe.
\newblock Quantum singular value transformation and beyond: exponential
  improvements for quantum matrix arithmetics.
\newblock In {\em Proceedings of the 51st Annual ACM SIGACT Symposium on Theory
  of Computing}, pages 193--204, 2019.

\bibitem[GTKL{\etalchar{+}}20]{giurgica2020low}
Tudor Giurgica-Tiron, Iordanis Kerenidis, Farrokh Labib, Anupam Prakash, and
  William Zeng.
\newblock Low depth algorithms for quantum amplitude estimation.
\newblock {\em arXiv preprint arXiv:2012.03348}, 2020.

\bibitem[GWOB19]{grant2019initialization}
Edward Grant, Leonard Wossnig, Mateusz Ostaszewski, and Marcello Benedetti.
\newblock An initialization strategy for addressing barren plateaus in
  parametrized quantum circuits.
\newblock {\em Quantum}, 3:214, 2019.

\bibitem[HAY{\etalchar{+}}20]{holmes2020barren}
Zo{\"e} Holmes, Andrew Arrasmith, Bin Yan, Patrick~J Coles, Andreas Albrecht,
  and Andrew~T Sornborger.
\newblock Barren plateaus preclude learning scramblers.
\newblock {\em arXiv preprint arXiv:2009.14808}, 2020.

\bibitem[HD15]{hamerly2015accelerating}
Greg Hamerly and Jonathan Drake.
\newblock Accelerating lloyd’s algorithm for k-means clustering.
\newblock In {\em Partitional clustering algorithms}, pages 41--78. Springer,
  New York, NY, USA, 2015.

\bibitem[HHL09]{HHL}
Aram~W Harrow, Avinatan Hassidim, and Seth Lloyd.
\newblock Quantum algorithm for linear systems of equations.
\newblock {\em Physical review letters}, 103(15):150502, 2009.

\bibitem[HLGJ21]{hann2021resilience_qram}
Connor~T Hann, Gideon Lee, SM~Girvin, and Liang Jiang.
\newblock Resilience of quantum random access memory to generic noise.
\newblock {\em PRX Quantum}, 2(2):020311, 2021.

\bibitem[HN92]{hecht1992theory}
Robert Hecht-Nielsen.
\newblock Theory of the backpropagation neural network.
\newblock In {\em Neural networks for perception}, pages 65--93. Elsevier,
  1992.

\bibitem[HRS18]{haner2018optimizing}
Thomas H{\"a}ner, Martin Roetteler, and Krysta~M Svore.
\newblock Optimizing quantum circuits for arithmetic.
\newblock {\em arXiv preprint arXiv:1805.12445}, 2018.

\bibitem[HSG{\etalchar{+}}21]{gravitationallenses}
Xiaosheng Huang, Christopher Storfer, A~Gu, V~Ravi, A~Pilon, W~Sheu,
  R~Venguswamy, S~Banka, A~Dey, M~Landriau, et~al.
\newblock Discovering new strong gravitational lenses in the desi legacy
  imaging surveys.
\newblock {\em The Astrophysical Journal}, 909(1):27, 2021.

\bibitem[JDM{\etalchar{+}}20]{johri2020nearest_dataloaders}
Sonika Johri, Shantanu Debnath, Avinash Mocherla, Alexandros Singh, Anupam
  Prakash, Jungsang Kim, and Iordanis Kerenidis.
\newblock Nearest centroid classification on a trapped ion quantum computer.
\newblock {\em arXiv preprint arXiv:2012.04145}, 2020.

\bibitem[JLW{\etalchar{+}}19]{jia2019orthogonal}
Kui Jia, Shuai Li, Yuxin Wen, Tongliang Liu, and Dacheng Tao.
\newblock Orthogonal deep neural networks.
\newblock {\em IEEE transactions on pattern analysis and machine intelligence},
  2019.

\bibitem[KH09]{krizhevsky2009learning}
Alex Krizhevsky and Geoffrey Hinton.
\newblock Learning multiple layers of features from tiny images.
\newblock Technical report, Citeseer, 2009.

\bibitem[Kit95]{kitaev1995quantum}
A~Yu Kitaev.
\newblock Quantum measurements and the abelian stabilizer problem.
\newblock {\em arXiv preprint quant-ph/9511026}, 1995.

\bibitem[KL97]{knill1997theory}
Emanuel Knill and Raymond Laflamme.
\newblock Theory of quantum error-correcting codes.
\newblock {\em Physical Review A}, 55(2):900, 1997.

\bibitem[KL20]{kerenidis2020classification_QSFA}
Iordanis Kerenidis and Alessandro Luongo.
\newblock Classification of the {MNIST} data set with quantum slow feature
  analysis.
\newblock {\em Physical Review A}, 101(6):062327, 2020.

\bibitem[KL21]{quantumspectralclustering}
Iordanis Kerenidis and Jonas Landman.
\newblock Quantum spectral clustering.
\newblock {\em Physical Review A}, 103(4):042415, 2021.

\bibitem[KLLP19]{qmeans}
Iordanis Kerenidis, Jonas Landman, Alessandro Luongo, and Anupam Prakash.
\newblock q-means: A quantum algorithm for unsupervised machine learning.
\newblock In {\em Advances in Neural Information Processing Systems},
  volume~32. Curran Associates, Inc., 2019.

\bibitem[KLM{\etalchar{+}}07]{kaye2007introduction}
Phillip Kaye, Raymond Laflamme, Michele Mosca, et~al.
\newblock {\em An introduction to quantum computing}.
\newblock Oxford University Press on Demand, 2007.

\bibitem[KLM21]{Quantum_OrthoNN}
Iordanis Kerenidis, Jonas Landman, and Natansh Mathur.
\newblock Classical and quantum algorithms for orthogonal neural networks.
\newblock 2021.

\bibitem[KLP20a]{QCNN}
Iordanis Kerenidis, Jonas Landman, and Anupam Prakash.
\newblock Quantum algorithms for deep convolutional neural networks.
\newblock In {\em Proceedings of the International Conference on Learning
  Representations ({ICLR})}, 2020.

\bibitem[KLP20b]{kerenidis2020quantum_guassianmixture}
Iordanis Kerenidis, Alessandro Luongo, and Anupam Prakash.
\newblock Quantum expectation-maximization for gaussian mixture models.
\newblock In {\em International Conference on Machine Learning}, pages
  5187--5197. PMLR, 2020.

\bibitem[KP16]{kerenidis_recommendation_system}
Iordanis Kerenidis and Anupam Prakash.
\newblock Quantum recommendation systems.
\newblock {\em arXiv preprint arXiv:1603.08675}, 2016.

\bibitem[KP20a]{kerenidis2020_gradient_descent}
Iordanis Kerenidis and Anupam Prakash.
\newblock Quantum gradient descent for linear systems and least squares.
\newblock {\em Physical Review A}, 101(2):022316, 2020.

\bibitem[KP20b]{kerenidis2020quantum_IPM}
Iordanis Kerenidis and Anupam Prakash.
\newblock A quantum interior point method for {LP}s and {SDP}s.
\newblock {\em ACM Transactions on Quantum Computing}, 1(1):1--32, 2020.

\bibitem[KPS21]{kerenidis2021quantum_SOCP}
Iordanis Kerenidis, Anupam Prakash, and D{\'a}niel Szil{\'a}gyi.
\newblock Quantum algorithms for second-order cone programming and support
  vector machines.
\newblock {\em Quantum}, 5:427, 2021.

\bibitem[KSH12]{krizhevsky2012imagenet}
Alex Krizhevsky, Ilya Sutskever, and Geoffrey~E Hinton.
\newblock Imagenet classification with deep convolutional neural networks.
\newblock In {\em Advances in neural information processing systems}, pages
  1097--1105, 2012.

\bibitem[LBBH98]{lecun1998gradient}
Yann LeCun, L{\'e}on Bottou, Yoshua Bengio, and Patrick Haffner.
\newblock Gradient-based learning applied to document recognition.
\newblock {\em Proceedings of the IEEE}, 86(11):2278--2324, 1998.

\bibitem[LBH15]{lecun2015deep}
Yann LeCun, Yoshua Bengio, and Geoffrey Hinton.
\newblock Deep learning.
\newblock {\em nature}, 521(7553):436--444, 2015.

\bibitem[LCMR19]{lezcano2019cheap}
Mario Lezcano-Casado and David Mart{\i}nez-Rubio.
\newblock Cheap orthogonal constraints in neural networks: A simple
  parametrization of the orthogonal and unitary group.
\newblock In {\em International Conference on Machine Learning}, pages
  3794--3803. PMLR, 2019.

\bibitem[LGZ16]{Lloyd_topological_ml}
Seth Lloyd, Silvano Garnerone, and Paolo Zanardi.
\newblock Quantum algorithms for topological and geometric analysis of data.
\newblock {\em Nature communications}, 7(1):1--7, 2016.

\bibitem[LJL19]{lee2019hybrid}
Yonghae Lee, Jaewoo Joo, and Soojoon Lee.
\newblock Hybrid quantum linear equation algorithm and its experimental test on
  ibm quantum experience.
\newblock {\em Scientific reports}, 9(1):1--12, 2019.

\bibitem[LLKL11]{li2011time}
Mu~Li, Xiao-Chen Lian, James~T Kwok, and Bao-Liang Lu.
\newblock Time and space efficient spectral clustering via column sampling.
\newblock In {\em CVPR 2011}, pages 2297--2304. IEEE, Colorado Springs, CO,
  USA, 2011.

\bibitem[Llo82]{lloyd1982least}
Stuart Lloyd.
\newblock Least squares quantization in pcm.
\newblock {\em IEEE transactions on information theory}, 28(2):129--137, 1982.

\bibitem[LLPS93]{leshno1993multilayer}
Moshe Leshno, Vladimir~Ya Lin, Allan Pinkus, and Shimon Schocken.
\newblock Multilayer feedforward networks with a nonpolynomial activation
  function can approximate any function.
\newblock {\em Neural networks}, 6(6):861--867, 1993.

\bibitem[LMR13]{LMR13}
Seth Lloyd, Masoud Mohseni, and Patrick Rebentrost.
\newblock {Quantum algorithms for supervised and unsupervised machine
  learning}.
\newblock {\em arXiv}, 1307.0411:1--11, 7 2013.

\bibitem[LMR14]{Lloyd_PCA_quantum}
Seth Lloyd, Masoud Mohseni, and Patrick Rebentrost.
\newblock Quantum principal component analysis.
\newblock {\em Nature Physics}, 10(9):631--633, 2014.

\bibitem[LW18]{lloyd2018quantum}
Seth Lloyd and Christian Weedbrook.
\newblock Quantum generative adversarial learning.
\newblock {\em Physical review letters}, 121(4):040502, 2018.

\bibitem[MBS{\etalchar{+}}18]{mcclean2018barren}
Jarrod~R McClean, Sergio Boixo, Vadim~N Smelyanskiy, Ryan Babbush, and Hartmut
  Neven.
\newblock Barren plateaus in quantum neural network training landscapes.
\newblock {\em Nature communications}, 9(1):1--6, 2018.

\bibitem[MLL{\etalchar{+}}21]{QMedicalImagingRoche}
Natansh Mathur, Jonas Landman, Yun~Yvonna Li, Martin Strahm, Skander Kazdaghli,
  Anupam Prakash, and Iordanis Kerenidis.
\newblock Medical image classification via quantum neural networks.
\newblock {\em arXiv preprint arXiv:2109.01831}, 2021.

\bibitem[MLM17]{moylett2017quantum}
Dominic~J Moylett, Noah Linden, and Ashley Montanaro.
\newblock Quantum speedup of the traveling-salesman problem for bounded-degree
  graphs.
\newblock {\em Physical Review A}, 95(3):032323, 2017.

\bibitem[MNKF18]{mitarai2018quantum}
Kosuke Mitarai, Makoto Negoro, Masahiro Kitagawa, and Keisuke Fujii.
\newblock Quantum circuit learning.
\newblock {\em Physical Review A}, 98(3):032309, 2018.

\bibitem[NC02]{nielsen2002quantum}
Michael~A Nielsen and Isaac Chuang.
\newblock Quantum computation and quantum information, 2002.

\bibitem[NJW02]{ng2002spectral}
Andrew~Y Ng, Michael~I Jordan, and Yair Weiss.
\newblock On spectral clustering: Analysis and an algorithm.
\newblock In {\em Advances in neural information processing systems}, pages
  849--856, 2002.

\bibitem[NM14]{nakata2014diagonal}
Yoshifumi Nakata and Mio Murao.
\newblock Diagonal quantum circuits: their computational power and
  applications.
\newblock {\em The European Physical Journal Plus}, 129(7):1--14, 2014.

\bibitem[OMA{\etalchar{+}}17]{Otterbach17}
JS~Otterbach, R~Manenti, N~Alidoust, A~Bestwick, M~Block, B~Bloom, S~Caldwell,
  N~Didier, E~Schuyler Fried, S~Hong, et~al.
\newblock Unsupervised machine learning on a hybrid quantum computer.
\newblock {\em arXiv preprint arXiv:1712.05771}, 2017.

\bibitem[PCW{\etalchar{+}}20]{pesah2020absence}
Arthur Pesah, M~Cerezo, Samson Wang, Tyler Volkoff, Andrew~T Sornborger, and
  Patrick~J Coles.
\newblock Absence of barren plateaus in quantum convolutional neural networks.
\newblock {\em arXiv preprint arXiv:2011.02966}, 2020.

\bibitem[PGC{\etalchar{+}}17]{paszke2017automatic}
Adam Paszke, Sam Gross, Soumith Chintala, Gregory Chanan, Edward Yang, Zachary
  DeVito, Zeming Lin, Alban Desmaison, Luca Antiga, and Adam Lerer.
\newblock Automatic differentiation in pytorch.
\newblock 2017.

\bibitem[PMS{\etalchar{+}}14]{peruzzo2014variational}
Alberto Peruzzo, Jarrod McClean, Peter Shadbolt, Man-Hong Yung, Xiao-Qi Zhou,
  Peter~J Love, Al{\'a}n Aspuru-Guzik, and Jeremy~L O’brien.
\newblock A variational eigenvalue solver on a photonic quantum processor.
\newblock {\em Nature communications}, 5(1):1--7, 2014.

\bibitem[PPR19]{park2019circuit}
Daniel~K Park, Francesco Petruccione, and June-Koo~Kevin Rhee.
\newblock Circuit-based quantum random access memory for classical data.
\newblock {\em Scientific reports}, 9(1):1--8, 2019.

\bibitem[Pra14]{prakash2014quantum}
Anupam Prakash.
\newblock {\em Quantum algorithms for linear algebra and machine learning.}
\newblock PhD thesis, UC Berkeley, 2014.

\bibitem[Pre18]{NISQpreskill}
John Preskill.
\newblock Quantum computing in the nisq era and beyond.
\newblock {\em Quantum}, 2:79, 2018.

\bibitem[PSCLGFL20]{perez2020data}
Adri{\'a}n P{\'e}rez-Salinas, Alba Cervera-Lierta, Elies Gil-Fuster, and
  Jos{\'e}~I Latorre.
\newblock Data re-uploading for a universal quantum classifier.
\newblock {\em Quantum}, 4:226, 2020.

\bibitem[PVG{\etalchar{+}}11]{scikit-learn}
F.~Pedregosa, G.~Varoquaux, A.~Gramfort, V.~Michel, B.~Thirion, O.~Grisel,
  M.~Blondel, P.~Prettenhofer, R.~Weiss, V.~Dubourg, J.~Vanderplas, A.~Passos,
  D.~Cournapeau, M.~Brucher, M.~Perrot, and E.~Duchesnay.
\newblock Scikit-learn: Machine learning in {P}ython.
\newblock {\em Journal of Machine Learning Research}, 12:2825--2830, 2011.

\bibitem[PZ21]{pan2021simulating}
Feng Pan and Pan Zhang.
\newblock Simulating the sycamore quantum supremacy circuits.
\newblock {\em arXiv preprint arXiv:2103.03074}, 2021.

\bibitem[QSW{\etalchar{+}}20]{qian2020orchestrating}
Bin Qian, Jie Su, Zhenyu Wen, Devki~Nandan Jha, Yinhao Li, Yu~Guan, Deepak
  Puthal, Philip James, Renyu Yang, Albert~Y Zomaya, et~al.
\newblock Orchestrating the development lifecycle of machine learning-based iot
  applications: A taxonomy and survey.
\newblock {\em ACM Computing Surveys (CSUR)}, 53(4):1--47, 2020.

\bibitem[RBWL18]{Lloyd_hopfield_nn}
Patrick Rebentrost, Thomas~R Bromley, Christian Weedbrook, and Seth Lloyd.
\newblock Quantum hopfield neural network.
\newblock {\em Physical Review A}, 98(4):042308, 2018.

\bibitem[RDK{\etalchar{+}}19]{climatechangeAI}
David Rolnick, Priya~L Donti, Lynn~H Kaack, Kelly Kochanski, Alexandre Lacoste,
  Kris Sankaran, Andrew~Slavin Ross, Nikola Milojevic-Dupont, Natasha Jaques,
  Anna Waldman-Brown, et~al.
\newblock Tackling climate change with machine learning.
\newblock {\em arXiv preprint arXiv:1906.05433}, 2019.

\bibitem[RHW86]{rumelhart1986learning}
David~E Rumelhart, Geoffrey~E Hinton, and Ronald~J Williams.
\newblock Learning representations by back-propagating errors.
\newblock {\em nature}, 323(6088):533--536, 1986.

\bibitem[Roj96]{rojas1996backpropagation}
Raul Rojas.
\newblock The backpropagation algorithm.
\newblock In {\em Neural networks}, pages 149--182. Springer, 1996.

\bibitem[RR08]{rahimi2008random}
Ali Rahimi and Benjamin Recht.
\newblock Random features for large-scale kernel machines.
\newblock In {\em Advances in neural information processing systems}, pages
  1177--1184, 2008.

\bibitem[RSML18]{rebentrost2018quantum_svm}
Patrick Rebentrost, Adrian Steffens, Iman Marvian, and Seth Lloyd.
\newblock Quantum singular-value decomposition of nonsparse low-rank matrices.
\newblock {\em Physical review A}, 97(1):012327, 2018.

\bibitem[SBG{\etalchar{+}}19]{schuld2019evaluating}
Maria Schuld, Ville Bergholm, Christian Gogolin, Josh Izaac, and Nathan
  Killoran.
\newblock Evaluating analytic gradients on quantum hardware.
\newblock {\em Physical Review A}, 99(3):032331, 2019.

\bibitem[Sch21]{schuld2021quantum}
Maria Schuld.
\newblock Quantum machine learning models are kernel methods.
\newblock {\em arXiv preprint arXiv:2101.11020}, 2021.

\bibitem[SEJ{\etalchar{+}}20]{deepmindalphafold}
Andrew~W Senior, Richard Evans, John Jumper, James Kirkpatrick, Laurent Sifre,
  Tim Green, Chongli Qin, Augustin {\v{Z}}{\'\i}dek, Alexander~WR Nelson, Alex
  Bridgland, et~al.
\newblock Improved protein structure prediction using potentials from deep
  learning.
\newblock {\em Nature}, 577(7792):706--710, 2020.

\bibitem[SHM{\etalchar{+}}16]{silver2016mastering}
David Silver, Aja Huang, Chris~J Maddison, Arthur Guez, Laurent Sifre, George
  Van Den~Driessche, Julian Schrittwieser, Ioannis Antonoglou, Veda
  Panneershelvam, Marc Lanctot, et~al.
\newblock Mastering the game of go with deep neural networks and tree search.
\newblock {\em nature}, 529(7587):484, 2016.

\bibitem[Sho99]{shor1999polynomial}
Peter~W Shor.
\newblock Polynomial-time algorithms for prime factorization and discrete
  logarithms on a quantum computer.
\newblock {\em SIAM review}, 41(2):303--332, 1999.

\bibitem[Sim97]{simon1997power}
Daniel~R Simon.
\newblock On the power of quantum computation.
\newblock {\em SIAM journal on computing}, 26(5):1474--1483, 1997.

\bibitem[SM21]{shao2021faster}
Changpeng Shao and Ashley Montanaro.
\newblock Faster quantum-inspired algorithms for solving linear systems.
\newblock {\em arXiv preprint arXiv:2103.10309}, 2021.

\bibitem[SSM21]{schuld2021effect}
Maria Schuld, Ryan Sweke, and Johannes~Jakob Meyer.
\newblock Effect of data encoding on the expressive power of variational
  quantum-machine-learning models.
\newblock {\em Physical Review A}, 103(3):032430, 2021.

\bibitem[SSP14]{schuld2014quest}
Maria Schuld, Ilya Sinayskiy, and Francesco Petruccione.
\newblock The quest for a quantum neural network.
\newblock {\em Quantum Information Processing}, 13(11):2567--2586, 2014.

\bibitem[SUR{\etalchar{+}}20]{suzuki2020amplitude}
Yohichi Suzuki, Shumpei Uno, Rudy Raymond, Tomoki Tanaka, Tamiya Onodera, and
  Naoki Yamamoto.
\newblock Amplitude estimation without phase estimation.
\newblock {\em Quantum Information Processing}, 19(2):1--17, 2020.

\bibitem[SWM17]{samek2017explainable}
Wojciech Samek, Thomas Wiegand, and Klaus-Robert M{\"u}ller.
\newblock Explainable artificial intelligence: Understanding, visualizing and
  interpreting deep learning models.
\newblock {\em arXiv preprint arXiv:1708.08296}, 2017.

\bibitem[SZ14]{simonyan2014very}
Karen Simonyan and Andrew Zisserman.
\newblock Very deep convolutional networks for large-scale image recognition.
\newblock {\em arXiv preprint arXiv:1409.1556}, 2014.

\bibitem[Tan18]{tang2018quantum}
Ewin Tang.
\newblock Quantum-inspired classical algorithms for principal component
  analysis and supervised clustering.
\newblock {\em arXiv preprint arXiv:1811.00414}, 2018.

\bibitem[Tan19]{tang2019quantum}
Ewin Tang.
\newblock A quantum-inspired classical algorithm for recommendation systems.
\newblock In {\em Proceedings of the 51st Annual ACM SIGACT Symposium on Theory
  of Computing}, pages 217--228, 2019.

\bibitem[TL20]{tremblay2020approximating}
Nicolas Tremblay and Andreas Loukas.
\newblock Approximating spectral clustering via sampling: a review.
\newblock In {\em Sampling Techniques for Supervised or Unsupervised Tasks},
  pages 129--183. Springer, New York, NY, USA, 2020.

\bibitem[TMGB19]{tacchino2019artificial}
Francesco Tacchino, Chiara Macchiavello, Dario Gerace, and Daniele Bajoni.
\newblock An artificial neuron implemented on an actual quantum processor.
\newblock {\em npj Quantum Information}, 5(1):1--8, 2019.

\bibitem[TS13]{T13}
Amnon Ta-Shma.
\newblock Inverting well conditioned matrices in quantum logspace.
\newblock In {\em Proceedings of the forty-fifth annual ACM symposium on Theory
  of computing}, pages 881--890. ACM, 2013.

\bibitem[TYL17]{tai2017image}
Ying Tai, Jian Yang, and Xiaoming Liu.
\newblock Image super-resolution via deep recursive residual network.
\newblock In {\em Proceedings of the IEEE conference on computer vision and
  pattern recognition}, pages 3147--3155, 2017.

\bibitem[vAG18]{van2018improvements}
Joran van Apeldoorn and Andr{\'a}s Gily{\'e}n.
\newblock Improvements in quantum sdp-solving with applications.
\newblock {\em arXiv preprint arXiv:1804.05058}, 2018.

\bibitem[VAGGdW17]{van2017quantum}
Joran Van~Apeldoorn, Andr{\'a}s Gily{\'e}n, Sander Gribling, and Ronald
  de~Wolf.
\newblock Quantum sdp-solvers: Better upper and lower bounds.
\newblock In {\em 2017 IEEE 58th Annual Symposium on Foundations of Computer
  Science (FOCS)}, pages 403--414. IEEE, 2017.

\bibitem[Vat09]{vattani2009hardness}
Andrea Vattani.
\newblock The hardness of k-means clustering in the plane.
\newblock {\em Manuscript, accessible at http://cseweb. ucsd.
  edu/avattani/papers/kmeans\_hardness. pdf}, 617, 2009.

\bibitem[VBB17]{verdon2017quantum}
Guillaume Verdon, Michael Broughton, and Jacob Biamonte.
\newblock A quantum algorithm to train neural networks using low-depth
  circuits.
\newblock {\em arXiv preprint arXiv:1712.05304}, 2017.

\bibitem[Wal16]{mooreslawend}
M~Mitchell Waldrop.
\newblock The chips are down for moore’s law.
\newblock {\em Nature News}, 530(7589):144, 2016.

\bibitem[WCCY20]{wang2020orthogonal}
Jiayun Wang, Yubei Chen, Rudrasis Chakraborty, and Stella~X Yu.
\newblock Orthogonal convolutional neural networks.
\newblock In {\em Proceedings of the IEEE/CVF Conference on Computer Vision and
  Pattern Recognition}, pages 11505--11515, 2020.

\bibitem[WGM19]{wang2019scalable}
Shusen Wang, Alex Gittens, and Michael~W Mahoney.
\newblock Scalable kernel k-means clustering with nystr{\"o}m approximation:
  relative-error bounds.
\newblock {\em The Journal of Machine Learning Research}, 20(1):431--479, 2019.

\bibitem[WKS14a]{wiebe_nearest_neigbhors}
Nathan Wiebe, Ashish Kapoor, and Krysta Svore.
\newblock Quantum algorithms for nearest-neighbor methods for supervised and
  unsupervised learning.
\newblock {\em arXiv preprint arXiv:1401.2142}, 2014.

\bibitem[WKS14b]{wiebe2014quantum_deeplearning}
Nathan Wiebe, Ashish Kapoor, and Krysta~M Svore.
\newblock Quantum deep learning.
\newblock {\em arXiv preprint arXiv:1412.3489}, 2014.

\bibitem[Wu17]{CNNIntro}
J~Wu.
\newblock Introduction to convolutional neural networks.
\newblock 2017.

\bibitem[XRV17]{xiao2017fashion}
Han Xiao, Kashif Rasul, and Roland Vollgraf.
\newblock Fashion-mnist: a novel image dataset for benchmarking machine
  learning algorithms.
\newblock {\em arXiv preprint arXiv:1708.07747}, 2017.

\bibitem[YHJ09]{yan2009fast}
Donghui Yan, Ling Huang, and Michael~I Jordan.
\newblock Fast approximate spectral clustering.
\newblock In {\em Proceedings of the 15th ACM SIGKDD international conference
  on Knowledge discovery and data mining}, pages 907--916. ACM, New York, NY,
  USA, 2009.

\bibitem[YLC14]{yoder2014fixed}
Theodore~J Yoder, Guang~Hao Low, and Isaac~L Chuang.
\newblock Fixed-point quantum search with an optimal number of queries.
\newblock {\em Physical review letters}, 113(21):210501, 2014.

\bibitem[ZDF{\etalchar{+}}18]{zhao2018note}
Zhikuan Zhao, Vedran Dunjko, Jack~K Fitzsimons, Patrick Rebentrost, and
  Joseph~F Fitzsimons.
\newblock A note on state preparation for quantum machine learning.
\newblock {\em arXiv preprint arXiv:1804.00281}, 2018.

\bibitem[ZHLT20]{zhang2020toward}
Kaining Zhang, Min-Hsiu Hsieh, Liu Liu, and Dacheng Tao.
\newblock Toward trainability of quantum neural networks.
\newblock {\em arXiv preprint arXiv:2011.06258}, 2020.

\bibitem[ZNL21]{zlokapa2021quantum}
Alexander Zlokapa, Hartmut Neven, and Seth Lloyd.
\newblock A quantum algorithm for training wide and deep classical neural
  networks.
\newblock {\em arXiv preprint arXiv:2107.09200}, 2021.

\bibitem[ZYST19]{zhang2019deep}
Shuai Zhang, Lina Yao, Aixin Sun, and Yi~Tay.
\newblock Deep learning based recommender system: A survey and new
  perspectives.
\newblock {\em ACM Computing Surveys (CSUR)}, 52(1):1--38, 2019.

\end{thebibliography}

%%%%---------------APPENDIX------------%%%%
%\appendix
%\chapter{Appendix}
%\input{chapters/appendix}

\end{document}